\documentclass[aps,prc,showpacs,twocolumn,superscriptaddress,nofootinbib,preprintnumbers,floatfix]{revtex4-1}
\usepackage{amsmath,amssymb,graphicx,bm,slashed,tikz,mathrsfs,cancel,epstopdf,esint}
\usepackage[left=1in,right=1in,top=1in,bottom=1in]{geometry}
\usetikzlibrary{decorations.pathmorphing}
\usepackage{longtable}
\usepackage{subfigure}
\usepackage{graphics}
\usepackage{hyperref}
\usepackage{array}
\usepackage{dcolumn}
\usepackage{tensor}
\usepackage{gensymb}
\usepackage{color}
\usepackage{float}

\usepackage[normalem]{ulem}  % \sout{old text} for strikeout

\renewcommand\sout{\bgroup \color{red} \ULdepth=-.5ex \ULset}
\begin{document}
%%%%%%%%%%%%%%%%%%%%%%%%%%%%%%%%%%%%%%%%%%%%%%%%%%%%%%%%%%%%%%%%%%%%%%%%%%%%%%%%%%%%%%%%%%%%%
\title{Radiative and chiral corrections to elastic lepton-proton scattering 
in chiral perturbation theory}
%%%%%%%%%%%%%%%%%%%%%%%%%%%%%%%%%%%%%%%%%%%%%%%%%%%%%%%%%%%%%%%%%%%%%%%%%%%%%%%%%%%%%%%%%%%%% 
\author{Pulak Talukdar}
     \email[]{t.pulak@iitg.ac.in}
     \affiliation{Department of Physics, Indian Institute of Technology Guwahati, Guwahati - 781039, Assam, India.}%
     \affiliation{Department of Physics, S. B. Deorah College, Guwahati, Assam 781007.}% 
     
\author{Vanamali C. Shastry}
     \email[]{vanamalishastry@gmail.com}
     \affiliation{Department of Education in Science and Mathematics, Regional Institute of Education Mysuru, Mysore - 570006, India.}%
     \affiliation{Institute of Physics, Jan Kochanowski University, ul. Uniwersytecka  7, P-25-406, Kielce, Poland}%

\author{Udit Raha}
     \email[]{udit.raha@iitg.ac.in}
     \affiliation{Department of Physics, Indian Institute of Technology Guwahati, Guwahati - 781039, Assam, India.}%
          
\author{Fred Myhrer}
     \email[]{myhrer@mailbox.sc.edu}
     \affiliation{Department of Physics and Astronomy, University of South Carolina, Columbia, SC 29208, USA.}%
                 
%%%%%%%%%%%%%%%%%%%%%%%%%%%%%%%%%%%%%%%%%%%%%%%%%%%%%%%%%%%%%%%%%%%%%%%%%%%%%%%%%%%%%%%%%%%%%%

\begin{abstract}  
A unified treatment of both chiral and radiative corrections to the low-energy elastic lepton-proton 
scattering processes is presented in Heavy Baryon Chiral Perturbations Theory. The proton hadronic 
chiral corrections include the next-to-next-to leading order corrections whereas the radiative 
corrections include the next-to-leading order terms in our novel power counting scheme. We find that 
the net fractional well-defined chiral corrections with respect to the leading order Born cross section   
can be as large as $10\%$ ($20\%$) for electron (muon) scattering process for MUon proton Scattering 
Experiment (MUSE) kinematics. We show {\it via} our model-independent treatment of the low-energy 
lepton-proton kinematics, that the largest theoretical uncertainty is due to the recent different 
published values of the proton's rms radius while, e.g., the next higher order hadronic chiral terms are
expected to give rather nominal errors. For the radiative corrections we demonstrate a systematic order 
by order cancellation of all infrared singularities and present our finite  ultraviolet regularization 
results. We find that the radiative corrections for muon-proton scattering is of the order of $2\%$, 
whereas for electron scattering the radiative corrections could be as large as $25\%$. We attribute such 
a contrasting result partially to the fact that in muon scattering the leading radiative order correction 
goes through zero in some intermediate low-momentum transfer region, leaving the sub-leading radiative 
chiral order effects to play a dominant role in this particular kinematic region. For the low-energy MUSE 
experiment, the often neglected lepton mass as well as the Pauli form factor contributions of the 
relativistic leptons are incorporated in all our computations. 
\end{abstract}

\maketitle

%%%%%%%%%%%%%%%%%%%%%%%%%%%%%%%%%%%%%%%%%%%%%%%%%%%%%%%%%%%%%%%%%%%%%%%%%%%%%%%%%%%%%%%%%%%%%%
\section{Introduction}\label{intro}
%%%%%%%%%%%%%%%%%%%%%%%%%%%%%%%%%%%%%%%%%%%%%%%%%%%%%%%%%%%%%%%%%%%%%%%%%%%%%%%%%%%%%%%%%%%%%%
Scattering processes involving charged particles, like the lepton-proton ($\ell$-p) elastic 
scattering, involve arbitrary number of real and virtual photons. The inelastic bremsstrahlung 
process, $\ell+{\rm p}\to \ell +{\rm p}+ \gamma^*$, where $\ell\equiv e^\pm,\mu^\pm$, 
constitutes the most significant undetected background radiative process.\footnote{The $\gamma^*$ 
symbol denotes an emitted real bremsstrahlung photon to distinguish it from the virtual loop-photon 
$\gamma$.} Many prominent works have estimated radiative corrections nearly as large as $30\%$, for 
electron scattering, based on analyses over a wide range of momentum transfers and a variety of 
experimental conditions, {\it viz.}  detector designs and resolutions~\cite{Maximon:2000hm}. 
Especially for soft (low-energy) photons these corrections must be theoretically evaluated as 
they are inaccessible to direct experimental probes. 

%\vspace{0.1cm}

The $\ell$-p elastic scattering process has particularly engendered extensive interests in the 
scientific community over the past two decades because of its significant role in bringing forth 
various discrepancies in our basic understanding of the electromagnetic properties of the proton. 
An accurate experimental determination of the proton's electromagnetic form factors can shed
much needed light on the proton's basic hadronic structure and internal dynamics. The original 
discrepancies in the measurements of the electric ($G^p_E$) and magnetic ($G^p_M$) proton form 
factors (so-called the ``proton form factor ratio puzzle") that emerged about two decades ago, 
stemmed from the utilization of the novel experimental recoil polarization transfer 
technique~\cite{Jones:1999rz,Perdrisat:2006hj,Punjabi:2015bba,Puckett:2010}. Such measurements 
led us not only to question the validity of the conventional Rosenbluth separation technique but 
also raised serious concerns regarding our basic understanding of the proton structure itself. In 
order to resolve these problems a flurry of ingenious ideas and methodologies ensued which were 
extensively discussed in numerous published works as well as reviews, see e.g.,
Refs.~\cite{Arrington:2003,Guichon:2003,Blunden:2003,Arrington:2011}. Furthermore, the 2013 
revelation of the so-called ``proton radius 
puzzle"~\cite{Pohl:2010zza,Pohl:2013,Antognini:1900ns,Bernauer:2014,Carlson:2015} concerns the 
irreconcilable inconsistency amongst the different measurement techniques in determining the 
proton's {\it root-mean-squared} (rms) charge radius. Subsequently, unremitting efforts geared 
towards the development of high-precision
experimental~\cite{ISR_2017,Beyer_2017,Fleurbaey_2019,ISR_2019,Adams_2018,Bezginov_2019,Xiong_2019,Grinin2020} 
and novel theoretical~\cite{Rejula_2010,Rejula_2011,Bea_2011,Smith_2011,Barger_2011,Barger_2012,Carlson_2012,Wang_2013,Karshenboim_2014,Lorentz_2015,Bernauer_2016,Barnauer_2020,Hammer_2019,Horbatsch:2016ilr,Alarcon:2018zbz,Lee:2015jqa,Arrington:2015yxa,Sick:2018fzn,Tomalak:2014dja,Tomalak:2015aoa,Tomalak:2015hva,Koshchii2017,Tomalak:2018jak} techniques, have been pursued over the last nine 
years seeking a definitive answer to the conundrum. However, despite the efforts such discrepancies 
are yet to be conclusively resolved, requiring further improved approaches on either fronts.    

%\vspace{0.1cm}
 
In contrast to the previously designed experiments, the MUon proton Scattering Experiment (MUSE), 
currently underway at the Paul Scherrer Institute (PSI), aims at a resolution of the proton radius 
puzzle. The MUSE Collaboration plans to carry out simultaneous high-precision measurements of 
low-energy electron-proton ($e^\pm$p) and muon-proton ($\mu^\pm$p) scattering cross 
sections~\cite{Gilman:2013eiv,Gilman:2017hdr}. MUSE's goal is to measure the proton's rms radius at a 
%projected accuracy 
better than $1\%$ precision~\cite{PComm}. On the theoretical side an improved assessment of the 
systematic uncertainties is needed in order to meet the expected level of accuracy of future MUSE data
analysis. In this work we utilize a systematic {\it model-independent perturbative} procedure to 
determine higher-order corrections of the {\it leading order} (LO) Born contribution (i.e., with 
point-like lepton and proton) for unpolarized elastic $\ell$-p scattering cross section. In particular, 
our analysis demonstrates how we handle both the strong interaction as well as the standard QED 
radiative corrections in our formalism. We work in the framework of a  low-energy effective field 
theory (EFT), namely, {\it Chiral Perturbation Theory} ($\chi$PT)~\cite{Scherer:2003}, which reflects 
the inherent low-energy non-perturbative features of QCD manifested in hadrons, where  chiral symmetry 
and its violations play decisive roles in determining the observables. The rationale for using $\chi$PT 
to evaluate the sub-leading corrections in an essential perturbative framework is that the methodology  
allows us to systematically extend the theoretical predictions to higher levels of accuracy through a 
well-defined power counting scheme. This is a distinctive feature of our approach which sets our 
analysis in contrast to the existing conventional approaches where hadron structure effects are 
empirically modeled through the use of phenomenological proton form factors.   

%\vspace{0.1cm}

Our evaluations are based on the well-established non-relativistic version of $\chi$PT, namely, Heavy 
Baryon $\chi$PT (HB$\chi$PT), e.g., Ref.~\cite{Bernard:1992qa,Bernard:1995dp}. HB$\chi$PT provides a 
convenient tool to study processes like the low-energy $\ell$-p scattering, where non-relativistic 
baryons and relativistic mesons and leptons are the fundamental degrees of freedom. Furthermore, all 
these particles naturally couple to the photon in a gauge invariant manner. Here we adopt the so-called 
"SU(2) isospin" scheme  which is tailor-made to deal with the low-energy hadron dynamics of the nucleon. 
The HB$\chi$PT power counting incorporates a chiral expansion in terms of powers of the ratio of a generic 
small momentum ${\cal Q}$ over the large chiral scale $\Lambda_\chi\simeq 4\pi f_\pi \simeq 1 {\rm GeV}$ 
($f_\pi\approx 92.4$ MeV is the pion decay constant), plus a {\it recoil expansion} in powers of the 
typical momentum scale of the process, ${\cal Q}\simeq 0.2$ GeV/c (in regard to MUSE kinematics), over 
the ``heavy'' proton mass $M \simeq 1 {\rm GeV}$, 
\begin{equation*}
\frac{{\cal Q} }{\Lambda_\chi}\sim \frac{{\cal Q} }{M} \ll 1.   
\end{equation*}

%\vspace{0.1cm}

Apart from the above  mentioned  chiral momentum expansion, the counting scheme also includes the standard
QED perturbative expansion where the effective Born amplitude (see next section), including all its 
non-radiative chiral effects,  count as order $\alpha = \frac{e^2}{4\pi}\simeq 1/137\approx 0.007$. In fact, 
for the evaluation of our  ``effective Born" contribution, we make a chiral expansion to include corrections 
up-to-and-including ${\cal Q}^2/M^2\simeq 0.04$ in this work. For the sake of transparency we assign distinct
nomenclatures to the various chiral corrections. For example, we denote the leading order Born amplitude 
${\mathcal M}^{(0)}_\gamma \sim {\cal O}(e^2)$ as ``LO$_\chi$" or simply ``LO Born", the {\it next-to-leading 
order} chiral Born amplitude ${\mathcal M}^{(1)}_\gamma\sim {\cal O}(e^2{\cal Q}/M)\simeq {\cal O}(e^2 \cdot 0.2)$ 
is denoted as ``NLO$_\chi$", and the {\it next-to-next-to-leading order} chiral Born amplitude
${\mathcal M}^{(2)}_\gamma \sim {\cal O}(e^2{\cal Q}^2/M^2)\simeq {\cal O}(e^2 \cdot 0.04)$ is labelled 
NNLO$_\chi$, (cf. Fig.~\ref{LN:tree}). We note that at the order ${\cal Q}^2/M^2$ and higher, the ${\cal Q}^2$ 
dependence of the proton form factors enters naturally in the chiral expansion. At NNLO$_\chi$ the 
{\it low-energy constants} (LECs) of HB$\chi$PT parametrize the short-distance physics and, in addition, 
regulate the {\it ultraviolet divergences} (UV) in diagrams with 
pion-loops~\cite{Bernard:1992qa,Bernard:1995dp,Fettes2000}.  The $\chi$PT evaluation of these NNLO$_\chi$ 
contributions are well-known, and yields analytical expressions for the proton's charge and magnetic
radii~\cite{Bernard:1992qa,Bernard:1995dp,Fearing1998}. In this work, however, we do not repeat  such 
evaluations. Instead, we use the analytical expression of the $\chi$PT renormalized ``effective Born" amplitude
${\mathcal M}^{(2);{\rm rms}}_\gamma\sim {\cal O}(e^2{\cal Q}^2/M^2)$. In this work we shall denote the chirally
corrected NNLO$_\chi$ fractional contributions as $\delta^{(2)}_\chi\sim {\mathcal O}({\cal Q}^2/M^2)$, with 
respect to the LO Born cross section of ${\cal O}(\alpha^2)$ (cf. Sec.~\ref{sec:formalism} for details). 
Furthermore, it should be noted that the {\it next-to-next-to-next-to-leading order} (N$^3$LO$_\chi$) 
contributions from terms of ${\cal O}(e^2{\cal Q}^3/M^3)\simeq {\cal O}(e^2\cdot 0.008)$,  which are not included
in this work, constitute an important theoretical uncertainty. We remark that including possible contributions 
arising from spin-3/2 $\Delta(1232)$ and other excited nucleon resonances, would be a significant extension of our
present analysis, and are therefore not included in this work. In particular, the inclusion of $\Delta(1232)$ in 
$\chi$PT requires the chiral counting to be supplemented by the so-called {\it small scale} or 
$\delta$-{\it expansion}~\cite{Bernard:1992qa,Bernard:1995dp,Hemmert:1996xg,Hemmert:1997ye}. In addition, we refer 
to Refs.~\cite{Lorentz_2015,Lee:2015jqa,Alarcon:2018zbz,Horbatsch:2016ilr} regarding some recently developed 
techniques of the so-called {\it Dispersively-Improved} $\chi$PT, for a consistent inclusion of resonance 
contributions. 

%\vspace{0.1cm}

The predominant portion of this work deals with the evaluation the first two orders of our ``chiral-radiative" 
corrections, namely, the {\it radiative leading order} whose amplitudes are denoted as ``LO$_\alpha$", i.e., 
${\cal O}(e^2\alpha)\simeq {\cal O}(e^2\cdot 0.007)$, and the {\it radiative next-to-leading order} corrections, 
denoted as ``NLO$_\alpha$", i.e., ${\cal O}(e^2\alpha{\cal Q}/M)\simeq {\cal O}(e^2\cdot 0.0015)$.  The next 
higher order chiral-radiative amplitudes, i.e.,  
${\cal O}(e^2\alpha{\cal Q}^2/M^2)\simeq {\cal O}(e^2\cdot 0.0003)$ are expected to yield only a tiny 
contribution.  

%\vspace{0.1cm}

To the best of our knowledge, the current work represents the first attempt using a model-independent EFT 
framework, namely HB$\chi$PT, to simultaneously evaluate the chiral and the radiative corrections in an unified 
framework. The EFT radiative evaluation include all one-loop virtual correction, {\it viz.} the self-energies 
(SE), vertex corrections (VC), vacuum polarization (VP), and the {\it two-photon exchange} (TPE) contributions 
to $\ell$-p elastic scattering process. Moreover, in our work the contributions from single soft photon 
($\gamma^*_{\rm soft}$) emission are required in order to demonstrate the cancellation of the {\it infrared} (IR)
divergences arising from the virtual processes. The {\it modus operandi} adopted in this paper is reminiscent of 
the seminal work of Yennie, Frautschi and Suura (YFS) of Ref.~\cite{Yennie:1961ad},  developed within a QED 
approach with relativistic point-like Dirac particles. 

\vspace{0.1cm} 
  
In order to render the radiative corrections IR-finite, we need to include the LO$_\alpha$ and NLO$_\alpha$ 
soft photon bremsstrahlung amplitudes of $\mathcal{O}(e^3)$ and $\mathcal{O}(e^3{\cal Q}/M)$ respectively. In 
the ensuing analysis, the LO$_\alpha$ fractional radiative corrections are denoted by 
$\delta^{(0)}_{2\gamma}\sim {\cal O}(\alpha)$, and likewise the NLO$_\alpha$ corrections by 
$\delta^{(1)}_{2\gamma}\sim {\cal O}(\alpha{\cal Q}/M)$, with respect to the elastic LO (Born) cross section 
of ${\cal O}(\alpha^2)$. The interference of the NNLO$_\chi$ terms with the LO$_\alpha$ radiative corrections 
is included together with other non-factorizable NNLO$_\alpha$ corrections in  
$\delta^{(2)}_{2\gamma}\sim {\cal O}(\alpha{\cal Q}^2/M^2)$, and $\delta^{(2)}_{2\gamma}$ 
will be used in our uncertainty assessment, see Eq.~\eqref{eq:full}. Thus, our result for the total fractional 
radiative correction to the $\ell$-p elastic LO (Born) cross section can be symbolically expressed in the form 
$\delta_{2\gamma}\! =\!\delta^{(0)}_{2\gamma}\! +\!\delta^{(1)}_{2\gamma}\! +\!\delta^{(2)}_{2\gamma}$\,.

%\vspace{0.1cm}

In one of our previous work, Ref.~\cite{Talukdar:2019dko}, we evaluated the {\it two-photon-exchange} (TPE) 
contributions to the $\ell$-p elastic process at NLO$_\alpha$ accuracy in HB$\chi$PT invoking a soft photon 
approximation (SPA), e.g., the approach as pursued in Ref.~\cite{Maximon:2000hm} (see discussion relating to 
the use of SPA later in this paper). We demonstrated that the TPE amplitudes diverge in the vanishing limit
of the photon momenta, {\it vis-a-vis}, IR divergences. The present work is an essential follow-up of that 
analysis~\cite{Talukdar:2019dko}. Here we shall detail the systematical step-wise evaluation of the radiative 
corrections at LO$_\alpha$  and NLO$_\alpha$. In this work the NNLO$_\alpha$ corrections are only partially 
included (for brevity the analytical NNLO$_\alpha$ results are not displayed explicitly) and contribute to our 
estimate of the theoretical error. We explicitly demonstrate how the chiral power counting allows an order by 
order cancellation of all the IR-divergences arising from the one-loop virtual and the single photon 
bremsstrahlung processes. 

%\vspace{0.1cm}

We assign the real bremsstrahlung photon as being either ``soft" or ``hard" in comparison to some fixed but 
frame dependent small energy scale $\Delta E$.  $\Delta E$ is associated with the outgoing detected lepton, 
which practically fixes the upper limit of the energy integration of the undetected emitted {\it soft} photon 
when evaluating the bremsstrahlung cross section. Specially, in the context of laboratory ({\it lab.}) frame 
kinematics, $\Delta E=\Delta_{\gamma^*}$, where  $\Delta_{\gamma^*}$ is the so-called detector 
{\it acceptance}.\footnote{In the present radiative analysis of $\ell$-p elastic scattering in the 
{\it lab.}-frame, we include only the {\it soft} photons whose energies lie below the detector threshold 
$\Delta_{\gamma^*}$. The {\it hard} photons with energies larger than $\Delta_{\gamma^*}$ will not be considered 
since they contribute to the inelastic radiative process, $\ell+{\rm p}\to \ell +{\rm p}+ \gamma^*$. The soft 
photons go undetected in a typical experiment and therefore it is necessary to integrate the part of the elastic
radiative tail distribution for photon energies between $0$ and $\Delta_{\gamma^*}$, in order to compare with 
the measured differential cross section.} In particular, to integrate over this ``soft" elastic radiative tail,
we adopt Tsai's formalism~\cite{Tsai:1961zz,Mo:1968cg} of boosting to the so-called $S$-frame, where the otherwise 
complicated phase-space integration involving the photon emission angles becomes rather simplified, as 
demonstrated in the Appendices. Our results for the radiative corrections will depend on the resolution factor
$\Delta_{\gamma^*}$, which constitutes a free parameter in our theoretical framework specified by the design of a
given experimental arrangement. In this work for the sake of numerical evaluations, we have chosen the value of 
this parameter to be $\sim 1\%$ of the incoming lepton energy. 

%\vspace{0.1cm}

The evaluation of the radiative correction diagrams involve virtual or real photons,  which yield matrix elements 
containing UV and/or IR divergences. In this work both the chiral and the QED divergences will be treated  by 
employing {\it dimensional regularization} (DR).\footnote{Most works in the literature prefer to use a non-zero 
photon mass, $\lambda$ (see, e.g., Refs.~\cite{Mo:1968cg,Maximon:2000hm}) which leads to IR-divergent terms in the 
form of logarithms, e.g., $\ln\left(\frac{\lambda^2}{-Q^2}\right)$. We follow the DR treatment of IR divergences, 
e.g., Refs.~\cite{Gastmans:1973uv,Marciano:1974tv,Vanderhaeghen:2000ws,Bucoveanu:2018soy}). A naive comparison of 
the IR treatments leads to the correspondence: 
\begin{equation*}
\frac{1}{|\epsilon_{\rm IR}|}+\gamma_E-\ln\left(\frac{4\pi\mu^2}{-Q^2}\right)\leftrightarrow 
-\ln\left(\frac{\lambda^2}{-Q^2}\right)\,,
\end{equation*} 
where $\mu$ is the subtraction scale (typically chosen as the momentum scale associated with the scattering 
process).} While the UV divergent terms are renormalized using LECs and QED counter-terms in the Lagrangian, the 
IR divergences, as demonstrated in this paper, systematically cancel at each order in the chiral expansion. The 
ultimate objective of this paper is to obtain a finite analytical expression for the fractional chiral-radiative 
correction $\delta_{2\gamma}$ to the elastic $\ell$-p differential cross section in the {\it lab.}-frame, namely,
\begin{equation}
\frac{{\rm d}\sigma_{el}}{{\rm d}\Omega^\prime_l}\Bigg|_{\rm lab}=
\left[\frac{{\rm d}\sigma_{el}}{{\rm d}\Omega^\prime_l}\right]_\gamma\, 
\Big\{1+\delta_{2\gamma}(Q^2)\Big\}\,,
\end{equation}
where the pre-factor on the right side includes all our ${\cal O}(\alpha^2{\cal Q}^2/M^2)$ (i.e., NNLO$_\chi$) 
hadronic chiral corrections to elastic Born differential cross section (cf. Sec.~\ref{sec:formalism} for details).   

%\vspace{0.1cm} 

The paper is outlined as follows. The details of our methodology are presented in Sec.~\ref{sec:formalism} where 
we display the pertinent terms of the effective Lagrangian (QED + HB$\chi$PT). Based on the aforementioned chiral
power counting scheme, we determine the analytical expressions for the two chiral corrections to the LO  Born 
cross section, namely, the NLO$_\chi$ and NNLO$_\chi$ terms in HB$\chi$PT. The details of the chiral-radiative 
corrections, namely, at LO$_\alpha$ and NLO$_\alpha$, involving evaluations of the corresponding one-loop virtual 
and single real soft photon emission diagrams, are presented in Sec.~\ref{sec:LO} and Sec.~\ref{sec:NLO} 
respectively. In Sec.~\ref{sec:results}, we discuss and compare the numerical estimates of the various 
contributions in regard to the MUSE kinematical region~\cite{Gilman:2013eiv,Gilman:2017hdr}. We also outline the 
major sources of theoretical uncertainties in this section. Finally, Sec.~\ref{sec:summary} summarizes the key 
features of our analysis and prospects of possible future extensions of this work. Technical details regarding the 
utilization of the $S$-frame kinematics to perform the bremsstrahlung phase-space integrals are relegated to the 
Appendices.  

%%%%%%%%%%%%%%%%%%%%%%%%%%%%%%%%%%%%%%%%%%%%%%%%%%%%%%%%%%%%%%%%%%%%%%%%%%%%%%%%%%%%%%%%%%%%%%
\section{HB$\chi$PT: Formalism}\label{sec:formalism}  
%%%%%%%%%%%%%%%%%%%%%%%%%%%%%%%%%%%%%%%%%%%%%%%%%%%%%%%%%%%%%%%%%%%%%%%%%%%%%%%%%%%%%%%%%%%%%% 
The most general effective Lagrangian consistent with the low-energy symmetries,  is the sum of the QED Lagrangian 
for the lepton fields $\psi_l$ and the hadronic $\pi N$ effective Lagrangian, namely, 
\begin{eqnarray}
\mathscr{L}_{lp\gamma} &=& -\frac{1}{4}F_{\mu\nu}F^{\mu\nu}+\frac{1}{2\xi_A}(\partial\cdot A)^2
\nonumber\\
&&+\,\sum_{l=e,\mu}\bar{\psi}_l\left(i\slashed{D}-m_l\right)\psi_l + \mathcal{L}^{\rm eff}_{\pi N}\,,   
\label{eq:full_Lagrangian}
\end{eqnarray}
where $F_{\mu\nu}=\partial_\mu A_\nu - \partial_\nu A_\mu$ is the electromagnetic field tensor with $A_\mu$ being 
the photon field, and $\xi_A$ is the gauge parameter, which in the Feynman gauge is $\xi_A=1$. The gauge covariant 
derivatives appearing in the Lagrangian is defined as $D_\mu = \partial_\mu - ieA_\mu$. The hadronic part of the
effective Lagrangian is here expressed as the sum of the lowest order pionic Lagrangian $\mathscr{L}^{(2)}_\pi$ 
and the $\pi N$ Lagrangian expanded in an infinite sequence of operator terms characterized by the chiral dimension 
$\nu=0,1,2,...$, namely, 
\vspace*{-1mm}
\begin{eqnarray}
\mathcal{L}^{\rm eff}_{\pi N} = \mathscr{L}^{(2)}_\pi + \sum_{\nu=0}\mathcal{L}_{\pi N}^{(\nu)} \, , 
\nonumber
\end{eqnarray}
where\footnote{In principle the LECs in the  Lagrangian should be taken in the chiral limit, e.g.,  
$\stackrel{\circ}{f}$ and $\stackrel{\circ}{g_A}$.  However, they will be renormalized to their respective physical 
values, $f_\pi$ and  $g_A$, at  a given chiral order.} 
\begin{eqnarray}
\mathscr{L}^{(2)}_\pi &=& 
\frac{f_\pi}{4}{\rm Tr}\left[\nabla_\mu U^\dagger \nabla^\mu U+\chi^\dagger U+\chi U^\dagger\right]\,;
\nonumber\\
U&=&\sqrt{1 - \frac{\vec{\pi}^2}{f_\pi^2}} + \frac{i}{f_\pi} \vec{\tau}\cdot\vec{\pi}\,, 
\nonumber\\
\chi &=& 2B  
\begin{pmatrix}
m_u & 0\\ 
0   & m_d
\end{pmatrix}\,.
\label{eq:LpiN}
\end{eqnarray}
Here we use the so-called {\it sigma gauge} parametrization of the non-linear pion field $U$, and the constant $B$ 
is related to the scalar current quark condensate $\langle 0|{\bar q}q|0\rangle$, the order parameter of spontaneously
broken chiral symmetry. The chiral covariant derivative $\nabla_{\!\mu}$ is given in Eq.~\eqref{eq:covariant}. Below we 
explicitly specify only the $\nu=0$ (LO$_\chi$) and $\nu=1$ (NLO$_\chi$) terms of the $\pi N$ Lagrangian. The $\nu=2$ 
(NNLO$_\chi$) chiral Lagrangian contains many additional LECs as well as $1/M^{2}$ order ``fixed" terms and 
counterterms, some of which contribute to the lowest order proton's form factors, including the rms radii (see e.g.,
Ref.~\cite{Fearing1998}). Owing to the large number of operator terms in ${\mathcal L}^{(2)}_{\pi N}$, we just refer 
the reader to, e.g., Refs.~\cite{Bernard:1995dp,Fettes2000} where complete expression can be found. The lowest chiral 
order $\pi N$ Lagrangian is given as
\begin{eqnarray}
\mathcal{L}_{\pi N}^{(0)} &=& {\bar N}_v (i v\cdot {\mathcal D} + g_A S \cdot u) N_v\,; \,\,\,
N_v \equiv \binom{{\rm p}_v}{{\rm n}_v}\,,\,\,\,
%\nonumber\\
\end{eqnarray} 
where in HB$\chi$PT the nucleon velocity and spin four-vectors can be chosen as $v_\mu=(1,\vec{0})$ and 
$S_\mu=(0,\vec{\sigma}/2)$ respectively, satisfying the condition, $v\cdot S=0$. The pion fields enter 
$\mathcal{L}_{\pi N}^{(0)}$ through the term $u=\sqrt U$ where $u_\mu = i u^\dagger \nabla_{\!\mu} U u^\dagger$ is the  
chiral vielbein. The LEC $g_A\simeq 1.26$ is the axial-vector coupling constant. The velocity-dependent heavy nucleon 
field $N_v$ represents the ``large" components of the nucleon iso-spinor field with the proton (p${}_v$) and neutron 
(n${}_v$) projections. The  next chiral order $\pi N$ Lagrangian 
\begin{eqnarray}
\mathcal{L}_{\pi N}^{(1)} \!&=&\! {\bar N}_v\! \left[\left(\frac{ (v\cdot {\mathcal D})^2 
- {\mathcal D}\cdot {\mathcal D}}{2M}\right) 
- \frac{i g_A}{2M}\big\{S\cdot {\mathcal D} , v\cdot u\big\}  \right.
\nonumber\\
&& \left. \!+\, c_1 {\rm Tr}(\chi_+) + \left( c_2 - \frac{g_A^2}{8M}\right) ( v \cdot u)^2 
+ c_3 u \cdot u  \right.
\nonumber\\
&& \left. \!+\, \left( c_4 +\frac{1}{4M}\right) [S^\mu , S^\nu]u_\mu u_\nu 
+ c_5 {\rm Tr}(\tilde{\chi}_+) \right.
\nonumber\\
&& \left. \!-\, \frac{i}{4M}[S^\mu , S^\nu] \left[(1 +c_6)f^+_{\mu \nu} 
+ c_7 {\rm Tr}(f^+_{\mu\nu}) \right] \right]\!\! N_v,  
\nonumber\\
\end{eqnarray}
where in our case we have 
%where
\begin{eqnarray}
\chi_+ &=& u^\dagger \chi u + u \chi^\dagger u\,, \hspace*{2mm} %{\rm where} \hspace*{1mm} 
%\nonumber\\
\tilde{\chi}_+ %&=&
=  \chi_+ - \frac{\mathbb I}{2}{\rm Tr} (\chi_+)\,,
\nonumber\\
f^{+}_{\mu\nu} &=& u^\dagger \left(f^R_{\mu\nu}+v^{(s)}_{\mu\nu}\right) u 
+ u \left(f^L_{\mu\nu}+v^{(s)}_{\mu\nu}\right) u^\dagger 
\nonumber\\
&=& eF_{\mu\nu}(u {\mathscr Q} u^\dagger + u^\dagger{\mathscr Q} u)\,,
\nonumber\\
f^{R}_{\mu\nu} &=& \partial_\mu r_\nu -\partial_\nu r_\mu - i[r_\mu,r_\nu]
=e\frac{\tau^3}{2} F_{\mu\nu}\,,
\nonumber\\
f^{L}_{\mu\nu} &=& \partial_\mu l_\nu -\partial_\nu l_\mu - i[l_\mu,l_\nu]\,,
= e\frac{\tau^3}{2} F_{\mu\nu}\,,
\nonumber\\
v^{(s)}_{\mu\nu} &=& 
e\frac{\mathbb I}{2} \left(\partial_\mu A_\nu -\partial_\nu A_\mu\right)\,, \hspace*{3mm} {\rm and} 
\nonumber\\
{\mathscr Q}&=& \frac{1}{2}({\mathbb I}+\tau_3)
= \begin{pmatrix}
1 & 0\\ 
0   & 0
\end{pmatrix}\,. 
\label{eq:seven}
\end{eqnarray}
Apart from the $1/M$ order terms, the $\nu=1$ chiral dimension Lagrangian contains the seven LECs, 
$c_i$, $i = 1,2,...,7$, whose values are phenomenologically determined~\cite{Scherer:2003,Fettes2000}. The 
speciality of these dimension zero and one LECs are that they are finite and unaffected by pion-loop effects, %to these orders 
which start at chiral order %dimension 
two. In particular, the LECs $c_6 =\kappa_v$ and 
$c_7=(\kappa_s-\kappa_v)/2$, where the nucleon iso-vector and iso-scalar anomalous magnetic moments are 
$\kappa_v=3.71$ and $\kappa_s=-0.12$ respectively~\cite{Fearing1998}. The field tensor $f^+_{\mu \nu}$ represents 
the external iso-scalar and iso-vector sources. As shown in Eq.~(\ref{eq:seven}) the external iso-scalar field is
$v_\mu^{(s)}=-e\frac{\mathbb I}{2}A_\mu$, and the iso-vector right-handed ($r_\mu$) and left-handed ($l_\mu$) 
external sources are given as $l_\mu=r_\mu=-e\frac{\tau^3}{2}A_\mu$, where $\mathbb I$ is the identity matrix and 
$\tau^3$ is the diagonal Pauli matrix in isospin SU(2). In this work we have ignored all sources of isospin 
violation. Consequently, there is no contribution from the term proportional to $c_5$, since $\tilde{\chi}_+\to 0$ 
in the limit $m_d\to m_u$. Finally, the covariant derivatives used in the HB$\chi$PT Lagrangian are  
\begin{eqnarray}
{\mathcal D}_\mu &=& \partial_\mu + \Gamma_\mu - i v_\mu^{(s)}\,, \hspace*{3mm} {\rm and} 
\nonumber\\
\nabla_{\!\mu} U &=& \partial_\mu U-ir_\mu U+iUl_\mu\,, 
\label{eq:covariant}
\end{eqnarray}
where
\begin{eqnarray}
\Gamma_\mu = \frac{1}{2}\left[u^\dagger(\partial_\mu-i r_\mu)u+u(\partial_\mu-i l_\mu)u^\dagger\right]\,.
\end{eqnarray}

%\vspace{0.1cm}

In Fig.~\ref{LN:tree} we display the LO$_\chi$, NLO$_\chi$ and NNLO$_\chi$ Born amplitudes (diagrams B${}^{(0,1,2)}$) 
for the elastic lepton-proton scattering process. As mentioned, we prefer to represent the proton form factor (rms 
radii) contributions at NNLO$_\chi$ {\it via} the effective Born amplitude ${\mathcal M}^{(2);{\rm rms}}_\gamma$ 
(diagram B${}^{(2);\,{\rm rms}}$). The LO Born amplitude is given as
\begin{equation}
\mathcal{M}^{(0)}_{\gamma} = -\,\frac{e^2}{ Q^2}[{\bar u}_l(p^\prime)\gamma^\mu u_l(p)]\,
\left[\chi^\dagger(p_p^\prime)\,v_\mu \,\chi(p_p)\right]\,,
\label{M0_gamma}
\end{equation} 
where  $\chi$ is the two-component Pauli spinor for the non-relativistic proton, while $u_l$ is the Dirac spinors for 
the relativistic leptons. The sub-leading Born amplitudes needed in this work are given as 
\begin{eqnarray} 
\mathcal{M}^{(1)}_{\gamma} &=& \mathcal{M}^{(1);\,{\rm a}}_\gamma + \mathcal{M}^{(1);\,{\rm b}}_\gamma\,;
\nonumber\\
\mathcal{M}^{(1);\,{\rm a}}_\gamma &=& -\,\frac{e^2}{2 M Q^2}[{\bar u}_l(p^\prime)\gamma^\mu\,u_l(p)]\,
\left[\chi^\dagger(p_p^\prime)\right.
\nonumber\\
&& \left. \times\, \left\{(p_p + p_p^\prime)_\mu - v_\mu v \cdot (p_p + p_p^\prime) \right\} \chi(p_p)\right]\,,
\nonumber\\
\mathcal{M}^{(1);\,{\rm b}}_\gamma &=& -\,\frac{e^2}{2 M Q^2}[{\bar u}_l(p^\prime)\gamma^\mu\,u_l(p)]\,
\left[\chi^\dagger(p_p^\prime)\right.
\nonumber\\
&& \left. \times\, (2+\kappa_s +\kappa_v) \left[S_\mu , S\cdot Q\right]\,\chi(p_p)\,\right]\,,
\label{eq:M1_gamma}
\end{eqnarray} 
where $\mathcal{M}^{(1);\,{\rm a}}_\gamma$ and $\mathcal{M}^{(1);\,{\rm b}}_\gamma$ are the spin-independent and 
spin-dependent parts of the NLO$_\chi$ amplitude $\mathcal{M}^{(1)}_\gamma$ respectively. Including the $\nu=2$ chiral order 
interactions, the $1/M^{2}$ order Born amplitude is given as
\begin{eqnarray}
\mathcal{M}^{(2)}_{\gamma} &=& -\frac{e^2}{ 8M^2 Q^2}[{\bar u}_l(p^\prime)\gamma^\mu \,u_l(p)]\, 
\left[\chi^\dagger(p_p^\prime)\right.
\nonumber\\ 
&& \left. \times \left\{\left(2(v\cdot Q)^2 -  Q^2\right) \!v_\mu  - Q_\mu v\cdot Q \right\}\! \chi(p_p) \right].
\nonumber\\
\label{M1M2_gamma}
\end{eqnarray}
Furthermore, the proton's Dirac and Pauli form factors up to ${\mathcal O}({\cal Q}^2/M^2)$ order get 
contributions from the pion-loops and various   %higher order 
LECs. Expressed in terms of the mean square radii 
$\langle r^2_{1} \rangle$ and $\langle r^2_{2} \rangle$ respectively, they are viable in %to 
the following low-energy expansions in $Q^2$:
%%%%%%%%%%%%%%%%%%%%%%%%%%%%%%%%%%%%%%%%%%FIGURE%%%%%%%%%%%%%%%%%%%%%%%%%%%%%%%%%%%%%%%%%%%%%	
\begin{figure}[tbp]
\centering
       \includegraphics[scale=0.58]{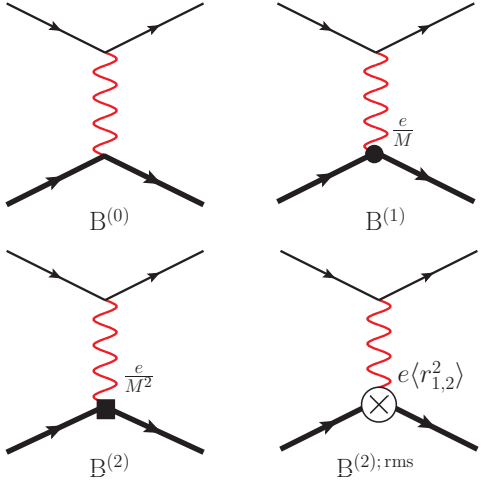}	
       \caption{Born diagrams B${}^{(0,1,2)}$ for $\ell$-p elastic scattering at leading order 
                [i.e., $\mathcal{O}(e^2)$], next-to-leading order [i.e., $\mathcal{O}(e^2{\cal Q}/M)$] and 
                next-to-next-to-leading order [i.e., $\mathcal{O}(e^2{\cal Q}^2/M^2)$] in chiral expansion 
                [see Eqs.~\eqref{M0_gamma} - \eqref{M1M2_gamma}]. The effective Born diagram 
                $B^{(2);\,{\rm rms}}$ also contributing at next-to-next-to-leading order [see 
                Eqs.~\eqref{M2rms_gamma} and \eqref{Vhadron}] parametrizes the proton's structure effects 
                (pion-loops and higher-order LECs) {\it via}  insertions of the Dirac and Pauli mean-squared 
                radii $\langle r^2_{1,2} \rangle$ (crossed blob). The dark filled blob and square represent 
                the proton-photon vertex insertion of order $1/M$ and $1/M^2$ respectively.}
       \label{LN:tree}
\end{figure}
%%%%%%%%%%%%%%%%%%%%%%%%%%%%%%%%%%%%%%%%%%%%%%%%%%%%%%%%%%%%%%%%%%%%%%%%%%%%%%%%%%%%%%%%%%%%
\begin{eqnarray}
F^p_1(Q^2) &=& 1 + \frac{Q^2}{6} \langle r_1^2\rangle +{\cal O}(M^{-3}) \,, \quad {\rm and}
\nonumber\\
F^p_2(Q^2)\!\! &=& \kappa_p\! +\!\frac{Q^2}{6} \langle r_2^2\rangle +{\cal O}(M^{-3}) \,. 
\end{eqnarray}
Including these form factors we derive the general form for the effective Born amplitude, 
diagram B${}^{(2);\,{\rm rms}}$, namely,   
\begin{eqnarray}
\mathcal{M}^{(2);\,{\rm rms}}_{\gamma} \!\!=\! - \frac{e^2}{ Q^2}[{\bar u}_l(p^\prime)\gamma^\mu u_l(p)]
\left[\chi^\dagger(p_p^\prime)\,{\cal V}^{(2)}_\mu \,\chi(p_p)\right],
\nonumber\\
\label{M2rms_gamma}
\end{eqnarray}
where the %renormalized 
effective proton-photon vertex, due to form factor corrections from pion-loops and LECs at 
NNLO$_\chi$, is given as
\begin{eqnarray}
{\cal V}^{(2)}_\mu \!&=&   (F^p_1-1)v_\mu  \!+\!\frac{1}{M}\Bigg\{(F^p_1-1)\left(Q_\mu + \frac{Q^2}{2M}v_\mu \right)  
\nonumber\\
&& +\, 2(F^p_1+F^p_2-1-\kappa_p)\left[S_\mu , S\cdot Q\right]\,\Bigg\}
\nonumber\\
&& -\, \frac{Q^2}{8M^2}(F^p_1 - 2F^p_2 - 1)v_\mu  + {\mathcal O}\left(\frac{1}{M^3}\right). \,\quad
\label{Vhadron}
\end{eqnarray}
Here $\kappa_p=(\kappa_v+\kappa_s)/2=1.795$ is the anomalous magnetic moment of the proton. Analytical expressions 
for the form factors evaluated to NNLO$_\chi$ in HB$\chi$PT already exist in the literature~\cite{Fearing1998} and 
may be used to determine the scattering cross section. However, in this work we shall only consider some 
representative input for the proton's rms charge radius amongst the recently measured values from scattering as well
as atomic-spectroscopy 
measurements~\cite{Pohl:2010zza,Antognini:1900ns,ISR_2017,Beyer_2017,Fleurbaey_2019,ISR_2019,Adams_2018,Bezginov_2019,Xiong_2019,Grinin2020}. 
%\vspace{1.5mm}

In this paper we define the four-momentum transfer $Q_\mu=p_\mu-p^\prime_\mu= (p^\prime_p)_\mu-(p_p)_\mu$ of the 
$\ell$-p elastic scattering process, $\ell(p)+{\rm p}(p_p)\to\ell(p^\prime)+{\rm p}(p^\prime_p)$, where $Q^2 < 0$. The
incoming and outgoing lepton four-momenta are $p=(E,\vec{p})$ and $p^\prime=(E^\prime,\vec{p}^{\,\prime})$. The initial
and final proton four-momenta in the {\it lab.}-frame are $P=(M,0)\,,\,P^\prime=(E^\prime_p,\vec{p}^{\,\prime}_p)$. In 
the heavy baryon formalism the initial and final state proton four-momenta are $P^\mu=Mv^\mu+p^\mu_p$ and 
$P^{\prime \mu}=Mv^\mu+p^{\prime\mu}_p$ respectively. We have in the {\it lab.}-frame, $v\cdot p_p=0$, and 
$v\cdot p_p^\prime =-\frac{(p_p^\prime)^2}{2M}+{\mathcal O}(M^{-2})$. 

%\vspace{0.15cm}

\begin{figure*}[tbp]
 \centering
	\includegraphics[scale=0.52]{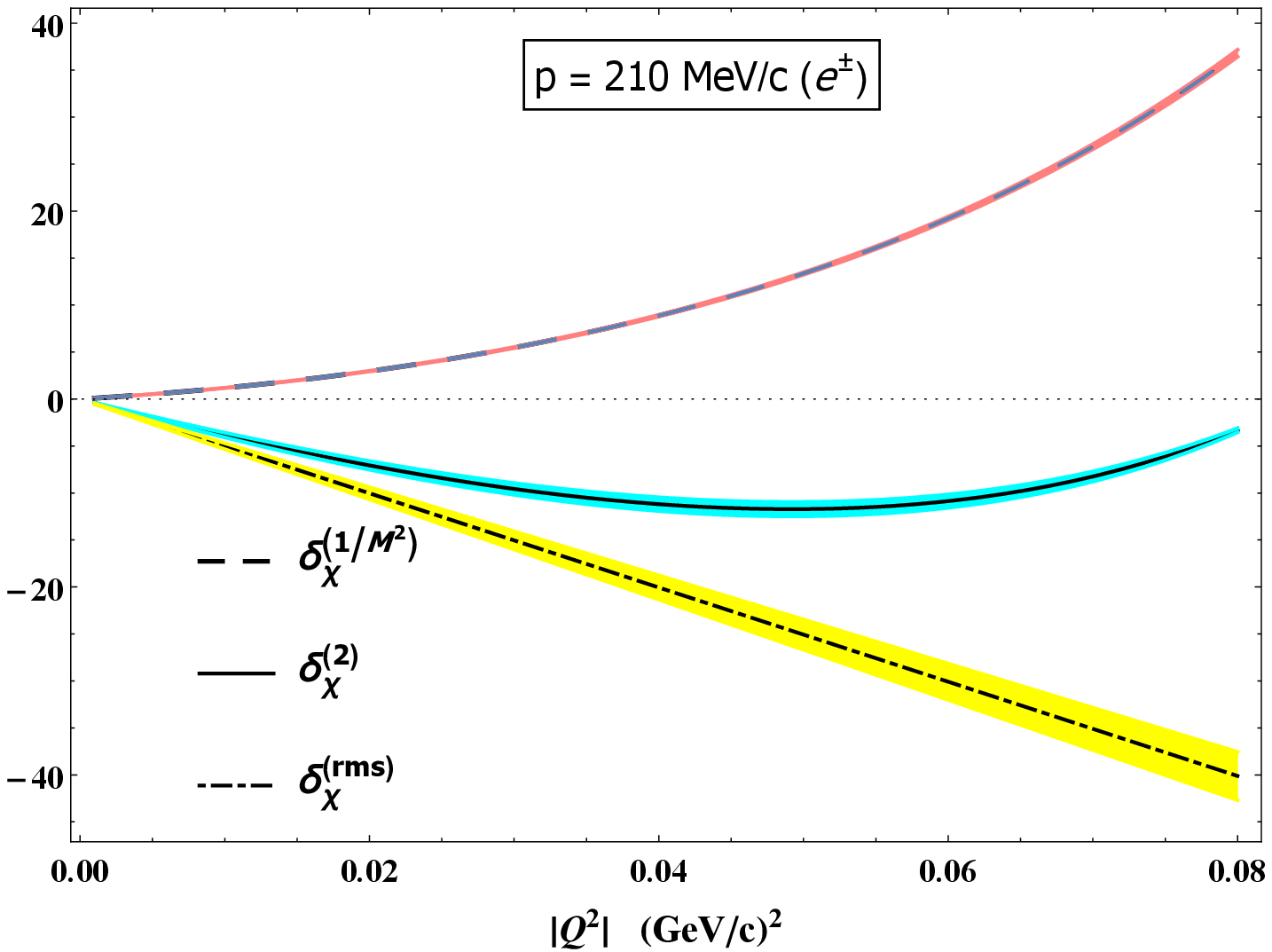}\qquad
	\includegraphics[scale=0.52]{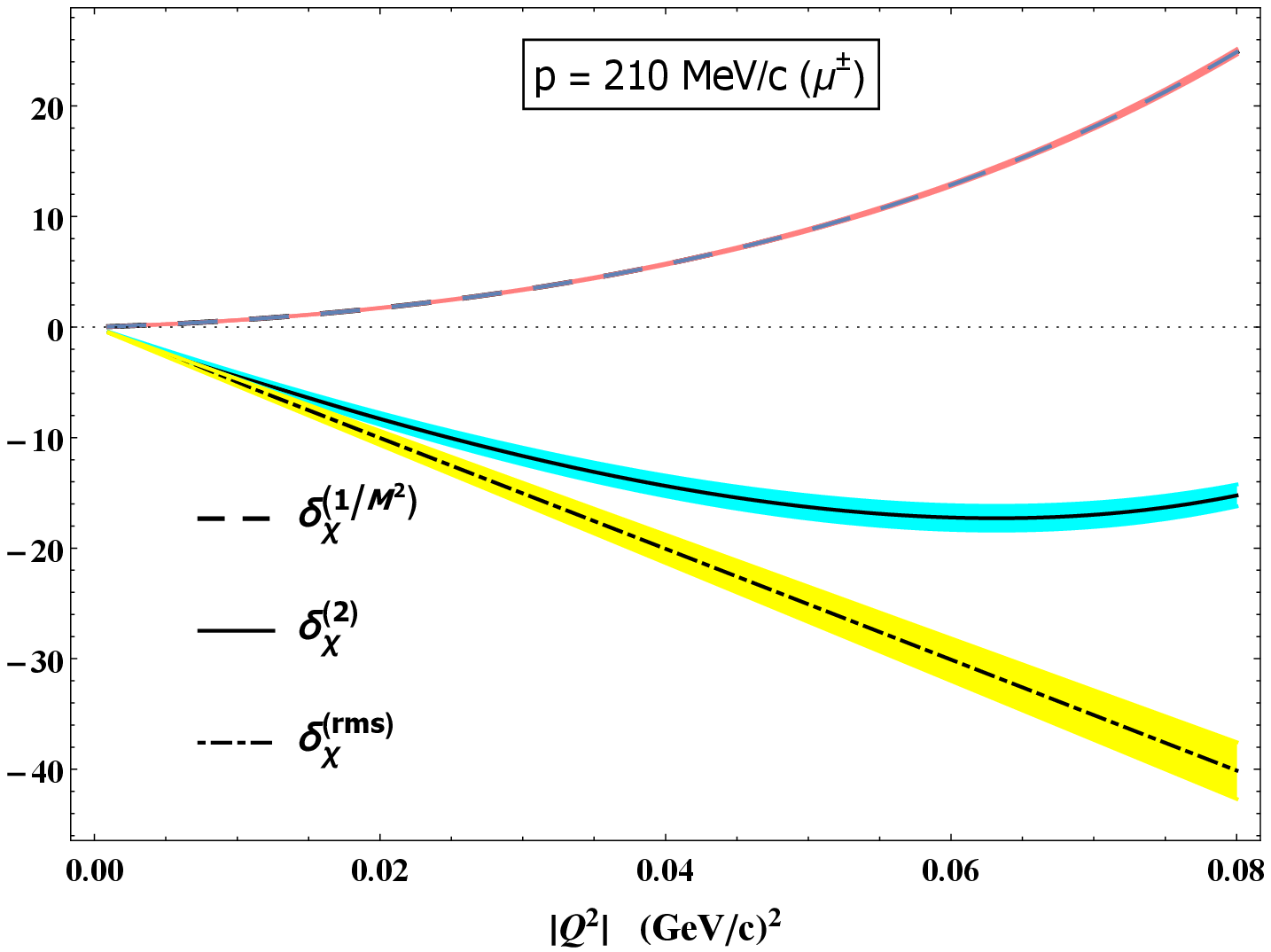}

    \vspace{0.7cm}

    \includegraphics[scale=0.52]{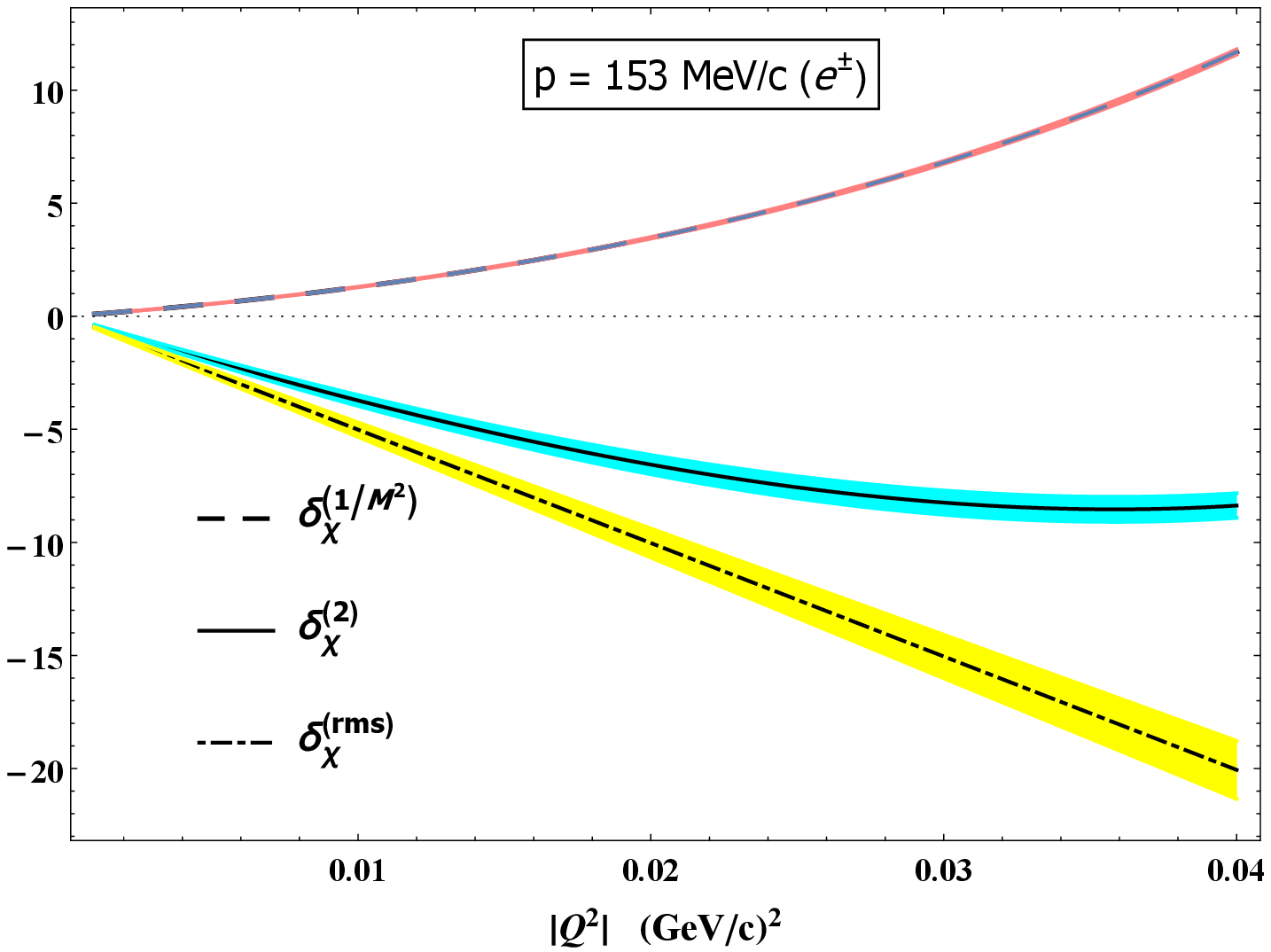}\qquad 
    \includegraphics[scale=0.52]{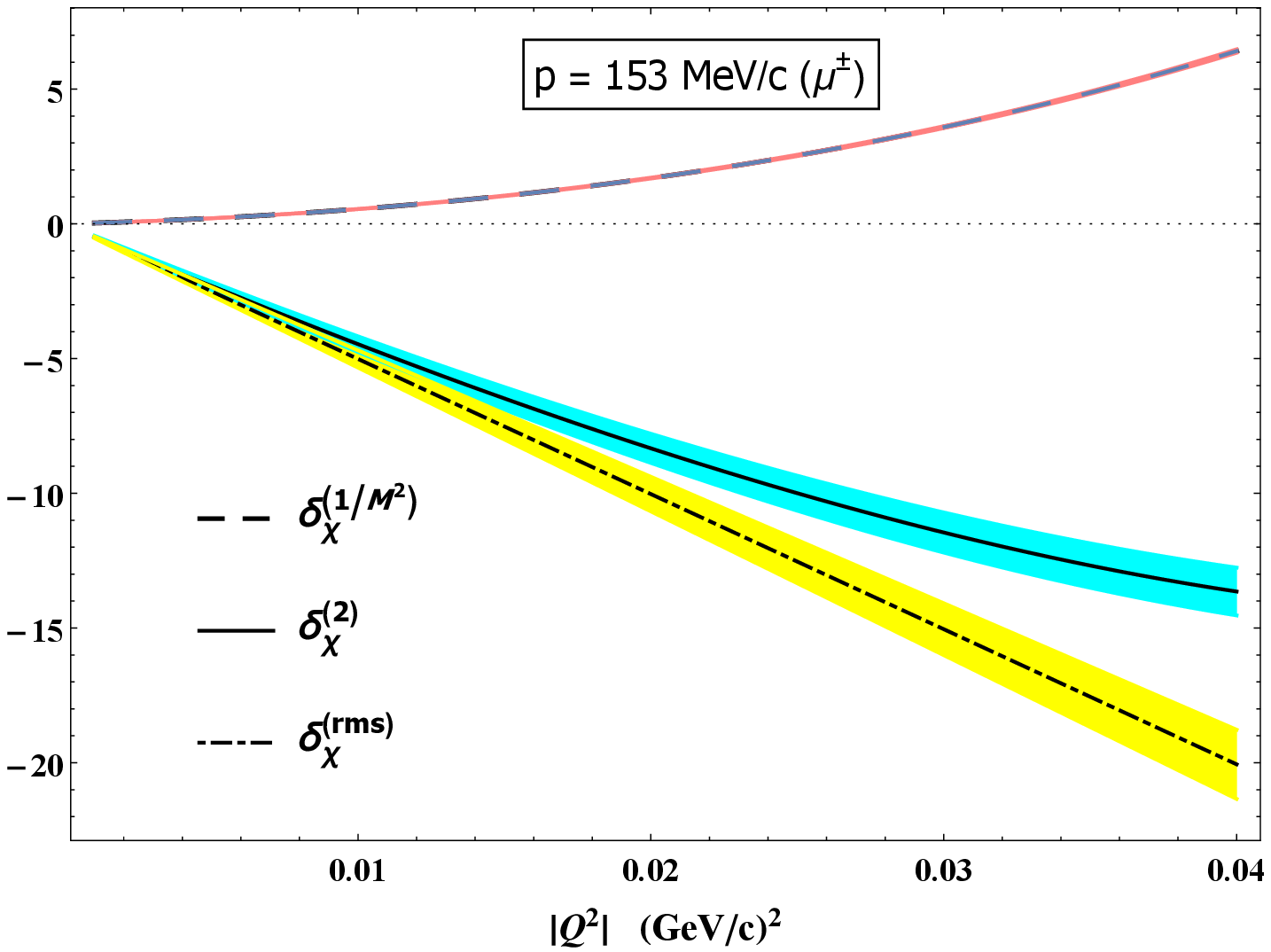}

    \vspace{0.7cm}

    \includegraphics[scale=0.52]{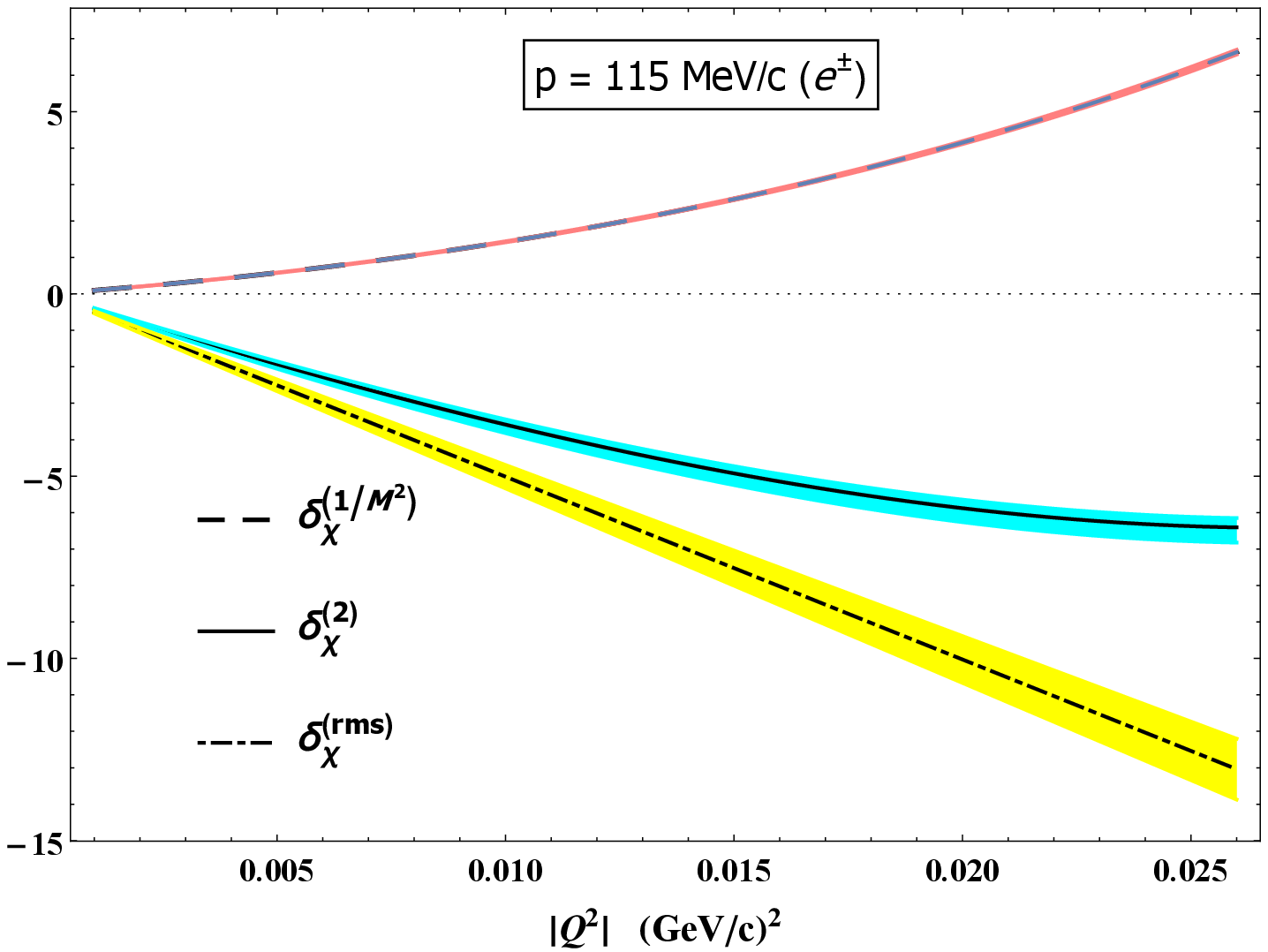}\qquad 
    \includegraphics[scale=0.52]{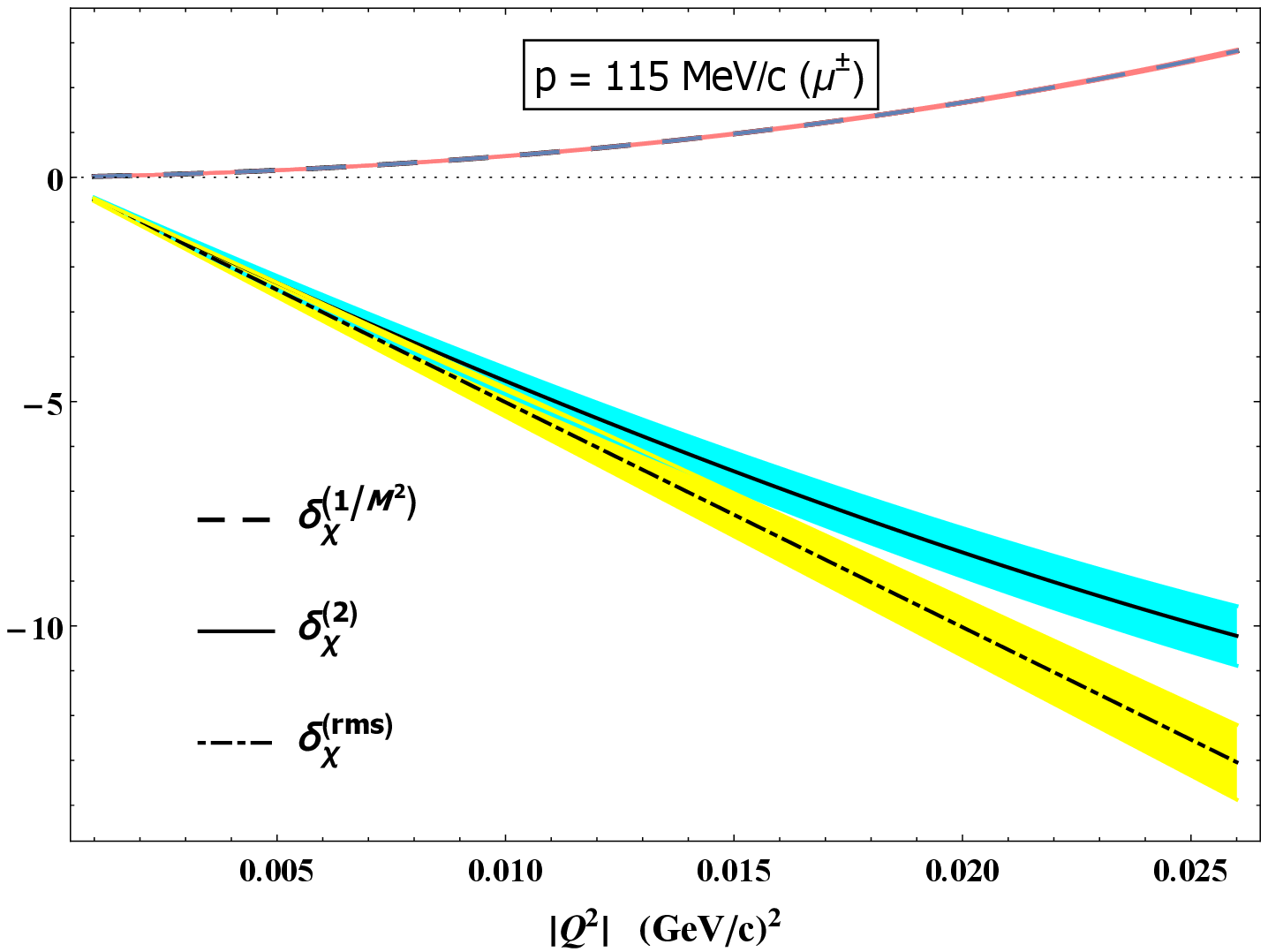}
	\caption{The fractional chiral corrections (in percentage) to the leading order Born elastic cross section 
	         [Eq.~\eqref{dsigmaBornL0}] for $e$-p (left panel) and $\mu$-p  (right panel) at NNLO$_\chi$ in 
	         HB$\chi$PT, see Eq.~\eqref{eq:delta_chi1}, as a function of the squared four-momentum transfer 
	         $|Q^2|$. The contributions due to the proton's rms radius $\delta^{(rms)}_\chi$ [dashed-dotted 
	         (yellow) curve], recoil contributions $\delta^{(1/M^2)}_\chi$ [dashed (red) curve], and their sum 
	         $\delta^{(2)}_\chi$ [solid (cyan) curve], are separately displayed. Each plot covers the MUSE 
	         kinematic range of $|Q^2|$ where the scattering angle lies within the range, 
	         $\theta\in[20^\circ,100^\circ]$, at specific incoming lepton momenta, 
	         $|\vec p\,|=p=115,\, 153,\, 210$~MeV/c. The curves for the proton's rms charge radius in the 
	         above plots are varied within the range corresponding to the extracted values from the recent 
	         precision $e$-p scattering measurements by the PRad Collaboration~\cite{Xiong_2019}, and that 
	         from the erstwhile high-precision muonic hydrogen atomic-spectroscopy measurements by the CREMA 
	         Collaboration~\cite{Pohl:2013,Antognini:1900ns}. The theoretical uncertainty due to the input 
	         variation of the rms radius along with the N$^3$LO error are depicted by the widths of the 
	         colored bands.}
	\label{plot:deltachi}
\end{figure*}

The full (chirally corrected) effective Born cross section in the {\it lab.}-frame is determined by evaluating 
the phase-space integral of the expression 
\begin{eqnarray}
\left[{\rm d}\sigma_{el}\right]_\gamma&=&\frac{(2\pi)^4\delta^4\left(p+P-p^\prime-P^\prime\right)}{4ME}
\nonumber\\
&&\times\frac{{\rm d}^3\vec{p}^{\,\prime}}{(2\pi)^3 2E^\prime}\frac{{\rm d}^3\vec{P}^{\prime}}{(2\pi)^3
2E^\prime_p}\,\frac{1}{4}\sum_{spins}|\mathcal{M}_\gamma|^2\,,\,\,\quad
\end{eqnarray}
where the squared amplitude is  
\begin{eqnarray}
\left|\mathcal{M}_\gamma\right|^2 = \left|\mathcal{M}^{(0)}_\gamma + \mathcal{M}^{(1)}_\gamma 
+ \mathcal{M}^{(2)}_{\gamma} + \mathcal{M}^{(2);\,{\rm rms}}_{\gamma}\right|^2.\,\quad 
\end{eqnarray}
It is notable that due to the sum over spins, only the spin-independent parts of $\mathcal{M}^{(1)}_\gamma$
and $\mathcal{M}^{(2);\,{\rm rms}}_{\gamma}$ contribute to the interference with the LO Born amplitude 
$\mathcal{M}^{(0)}_\gamma$. The lowest order Born contribution is well known, namely,
\begin{eqnarray}
\frac{1}{4}\sum_{spins}\!\!|\mathcal{M}^{(0)}_\gamma|^2=\frac{4e^4}{Q^2} M^2 \left(1-\frac{Q^2}{4M^2}\right)
\left[1+\frac{4EE^\prime}{Q^2}\right]\,.
\nonumber\\
\label{M0gamma2}
\end{eqnarray} 
The following are the relevant fractional contributions needed at $1/M^2$ order accuracy involving the 
NLO$_\chi$ amplitudes, where we introduce the compact notation $\mathscr{R}_Q \equiv \frac{Q^2}{2M^2}$ 
for later convenience:
%%%%%%%%%%%%%%%%%%%%%%%%%%%%%%%%%%%%%%%%%%%%%%%%%%%%
\begin{widetext}
\begin{eqnarray}
&&\frac{2{\mathcal R}{\rm e} \sum\limits_{spins}\left({\mathcal{M}}^{(0)\dagger}_\gamma
\mathcal{M}^{(1)}_\gamma\right)}{\sum\limits_{spins}|\mathcal{M}^{(0)}_\gamma|^2} 
=\frac{2{\mathcal R}{\rm e} \sum\limits_{spins}\!\left({\mathcal{M}}^{(0)\dagger}_\gamma
\mathcal{M}^{(1)\,;\,{\rm a}}_\gamma\right)}{\sum\limits_{spins}|\mathcal{M}^{(0)}_\gamma|^2} 
=  \frac{Q^2}{2M^2}\equiv 
\mathscr{R}_Q,
\nonumber\\
\nonumber\\
&&\frac{\sum\limits_{spins}|\mathcal{M}^{(1)}_\gamma|^2}{\sum\limits_{spins}|\mathcal{M}^{(0)}_\gamma|^2}
=\frac{\sum\limits_{spins}|\mathcal{M}^{(1);\,{\rm b}}_\gamma|^2}{\sum\limits_{spins}|\mathcal{M}^{(0)}_\gamma|^2}
= \frac{1}{2}(1+\kappa_p)^2\mathscr{R}_Q\,\left(\frac{Q^2+4(m^2_l-EE^\prime)}{Q^2+4EE^\prime}\right)
+ {\mathcal O}\left(\frac{{\cal Q}^4}{M^4}\right)\,,
\label{Ratio_NLO}
\end{eqnarray}
\end{widetext}
%%%%%%%%%%%%%%%%%%%%%%%%%%%%%%%%%%%%%%%%%%%%%%%%%%%%
and those involving the NNLO$_\chi$ amplitudes:
\begin{eqnarray}
&&\frac{2{\mathcal R}{\rm e}\!\sum\limits_{spins}\!\left({\mathcal{M}}^{(0)\dagger}_\gamma
\mathcal{M}^{(2);\,{\rm rms}}_\gamma\right)}{\sum\limits_{spins}|\mathcal{M}^{(0)}_\gamma|^2}
=\frac{Q^2}{3}\langle r^2_1 \rangle + {\mathcal O}\left(\!\frac{{\cal Q}^3}{M^3}\!\right), 
\nonumber\\
&&\frac{2{\mathcal R}{\rm e}\!\sum\limits_{spins}\!\left({\mathcal{M}}^{(0)\dagger}_\gamma
\mathcal{M}^{(2)}_\gamma\right)}{\sum\limits_{spins}|\mathcal{M}^{(0)}_\gamma|^2}
=-\frac{1}{2}\mathscr{R}_Q + {\mathcal O}\left(\!\frac{{\cal Q}^4}{M^4}\!\right).\,\quad
\label{Ratio_NNLO}
\end{eqnarray}
In fact, the effective Born differential cross section including up to NNLO$_\chi$ corrections may be expressed in 
terms of the proton form factors $F^p_{1,2}$ and the incoming and outgoing lepton velocities, 
\begin{eqnarray*}
\beta&=&\frac{|\vec{p}\,|}{E}\,\, \hspace*{2mm} {\rm and} 
\\
\beta^\prime&=&\frac{|\vec{p}^{\,\prime}|}{E^\prime}=\sqrt{1-\left(\frac{\eta m_l}{E}\right)^2}
\end{eqnarray*}
respectively, in the generic form
%%%%%%%%%%%%%%%%%%%%%%%%%%%%%%%%%%%%%%%%%%%%%%%%%%%%
\begin{widetext} 
\begin{eqnarray}
\left[\frac{{\rm d}\sigma_{el}(Q^2)}{{\rm d} \Omega^\prime_l}\right]_\gamma 
=\int\!\frac{|\vec{p}^{\,\prime}\,|\,{\rm d}E^\prime}{(2\pi)^2 16M^2E}\,\delta\!\left\{E\! +\! M\! -\! E^\prime\! -\! E_p^\prime \right\} \frac{1}{4}\!\sum_{spins}|\mathcal{M}_\gamma|^2 
= \frac{1}{64\pi^2M^2}\left(\frac{\beta^\prime}{\eta\beta}\right)\frac{1}{4}\sum_{spins}|\mathcal{M}_\gamma|^2\,,
\label{dsigmaBornformfactor}
\end{eqnarray} 
where 
\begin{eqnarray}
\eta \equiv \frac{E}{E^{\prime}}=\frac{E}{E+Q^2/2M} = 1+\frac{E}{M}\left(1-\cos\theta\right)\,,\,\quad
\label{eta}
\end{eqnarray}
is the proton recoil factor which may either be expressed in terms the four momentum transfer $Q^2$, or equivalently, 
in terms of the {\it lab\,}-frame lepton scattering angle $\theta$. The spin-averaged squared amplitude is 
\begin{eqnarray}
\frac{1}{4}\sum_{spins}|\mathcal{M}_\gamma|^2 &=&\frac{1}{4}\sum_{spins}\!
\Bigg[ \!\Big\{ |\mathcal{M}^{(0)}_\gamma|^2\! +\! |\mathcal{M}^{(1);\,{\rm a}}_{\gamma}|^2\! +
 \! 2{\mathcal R}{\rm e} \left(\mathcal{M}^{(0)\,\dagger}_\gamma\mathcal{M}^{(1);\,{\rm a}}_{\gamma}\right)\!\Big\} 
 \!\left( F^p_1(Q^2)\right)^2 \!+\!\frac{1}{(1+\kappa_p)^2} |\mathcal{M}^{(1);\,{\rm b}}_{\gamma}|^2 
\nonumber\\ 
&&\!\!\times\left( F^p_1(Q^2) \!+\! F^p_2(Q^2)\right)^2 
+ 2{\mathcal R}{\rm e}\left(\mathcal{M}^{(0)\,\dagger }_\gamma\mathcal{M}^{(2)}_\gamma\right) 
\left( F^p_1(Q^2)\! -\! 2 F^p_2(Q^2)\right)\! F^p_1(Q^2)
\!+\! {\mathcal O}\left(\!\alpha^2\frac{{\cal Q}^3}{M^3}\!\right)\!\Bigg],\,\,\quad
\end{eqnarray}
\end{widetext}  
%%%%%%%%%%%%%%%%%%%%%%%%%%%%%%%%%%%%%%%%%%%%%%%%%%%%
The pre-factor $\eta$, which arises from the phase-space integration over the energy conservation 
$\delta$-function, exactly cancels out while considering the different ratios of the chirally or radiatively 
corrected $\ell$-p scattering cross sections to the LO Born contribution, 
\begin{equation}
\left[\frac{{\rm d}\sigma_{el}(Q^2)}{{\rm d} \Omega^\prime_l}\right]_0
=\frac{\alpha^2\beta^\prime}{\eta\beta Q^2}
\left(1-\frac{Q^2}{4M^2}\right)\left[1+\frac{4E^2}{\eta Q^2}\right]\,.
\label{dsigmaBornL0}
\end{equation}  
Firstly, we note that $|\mathcal{M}^{(0)}_\gamma|^2$ contains terms up-to-and-including ${\cal O}(M^{-2})$. 
Retaining its complete expression generates the above LO Born cross section. Secondly, the term 
$\sum |\mathcal{M}^{(1);\,{\rm a}}_{\gamma}|^2 \sim \sum |\mathcal{M}^{(0)}_\gamma|^2 {\cal O}(M^{-4})$, 
and is therefore ignored. Thus, we obtain the full chirally corrected NNLO$_\chi$ result for the elastic 
differential cross section of the form 
\begin{eqnarray} 
\left[\frac{{\rm d}\sigma_{el}(Q^2)}{{\rm d} \Omega^\prime_l}\right]_\gamma 
&=&\left[\frac{{\rm d}\sigma_{el}(Q^2)}{{\rm d} \Omega^\prime_l}\right]_0 \left\{1+\delta^{(2)}_\chi(Q^2) \right\}\,.\,\quad
\label{eq:dsigmaBornNNLO}
\end{eqnarray} 
where the  ${\mathcal O}({\cal Q}^2/M^2)$ fractional contributions due to the pure hadronic chiral effects are 
represented as 
\begin{equation}
\delta^{(2)}_\chi(Q^2) =\delta_\chi^{(rms)}(Q^2)+\delta_\chi^{(1/M^2)}(Q^2)\, . 
\label{eq:delta_chi1}
\end{equation}
Here   
\begin{eqnarray}
\delta_\chi^{(rms)}(Q^2) &=&\frac{Q^2}{3}\langle r_1^2\rangle + \mathcal{O}\left(\frac{{\cal Q}^3}{M^3}\right)
\nonumber\\ 
&=&\frac{Q^2}{3}\left[\langle r_E^2\rangle-\frac{3\kappa_p}{2M^2}\right] 
+ \mathcal{O}\left(\frac{{\cal Q}^3}{M^3}\right)\,,\quad\,
\label{eq:delta_chi2}
\end{eqnarray}   
stands for the NNLO$_\chi$ corrections due to the proton's rms {\it electric} charge radius, 
$r_p=\sqrt{\langle r^2_E\rangle}$, and
\begin{eqnarray}
\delta_\chi^{(1/M^2)}(Q^2) &=& \frac{1}{2}{\mathscr R}_Q \!
\Bigg[1+2\kappa_p+(1+\kappa_p)^2
\nonumber\\ \!\!
&\times&\!\!\left(\!\frac{\eta Q^2\! +\! 4(\eta m_l^2\! -\! E^2)}{\eta Q^2+4E^2}\!\right)\!\!\Bigg]
\!\!+\!\mathcal{O}\left(\frac{{\cal Q}^3}{M^3}\right)\, .
\nonumber\\
\label{eq:delta_chi3}
\end{eqnarray} 
are ${\mathcal O}({\cal Q}^2/M^2)$ contributions. Figure~\ref{plot:deltachi} displays separately 
the  proton's NNLO$_\chi$ recoil and structure dependent corrections to the LO Born $e$-p and $\mu$-p elastic 
scattering cross sections in the MUSE kinematic region~\cite{Gilman:2013eiv,Gilman:2017hdr}. These two 
NNLO$_\chi$  terms are the largest of the corrections to the LO Born contributions, each increasing to about 
$40\%$ at the largest $Q^2$ values. However, due to their opposite signs there are large cancellation as 
observed in Fig.~\ref{plot:deltachi}. Although the proton rms radius dependent effects are independent of the 
lepton mass, the $1/M^2$ order effects are about one half times smaller in muon as compared to electron 
scattering. The overall contributions are only somewhat sensitive to lepton mass dependence. To check the 
sensitivity of our chiral corrections to the input $r_p$ value, we vary $r_p$ within the range corresponding 
to the extracted value from recent precision $e$-p scattering measurements at the Jefferson Laboratory (PRad 
Collaboration)~\cite{Xiong_2019}, and that from the high-precision muonic hydrogen atomic-spectroscopy 
measurements at PSI (CREMA Collaboration)~\cite{Pohl:2013,Antognini:1900ns}. The resulting plots for 
$\delta_\chi^{(rms)}$ show a sensitivity of about $\pm 6.4\%$. We have included both the experimental and 
theoretical uncertainties in quadrature (see footnote 8). These uncertainties are represented by the error 
bands in yellow (color online). For the purpose of estimating the theoretical error, we have, in addition, 
varied each of the two chiral corrections in Fig.~\ref{plot:deltachi} by $\pm 1$\% to incorporate the 
uncertainties due to the N$^3$LO$_\chi$ [i.e., ${\mathcal O}({\cal Q}^3/M^3)\sim 0.008$] effects excluded in 
our analysis. The error bands in cyan (color online) associated with the total chiral corrections 
$\delta^{(2)}_\chi$ [with the error bands in pink representing the $\delta_\chi^{(1/M^2)}$ terms] yield about
$7\%$ uncertainty relative to their central estimates, after combining the two errors. 

%\vspace{0.15cm}

In the next section we demonstrate that the radiative (QED) contributions are smaller in comparison to the 
individual chiral corrections. However, owing to the subtlety of the large chiral cancellations noted in case
of $e$-p scattering, the otherwise power counting sub-dominant QED effects get effectively promoted as a 
serious correction to the Born cross section. Thus, estimating the crucial radiative effects along with the 
chiral corrections becomes a necessary precursor before attempting a precision extraction of the proton's 
rms charge radius.  
%%%%%%%%%%%%%%%%%%%%%%%%%%%%%%%%%%%%%%%%%%%FIGURE%%%%%%%%%%%%%%%%%%%%%%%%%%%%%%%%%%%%%%%%%%%
%%%%%%%%%%%%%%%%%%%%%%%%%%%%%%%%%%%%%%%%%%%%%%%%%%%%%%%%%%%%%%%%%%%%%%%%%%%%%%%%%%%%%%%%%%%% 

%\vspace{-0.1cm}

%%%%%%%%%%%%%%%%%%%%%%%%%%%%%%%%%%%%%%%%%%%%%%%%%%%%%%%%%%%%%%%%%%%%%%%%%%%%%%%%%%%%%%%%%%%%%%
\section{Radiative Correction at LO$_\alpha$}\label{sec:LO}
%%%%%%%%%%%%%%%%%%%%%%%%%%%%%%%%%%%%%%%%%%%%%%%%%%%%%%%%%%%%%%%%%%%%%%%%%%%%%%%%%%%%%%%%%%%%%%
The lowest order radiative (LO$_\alpha$) corrections  to the $\ell$-p elastic scattering process constitute 
diagrams with amplitudes either of $\mathcal{O}(e^2\alpha)$, which arise from one-loop virtual corrections, or 
of $\mathcal{O}(e^3)$, associated with the emission of a single undetectable real soft photon. In this section 
we outline all UV and IR divergences in DR arising in the analytical evaluation of the LO$_\alpha$ 
contributions, $\delta^{(0)}_{2\gamma}\sim\mathcal{O}(\alpha)$.   

%\vspace{0.1cm}

The UV-divergent terms are renormalized by the Bogoliubov, Parasiuk, Hepp, Zimmermann (BPHZ) renormalization 
method~\cite{BPHZ,akhiezer1965quantum} using Lagrangian counter-terms to render UV-finite results. To this end, all bare Lagrangian 
masses, charges, coupling constants etc. are replaced by the corresponding physical ones in the standard way. As
in our previous work~\cite{Talukdar:2019dko}, we analytically evaluate the one-loop virtual diagrams in order to 
project out the complete IR-singularity structures ensuring exact cancellation with the soft bremsstrahlung IR 
divergences. Note that unlike the {\it soft photon approximation} (SPA) which was invoked to allow analytical 
evaluation of the TPE {\it box amplitudes} in Ref.~\cite{Talukdar:2019dko}, the other one-loop virtual amplitudes 
in this paper will be analytically performed without any approximations. However, for the purpose of extracting 
the IR-singularities from the soft photon bremsstrahlung diagrams, we need to rely on SPA as a basic precept of 
the YFS methodology~\cite{Yennie:1961ad}. In what follows we consider each of the {\it virtual} and {\it real} 
(bremsstrahlung) contributions separately.
    
\vspace{-0.1cm}
    
%%%%%%%%%%%%%%%%%%%%%%%%%%%%%%%%%%%%%%%%%%%%%%%%%%%%%%    
\subsection{One-loop Virtual Corrections at LO$_\alpha$}
\begin{figure*}[tbp]
 \centering
       \includegraphics[scale=0.6]{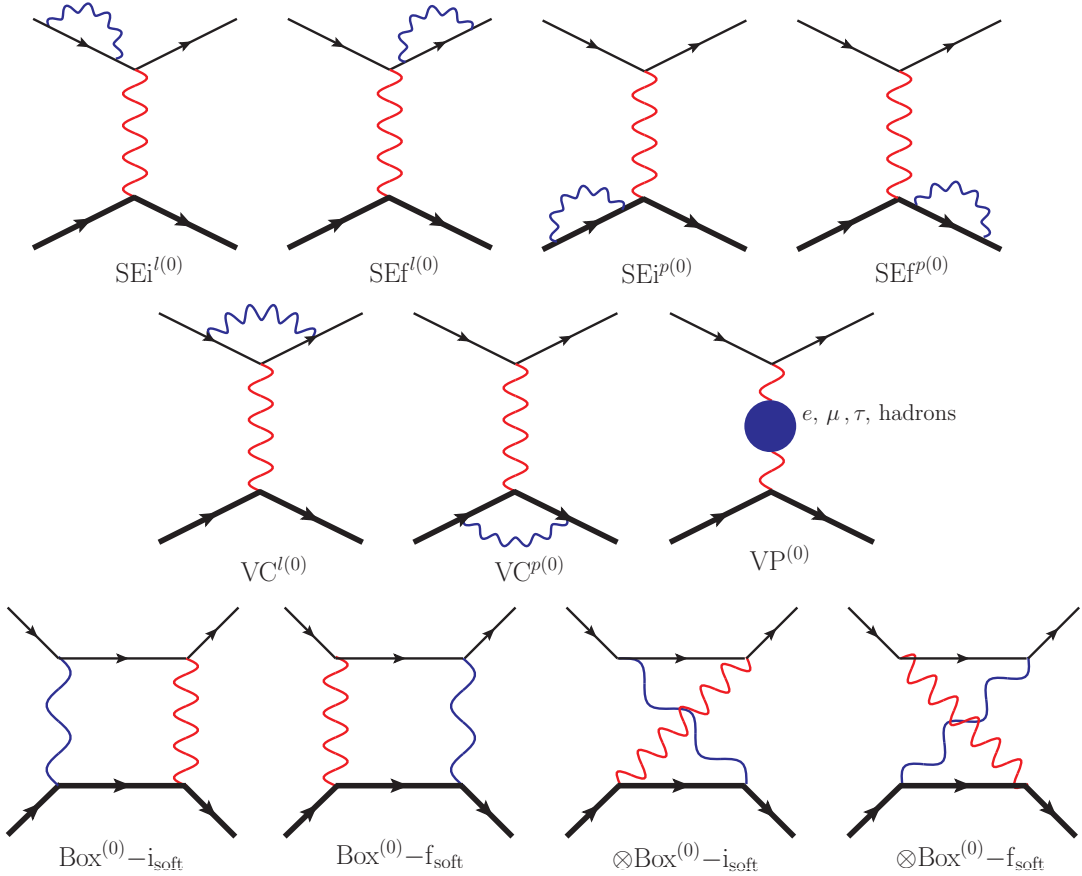}
       \caption{The one-loop $\mathcal{O}(e^2\alpha)$ diagrams at LO$_\alpha$ in HB$\chi$PT, contributing 
                to the virtual radiative corrections to the elastic leading order Born $\ell$-p scattering 
                amplitude [see Eq.~\eqref{M0_gamma}]. The blob in the diagram VP$^{(0)}$ represents one-loop 
                leptonic and hadronic vacuum polarization contributions. For the sake of illustration, each 
                leading order two-photon exchange (``direct" and ``crossed") box diagram is shown with one 
                hard photon (red color online) and one soft photon (blue color online) exchange.}
        \label{1loop}
\end{figure*}

%%%%%%%%%%%%%%%%%%%%%%%%%%%%%%%%%%%%%%%%%%%%%%%%%%%%%%
The one-loop diagrams contributing to the virtual radiative corrections are displayed in Fig.~\ref{1loop}. It is 
notable that the SE loops renormalize the bare masses and wave functions of the external lepton and proton, but on 
the mass-shell such SE diagrams themselves vanish upon renormalization (these corrections are non-vanishing for 
internal off-shell propagator lines). Nonetheless, their expressions are needed to determine the respective 
wave-function renormalization constants $Z^{l,p}_2$, which by virtue of {\it Ward-Takahashi identity} in QED is 
equal to the corresponding vertex renormalization constants $Z^{l,p}_1$. Furthermore, as previously discussed in 
Ref.~\cite{Talukdar:2019dko}, the sum of the real parts of TPE ``direct" and ``crossed" box diagrams at LO$_\alpha$ 
in HB$\chi$PT (cf. diagrams in the last row of Fig.~\ref{1loop}) vanishes with or without SPA.\footnote{The imaginary
part of the sum of the LO$_\alpha$ TPE box amplitudes is, however, non-vanishing even after invoking SPA, but it 
is irrelevant in this work since it does not contribute to the unpolarized elastic cross section.} 
The remaining one-loop contributions listed in Fig.~\ref{1loop} are evaluated without invoking SPA, and will be 
discussed in the following: 

%\vspace{0.1cm}

%%%%%%%%%%%%%%%%%%%%%%%%%%%%%%%%%%%%%%%%%%%%%%%%%%%%%%
\noindent {\it 1. Lepton-Photon Vertex Correction.\,}
%%%%%%%%%%%%%%%%%%%%%%%%%%%%%%%%%%%%%%%%%%%%%%%%%%%%%%
The one-loop lepton-photon VC amplitude from the VC$^{l(0)}$ diagram in Fig.~\ref{1loop}  using DR is well known 
(see, e.g., Refs.~\cite{Vanderhaeghen:2000ws,akhiezer1965quantum}). The VC amplitude VC$^{l(0)}$ is 
\begin{eqnarray}
\mathcal{M}^{l(0)}_{\gamma\gamma;\,{\rm vertex}} &=& -\frac{e^2}{Q^2}\,\Big[ {\bar u}_l(p^\prime)\,
\delta\Gamma_l^\mu(p,p^\prime)\, u_l(p)\Big] 
\nonumber\\
&&\times\,\left[\chi^\dagger(p^\prime_p) v_\mu \chi(p_p)\right]\,,  
\end{eqnarray}  
and the radiative corrections to the lepton-photon vertex in terms of the Dirac and Pauli form factors, 
$F^l_1=1+\delta F^l_1$ and Pauli $F^l_2$ respectively, is expressed in the general form 
\begin{eqnarray}
\delta\Gamma_l^\mu(p,p^\prime)=\gamma^\mu \delta F^l_1(Q^2)+\frac{i\sigma^{\mu\nu}Q_\nu}{2m_l}F^l_2(Q^2)\,.  
\end{eqnarray}  
The evaluation of diagram VC$^{l(0)}$ yields both UV and IR divergences for the form factor $F^l_1$, while $F^l_2$ 
is finite at $\mathcal O(\alpha)$. It is noteworthy that while the ``Dirac'' part of amplitude factorizes into the 
Born amplitude $\mathcal{M}^{(0)}_\gamma$, the ``Pauli" part does not manifest itself in the same way, namely, 
\begin{eqnarray}
\mathcal{M}^{l (0)}_{\gamma\gamma;{\rm vertex}} &=& \mathcal{M}^{(0)}_\gamma \delta F^l_1(Q^2) 
+ \overline{\mathcal{M}}^{(0)}_\gamma F^l_2(Q^2)\,,
\nonumber
\end{eqnarray}

\vspace{-0.8cm}

\begin{eqnarray}
{\rm with}\quad \overline{\mathcal{M}}^{(0)}_\gamma &=& -\frac{e^2}{2m_lQ^2}\left[\bar u_l(p^\prime) 
i\sigma^{\mu\nu}Q_\nu u_l(p)\right]
\nonumber\\
& & \times\left[\chi^\dagger(p^\prime_p)v_\mu\chi(p_p)\right]\,.
\label{Ml0gamgam-vertex}
\end{eqnarray} 
As discussed, we use DR in order to simultaneously extract the UV and IR divergences from the loop diagrams. The UV 
divergence in space-time dimensions $d=4-2\epsilon_{UV}$ (with $\epsilon_{UV} > 0$) is characterized by the 
pole-term proportional to $1/\epsilon_{UV}$ and a log-dependent subtraction scale $\mu$. Likewise, for space-time 
dimensions $d=4-2\epsilon_{IR}$ (with $\epsilon_{IR} < 0$), the IR-singularity appear as a pole term proportional to 
$1/\epsilon_{IR}$. The one-loop LO$_\alpha$ expressions for the Dirac and Pauli form factor evaluated using DR are 
respectively given as~\cite{Vanderhaeghen:2000ws,Bucoveanu:2018soy} 
\begin{eqnarray}
\delta F^l_1(Q^2) &=& \frac{\alpha}{4\pi}\Bigg[\left[\frac{1}{\epsilon_{\rm UV}}-\gamma_E
+\ln\left(\frac{4\pi\mu^2}{m^2_l}\right)\right]
\nonumber\\
&& -\,\left[\frac{1}{|\epsilon_{\rm IR}|}+\gamma_E-\ln\left(\frac{4\pi\mu^2}{m^2_l}\right)\right]
\frac{\nu^2+1}{\nu}
\nonumber\\
&&\times\,\ln\left[\frac{\nu+1}{\nu-1}\right]+\frac{\nu^2+1}{2\nu}\ln\left[\frac{\nu+1}{\nu-1}\right]
\nonumber\\
&&\times\,\ln\left[\frac{\nu^2-1}{4\nu^2}\right]+\frac{2\nu^2+1}{\nu}\ln\left[\frac{\nu+1}{\nu-1}\right]
\nonumber\\
&&-\,\frac{\nu^2+1}{\nu}\!\left\{\text{Sp}\!\left(\frac{\nu+1}{2\nu}\right)\!
-\!\text{Sp}\!\left(\frac{\nu-1}{2\nu}\right)\right\}\Bigg]\,,
\nonumber
\end{eqnarray}
and
\begin{eqnarray}
F^l_2(Q^2) = \frac{\alpha}{4\pi}\frac{\nu^2-1}{\nu}\ln\left[\frac{\nu+1}{\nu-1}\right]\,,
\label{Fl1Fl2unren}
\end{eqnarray}
where $\displaystyle{\nu=\sqrt{1-4m_l^2/Q^2}}$ and ``Sp" denotes the Spence function, defined as
\begin{equation}
{\rm Sp}(x)=\int^x_0 {\rm d}t\, \frac{\ln|1-t|}{t}\,\,;\,\, x<1\, ,  
\end{equation}  \\
and $\gamma_E=0.577216...$  is the Euler-Mascheroni constant. Note that our definition of the Spence function 
differs from the standard one (e.g., as used in Ref.~\cite{Vanderhaeghen:2000ws}) by an overall sign.

%\vspace{0.1cm}

The UV divergence is renormalized in the standard way by adding the counter-term vertex $(Z^l_1-1)\gamma^\mu$ 
to the vertex function $\Gamma^\mu_l$, requiring that the total vertex function, 
$\tilde{\Gamma}^\mu_l=\Gamma^\mu_l+(Z^l_1-1)\gamma^\mu$, defines the physical charge at $Q^2=0$ according to 
the renormalization condition, i.e., $\tilde{\Gamma}^\mu(Q^2=0)=\gamma^\mu$. The wave function renormalization 
constant $Z^l_2$ is defined by the derivative of the lepton SE function $\Sigma_l(p)$ in the on-shell limit, 
namely, 
\begin{eqnarray}
Z^l_2 &=& 1+\frac{\partial\Sigma^l(p)}{\partial\slashed{p}}\Bigg|_{\slashed{p}\, =\,m_l}+{\mathcal O}(\alpha^2) 
\nonumber\\
&\equiv& Z^l_1=1-\delta F^l_1(Q^2=0)\,.
\end{eqnarray}\\
Taking the limit  $Q^2\to 0$ (or $\nu\to \infty$), i.e.,
\begin{eqnarray}
\lim_{\nu \to \infty} \frac{\nu^2\pm 1}{\nu}\ln\left[\frac{\nu+1}{\nu-1}\right]=2\,,
\end{eqnarray}
we obtain
\begin{eqnarray}
\frac{\partial\Sigma^l(p)}{\partial \slashed{p}}\Bigg|_{\slashed{p}\,=\, m_l}
\hspace{-0.3cm}&=&-\frac{\alpha}{4\pi}\left[\frac{1}{\epsilon_{\rm UV}}-\gamma_E
+\ln\left(\frac{4\pi\mu^2}{m^2_l}\right)\right]
\nonumber\\
&&+\, \frac{\alpha}{2\pi}\!\left[\frac{1}{|\epsilon_{\rm IR}|}
+\gamma_E\!-\!\ln\left(\frac{4\pi\mu^2}{m^2_l}\right)\!-\!2\right]\,.
\nonumber\\
\label{senergy}
\end{eqnarray}  
Thus, by adding the counter-term $-\delta F^l_1(0)\gamma^\mu$, the renormalized amplitude is given by 
\begin{eqnarray}
 \left[\mathcal{M}^{l (0)}_{\gamma\gamma;\,{\rm vertex}}\right]_{\rm ren}&=& 
 \mathcal{M}^{(0)}_\gamma \left[\delta F^l_1(Q^2)-\delta F^l_1(0)\right] 
 \nonumber\\
 && +\, \overline{\mathcal{M}}^{(0)}_\gamma F^l_2(Q^2)\,,
\end{eqnarray}
where $\overline{\mathcal{M}}^{(0)}_\gamma$ is given in Eq.~\eqref{Ml0gamgam-vertex}, and the renormalized 
one-loop expression for the Dirac form factor of the lepton is given 
as~\cite{Vanderhaeghen:2000ws,Bucoveanu:2018soy}
%%%%%%%%%%%%%%%%%%%%%%%%%%%%%%%%%%%%%%%%%FIGURE%%%%%%%%%%%%%%%%%%%%%%%%%%%%%%%%%%%%%%%%%%%%%	  
%%%%%%%%%%%%%%%%%%%%%%%%%%%%%%%%%%%%%%%%%%%%%%%%%%%%%%%%%%%%%%%%%%%%%%%%%%%%%%%%%%%%%%%%%%%%
%%%%%%%%%%%%%%%%%%%%%%%%%%%%%%%%%%%%%%%%%%%%%%%%%%%%
\begin{widetext}
\begin{eqnarray}
F^{l;\,{\rm ren}}_1(Q^2)&=&1+\delta F^l_1(Q^2)-\delta F^l_1(0)
\nonumber\\
&=& 1+\frac{\alpha}{2\pi}\Bigg[-\left[\frac{1}{|\epsilon_{\rm IR}|}
+\gamma_E-\ln\left(\frac{4\pi\mu^2}{m^2_l}\right)\right]
\left[\frac{\nu^2+1}{2\nu}\ln\left[\frac{\nu+1}{\nu-1}\right]-1\right]
+\frac{\nu^2+1}{4\nu}\ln\left[\frac{\nu+1}{\nu-1}\right]\ln\left[\frac{\nu^2-1}{4\nu^2}\right]
\nonumber\\
&& +\,\frac{2\nu^2+1}{2\nu}\ln\left[\frac{\nu+1}{\nu-1}\right]-2-\frac{\nu^2+1}{2\nu}
\left\{\text{Sp}\left(\frac{\nu+1}{2\nu}\right)-\text{Sp}\left(\frac{\nu-1}{2\nu}\right)\right\}\Bigg].
\label{Fl1ren}
\end{eqnarray}
\end{widetext} 
%%%%%%%%%%%%%%%%%%%%%%%%%%%%%%%%%%%%%%%%%%%%%%%%%%%%
Besides, the finite Pauli form factor $F_2^l$ contributes to the lepton's spin magnetic moment 
as~\cite{Vanderhaeghen:2000ws} 
\begin{eqnarray}
\vec{\mu}^{\,l}_S =   % -
\frac{e\vec{S}}{2m_l}[1 + F_2^l(Q^2 = 0)]
=\frac{e\vec{S}}{2m_l}\!\left(1 +\frac{\alpha}{2\pi}\!\right) \,.\,\quad\,
\end{eqnarray} 
For $|Q^2|\gg m_l^2$, i.e., $\nu\to 1$, implies that $F^l_2\to 0$. Hence, $F^l_2$'s contribution to the unpolarized 
scattering cross section can safely be ignored relative to $F^l_1$ for the case of electron scatterings. However, 
for low-energy muon scattering, e.g., at MUSE kinematics, the Pauli form factor could give significant contributions.   

%\vspace{0.1cm}

%%%%%%%%%%%%%%%%%%%%%%%%%%%%%%%%%%%%%%%%%%%%%%%%%%%%%%
\noindent {\it 2. Proton-Photon Vertex Correction.\,} 
%%%%%%%%%%%%%%%%%%%%%%%%%%%%%%%%%%%%%%%%%%%%%%%%%%%%%%
In literature, the proton-photon vertex has often been modeled using phenomenological form factors, e.g., as done 
in Refs.~\cite{Mo:1968cg,Maximon:2000hm}. As discussed earlier, HB$\chi$PT allows a systematic order by order 
estimation of this vertex using the gauge invariant couplings of the photon with the hadrons involved. In general, 
one  parametrizes the proton-photon vertex in terms of the non-relativistic electric $G^p_E$ and magnetic $G^p_M$ 
Sachs form factors which are related to the standard relativistic Dirac and Pauli form factors {\it via} 
\begin{eqnarray}
G^p_E(Q^2)&=& F^p_1(Q^2)+\frac{Q^2}{4M^2} F^p_2(Q^2) \,,
\nonumber\\
G^p_M(Q^2)&=&  F^p_1(Q^2) +  F^p_2(Q^2)\,.
\end{eqnarray}
The matrix element of the electromagnetic quark current between proton states is given by
\begin{eqnarray}
\Big\langle \hspace{-0.5cm}&&{\rm p}(P^\prime)\Big|\bar{q}\gamma^\mu q\Big|{\rm p}(P)\Big\rangle
\\
&\rightarrow &  \chi^\dagger(p^\prime_p)\!\left[v^\mu G^p_E(Q^2)
+\frac{[S^\mu,S\cdot Q]}{M}G^p_M(Q^2)\right]\!\chi(p_p)\,,
\nonumber
\end{eqnarray} 
In our heavy baryon formulation when including ${\mathcal O}(\alpha)$ radiative corrections, only the proton's 
electric form factor $G^{p}_E$ is expected to contribute, while the magnetic form factor $G^{p}_M$ contributes 
at a higher chiral order. This is already apparent in our ${\mathcal O}({\cal Q}^2/M^2)$ chiral corrections to 
the LO Born cross section presented in Sec.~\ref{sec:formalism} (also see Ref.~\cite{Fearing1998} for details 
regarding $\chi$PT expressions of the proton form factors). As revealed in our analysis in the next section, 
even the inclusion of the chiral-radiative corrections of ${\mathcal O}(\alpha {\cal Q}/M)$ does not renormalize
the form factors. To this end, we write $G^{p}_E=G^{p(0)}_E +\delta G^{p(0)}_E$, where $\delta G^{p(0)}_E$ 
incorporates the radiative corrections to the electric form factor. Subsequently, the possible UV divergences, 
arising from the LO$_\alpha$ one-loop photon corrections to the vertex function, namely, $v^\mu G^{p(0)}_E$, are 
renormalized by adding the counter-term $(Z^{p}_1-1)v^\mu$, with the requirement that the total renormalized 
vertex function, ${\mathcal V}^\mu_p=v^\mu_l G^{p(0)}_E +(Z^{p}_1-1)v^\mu$, defines the physical proton charge 
at $Q^2=0$, i.e.,  ${\mathcal V}^\mu_p(Q^2=0)=v^\mu$. Similarly to the lepton counterpart, the proton the 
wave-function renormalization constant $Z^{p}_2$ can be defined as
\begin{eqnarray}
Z^{p(0)}_2&=& 1+\frac{\partial\Sigma^p(p_p)}{\partial(v\cdot p_p)}\Bigg|_{v\cdot p_p=0}
+{\mathcal O}\left(\alpha^2,\frac{1}{M}\right)
\nonumber\\
&\equiv& Z^{p(0)}_1=1-\delta G^{p(0)}_E(Q^2=0)\,, 
\nonumber
\end{eqnarray}
where
\begin{eqnarray}
\frac{\partial\Sigma^p(p_p)}{\partial(v\cdot p_p)}\Big|_{v\cdot p_p=\,0}=
-\delta G^{p(0)}_E(Q^2=0)\,.\,\,\quad\,\,
\end{eqnarray}	
With the on-shell conditions for the external protons, $v\cdot p_p=0$ and 
$v\cdot p_p^\prime=-\frac{(p^\prime_p)^2}{2M}+{\mathcal O}(M^{-2})$, the amplitude of the diagram 
VC$^{p(0)}$ in Fig.~\ref{1loop} is 
\begin{eqnarray}
\mathcal{M}^{p(0)}_{\gamma\gamma;\,{\rm vertex}} 
&=& \frac{ie^4}{Q^2}\,\Big[ {\bar u}_l(p^\prime)\gamma^\mu(p,p^\prime) u_l(p)\Big]
\nonumber\\
&&\times\,\int \frac{{\rm d}^4k}{(2\pi)^4}
\frac{\left[\chi^\dagger(p^\prime_p) v_\mu \chi(p_p)\right]}{(k^2+i0)(-v\cdot k+i0)^2}
\nonumber\\
&&\times\,\left(1-\frac{(p^\prime_p)^2}{2M (v\cdot k)}+\cdots\right)
\nonumber\\
&\equiv& \mathcal{M}^{(0)}_\gamma\delta G^{p(0)}_E \stackrel{\rm DR}{\longmapsto} 0\,.
\end{eqnarray}	
Here we used the fact that all scaleless loop integrals of the type
\begin{eqnarray}
{\mathcal I}(m,n)=\int \frac{{\rm d}^d k}{(2\pi)^d}\frac{(k^2)^m}{(-v\cdot k+i0)^n}
\label{Ipvertex}
\end{eqnarray}
vanish in DR (see, e.g., Ref.~\cite{itzykson2012quantum}). Consequently, there is no contribution to the 
proton VC at LO$_\alpha$. In fact, this result is intuitively anticipated from the fact that in HB$\chi$PT 
there is no proton bremsstrahlung at this  order~\cite{Talukdar:2018hia}. In other words, at LO$_\alpha$ 
the proton is static and unaffected by radiative correction. 

%\vspace{0.1cm}

%%%%%%%%%%%%%%%%%%%%%%%%%%%%%%%%%%%%%%%%%%%%%%%%%%%%%%
\noindent{\it 3. Vacuum Polarization.\,}
%%%%%%%%%%%%%%%%%%%%%%%%%%%%%%%%%%%%%%%%%%%%%%%%%%%%%% 
%
%%%%%%%%%%%%%%%%%%%%%%%%%%%%%%%%%%%%%%%FIGURE%%%%%%%%%%%%%%%%%%%%%%%%%%%%%%%%%%%%%%%%%%%%%%%	  
\begin{figure*}[tbp]
 \centering
	\includegraphics[scale=0.58]{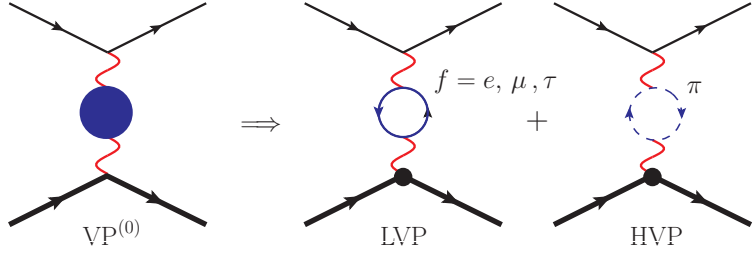}
        \caption{The one-loop vacuum polarization diagram at LO$_\alpha$ receives contributions 
                 from both leptonic (LVP) and hadronic (HVP) particle-antiparticle pairs.
                 We only consider the dominant HVP due to structureless pions.}
    \label{VP_blob}
\end{figure*}
%%%%%%%%%%%%%%%%%%%%%%%%%%%%%%%%%%%%%%%%%%%%%%%%%%%%%%%%%%%%%%%%%%%%%%%%%%%%%%%%%%%%%%%%%%%%
The one-loop VP contribution from diagram VP$^{(0)}$ in Fig.~\ref{1loop} at LO$_\alpha$ in HB$\chi$PT is 
IR-finite. However, it  contains a logarithmic UV divergence. Its unrenormalized amplitude in terms of 
the bare electric charge $e_0$ is given by
\begin{eqnarray}
{\mathcal M}^{(0)}_{\gamma\gamma;\,{\rm v.p.}}&=&\left[ {\bar u}_l(p^\prime)\,
e_0\gamma_\mu \,u_l(p)\right]\,D^{\mu\nu}(Q)
\nonumber\\
& &\times \left[\chi^\dagger(p^\prime_p) e_0 v_\nu \chi(p_p)\right]\,.
\end{eqnarray}
The full (interacting) photon propagator expressed in terms of the polarization tensor, 
$\Pi^{\mu\nu}=(Q^2g^{\mu\nu}-Q^{\mu}Q^{\nu})\,\Pi(Q^2)$, is
\begin{eqnarray}
iD^{\mu\nu}(Q)&=&\frac{-ig^{\mu\nu}}{Q^2}+iD^{\mu\rho}(Q)\,
i\Pi_{\rho\sigma}(Q)\,\left(\frac{-ig^{\sigma\nu}}{Q^2}\right)
\nonumber\\
&=&\frac{-ig^{\mu\nu}}{Q^2[1-\Pi(Q^2)]}+ \text{terms with\, }Q^\mu Q^\nu
\nonumber\\
&\simeq& \frac{-ig^{\mu\nu}}{Q^2\!\left[1\!-\!\Pi(0)\right]\!\left[1-\left(\Pi(Q^2)\!
-\!\Pi(0)\right)\right]}+\cdots\,.
\nonumber\\
\label{eq:PhotonD}
\end{eqnarray}
The UV divergence is as usual renormalized by adding the counter-term $-(Q^2g^{\mu\nu}-Q^{\mu}Q^{\nu})(Z_3-1)$ 
to $\Pi^{\mu\nu}$ which renormalizes the photon propagator:
\begin{eqnarray}
i\tilde{D} ^{\mu\nu}(Q)&=&\frac{-ig^{\mu\nu}}{Q^2[1-\Pi(Q^2)+(Z_3-1)]}+\cdots\,\,,\quad\,
\end{eqnarray}
where the ellipses denote the ``gauge terms" containing $Q_\mu Q_\nu$ which do not contribute in any gauge 
invariant result~\cite{Vanderhaeghen:2000ws}. The requirement that $\tilde{D}^{\mu\nu}$ has a pole at $Q^2=0$ 
with residue 1, yields $Z_3=1+\Pi(0)$, which renormalizes the bare QED coupling, $\alpha_0=\alpha/Z_3$, where 
$\alpha=e^2/(4\pi)\approx 1/137$ is the physical QED coupling. Finally, the renormalized amplitude factorizes 
into the LO Born amplitude as  
\begin{eqnarray}
\left[{\mathcal M}^{(0)}_{\gamma\gamma;\,{\rm v.p.}}\right]_{\rm ren}=\mathcal{M}^{(0)}_\gamma\Delta\Pi(Q^2)\,, 
\label{eq:vac_ren}
\end{eqnarray}
with the renormalized polarization function 
\begin{eqnarray}
\Delta\Pi(Q^2)&=&\Pi(Q^2)-\Pi(0)
\nonumber\\
&=&\Delta\Pi_{\rm lepton}(Q^2)+\Delta\Pi_{\rm hadron}(Q^2)\,,
\end{eqnarray}
receiving both leptonic $\Delta\Pi_{\rm lepton}$ and hadronic $\Delta\Pi_{\rm hadron}$ contributions at the 
one-loop level. Using DR one can readily obtain the well-known expression for the one-loop leptonic vacuum 
polarization (LVP) contribution~\cite{Vanderhaeghen:2000ws,Feynman:1949zx,Wheeler:1956}: 
\begin{eqnarray}
\Delta\Pi_{\rm lepton}(Q^2)
&=&\frac{\alpha}{2\pi}\!\!\sum_{f\,=\,e,\mu,\tau}
\left\{\frac{2}{3}\left(\nu^2_f-\frac{8}{3}\right)\right.
\nonumber\\
&&\!\!\!\left. +\,\nu_f\left(\frac{3-\nu^2_f}{3}\right)\ln\left[\frac{\nu_f+1}{\nu_f-1}\right]\right\}\,.\quad\,
\label{vac0_l}
\end{eqnarray}
Here $\nu_f=\sqrt{1-4m_f^2/Q^2}$, with index $f=e,\mu,\tau$ that is used to distinguish between the different 
lepton flavors contributing to the fermion loop. The hadronic vacuum polarization (HVP) contribution is 
illustrated in Fig.~\ref{VP_blob}. It only shows the contribution arising from structure-less, non-interacting 
pions in the loop. There is no unique method to determine the contributions for the HVP contributions and we 
consider a simplistic one-loop {\it estimate} of HVP that arises due to a $\pi^+\pi^-$ pair, evaluated using 
scalar QED.  In this regard, we quote the renormalized expression obtained by Tsai~\cite{Tsai:1960zz}: 
%%%%%%%%%%%%%%%%%%%%%%%%%%%%%%%%%%%%%%%%%FIGURE%%%%%%%%%%%%%%%%%%%%%%%%%%%%%%%%%%%%%%%%%%%%%
\begin{figure}[tbp]
 \centering
    \hspace{-0.6cm}
	\includegraphics[scale=0.53]{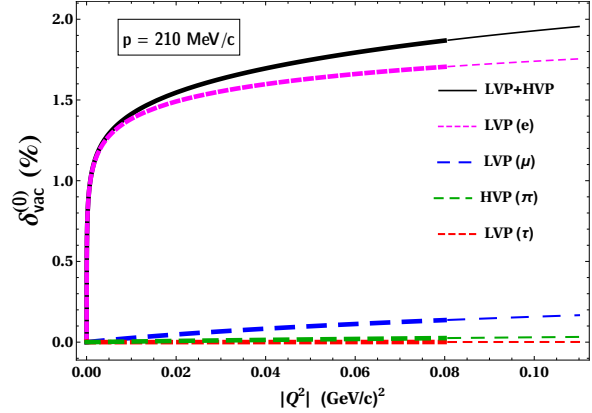}
	\caption{The one-loop leptonic and hadronic vacuum polarization corrections 
	         $\delta^{(0)}_{{\rm vac};\,e,\mu,\tau,\pi}$ and their sum $\delta^{(0)}_{\rm vac}$, 
	         contributing to the $\ell$-p elastic cross section at LO$_\alpha$, as a function 
	         of the squared four-momentum transfer $|Q^2|$. The plot covers the full kinematic 
	         scattering range, $0<|Q^2|<|Q^2_{\rm max}|$, when $\theta\in[0,\pi]$. The thickened 
	         section of each curve corresponds to the MUSE kinematic cut, where 
	         $\theta\in[20^\circ,100^\circ]$. }  
        \label{VP_delta}
\end{figure}
%%%%%%%%%%%%%%%%%%%%%%%%%%%%%%%%%%%%%%%%%%%%%%%%%%%%%%%%%%%%%%%%%%%%%%%%%%%%%%%%%%%%%%%%%%%%
\begin{eqnarray}
\Delta\Pi_{\rm hadron}(Q^2) &\to& \Delta\Pi_{\rm \pi^+\pi^-}(Q^2)
\nonumber \\
&&\hspace{-2cm}=\frac{\alpha}{2\pi}\left\{-\frac{2}{3}\left(\nu^2_\pi+\frac{1}{3}\right)
+\frac{\nu^3_\pi}{3}\ln\left[\frac{\nu_\pi+1}{\nu_\pi-1}\right]\right\},\,\quad\,\,
\label{vac0_h}
\end{eqnarray}
where $\displaystyle{\nu_\pi=\sqrt{1-4m_\pi^2/Q^2}}$. In Fig.~\ref{VP_delta} we display the 
${\mathcal O}(\alpha)$ fractional leptonic and pionic one-loop VP contributions at LO$_\alpha$, 
$\delta^{(0)}_{{\rm vac};\, f,\pi}=2\Delta\Pi_{f,\pi}(Q^2)$, with respect to the $\ell$-p elastic Born cross 
section, Eq.~\eqref{dsigmaBornL0}. The results are shown for the largest incoming momentum $p=210$ MeV/c for 
MUSE. We note that these corrections are  independent of the flavor of the incident lepton ($\ell=e,\mu$) and
the beam energy $E$. As expected, the $e^+e^-$ loop gives the dominant contribution due to the small electron 
mass and is an order larger than the other VP contributions combined. It amounts to about $1.7\%$ in the MUSE 
kinematic range. The $\mu^+\mu^-$, $\tau^+\tau^-$ and $\pi^+\pi^-$ pairs contribute about $0.15\%$, $0.002\%$ 
and $0.03\%$ respectively. Thus, the total UV finite VP contribution (i.e., LVP + HVP) at LO$_\alpha$ is   
\begin{eqnarray}
\delta^{(0)}_{\rm vac}=2\Delta\Pi (Q^2)=\sum_{f \,=\, e,\mu,\tau}\delta^{(0)}_{{\rm vac};f}
+\delta^{(0)}_{\rm vac;\,\pi}\,,
\label{delta0_vac}
\end{eqnarray}
which amounts to $\lesssim 2\%$ of the elastic Born differential cross section. 

%\vspace{0.2cm} 

\begin{figure*}[tbp]
 \centering
	\includegraphics[scale=0.53]{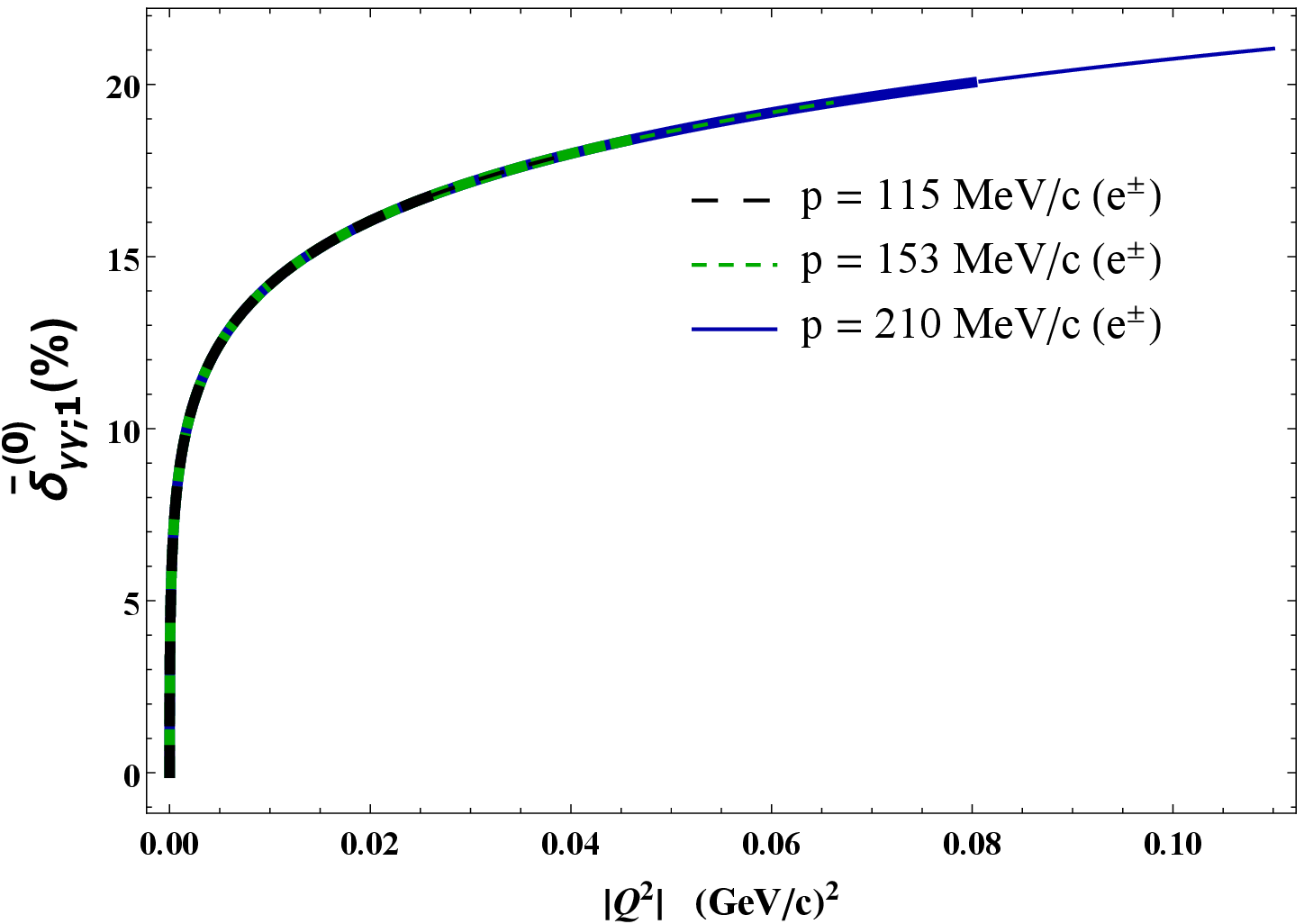} \qquad
	\includegraphics[scale=0.53]{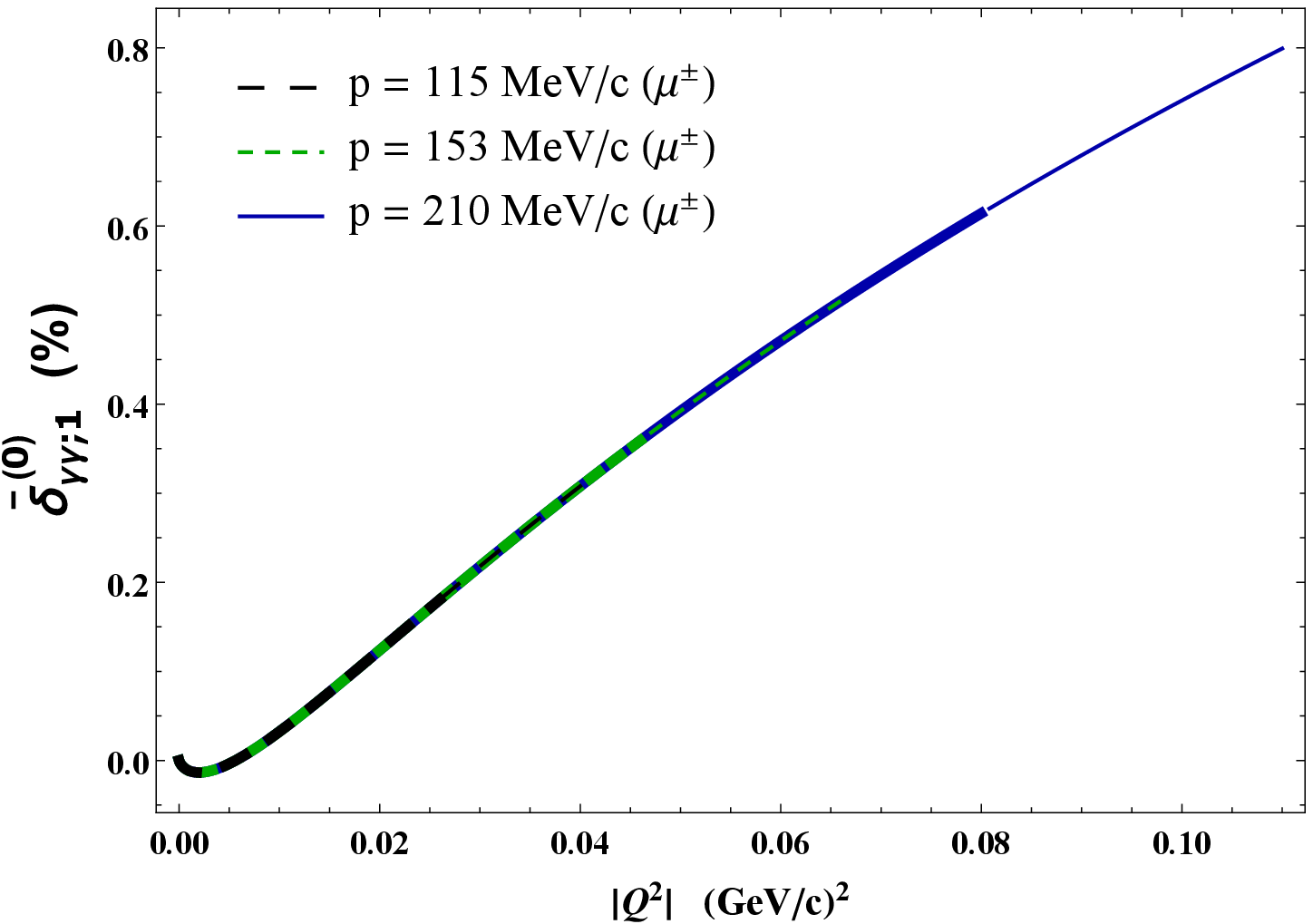}
	
	\vspace{1cm}
	
	\includegraphics[scale=0.53]{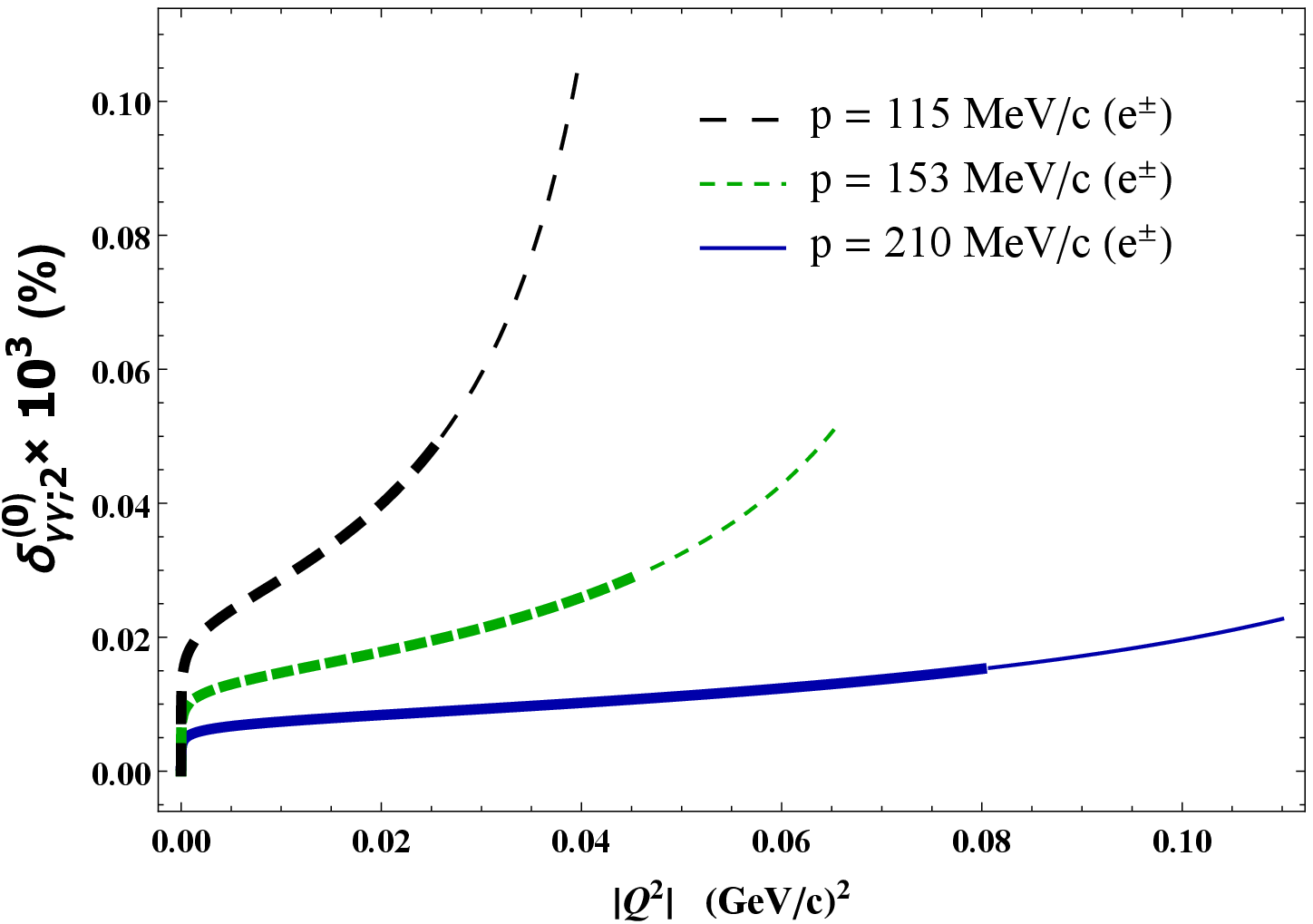} \qquad 
	\includegraphics[scale=0.53]{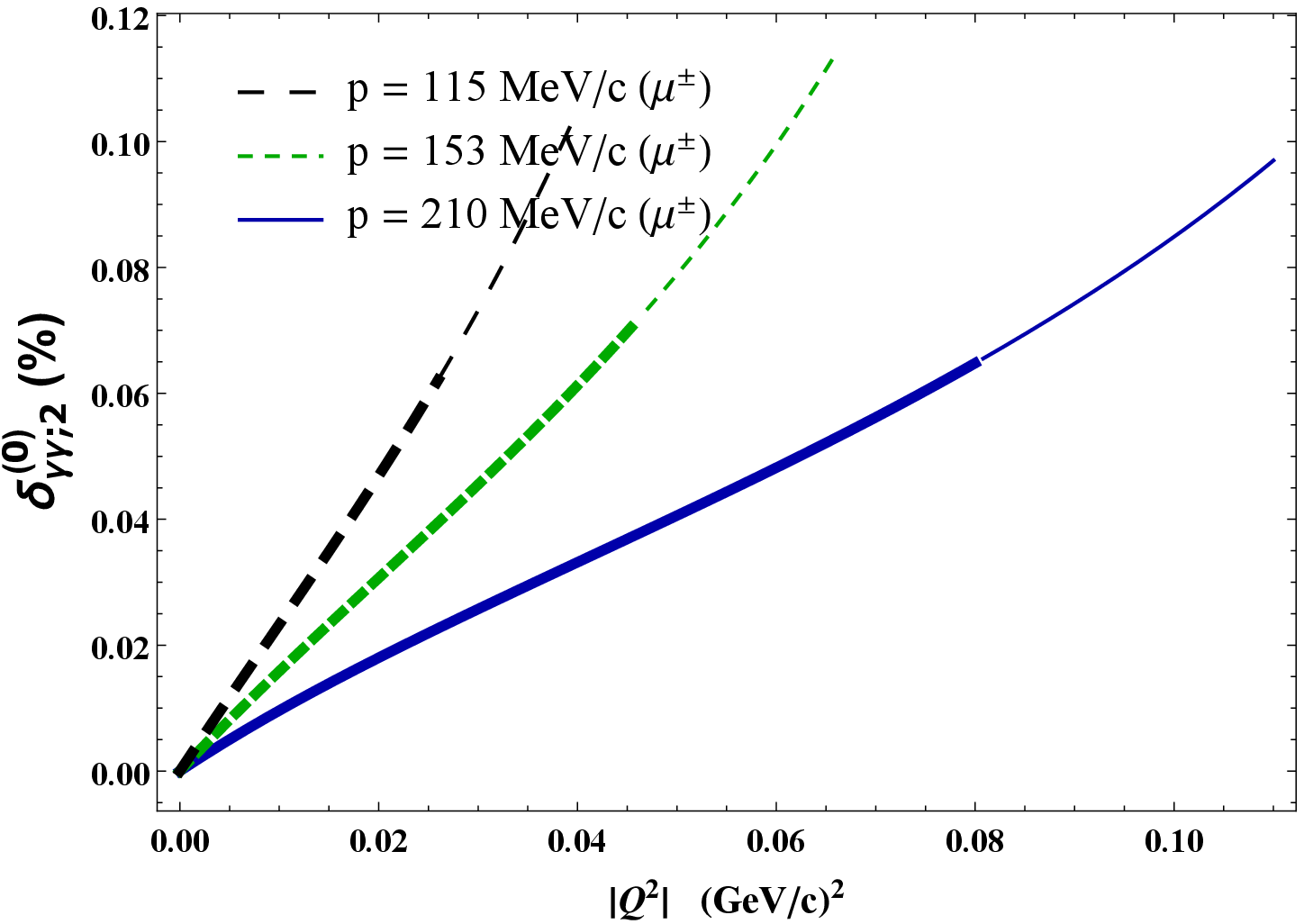}
	\caption{The one-loop LO$_\alpha$ contributions $\overline{\delta}^{(0)}_{\gamma\gamma;1}$ (upper panel) 
	         and $\delta^{(0)}_{\gamma\gamma;2}$ (lower panel) to the $e$-p (left panel) and $\mu$-p (right 
	         panel) elastic cross sections (in percentage) from the finite lepton-photon vertex corrections 
	         terms containing the (UV and IR-finite)  form factors $F^l_{1;{\rm ren}}$ and $F^l_{2}$. Each 
	         plot covers the full kinematic scattering range, $0<|Q^2|<|Q^2_{\rm max}|$, when 
	         $\theta\in[0,\pi]$ at the incoming lepton momenta, $|\vec p\,|=p=115,\, 153,\, 210$ MeV/c. The 
	         thickened section of each curve corresponds to the MUSE kinematic cut, where 
	         $\theta\in[20^\circ,100^\circ]$. }
	\label{plot:rad_f1f2}
\end{figure*}

%%%%%%%%%%%%%%%%%%%%%%%%%%%%%%%%%%%%%%%%%%%%%%%%%%%%%%%%%%%%%%
\noindent{\it 4. Complete one-loop Virtual Contribution\,}  
%%%%%%%%%%%%%%%%%%%%%%%%%%%%%%%%%%%%%%%%%%%%%%%%%%%%%%%%%%%%%%
Adding all the non-vanishing renormalized virtual contributions from the one-loop diagrams of Fig.~\ref{1loop} 
yield the total UV-finite elastic scattering amplitude at LO$_\alpha$:  
\begin{eqnarray}
\mathcal{M}^{(0)}_{\gamma\gamma}&=&\mathcal{M}^{(0)}_{\gamma}
+\left[\mathcal{M}^{l (0)}_{\gamma\gamma;\,{\rm vertex}}\right]_{\rm ren}
+\left[{\mathcal M}^{(0)}_{\gamma\gamma;\,{\rm v.p.}}\right]_{\rm ren}
\nonumber\\
&=&\mathcal{M}^{(0)}_\gamma + \mathcal{M}^{(0)}_\gamma \left[F^{l;\, {\rm ren}}_1(Q^2)-1 +\Delta\Pi(Q^2)\right]  
\nonumber\\
&& +\,\overline{\mathcal{M}}^{(0)}_\gamma F^l_2(Q^2)\,.\quad\,
\label{M0gamgam}
\end{eqnarray} 
Here $F^{l;\,{\rm ren}}_1$ is the UV renormalized one-loop leptonic Dirac form factor given in 
Eq.~\eqref{Fl1ren}, and $F^l_2$ is the one-loop {\it finite} leptonic Pauli form factor given in 
Eq.~\eqref{Fl1Fl2unren}. The renormalized VP corrections $\Delta\Pi$ are obtained from Eqs.~\eqref{vac0_l} 
and \eqref{vac0_h}, and  the amplitude $\overline{\mathcal{M}}^{(0)}_\gamma$ is given in 
Eq.~\eqref{Ml0gamgam-vertex}. The IR divergences arising from the photon loops are contained in the factor
multiplying the Born amplitude ${\mathcal{M}}^{(0)}_\gamma$. The {\it lab.}-frame LO$_\alpha$ radiative 
correction to the elastic differential cross section becomes: 
%%%%%%%%%%%%%%%%%%%%%%%%%%%%%%%%%%%%%%%%%%%%%%%%%%%%
\begin{widetext}
\begin{eqnarray}
\Delta\left[\frac{{\rm d}\sigma^{\rm (LO)}_{el}(Q^2)}{{\rm d}\Omega^\prime_l}\right]_{\gamma\gamma}
=\left[\frac{{\rm d}\sigma_{el}(Q^2)}{{\rm d}\Omega^\prime_l}\right]_{0}\delta^{(0)}_{\gamma\gamma}(Q^2)\,,\qquad
\end{eqnarray}
with
\begin{eqnarray}
&&\delta^{(0)}_{\gamma\gamma}(Q^2)= \frac{2{\mathcal R}\text{e}\sum\limits_{spins}\left(\mathcal{M}_\gamma^{(0)\dagger}
\mathcal{M}^{(0)}_{\gamma\gamma}\right)}{\sum\limits_{spins} |\mathcal{M}^{(0)}_\gamma|^2}-2
=\,\text{\bf IR}^{(0)}_{\gamma\gamma}(Q^2)+\overline{\delta}^{(0)}_{\gamma\gamma}(Q^2)\,, 
\label{eq:IR+finite}
\end{eqnarray}
\end{widetext}
%%%%%%%%%%%%%%%%%%%%%%%%%%%%%%%%%%%%%%%%%%%%%%%%%%%%
representing the ${\mathcal O}(\alpha)$ fractional contribution from the virtual photon-loops at LO$_\alpha$. 
The IR-divergent part is contained in $\text{\bf IR}^{(0)}_{\gamma\gamma}(Q^2)$. The corresponding finite part 
$\overline{\delta}^{(0)}_{\gamma\gamma}$ includes the LO$_\alpha$ lepton-photon VC contributions 
$\overline{\delta}^{(0)}_{\gamma\gamma;1,2}$, as extracted from Eq.~\eqref{M0gamgam}, namely, the contribution 
from the lepton Dirac form factor, 
\begin{equation}
\overline{\delta}^{(0)}_{\gamma\gamma;1}(Q^2)=
2\left[F^{l;\, {\rm ren}}_1(Q^2)-1\right]-\text{\bf IR}^{(0)}_{\gamma\gamma}(Q^2)\,,
\label{Eq:rad_electric}
\end{equation}
and that from the lepton Pauli form factor, 
%%%%%%%%%%%%%%%%%%%%%%%%%%%%%%%%%%%%%%%%%%%%%%%%%%%%
\begin{widetext}
\begin{eqnarray}
\delta^{(0)}_{\gamma\gamma;2}(Q^2)
&=&\frac{2{\mathcal R}\text{e}\sum\limits_{spins}\left(\mathcal{M}^{(0)\dagger}_\gamma\,\,
\overline{\mathcal{M}}^{(0)}_{\gamma}\right)}{\sum\limits_{spins} |\mathcal{M}^{(0)}_\gamma|^2}F_2^l(Q^2)
= -\left(\frac{2\eta Q^2}{\eta Q^2+4E^2}\right)\left(1-\frac{Q^2}{4M^2}\right)F^l_2(Q^2)
\nonumber\\ 
&=&\frac{\alpha}{\pi\nu}\left(\frac{2\eta m^2_l}{\eta Q^2+4E^2}\right)
\ln\left[\frac{\nu+1}{\nu-1}\right]+{\mathcal O}\left(\alpha\frac{{\cal Q}^2}{M^2}\right)\,.
\label{Eq:rad_magnetic}
\end{eqnarray} 
where the above $1/M^2$ order term is dropped from our central result presented below. 
However, these $1/M^2$ order terms will be included as a part of the theoretical error estimate. The finite 
part includes the total LO$_\alpha$ VP contribution, $\delta^{(0)}_{\rm vac}=2\Delta\Pi (Q^2)$, obtained 
earlier in this section. Thus, the total finite fractional virtual radiative corrections at LO$_\alpha$ 
reads 
\begin{eqnarray}
\overline{\delta}^{(0)}_{\gamma\gamma}(Q^2)&=&\overline{\delta}^{(0)}_{\gamma\gamma;1}(Q^2)
+\delta^{(0)}_{\gamma\gamma;2}(Q^2)+\sum_{f=e,\mu,\tau}\delta^{(0)}_{{\rm vac};f}(Q^2)
+\delta^{(0)}_{\rm vac;\,\pi}(Q^2)
\nonumber\\
&=&\frac{\alpha}{\pi}\Bigg[\frac{\nu^2+1}{4\nu}
\ln\left[\frac{\nu+1}{\nu-1}\right]\ln\left[\frac{\nu^2-1}{4\nu^2}\right]
+\ln\left(\frac{-Q^2}{m^2_l}\right)\left[\frac{\nu^2+1}{2\nu}\ln\left[\frac{\nu+1}{\nu-1}\right]-1\right]
+\frac{2\nu^2+1}{2\nu}\ln\left[\frac{\nu+1}{\nu-1}\right]
\nonumber\\
&&-2-\frac{\nu^2+1}{2\nu}\left\{\text{Sp}\left(\frac{\nu+1}{2\nu}\right)
-\text{Sp}\left(\frac{\nu-1}{2\nu}\right)\right\}
+\sum_{f=e,\mu,\tau}\left\{\frac{2}{3}\left(\nu^2_f-\frac{8}{3}\right)
+\nu_f\left(\frac{3-\nu^2_f}{3}\right)\ln\left[\frac{\nu_f+1}{\nu_f-1}\right]\right\}
\nonumber\\
&&\hspace{0.3cm}-\,\frac{2}{3}\left(\nu^2_\pi+\frac{1}{3}\right)
+\frac{\nu^3_\pi}{3}\ln\left[\frac{\nu_\pi+1}{\nu_\pi-1}\right]
+\frac{1}{\nu}\left(\frac{2\eta m^2_l}{\eta Q^2+4E^2}\right)
\ln\left[\frac{\nu+1}{\nu-1}\right]\Bigg]+{\mathcal O}\left(\alpha\frac{{\cal Q}^2}{M^2}\right)\,.
\label{delta:LO_vertex}
\end{eqnarray}
The IR-divergent part of Eqs.~\eqref{eq:IR+finite} and \eqref{Eq:rad_electric}, 
$\text{\bf IR}^{(0)}_{\gamma\gamma}(Q^2)$, which essentially stems from the ``Dirac" contribution to the 
one-loop lepton-photon VC at LO$_\alpha$, Eq.~\eqref{Fl1ren}, is given as
\begin{eqnarray}
\text{\bf IR}^{(0)}_{\gamma\gamma}(Q^2) & \equiv & \text{\bf IR}^{l(0)}_{\gamma\gamma;\,{\rm vertex}}(Q^2)
= -\,\frac{\alpha}{\pi}\left[\frac{1}{|\epsilon_{\rm IR}|}
+\gamma_E-\ln\left(\frac{4\pi\mu^2}{-Q^2}\right)\right]
\,\left[\frac{\nu^2+1}{2\nu}\ln\left[\frac{\nu+1}{\nu-1}\right]-1\right] \,.
\label{IR0_vert}
\end{eqnarray}
\end{widetext}
%%%%%%%%%%%%%%%%%%%%%%%%%%%%%%%%%%%%%%%%%%%%%%%%%%%%
In Sec. III B we show that this IR divergence is cancelled by the IR-divergence from the 
soft-bremsstrahlung process at LO$_\alpha$. Figure~\ref{plot:rad_f1f2} displays the LO$_\alpha$ 
fractional contributions from the lepton-photon VC terms, $\overline{\delta}^{(0)}_{\gamma\gamma;1}$ 
and $\delta^{(0)}_{\gamma\gamma;2}$, stemming from the form factors $F^{l;{\rm ren}}_1$  and $F^{l}_2$ 
respectively, for the full kinematic elastic scattering range $0<|Q^2|<|Q^2_{\rm max}|$ of the MUSE 
specified incoming lepton momenta. A summary of our observations are in order: 
\begin{itemize}
\item All the radiative corrections vanish in the limit $Q^2\rightarrow 0$, as dictated by gauge 
invariance.
\item There is no LO$_\alpha$ contribution from the TPE box diagrams, even without invoking SPA. 
\item The lepton Dirac form factor contribution is basically independent of the lepton beam energy. 
\item In contrast, the lepton Pauli form factor contribution depends strongly on the beam momentum. 
For electron scattering
in the MUSE momentum range, the contributions are practically negligible, $\sim\! 10^{-5}\%$. 
However, for muon scattering the relative corrections turn out to be much larger $\sim\! 10^{-1}\%$. 
\item The electronic and muonic Dirac form factor contributions in the region of low momentum 
transfers, $|Q^2|<0.1$~(GeV/c)$^2$, differ by almost two orders of magnitudes. The reason is that 
the electronic Dirac term $\overline{\delta}^{l(0)}_{\gamma\gamma;1}$ is enhanced in the {\it soft} 
and {\it collinear} region of the loop-momentum integration resulting from the so-called 
{\it Sudakov} double-logarithms, namely, 
\begin{eqnarray*}
\textcolor{white}{a}\hspace{0.7cm}\frac{\nu^2+1}{4\nu}\ln\left[\frac{\nu+1}{\nu-1}\right]&&
\ln\left[\frac{\nu^2-1}{4\nu^2}\right]
\\
&&\hspace{-2.7cm}+\frac{\nu^2+1}{2\nu}\ln\left(\frac{-Q^2}{m^2_l}\right)\!
\ln\left[\frac{\nu+1}{\nu-1}\right]\!\approx \!\frac{1}{2}\!\ln^2\!\left(\frac{-Q^2}{m^2_l}\right)\!,
\end{eqnarray*}
in the limit of a small lepton mass (i.e.,  $m_l^2\ll |Q^2|$). However, for muon scattering at MUSE 
kinematics where $m_\mu^2=0.01~{\rm GeV}^2\approx |Q^2|$, no such enhancements is manifest. 
\end{itemize}
%%%%%%%%%%%%%%%%%%%%%%%%%%%%%%%%%%%%%%%%%FIGURE%%%%%%%%%%%%%%%%%%%%%%%%%%%%%%%%%%%%%%%%%%%%%
%%%%%%%%%%%%%%%%%%%%%%%%%%%%%%%%%%%%%%%%%%%%%%%%%%%%%%%%%%%%%%%%%%%%%%%%%%%%%%%%%%%%%%%%%%%% 

%%%%%%%%%%%%%%%%%%%%%%%%%%%%%%%%%%%%%%%%%%%%%%%%%%%%%%
\subsection{Soft Bremsstrahlung Corrections at LO$_\alpha$}  
\begin{figure*}[tbp]
 \centering
	\includegraphics[scale=0.58]{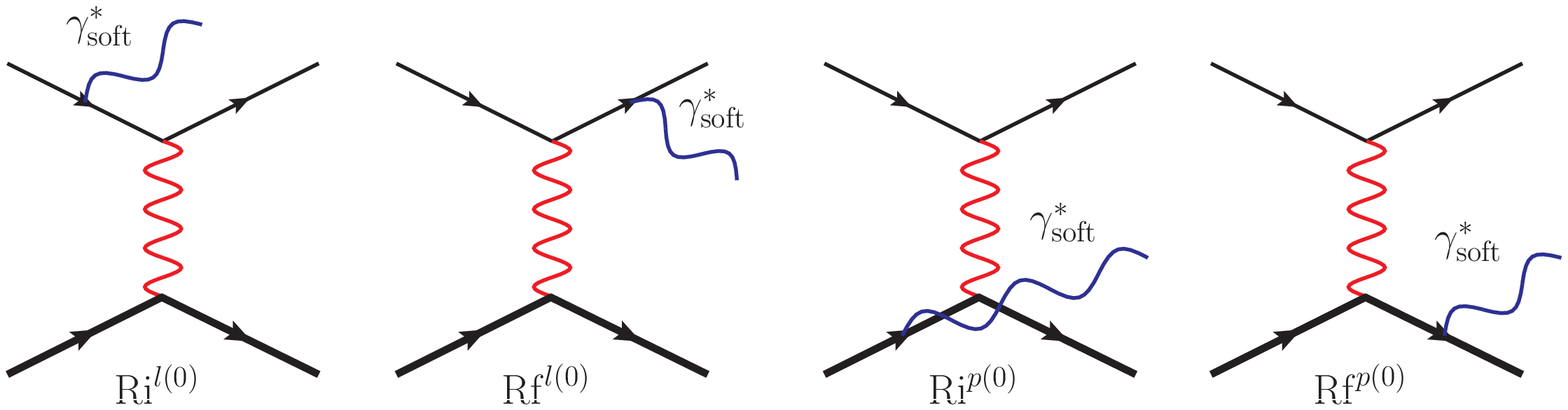}
	\caption{Soft bremsstrahlung diagrams at LO${_\alpha}$ [i.e., ${\mathcal O}(e^3)$] in HB$\chi$PT
	         contributing to the radiative corrections to the elastic leading order (Born) $\ell$-p 
	         scattering amplitude [see Eq~.\eqref{M0_gamma}]. The proton radiating diagrams vanish.}
	\label{lobrem}
\end{figure*}

%%%%%%%%%%%%%%%%%%%%%%%%%%%%%%%%%%%%%%%%%%%%%%%%%%%%%%
A review of known results using standard field theoretical techniques can be found in, e.g., 
Refs.~\cite{Tsai:1961zz,Mo:1968cg,Maximon:2000hm,Vanderhaeghen:2000ws,Bucoveanu:2018soy}).  We re-evaluated 
the bremsstrahlung process, $\ell {\rm p}\to \ell {\rm p} \gamma^*$, in our HB$\chi$PT work of 
Ref.~\cite{Talukdar:2018hia}. By virtue of tranversality of real photons with polarization four-vector 
$\varepsilon_\mu$, namely, $k\cdot \varepsilon=0$, the  Coulomb gauge condition, $v\cdot \varepsilon=0$, 
is naturally satisfied for the bremsstrahlung process. Consequently, with the LO$_\alpha$ proton-photon vertex 
in heavy baryon formalism proportional to $v\cdot \varepsilon^*$, the ``static" proton does not radiate at the 
leading chiral order. Therefore, in this case the lowest order soft bremsstrahlung process consists of a single
soft photon that is emitted from either the incoming lepton, diagram Ri$^{l(0)}$, or the outgoing one, diagram 
Rf$^{\,l(0)}$, as displayed in Fig.~\ref{lobrem}, with amplitudes given as 
%%%%%%%%%%%%%%%%%%%%%%%%%%%%%%%%%%%%%%%%%%%%%%%%%%%%
\begin{widetext}
\begin{eqnarray}
\mathcal{M}^{l(0);\,{\rm i}}_{\gamma\gamma^*} &=& 
-e^3\int \frac{{\rm d}^4k}{(2\pi)^4}\left[{\bar u}_l(p^{\prime\prime}) \, \gamma^\mu\,
\frac{(\slashed{p}-\slashed{k}+m)}{(p-k)^2-m^2_l} \,\slashed{\varepsilon}^*\, u_l(p)\right] 
\frac{1}{(Q-k)^2}\left[\chi^\dagger(p_p^{\prime}) \,v_\mu\, \chi(p_p)\right]\,, \quad {\rm and}
\nonumber\\
\nonumber\\
\mathcal{M}^{l(0);\,{\rm f}}_{\gamma\gamma^*} &=& 
-e^3\int \frac{{\rm d}^4k}{(2\pi)^4}\left[{\bar u}_l(p^{\prime\prime}) \, \slashed{\varepsilon}^*\,
\frac{(\slashed{p}^{\prime\prime}+\slashed{k}+m)}{(p^{\prime\prime}+k)^2-m^2_l}\,\gamma^\mu\,  u_l(p)\right] 
\frac{1}{(Q-k)^2}\left[\chi^\dagger(p_p^{\prime}) \,v_\mu\, \chi(p_p)\right]\,.\quad\,
\end{eqnarray}
\end{widetext} 
%%%%%%%%%%%%%%%%%%%%%%%%%%%%%%%%%%%%%%%%%%%%%%%%%%%%
In the above expressions, $k_\mu$ is the four-momentum of the bremsstrahlung photon and $p^{\prime\prime}_\mu$ 
is the four-momentum of the inelastically scattered outgoing lepton. In this paper we are  only concerned with 
the undetectable soft photon emissions. Thus, by the YFS methodology~\cite{Yennie:1961ad} the real photon 
emission amplitudes are evaluated using SPA, where SPA regards $p^{\prime\prime}_\mu$ as the physical 
four-momentum of the {\it elastically} scattered lepton. Henceforth, for reasons of brevity we drop all 
distinctions between the $p^\prime_\mu$ and $p^{\prime\prime}_\mu$, unless explicitly mentioned. In SPA the 
photon momentum of the propagator numerator is taken to be zero, i.e., $k_\mu \to 0$, with the crucial assumption 
that the soft emissions does not alter the elastic kinematics. In other words, in this limit the four-momentum 
transfer for the bremsstrahlung process, $q_\mu=(Q-k)_\mu$ is practically indistinguishable from its elastic 
counterpart, $Q_\mu=(p-p^\prime)_\mu=(P^\prime-P)_\mu$. Then, the matrix elements get factorized into the LO 
(Born) amplitude $\mathcal{M}^{(0)}_\gamma$, namely,
\begin{eqnarray*}
\mathcal{M}^{l(0);\,{\rm i}}_{\gamma\gamma^*} \,\,\stackrel{\gamma^*_{\rm soft}}{\leadsto}\,\,
\widetilde{\mathcal{M}^{l(0);\,{\rm i}}_{\gamma\gamma^*}} 
= e\mathcal{M}^{(0)}_\gamma\left(\frac{p\cdot\varepsilon^*} {p\cdot k}\right)\,,
\end{eqnarray*}
\begin{eqnarray}
\mathcal{M}^{l(0);\,{\rm f}}_{\gamma\gamma^*}\,\,\stackrel{\gamma^*_{\rm soft}}{\leadsto}\,\,
\widetilde{\mathcal{M}^{l(0);\,{\rm f}}_{\gamma\gamma^*}} 
= -e\mathcal{M}^{(0)}_\gamma\left(\frac{p^\prime\cdot\varepsilon^*} {p^\prime\cdot k}\right)\,.
\end{eqnarray}
Taking the square of the total LO$_\alpha$ bremsstrahlung matrix element, 
$\mathcal{M}_{\gamma\gamma^*}^{(0)}=\mathcal{M}^{l(0);\,{\rm i}}_{\gamma\gamma^*}
+\mathcal{M}^{l(0);\,{\rm f}}_{\gamma\gamma^*}$, in SPA yields a cross section in accordance with the 
well-known  Low's soft photon theorem~\cite{Low:1958sn}. This implies that in terms of the bremsstrahlung 
soft photon energy  the first two terms in the expansion of the unpolarized radiative cross section depend
only on the corresponding non-radiative unpolarized cross section. Thus, the {\it lab.}-frame differential
cross cross section for the LO$_\alpha$ bremsstrahlung process is given by the expression
%%%%%%%%%%%%%%%%%%%%%%%%%%%%%%%%%%%%%%FIGURE%%%%%%%%%%%%%%%%%%%%%%%%%%%%%%%%%%%%%%%%%%%%%%%%
%%%%%%%%%%%%%%%%%%%%%%%%%%%%%%%%%%%%%%%%%%%%%%%%%%%%%%%%%%%%%%%%%%%%%%%%%%%%%%%%%%%%%%%%%%%%
%%%%%%%%%%%%%%%%%%%%%%%%%%%%%%%%%%%%%%%%%%%%%%%%%%%%
\begin{widetext}
\begin{eqnarray}
\left[{\rm d}\sigma^{\rm (LO_\alpha)}_{br}\right]_{\gamma\gamma^*} = 
\frac{{\rm d}^3\vec{p}^{\,\prime}}{(2\pi)^3 2E^\prime}
\frac{{\rm d}^3\vec{P}^{\prime}}{(2\pi)^3 2E^\prime_p}
\frac{{\rm d}^3\vec{k}}{(2\pi)^3 2E_\gamma}\,
\frac{(2\pi)^4\delta^4\left(p+P-p^\prime-P^\prime-k\right)}{4ME}\,
\frac{1}{4}\sum_{spins}|\mathcal{M}^{(0)}_{\gamma\gamma^*}|^2\,,
\label{eq:dsigma_brem}
\end{eqnarray}
where
\begin{eqnarray}
\sum_{spin}|\mathcal{M}_{\gamma\gamma^*}^{(0)}|^2 \equiv |{\mathcal{M}^{l(0);\,{\rm i}}_{\gamma\gamma^*}} 
+{\mathcal{M}^{l(0);\,{\rm f}}_{\gamma\gamma^*}}|^2
&\stackrel{\gamma^*_{\rm soft}}{\leadsto}&\,\,|\widetilde{\mathcal{M}^{l(0);\,{\rm i}}_{\gamma\gamma^*}} 
+\widetilde{\mathcal{M}^{l(0);\,{\rm f}}_{\gamma\gamma^*}}|^2
\nonumber\\
&=&-\,e^2\!\sum_{spins}|\mathcal{M}^{(0)}_\gamma|^2 \Bigg(\frac{m^2_l}{(p\cdot k)^2}
+\frac{m^2_l}{(p^\prime\cdot k)^2}-\frac{2 p^\prime\cdot p}{(p\cdot k)(p^\prime\cdot k)}\Bigg)\,.
\label{M0gamgam*2}
\end{eqnarray}
\end{widetext}
%%%%%%%%%%%%%%%%%%%%%%%%%%%%%%%%%%%%%%%%%%%%%%%%%%%%%%%%%%%%%
The phase-space integrated cross section also factorizes into the Born cross section, Eq.~\eqref{dsigmaBornL0}, 
and reads:  
\begin{eqnarray}
\Delta\left[\frac{{\rm d}\sigma^{\rm (LO)}_{br}(Q^2)}{{\rm d}\Omega^\prime_l}\right]_{\gamma\gamma^*} \,\,
&\stackrel{\gamma^*_{\rm soft}}{\leadsto}& \,\, 
\frac{\alpha}{2\pi^2 }\left[\frac{{\rm d}\sigma_{el}(Q^2)}{{\rm d}\Omega^\prime_l}\right]_{0}
\nonumber\\
&&\times\, \left(- L_{\rm ii}-L_{\rm ff}+ L_{\rm if}\right).\,\quad\,\,\,
\label{LO:brem_cross}
\end{eqnarray} 
The integrals, $L_{\rm ii},\,L_{\rm ff}$ and $L_{\rm if}$, are three-momentum integrals involving the 
soft photons radiated by the leptons. They are evaluated in Appendix B using the method of, e.g.,  
Refs.~\cite{Tsai:1961zz,Mo:1968cg,Maximon:2000hm,Vanderhaeghen:2000ws,Bucoveanu:2018soy}. The integrals, 
$L_{\rm ii},\,L_{\rm ff}$ and $L_{\rm if}$ are all IR-divergent. As demonstrated in the Appendix B, we 
isolate the corresponding finite contributions, $\widetilde{L}_{\rm ii},\,\widetilde{L}_{\rm ff}$ and 
$\widetilde{L}_{\rm if}$, using DR. This yields the {\it lab.}-frame LO$_\alpha$ bremsstrahlung 
correction to the elastic differential cross section with all possible soft photon emissions with energies
less then $\Delta_{\gamma^*}$:  
\begin{eqnarray}
\Delta\!\!\left[\frac{{\rm d}\sigma^{\rm (LO)}_{br}(Q^2)}{{\rm d}
\Omega^\prime_l}\right]^{(E_{\gamma^*}<\Delta_{\gamma^*})}_{\gamma\gamma^*} 
\hspace{-0.5cm}&=&\left[\frac{{\rm d}\sigma_{el}(Q^2)}{{\rm d}
\Omega^\prime_l}\right]_{0}\!\!\!\delta^{(0)}_{\gamma\gamma^*}(Q^2),\,\quad\,\,
\end{eqnarray}
where the ${\mathcal O}(\alpha)$ fractional bremsstrahlung contribution $\delta^{(0)}_{\gamma\gamma^*}$ 
at LO$_\alpha$ reads~\cite{Maximon:2000hm,Vanderhaeghen:2000ws}
\begin{eqnarray}
\delta^{(0)}_{\gamma\gamma^*}(Q^2)=\text{\bf IR}^{(0)}_{\gamma\gamma^*}(Q^2)
+\overline{\delta}^{(0)}_{\gamma\gamma^*}(Q^2)\,,
\end{eqnarray}
with
%%%%%%%%%%%%%%%%%%%%%%%%%%%%%%%%%%%%%%%%%%%%%%%%%%%%
\begin{widetext}
\begin{eqnarray}
\text{\bf IR}^{(0)}_{\gamma\gamma^*}(Q^2) &\equiv& \text{\bf IR}^{l(0)}_{\gamma\gamma^*}(Q^2)
= \frac{\alpha}{\pi}\left[\frac{1}{|\epsilon_{\rm IR}|}+\gamma_E-\ln\left(\frac{4\pi\mu^2}{-Q^2}\right)\right]
\left[\frac{\nu^2+1}{2\nu}\ln\left[\frac{\nu+1}{\nu-1}\right]-1\right]\,,
\label{IR0_brem}
\end{eqnarray}
being the IR-divergent term, and the finite part of the LO$_\alpha$ bremsstrahlung contribution is 
represented by
\begin{eqnarray}
\overline{\delta}^{(0)}_{\gamma\gamma^*}(Q^2)&=&
\frac{\alpha}{\pi}\left(-\widetilde{L}_{\rm ii}-\widetilde{L}_{\rm ff}+\widetilde{L}_{\rm if}\right)
\nonumber\\
&=&\frac{\alpha}{\pi}\Bigg[\ln\left(\frac{4\eta^2\Delta^{2}_{\gamma^*}}{-Q^2}\right)
\left[\frac{\nu^2+1}{2\nu}\ln\left[\frac{\nu+1}{\nu-1}\right]-1\right]
+\frac{1}{4\beta}\ln\sqrt{\frac{1+\beta}{1-\beta}}
+\frac{1}{4\beta^\prime}\ln\sqrt{\frac{1+\beta^\prime}{1-\beta^\prime}}
\nonumber\\
&&\,\,\quad-\,\frac{\nu^2+1}{2\nu}\left\{\ln^2\sqrt{\frac{1+\beta}{1-\beta}}
-\ln^2\sqrt{\frac{1+\beta^\prime}{1-\beta^\prime}}
+\text{Sp}\left(1-\frac{\lambda_\nu-\eta}{(1-\beta^\prime)\xi_\nu}\right)
+\text{Sp}\left(1-\frac{\lambda_\nu-\eta}{(1+\beta^\prime)\xi_\nu}\right)\right.
\nonumber\\
&&\hspace{2.1cm} \left.-\,\text{Sp}\left(1-\frac{\lambda_\nu-\eta}{(1-\beta)\eta\lambda_\nu\xi_\nu}\right)
-\text{Sp}\left(1-\frac{\lambda_\nu -\eta}{(1+\beta)\eta\lambda_\nu\xi_\nu}\right)\right\}\Bigg]\,,
\label{delta:LO_brem}
\end{eqnarray}
\end{widetext}
%%%%%%%%%%%%%%%%%%%%%%%%%%%%%%%%%%%%%%%%%%%%%%%%%%%%%%%%%%%%%
where $\nu$ is defined below Eq.~\eqref{Fl1Fl2unren}, and  
$\xi_\nu=\frac{2\nu}{(\nu+1)(\nu-1)}$ and $\lambda_\nu=\frac{3\nu-1}{\nu-1}$, are $Q^2$ dependent kinematic 
variables. The IR-divergent term $\text{\bf IR}^{(0)}_{\gamma\gamma^*}$ from the LO$_\alpha$ bremsstrahlung 
diagrams, being equal and opposite to the LO$_\alpha$ one-loop IR-divergent counterpart 
$\text{\bf IR}^{(0)}_{\gamma\gamma}$ [cf. Eq.~\eqref{IR0_vert}], exactly cancels out in the sum of the 
LO$_\alpha$ real and virtual radiative contributions. Thus, the resulting finite contribution is 
\begin{eqnarray}
\delta^{(0)}_{2\gamma}(Q^2)&=&\delta^{(0)}_{\gamma\gamma}(Q^2)
+\delta^{(0)}_{\gamma\gamma^*}(Q^2)
\nonumber\\
&\equiv& \overline{\delta}^{(0)}_{\gamma\gamma}(Q^2)+\overline{\delta}^{(0)}_{\gamma\gamma^*}(Q^2)\,.
\end{eqnarray}

\begin{widetext}
%%%%%%%%%%%%%%%%%%%%%%%%%%%%%%%%%%%%%%%%%%%%%%%%%%%%%%%%%%%%%%%
\subsection{Total Radiative Corrections at LO$_\alpha$} 
%%%%%%%%%%%%%%%%%%%%%%%%%%%%%%%%%%%%%%%%%%%%%%%%%%%%%%%%%%%%%%% 
After eliminating all the UV and IR divergences we obtain the desired analytical result for the complete 
radiative contributions at LO$_\alpha$ in HB$\chi$PT. The finite one-loop ${\mathcal O}(\alpha)$ expression 
for the LO$_\alpha$ fractional radiative corrections to the $\ell$-p elastic differential cross section is 
given by the expression:
%%%%%%%%%%%%%%%%%%%%%%%%%%%%%%%%%%%%%%%%%%%%%%%%%%%%
\begin{eqnarray} 
\delta^{(0)}_{2\gamma}&=&\frac{\alpha}{\pi}\Bigg[\frac{\nu^2+1}{4\nu}
\ln\left[\frac{\nu+1}{\nu-1}\right]\ln\left[\frac{\nu^2-1}{4\nu^2}\right]
+\frac{2\nu^2+1}{2\nu}\ln\left[\frac{\nu+1}{\nu-1}\right]-\frac{\nu^2+1}{2\nu}
\left\{\text{Sp}\left(\frac{\nu+1}{2\nu}\right)
-\text{Sp}\left(\frac{\nu-1}{2\nu}\right)\right\}
\nonumber\\
&&\quad-\,2+\sum_{f=e,\mu,\tau}\left\{\frac{2}{3}\left(\nu^2_f-\frac{8}{3}\right)
+\nu_f\left(\frac{3-\nu^2_f}{3}\right)\ln\left[\frac{\nu_f+1}{\nu_f-1}\right]\right\}
-\frac{2}{3}\left(\nu^2_\pi+\frac{1}{3}\right)+\frac{\nu^3_\pi}{3}
\ln\left[\frac{\nu_\pi+1}{\nu_\pi-1}\right]
\nonumber\\
&&\quad +\,\frac{1}{\nu}\left(\frac{2\eta m^2_l}{\eta Q^2+4E^2}\right)
\ln\left[\frac{\nu+1}{\nu-1}\right]+
\ln\left(\frac{4\eta^2\Delta^{2}_{\gamma^*}}{m_l^2}\right)\left[\frac{\nu^2+1}{2\nu}
\ln\left[\frac{\nu+1}{\nu-1}\right]-1\right]
+\frac{1}{4\beta}\ln\sqrt{\frac{1+\beta}{1-\beta}}
\nonumber\\
&&\quad +\,\frac{1}{4\beta^\prime}\ln\sqrt{\frac{1+\beta^\prime}{1-\beta^\prime}}
-\frac{\nu^2+1}{2\nu}\left\{\ln^2\sqrt{\frac{1+\beta}{1-\beta}}
-\ln^2\sqrt{\frac{1+\beta^\prime}{1-\beta^\prime}}
+\text{Sp}\left(1-\frac{\lambda_\nu-\eta}{(1-\beta^\prime)\xi_\nu}\right)\right.
\nonumber\\
&&\quad \left. +\text{Sp}\left(1-\frac{\lambda_\nu-\eta}{(1+\beta^\prime)\xi_\nu}\right)
\!-\!\text{Sp}\left(1-\frac{\lambda_\nu -\eta}{(1-\beta)\eta\lambda_\nu\xi_\nu}\right)
\!-\!\text{Sp}\left(1-\frac{\lambda_\nu-\eta}{(1+\beta)\eta\lambda_\nu\xi_\nu}\right)\right\}\Bigg]
\!+\!{\mathcal O}\left(\frac{{\cal Q}^2}{M^2}\right).\,\,\quad\,
\label{delta02gamma}
\end{eqnarray}
\end{widetext} 
%%%%%%%%%%%%%%%%%%%%%%%%%%%%%%%%%%%%%%%%%%%%%%%%%%%%
Apart from the pionic VP contribution result of Tsai~\cite{Tsai:1960zz} that is included in above 
expression, the remaining expression is by and large identical to what is found in the    
literature~\cite{Tsai:1961zz,Mo:1968cg,Maximon:2000hm,Vanderhaeghen:2000ws,Bucoveanu:2018soy}.

%\vspace{0.1cm}

In Fig.~\ref{plot:delta02gamma} we summarize all the fractional radiative corrections to the $\ell$-p 
elastic cross section at LO$_\alpha$  in HB$\chi$PT, namely, the VP correction, 
$\delta^{(0)}_{\rm vac}\!\!
=\!\!\!\!\sum\limits_{f=e,\mu,\tau}\delta^{(0)}_{{\rm vac};f}+\delta^{(0)}_{\rm vac;\,\pi}$, the 
lepton-photon VC, $\overline{\delta}^{(0)}_{\gamma\gamma;\,{\rm vertex}}
=\overline{\delta}^{(0)}_{\gamma\gamma;1}+\delta^{(0)}_{\gamma\gamma;2}$, and the soft bremsstrahlung 
correction $\overline{\delta}^{(0)}_{\gamma\gamma^*}$, in the MUSE kinematic range. A {\it key feature} 
of these LO$_\alpha$ radiative corrections is that they are {\it charge-symmetric}, {\it viz.} the cross 
sections are identical for both $\ell^-$p and $\ell^+$p scatterings. We find that the negative 
bremsstrahlung contribution is the most dominant correction in this low-$|Q^2|$ range. In contrast, the 
lepton-photon VC and VP correction are both positive. The plots in essence suggest very little 
sensitivity of the radiative corrections to the incoming lepton beam momenta. While the VP contributions 
are identical in both electron and muon scatterings, the following observations depict the contrasting 
nature of the other two LO$_\alpha$ radiative corrections, {\it viz.} the lepton-photon vertex and soft 
bremsstrahlung corrections, associated with MUSE kinematics:
%%%%%%%%%%%%%%%%%%%%%%%%%%%%%%%%%%%%%%%%%%%%FIGURE%%%%%%%%%%%%%%%%%%%%%%%%%%%%%%%%%%%%%%%%%%%
\begin{figure*}[tbp]
 \centering
	\includegraphics[scale=0.53]{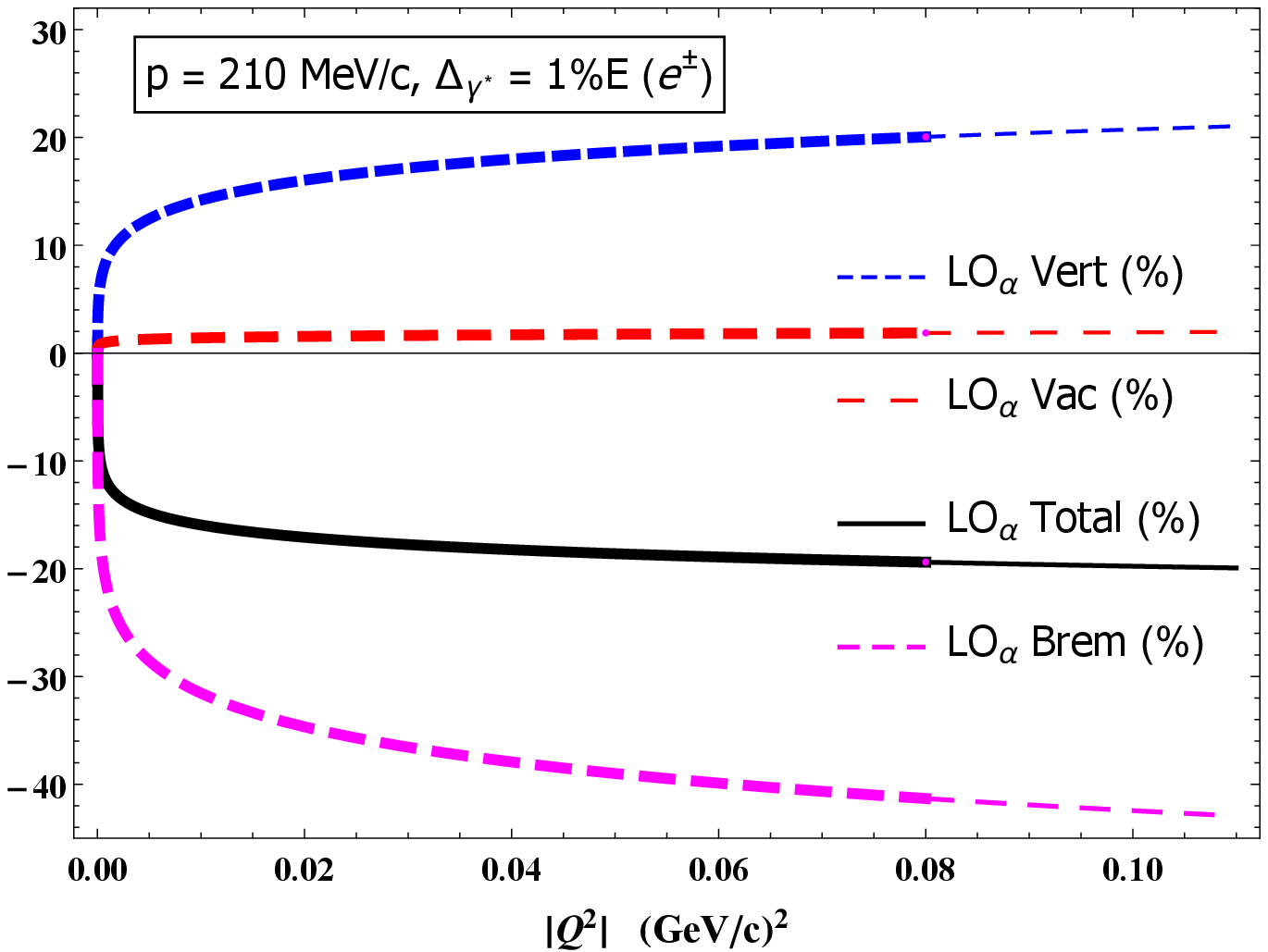}\qquad 
	\includegraphics[scale=0.53]{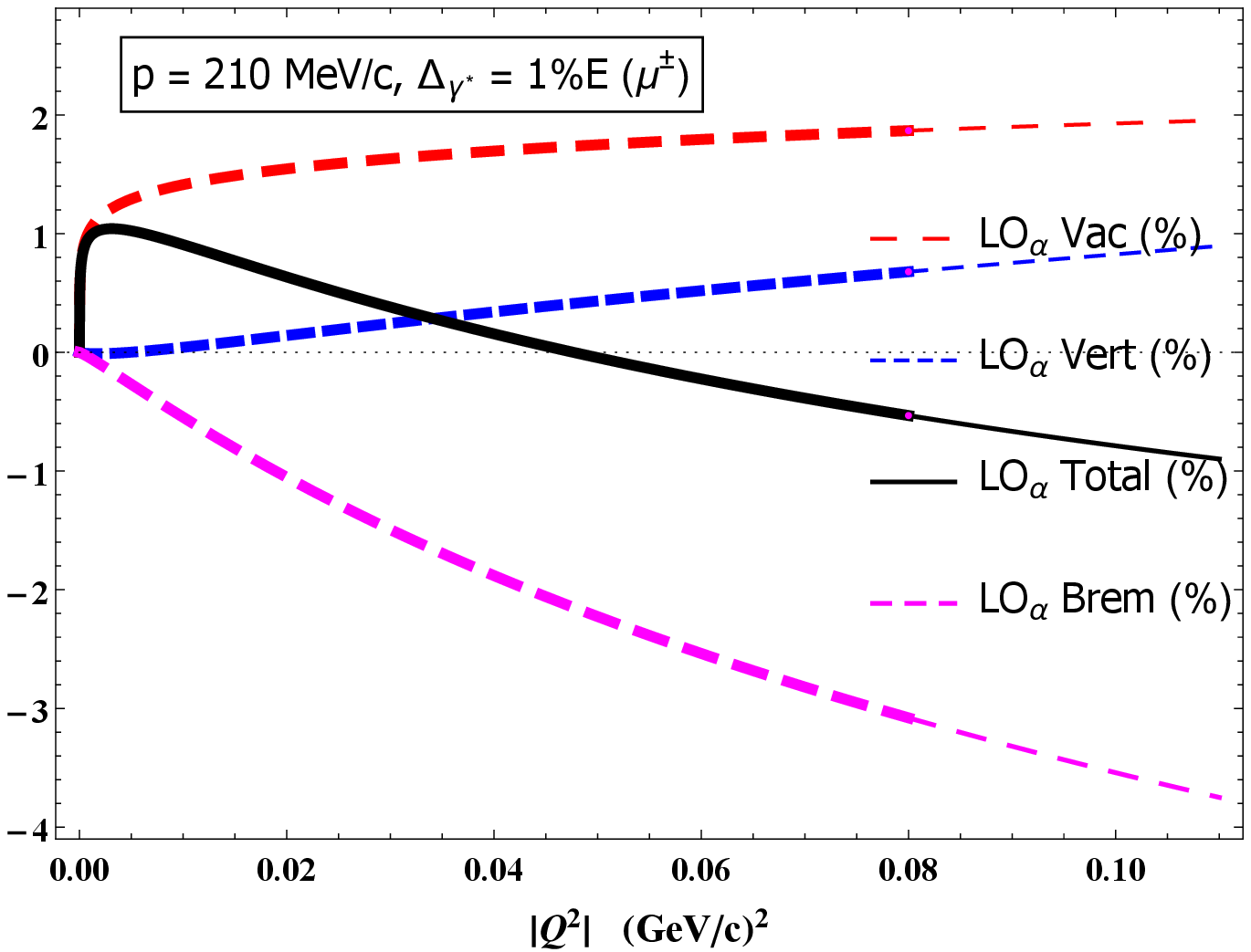}

    \vspace{1cm}

    \includegraphics[scale=0.53]{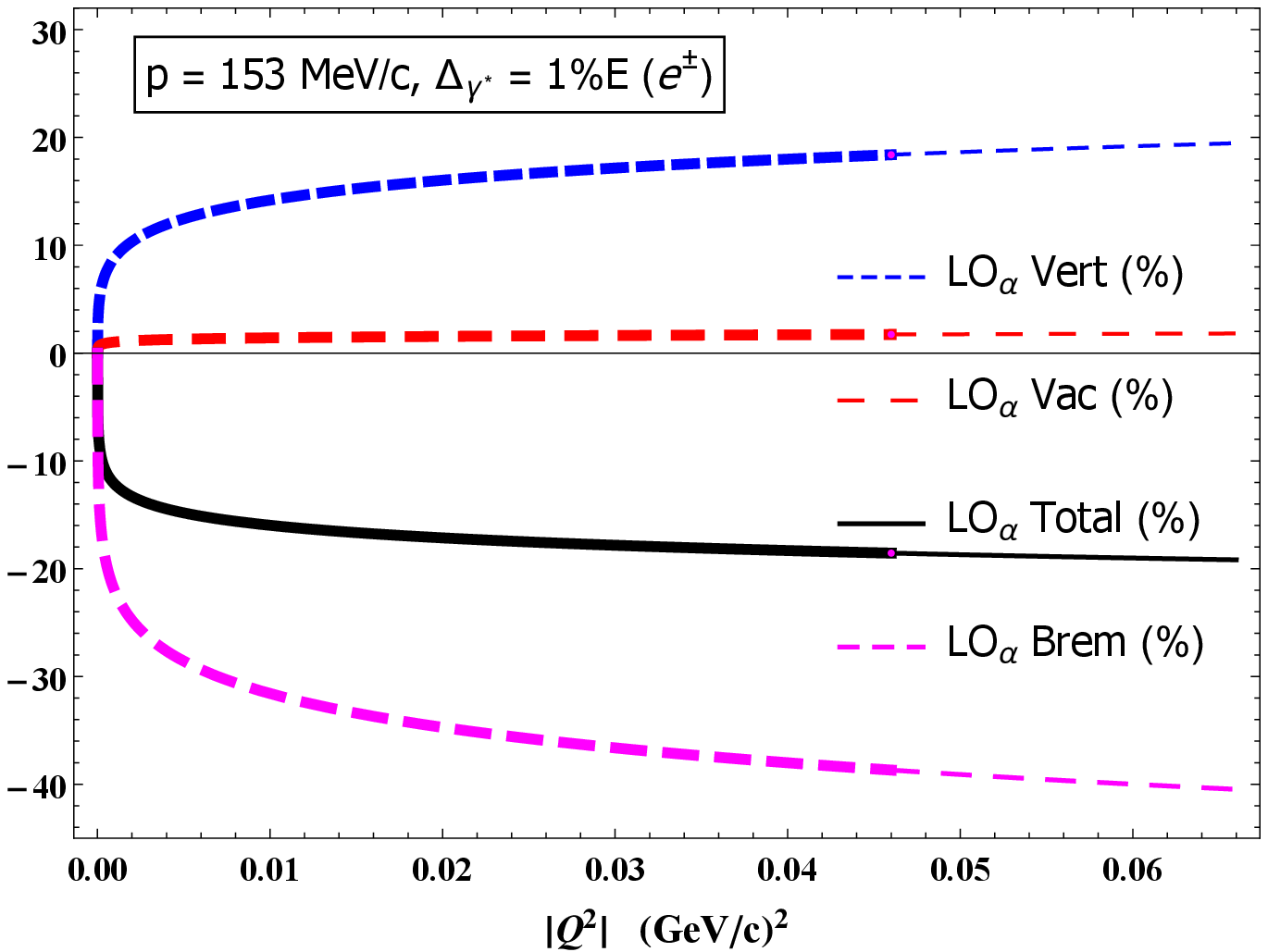}\qquad 
    \includegraphics[scale=0.53]{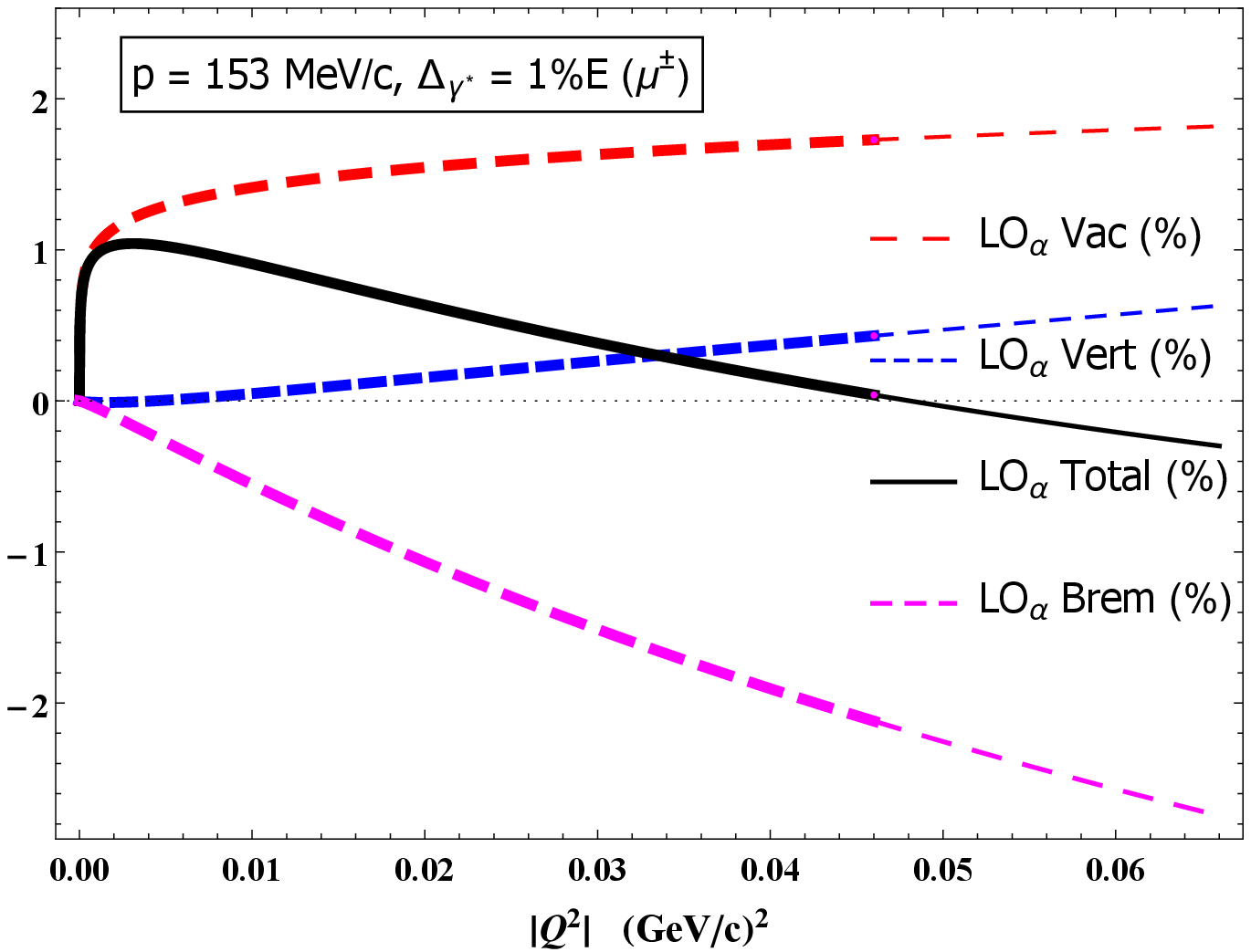}

    \vspace{1cm}

    \includegraphics[scale=0.53]{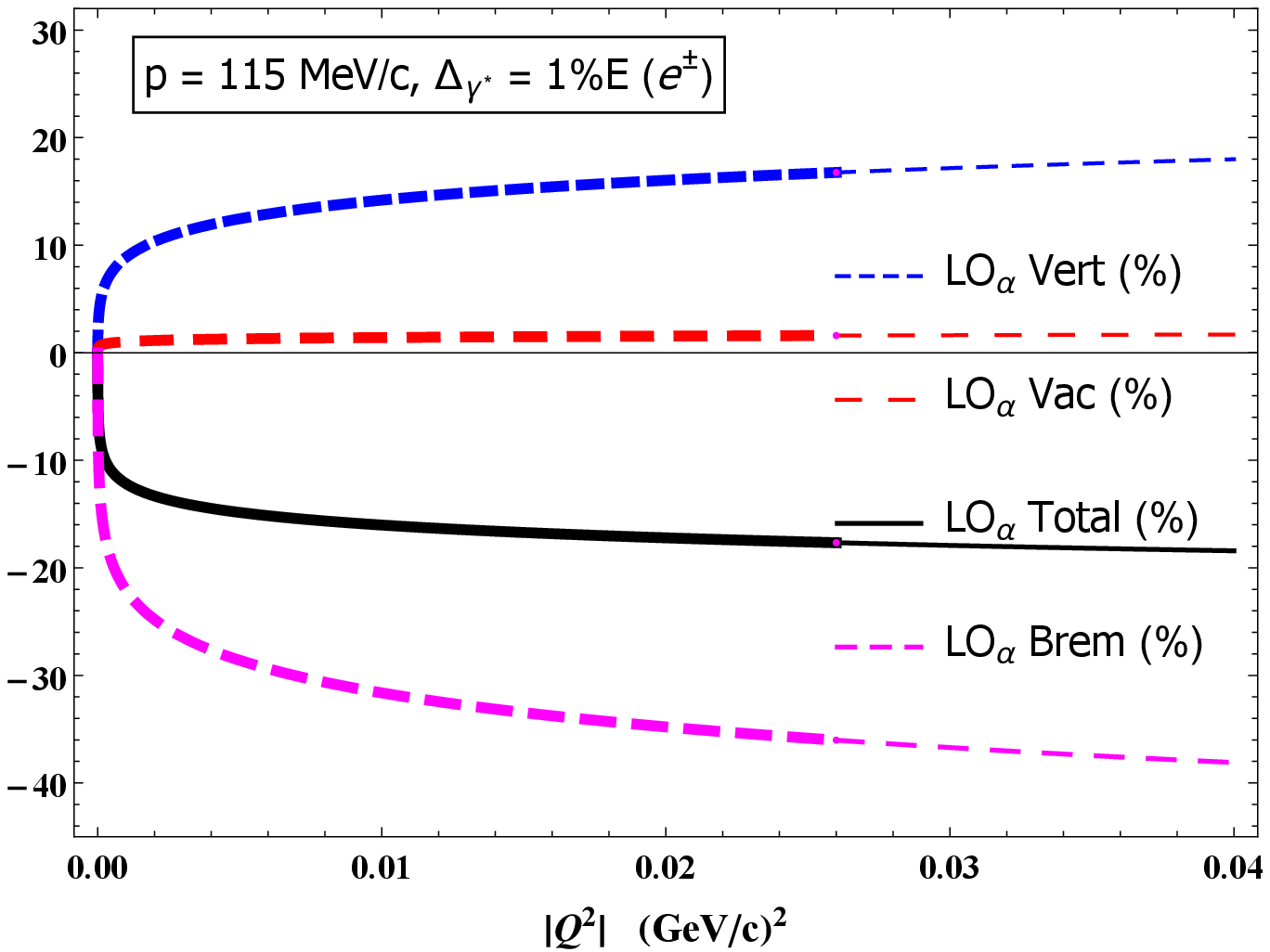}\qquad 
    \includegraphics[scale=0.53]{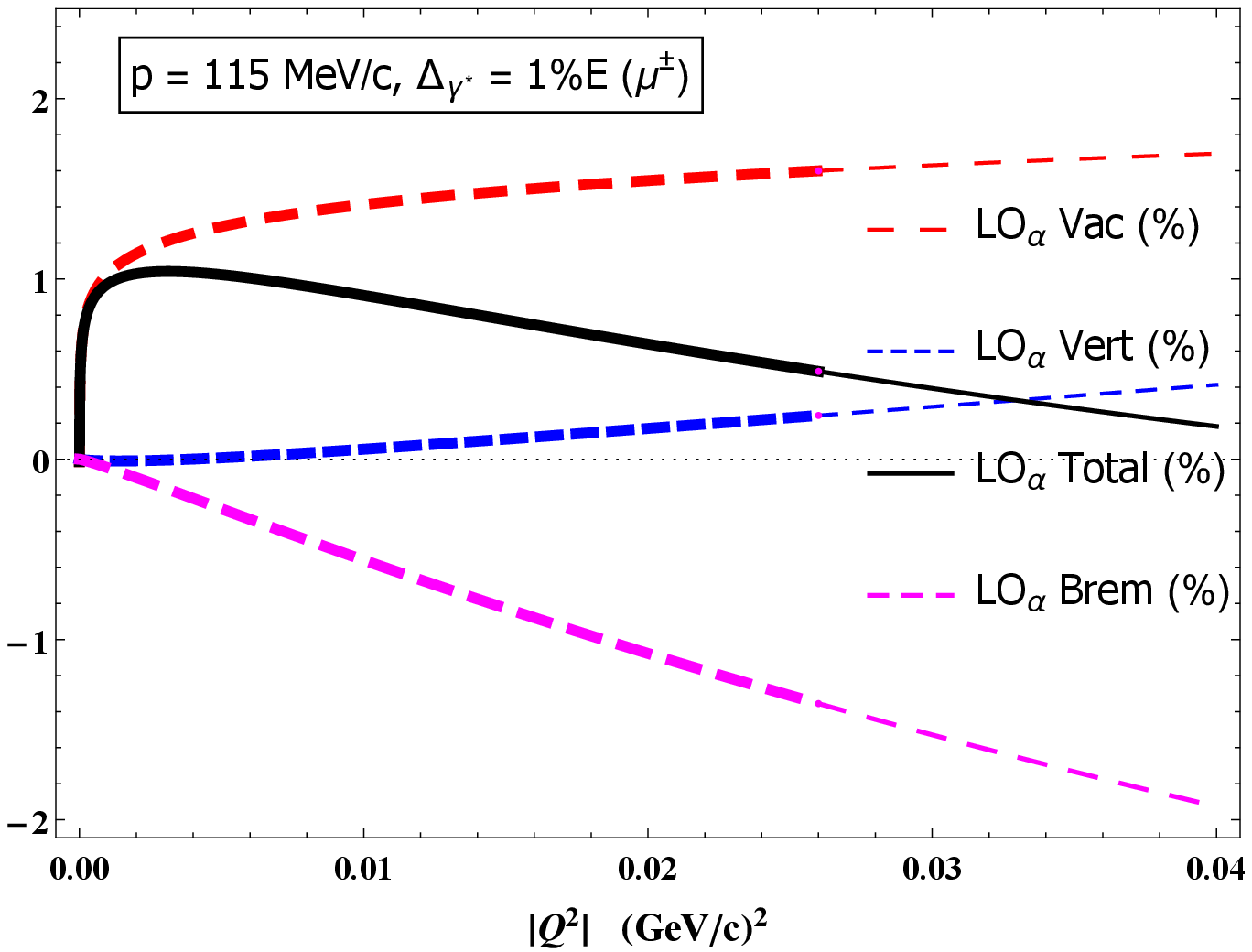}
	\caption{The individual one-loop LO$_\alpha$ contributions (in percentage), namely, 
	         the vacuum polarization correction $\delta^{(0)}_{\rm vac}$, lepton-photon 
	         vertex correction $\overline{\delta}^{(0)}_{\gamma\gamma;\,{\rm vertex}}$, 
	         and the soft photon bremsstrahlung correction 
	         $\overline{\delta}^{(0)}_{\gamma\gamma^*}$, to the $e^{\pm}$-p (left panels) 
	         and $\mu^{\pm}$-p (right panels) elastic cross sections as a function of the 
	         squared four-momentum transfer $|Q^2|$. The total LO$_\alpha$ contribution 
	         $\overline{\delta}^{(0)}_{\gamma\gamma}$ is displayed as the solid (black) 
	         line. Each plot covers the full kinematically allowed scattering range, 
	         $0<|Q^2|<|Q^2_{\rm max}|$, when $\theta\in[0,\pi]$ (cf. Table I of 
	         Ref.~\cite{Talukdar:2019dko}). The thickened portion of each curve 
	         corresponds to the MUSE kinematic cut, where $\theta\in[20^\circ,100^\circ]$. 
	         The {\it lab.}-frame detector acceptance $\Delta_{\gamma^*}$ is taken to be 
	         $1\%$ of the incident lepton beam energy $E$.}
	\label{plot:delta02gamma}
\end{figure*}
%%%%%%%%%%%%%%%%%%%%%%%%%%%%%%%%%%%%%%%%%%%%%%%%%%%%%%%%%%%%%%%%%%%%%%%%%%%%%%%%%%%%%%%%%%%%  
\begin{itemize}  
\item 
While both electronic VC and the soft photon bremsstrahlung contributions are very large and of comparable 
magnitudes, the muonic VC is roughly two orders of magnitude smaller. One reason for this contrast is 
evidently the absence of {\it Sudakov enhancement} in muonic scattering since $m_\mu^2\approx |Q^2|$, as 
mentioned earlier. 
\item
The electron and muon bremsstrahlung corrections are both negative, but the latter is over a magnitude 
smaller. Here too the Sudakov enhancement of the term $\overline{\delta}^{(0)}_{\gamma\gamma^*}$ plays a 
vital role which can be seen as follows. In the limit of small lepton mass (i.e., $m_l^2\ll |Q^2|$) we 
obtain
%%%%%%%%%%%%%%%%%%%%%%%%%%%%%%%%%%%%%%%%%%%%%%%%%%%%
\begin{widetext}

\vspace{-0.25cm}

\begin{eqnarray*}
\overline{\delta}^{(0)}_{\gamma\gamma^*}(Q^2) 
&\stackrel{Q^2\gg m_l^2}{\approx}& \frac{\alpha}{\pi}
\Bigg[\ln\left(\frac{\eta^3{\Delta}^{2}_{\gamma^*}}{E^2}\right)
\left[\ln\left(\frac{-Q^2}{m_l^2}\right)-1\right]
-\frac{1}{2}\ln^2\left(\frac{-Q^2}{m_l^2}\right)
+\ln\left(\frac{-Q^2}{m^2_l}\right)
\nonumber\\
&&\,\,\quad-\,\frac{1}{2}\ln^2\eta-\frac{\pi^2}{3}
-\text{Sp}\left(\cos^2\frac{\theta}{2}\right)\Bigg]\,.
\end{eqnarray*}

\vspace{-0.25cm}

\end{widetext} 
%%%%%%%%%%%%%%%%%%%%%%%%%%%%%%%%%%%%%%%%%%%%%%%%%%%%
In regard to the low-$|Q^2|$ MUSE kinematics, the ``high energy" approximations of 
$\overline{\delta}^{(0)}_{\gamma\gamma^*}$ for electron scattering is quite legitimate since 
$m^2_e\ll |Q^2|$ ($m_e^2=0.25 \times 10^{-6}$ GeV$^2$), which cannot be justified in case of the muon. With 
$\Delta_{\gamma^*}$ typically much smaller than the beam energies, the first two double-log terms containing 
the factor $-\ln \left(-Q^2/m^2_e\right)$ dominates in case of electron scattering, accounting for the large 
negative sign of the bremsstrahlung contribution. This contrasts the positive sign of electron-photon VC 
attributed to the dominant positive contributions from the Sudakov terms $\propto \ln^2 \left(-Q^2/m^2_e\right)$, 
as elucidated earlier.\footnote{It should be noted that the above expression differs by a factor
$\ln\left(\frac{-Q^2}{m_l^2}\right)\left[\ln\left(\frac{-Q^2}{m_l^2}\right)-1\right]$ compared to the 
standard expression known in existing literature (see, e.g., Refs.~\cite{Maximon:2000hm}). This difference is 
due to the DR scheme we have adopted in order to separate the IR-divergent part from the finite contribution. 
Specifically, in DR we prefer to retain the factor  $\ln\left(\frac{4\pi\mu^2}{-Q^2}\right)$ in the IR-singular 
part $\text{\bf IR}^{(0)}_{\gamma\gamma^*}$, instead of  $\ln\left(\frac{4\pi\mu^2}{m^2_l}\right)$,  which is 
a more standard representation in the literature. For the same reason our leptonic VC result [cf. 
Eq.~\eqref{delta:LO_vertex}] differs by the same factor.}
\item 
Large cancellations occur between the VC and bremsstrahlung contributions and lead to $\sim 20\%$ correction 
in electron scattering. For the muon cancellations between the comparable VP and bremsstrahlung contributions 
lead to $\sim 1\%$ correction only at the {\it largest} MUSE beam momenta.
\item 
At lowest order in chiral expansion the proton is essentially an infinitely heavy static object, i.e., leptons 
scatter off a static Coulomb potential. This naturally explains why all LO$_\alpha$ radiative effects on the 
proton vanish. 
\end{itemize}            

%%%%%%%%%%%%%%%%%%%%%%%%%%%%%%%%%%%%%%%%%%%%%%%%%%%%%%%%%%%%%%%%%%%%%%%%%%%%%%%%%%%%%%%%%%%%%%
\section{Radiative correction at NLO$_\alpha$}\label{sec:NLO}   
%%%%%%%%%%%%%%%%%%%%%%%%%%%%%%%%%%%%%%%%%%%%%%%%%%%%%%%%%%%%%%%%%%%%%%%%%%%%%%%%%%%%%%%%%%%%%% 
The next order radiative corrections are {\it dynamical} in nature, since they arise from the NLO$_\alpha$  
interactions in the HB$\chi$PT Lagrangian. Thus, the power counting scheme allows for diagrams containing 
either one NLO$_\alpha$  vertex or  one insertion of an NLO$_\chi$ proton propagator. The NLO$_\alpha$ 
diagrams that we consider are the ${\mathcal O}(e^2\alpha{\cal Q}/M)$ one-loop virtual correction 
amplitudes, along with those of the ${\mathcal O}(e^3{\cal Q}/M)$ soft photon bremsstrahlung amplitudes. 
Employing the DR scheme we extract the UV and IR divergences generated at NLO$_\alpha$ to obtain the 
corresponding fractional radiative corrections to the cross section, 
$\delta^{(1)}_{2\gamma}\sim\mathcal{O}\left(\alpha{\cal Q}/M\right)$. The NLO$_\alpha$  TPE box diagrams 
were evaluated analytically invoking SPA in Ref.~\cite{Talukdar:2019dko} (we shall make some pertinent 
comments regarding SPA and its validity in the evaluation of the TPE box diagrams in the discussion section 
of this paper). In this section the other virtual NLO$_\alpha$ one-loop diagrams shall be evaluated exactly, 
i.e., we make no approximations in our evaluation of the one-loop diagrams. Below we elucidate the details 
of the NLO$_\alpha$ virtual and real contributions and the subsequent cancellation of the IR divergences.

%%%%%%%%%%%%%%%%%%%%%%%%%%%%%%%%%%%%%%%%%%%%%%%%%%%%%%	 
\subsection{One-loop Virtual Corrections at NLO$_\alpha$} 
\begin{figure*}[tbp]
 \centering
	\includegraphics[scale=0.59]{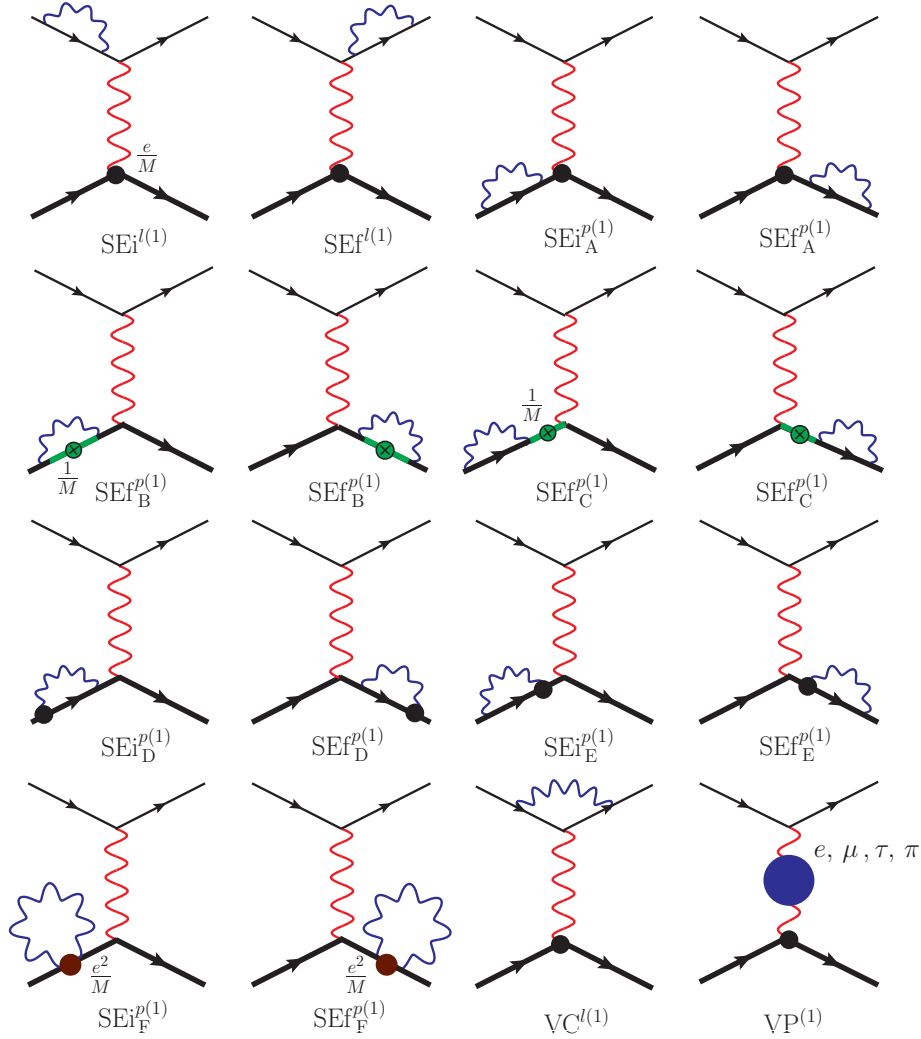}
	\caption{The one-loop lepton and proton self-energies, lepton-proton vertex corrections 
	         and the vacuum polarizations at NLO$_\alpha$ in HB$\chi$PT [i.e., 
	         ${\mathcal O}\left(e^2\alpha{\cal Q}/M\right)$], contributing to the radiative 
	         corrections to the LO (Born) $\ell$-p elastic scattering amplitude [see 
	         Eq.~\eqref{M0_gamma}]. The filled blobs represent $1/M$ order proton-photon 
	         vertex insertions. In particular, the proton self-energy ``tadpoles" (diagrams 
	         SEi,f$^{\,p(1)}_{\rm \,F}$) have $e^2/M$ order vertices. The proton propagators 
	         with the crossed blobs ``$\otimes$" represent $1/M$ order propagator insertions. 
	         While all the self-energy diagrams vanish in the on-shell limit of the external 
	         particles, the lepton vertex correction and vacuum polarization diagrams do not 
	         contribute at NLO$_\alpha$ since they are kinematically suppressed to $1/M^2$ 
	         order.} 
	\label{leptonic:NLO}
\end{figure*}

\begin{figure*}[tbp]
 \centering
	\includegraphics[scale=0.59]{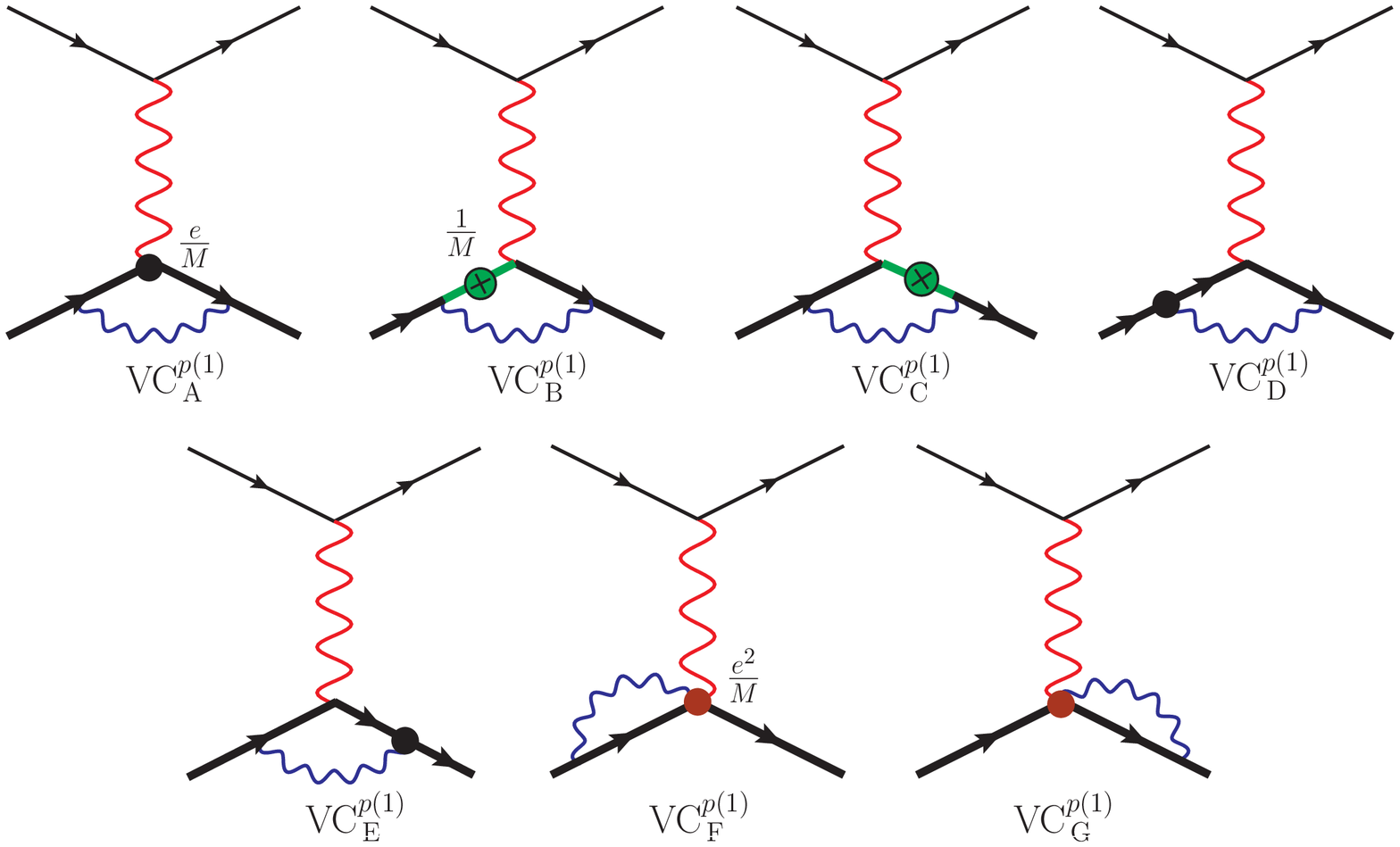}
	\caption{One-loop proton-photon vertex correction diagrams at NLO$_\alpha$ in HB$\chi$PT 
	         [i.e., ${\mathcal O}\left(e^2\alpha{\cal Q}/M\right)$], contributing to the 
	         radiative corrections to the LO (Born) $\ell$-p elastic scattering amplitude 
	         [see Eq.~\eqref{M0_gamma}]. The filled blobs represent $1/M$ order proton-photon 
	         vertex insertions. In particular, the two-photon proton vertices (diagrams 
	         VC$^{\,p(1)}_{\rm \,F,G}$) are of order $e^2/M$. The proton propagators with the 
	         crossed blobs ``$\otimes$" represent $1/M$ order propagator insertions (diagrams 
	         VC$^{\,p(1)}_{\rm \,B,C}$). None of these diagrams contribute to the cross 
	         section at NLO$_\alpha$ (see text).}
	\label{proton_vertex:NLO}
\end{figure*}

%%%%%%%%%%%%%%%%%%%%%%%%%%%%%%%%%%%%%%%%%%%%%%%%%%%%%% 
In case of the external on-shell particles the lepton and proton SE amplitude terms (diagrams 
SEi,f$^{\,l(1)}$ and  SEi,f$^{\,p(1)}_{\,{\rm A},\cdots,{\rm F}}$ shown in Fig.~\ref{leptonic:NLO}) do not 
contribute directly to the elastic scattering amplitude as they vanish due to the on-shell renormalization 
condition. The respective SE loops, however, renormalize the off-shell bare masses in the propagators. In 
addition, their derivatives contribute at NLO$_\alpha$ to the respective wave function renormalization 
constants $Z^{l,p(1)}_1$. In the following we discuss the evaluations of the other one-loop NLO$_\alpha$ 
amplitudes at ${\mathcal O}(e^2\alpha{\cal Q}/M)$, which include the lepton-photon VC diagram (VC$^{l(1)}$ 
in Fig.~\ref{leptonic:NLO}), the VP diagram (VP$^{(1)}$ in Fig.~\ref{leptonic:NLO}), and the proton VC 
diagrams (VC$^{p(1)}_{\rm A,..,G}$ in Fig.~\ref{proton_vertex:NLO}). Furthermore, the NLO$_\alpha$ TPE 
amplitudes which were already evaluated invoking SPA in our previous work~\cite{Talukdar:2019dko}, are used
in this work to determine the complete one-loop virtual radiative contribution at NLO$_\alpha$. 

%\vspace{0.1cm}

%%%%%%%%%%%%%%%%%%%%%%%%%%%%%%%%%%%%%%%%%%%%%%%%%%%%%%%%%%%%%%%%%%%%%%%%%%%%%
\noindent{\it 1. Lepton-Photon Vertex and Vacuum Polarization Corrections.\,} 
%%%%%%%%%%%%%%%%%%%%%%%%%%%%%%%%%%%%%%%%%%%%%%%%%%%%%%%%%%%%%%%%%%%%%%%%%%%%%
Formally, the only non-trivial NLO$_\alpha$ contributions in Fig.~\ref{leptonic:NLO} are expected to arise 
from the last two diagrams, namely, the lepton-photon VC (diagram VC$^{l(1)}$) and the VP (diagram VP${}^{(1)}$) 
contributions. The result closely resembles the corresponding LO$_\alpha$ result, Eq.~\eqref{M0gamgam}, in that
the amplitude of each of the above NLO$_\alpha$ diagram, apart from the vertex term $\propto F^l_2$, factorizes
into the NLO$_\chi$ Born amplitude $\mathcal{M}^{(1)}_{\gamma}$ [see Eq.~\eqref{M1M2_gamma}], namely, 
\begin{eqnarray}
\mathcal{M}^{l(1)}_{\gamma\gamma} &=& \mathcal{M}^{(1)}_\gamma+\mathcal{M}^{(1)}_\gamma 
\left[F^{l;\,{\rm ren}}_1(Q^2)-1 +\Delta\Pi(Q^2)\right] 
\nonumber\\
&& +\,\overline{\mathcal{M}}^{(1)}_\gamma  F^l_2(Q^2)\,,
\label{Ml1gamgam_1}
\end{eqnarray}
where
\begin{eqnarray}
\overline{\mathcal{M}}^{(1)}_\gamma &=&\overline{\mathcal{M}}^{(1);\,{\rm a}}_\gamma 
+ \overline{\mathcal{M}}^{(1);\,{\rm b}}_\gamma\,;
\nonumber\\
\overline{\mathcal{M}}^{(1);\,{\rm a}}_\gamma &=& 
-\frac{e^2}{4 m_l M Q^2}\left[{\bar u}_l(p^\prime)i\sigma^{\mu\nu}Q_\nu \,u_l(p)\right]\,
\left[\chi^\dagger(p_p^\prime)\right.
\nonumber\\
&& \left. \times\, \left\{(p_p + p_p^\prime)_\mu - v_\mu v \cdot (p_p + p_p^\prime) \right\} \chi(p_p)\right]\,,
\nonumber
\end{eqnarray}
\begin{eqnarray}
\overline{\mathcal{M}}^{(1);\,{\rm b}}_\gamma &=& 
-\frac{e^2}{4 m_l M Q^2}\left[{\bar u}_l(p^\prime)i\sigma^{\mu\nu}Q_\nu \,u_l(p)\right]\,
\left[\chi^\dagger(p_p^\prime)\right.
\nonumber\\
&& \left. \times\, (2+\kappa_s+\kappa_v)\left[S_\mu,\, S \cdot Q\right] \chi(p_p)\right]\,.
\label{Ml1gamgam_2}
\end{eqnarray} 
Here $F^{l;\, {\rm ren}}_1$ is the renormalized Dirac form factor of the lepton, Eq.~\eqref{Fl1ren}. In 
Sec.~\ref{sec:formalism} it was demonstrated that the interference of the LO and NLO$_\chi$ Born amplitudes 
is proportional to ${\mathscr R}_Q\sim {\mathcal O}(M^{-2})$ [see Eq.~\eqref{Ratio_NLO}]. Consequently, with 
$F^{l;\,{\rm ren}}_1(Q^2)-1$, $\Delta\Pi(Q^2)$ and $F^l_2(Q^2)$  of ${\mathcal O}(\alpha)$, the relevant
NLO$_\alpha$ terms which arise from the interference of the amplitudes, Eqs.~\eqref{M0gamgam} and 
\eqref{Ml1gamgam_1}, and which should formally contribute here, are {\it de facto} kinematically suppressed 
to ${\mathcal O}(\alpha{\cal Q}^2/M^2)$, i.e., in essence NNLO$_\alpha$ in HB$\chi$PT. Thus, we have  
%%%%%%%%%%%%%%%%%%%%%%%%%%%%%%%%%%%%%%%%%%%%%%%%%%%%
\begin{widetext}
\begin{eqnarray}
\delta^{l(1)}_{\gamma\gamma}(Q^2)&=&\frac{\sum\limits_{spins}\left[\left|{\mathcal M}^{(0)}_{\gamma\gamma}
+{\mathcal M}^{l(1)}_{\gamma\gamma}\right|^2 -|{\mathcal M}^{(1)}_\gamma|^2-2{\mathcal R}\text{e}
\left(\mathcal{M}_\gamma^{(0)\dagger}{\mathcal M}^{(1)}_\gamma
+\mathcal{M}_\gamma^{(0)\dagger}{\mathcal M}^{(0)}_{\gamma\gamma}\right)
\right]}{\sum\limits_{spins}|\mathcal{M}^{(0)}_\gamma|^2}+1
\nonumber\\
&=& 2\left[F^{l;\,{\rm ren}}_1(Q^2)-1 +\Delta\Pi(Q^2)\right]\left({\mathscr R}_Q
+\frac{\sum\limits_{spins}|{\mathcal M}^{(1)}_\gamma|^2}{\sum\limits_{spins}|\mathcal{M}^{(0)}_\gamma|^2}\right)
\nonumber\\
&&+\,\frac{2{\mathcal R}\text{e}\!\!\sum\limits_{spins}
\left(\mathcal{M}^{(0)\dagger}_\gamma\,\,\overline{\mathcal{M}}^{(1)}_\gamma 
+ {\mathcal M}^{(1)\dagger}_\gamma\,\,\overline{\mathcal{M}}^{(0)}_\gamma 
+ {\mathcal M}^{(1)\dagger}_\gamma\,\,\overline{\mathcal{M}}^{(1)}_\gamma \right)}{\sum\limits_{spins}
|\mathcal{M}^{(0)}_\gamma|^2}F_2^l(Q^2)
+ {\mathcal O}\left(\alpha^2\right)
\nonumber\\
&=&\, {\mathscr R}_Q\,\left[\text{\bf IR}^{(0)}_{\gamma\gamma}(Q^2)
+\bar{\delta}^{(0)}_{\gamma\gamma}(Q^2)\right]\,
\left\{1+\frac{1}{2}(1+\kappa_p)^2\left(\frac{Q^2+4(m^2_l-E^2)}{Q^2+4E^2}\right)\right\} 
\nonumber\\
&& +\, (1+\kappa_p)^2{\mathscr R}_Q\,{\delta}^{(0)}_{\gamma\gamma;2}(Q^2) 
\left\{1-\frac{1}{2}\left(\frac{Q^2+4(m^2_l-E^2)}{Q^2+4E^2}\right)\right\} 
+ {\mathcal O}\left(\alpha^2\right) 
\sim \mathcal{O}\left(\alpha \frac{{\cal Q}^2}{M^2}\right)\,. 
\label{eq:NNLO1}
\end{eqnarray}
\end{widetext}
%%%%%%%%%%%%%%%%%%%%%%%%%%%%%%%%%%%%%%%%%%%%%%%%%%%%
by using the results of Eqs.~\eqref{Ratio_NLO}, \eqref{Eq:rad_electric}, \eqref{Eq:rad_magnetic} and 
\eqref{delta:LO_vertex} as well as the following two estimates: 
%%%%%%%%%%%%%%%%%%%%%%%%%%%%%%%%%%%%%%%%%%%%%%%%%%%%
\begin{widetext}
\begin{eqnarray}
\frac{2{\mathcal R}\text{e}\!\!\sum\limits_{spins}
\left(\mathcal{M}^{(0)\dagger}_\gamma\,\,\overline{\mathcal{M}}^{(1);\,{\rm a}}_\gamma 
+ {\mathcal M}^{(1);\,{\rm a}\dagger}_\gamma\,\,
\overline{\mathcal{M}}^{(0)}_\gamma\right)}{\sum\limits_{spins}|\mathcal{M}^{(0)}_\gamma|^2} F^l_2(Q^2)
&=& {\mathscr R}_Q\,{\delta}^{(0)}_{\gamma\gamma;2}(Q^2)\,, \hspace{0.5cm} {\rm and} 
\nonumber\\
\nonumber\\
\frac{2{\mathcal R}\text{e}\!\!\sum\limits_{spins}
\left(\mathcal{M}^{(1);\,{\rm a}\dagger}_\gamma\,\,\overline{\mathcal{M}}^{(1);\,{\rm a}}_\gamma 
+ {\mathcal M}^{(1);\,{\rm b}\dagger}_\gamma\,\,
\overline{\mathcal{M}}^{(1);\,{\rm b}}_\gamma\right)}{\sum\limits_{spins}
|\mathcal{M}^{(0)}_\gamma|^2} F^l_2(Q^2)
&=& (1+\kappa_p)^2\,{\mathscr R}_Q\,{\delta}^{(0)}_{\gamma\gamma;2}(Q^2) 
+ o\left(\alpha\frac{{\cal Q}^4}{M^4}\right)\,, 
\label{eq:NNLO2}
\end{eqnarray}
\end{widetext}
%%%%%%%%%%%%%%%%%%%%%%%%%%%%%%%%%%%%%%%%%%%%%%%%%%%%
where the symbol $o(\alpha{\cal Q}^4/M^4)$ denote further terms of order  $1/M^4$ which we ignore. Please 
note that $\delta^{(0)}_{\gamma\gamma;2}$ already contains $1/M^2$ order terms [cf. 
Eq.~\eqref{Eq:rad_magnetic}]. As illustrated above, a notable feature regarding these sub-leading chiral 
order radiative corrections is that they generally do not completely factorize as a simple product of the 
${\mathcal O}({\cal Q}^2/M^2)$ pure chiral correction, e.g., $\delta^{(1/M^2)}_\chi$ [cf. 
Eq.~\eqref{eq:delta_chi1}], and the ${\mathcal O}(\alpha)$ leading virtual radiative correction 
$\bar{\delta}^{(0)}_{\gamma\gamma}$. At our accuracy of ${\mathcal O}(\alpha{\cal Q}/M)$ , however, we 
eventually drop all $1/M^2$ order terms in our final expressions for the fractional contributions. These 
NNLO$_\alpha$ terms of ${\mathcal O}(\alpha{\cal Q}^2/M^2)$ above, as mentioned earlier will be useful 
in estimating our theoretical uncertainties.  We then conclude that none of the diagrams displayed in 
Fig.~\ref{leptonic:NLO} contribute to the scattering cross section at NLO$_\alpha$.
%%%%%%%%%%%%%%%%%%%%%%%%%%%%%%%%%%%%%%%%%FIGURE%%%%%%%%%%%%%%%%%%%%%%%%%%%%%%%%%%%%%%%%%%%%%
%%%%%%%%%%%%%%%%%%%%%%%%%%%%%%%%%%%%%%%%%%%%%%%%%%%%%%%%%%%%%%%%%%%%%%%%%%%%%%%%%%%%%%%%%%%%
%%%%%%%%%%%%%%%%%%%%%%%%%%%%%%%%%%%%%%%%%%%%%%%%%%%%%%%%%%%%%%%%%%%%%%%%%%%%%%%%%%%%%%%%%%%%
%%%%%%%%%%%%%%%%%%%%%%%%%%%%%%%%%%%%%%%%%%%%%%%%%%%%%%%%%%%%%%%%%%%%%%%%%%%%%%%%%%%%%%%%%%%%

%\vspace{0.1cm}

%%%%%%%%%%%%%%%%%%%%%%%%%%%%%%%%%%%%%%%%%%%%%%%%%%%%%%%%%%%%%%%%%%%%%%%%%%%%%
\noindent{\it 2. Proton-Photon Vertex Corrections.\,}
%%%%%%%%%%%%%%%%%%%%%%%%%%%%%%%%%%%%%%%%%%%%%%%%%%%%%%%%%%%%%%%%%%%%%%%%%%%%%
As displayed in Fig.~\ref{proton_vertex:NLO} a total of seven diagrams, VC$^{p(1)}_{\rm A}$, $\cdots$, 
VC$^{p(1)}_{\rm G}$, contribute to the proton-photon VC at NLO$_\alpha$ in HB$\chi$PT.  Five of these 
diagrams involve $1/M$ order proton-photon vertex insertions, and the remaining two involve $1/M$ order 
``heavy" proton propagator insertions. The amplitudes of these diagrams 
$\mathcal{M}^{p;(1);A,...,G}_{\gamma\gamma;\,{\rm vertex}}$ can be expressed as follows:
%%%%%%%%%%%%%%%%%%%%%%%%%%%%%%%%%%%%%%%%%%%%%%%%%%%%
\begin{widetext}
\begin{eqnarray}
%%%%%%%%%%%%%%%%%%%%%%%%%%%%%%%%%
\mathcal{M}^{p(1);A}_{\gamma\gamma;\,{\rm vertex}} &=& 
\frac{ie^4}{2M Q^2}\left[{\bar u}_l(p^\prime)\gamma^\mu u_l(p)\right]\,
\int \frac{{\rm d}^4k}{(2\pi)^4}\frac{\left[\chi^\dagger(p_p^\prime)
\left\{(p_p+p_p^\prime-2k)_\mu-[v\cdot(p_p+p_p^\prime-2k)]v_\mu\right\} \chi(p_p)\right]}{(k^2+i0)\,
[v\cdot( p_p-k) +i0][v\cdot(p_p^\prime-k)+i0]}\,,
\nonumber\\
\nonumber\\
%%%%%%%%%%%%%%%%%%%%%%%%%%%%%%%%%
\mathcal{M}^{p(1);B}_{\gamma\gamma;\,{\rm vertex}} &=& 
\frac{ie^4}{2M Q^2}\left[{\bar u}_l(p^\prime)\gamma^\mu u_l(p)\right]\,
\int \frac{{\rm d}^4k}{(2\pi)^4}\frac{\left[\chi^\dagger(p_p^\prime)\,v_\mu \chi(p_p)\right]}{(k^2+i0)\,
[v\cdot(p_p^\prime-k) +i0]}\left(1+\frac{p_p^{2}}{(v\cdot k)^2}-\frac{(p_p-k)^2}{(v\cdot k)^2}\right)\,,
\nonumber\\
\nonumber\\
%%%%%%%%%%%%%%%%%%%%%%%%%%%%%%%%%
\mathcal{M}^{p(1);C}_{\gamma\gamma;\,{\rm vertex}} &=& 
\frac{ie^4}{2M Q^2}\left[{\bar u}_l(p^\prime)\gamma^\mu u_l(p)\right]\,
\int \frac{{\rm d}^4k}{(2\pi)^4} \frac{\left[\chi^\dagger(p_p^\prime)\,v_\mu \chi(p_p)\right]}{(k^2+i0)\,
[v\cdot(p_p-k) +i0]}\left(1+\frac{(p_p^\prime)^{2}}{(v\cdot k)^2}
-\frac{(p_p^\prime-k)^2}{(v\cdot k)^2}\right)\,,
\nonumber\\
\nonumber\\
%%%%%%%%%%%%%%%%%%%%%%%%%%%%%%%%%
\mathcal{M}^{p;(1);D}_{\gamma\gamma;\,{\rm vertex}} &=& 
\frac{ie^4}{2M Q^2}\left[{\bar u}_l(p^\prime)\gamma^\mu u_l(p)\right]\,
\int \frac{{\rm d}^4k}{(2\pi)^4}\frac{\left[\chi^\dagger(p_p^\prime)\,v_\mu\left\{[v\cdot(2p_p-k)]
- [v\cdot(2p_p-k)]v^2\right\} \chi(p_p)\right]}{(k^2+i0)\,
[v\cdot( p_p-k) +i0][v\cdot(p_p^\prime-k)+i0]}=0\,,
\nonumber\\
\nonumber\\
%%%%%%%%%%%%%%%%%%%%%%%%%%%%%%%%
\mathcal{M}^{p(1);E}_{\gamma\gamma;\,{\rm vertex}} &=& 
\frac{ie^4}{2M Q^2}\left[{\bar u}_l(p^\prime)\gamma^\mu u_l(p)\right]\,
\int \frac{{\rm d}^4k}{(2\pi)^4}\frac{\left[\chi^\dagger(p_p^\prime)\,v_\mu\left\{[v\cdot(2p_p^\prime-k)]
-[v\cdot(2p_p^\prime-k)]v^2\right\} \chi(p_p)\right]}{(k^2+i0)\,
[v\cdot( p_p-k) +i0][v\cdot(p_p^\prime-k)+i0]}=0\,,
\nonumber\\
\nonumber\\
%%%%%%%%%%%%%%%%%%%%%%%%%%%%%%%%%
\mathcal{M}^{p(1);F}_{\gamma\gamma;\,{\rm vertex}} &=& 
\frac{ie^4}{2M Q^2}\,\left[{\bar u}_l(p^\prime)\gamma^\mu u_l(p)\right]\,
\int \frac{{\rm d}^4k}{(2\pi)^4}\frac{\left[\chi^\dagger(p_p^\prime)
v_\mu(1-v^2) \chi(p_p)\right]}{(k^2+i0)\,[v\cdot(p_p-k) +i0]}=0\,,
\nonumber\\
\nonumber\\
%%%%%%%%%%%%%%%%%%%%%%%%%%%%%%%%%
\mathcal{M}^{p(1);G}_{\gamma\gamma;\,{\rm vertex}} &=&
\frac{ie^4}{2M Q^2}\,\left[{\bar u}_l(p^\prime)\gamma^\mu u_l(p)\right]\,
\int \frac{{\rm d}^4k}{(2\pi)^4}\frac{\left[\chi^\dagger(p_p^\prime)v_\mu(1-v^2) 
\chi(p_p)\right]}{(k^2+i0)\,[v\cdot(p_p^\prime-k) +i0]}=0\,.
%%%%%%%%%%%%%%%%%%%%%%%%%%%%%%%%%
\end{eqnarray}
\end{widetext}
%%%%%%%%%%%%%%%%%%%%%%%%%%%%%%%%%%%%%%%%%%%%%%%%%%%%
Since $v^2=1$, the last four amplitudes $\mathcal{M}^{p(1);D,E,F,G}_{\gamma\gamma;\,{\rm vertex}}$ 
vanish as indicated. To evaluate the remaining three amplitudes 
$\mathcal{M}^{p(1);A,B,C}_{\gamma\gamma;\,{\rm vertex}}$, as earlier we first use (in the 
{\it lab.}-frame) the on-shell relations, $v\cdot p_p=0$ and 
$v\cdot p_p^\prime= - \frac{(p_p^\prime)^2}{2M}+{\mathcal O}(M^{-2})$, in the denominators of the 
leading chiral order proton propagators. Second, we incorporate a $1/M$ expansion and retain terms up 
to ${\mathcal O}(e^2\alpha{\cal Q}/M)$ in the NLO$_\alpha$ proton-photon VC amplitudes. Using DM this 
eventually leads to vanishing contribution, namely,
%%%%%%%%%%%%%%%%%%%%%%%%%%%%%%%%%%%%%%%%%%%%%%%%%%%%
\begin{widetext}
\begin{eqnarray}
\mathcal{M}^{p(1)}_{\gamma\gamma;\,{\rm vertex}}\!\!\!\! &=& 
\!\!\mathcal{M}^{p(1);A}_{\gamma\gamma;\,{\rm vertex}} 
+ \mathcal{M}^{p(1);B}_{\gamma\gamma;\,{\rm vertex}}
+\mathcal{M}^{p(1);C}_{\gamma\gamma;\,{\rm vertex}}
\nonumber\\
&=& \!\!\frac{ie^4}{2M Q^2}\left[{\bar u}_l(p^\prime)\gamma^\mu u_l(p)\right]
\int \frac{{\rm d}^4k}{(2\pi)^4} 
\frac{\left[\chi^\dagger(p_p^\prime)(p_p+p_p^\prime-2k)_\mu\chi(p_p)\right]}{(k^2+i0)\,(-v\cdot k +i0)^2} 
\left(1-\frac{(p_p^\prime)^2}{2M (v\cdot k)}+\cdots\!\right)
\nonumber\\
&& \!\!+\,\frac{ie^4}{2M Q^2}\left[{\bar u}_l(p^\prime)\gamma^\mu u_l(p)\right]\,
\int \frac{{\rm d}^4k}{(2\pi)^4}\frac{\left[\chi^\dagger(p_p^\prime)
v_\mu\chi(p_p)\right]}{(k^2+i0)\,(-v\cdot k +i0)}
\left(1-\frac{k^2}{(v\cdot k)^2}\right)\!\left(2-\frac{(p_p^\prime)^2}{2M (v\cdot k)}
+\cdots\!\right)
\nonumber\\
&\stackrel{\rm DR}{\longmapsto}& 0\,. 
\end{eqnarray}
\end{widetext} 
%%%%%%%%%%%%%%%%%%%%%%%%%%%%%%%%%%%%%%%%%%%%%%%%%%%%
In the last step we again used the fact that all scale-less loop integrals of the type ${\mathcal I}(m,n)$, 
Eq.~\eqref{Ipvertex}, vanish in DR~\cite{itzykson2012quantum}. Consequently, none of the NLO$_\alpha$ 
proton-photon vertex correction diagrams shown in Fig.~\ref{proton_vertex:NLO} contributes to the radiative 
corrections. The non-vanishing radiative corrections  from NNLO$_\alpha$    % $1/M^2$ order 
vertices (excluded in this work) are potentially expected to renormalize the proton's Sachs form factors.   %at NNLO$_\alpha$. 
This leaves us only with the TPE diagrams which do contribute to the NLO$_\alpha$ one-loop radiative 
corrections for the LO Born lepton-proton elastic scattering cross section.

%\vspace{0.1cm}

%%%%%%%%%%%%%%%%%%%%%%%%%%%%%%%%%%%%%%%%%%%%%%%%%%%%%%%%%%%%%%%
\noindent{\it 3. Two-Photon Exchange Corrections.\,}  
%%%%%%%%%%%%%%%%%%%%%%%%%%%%%%%%%%%%%%%%%%%%%%%%%%%%%%%%%%%%%%%
The NLO${_\alpha}$ TPE diagrams of ${\mathcal O}\left(e^2\alpha{\cal Q}/M\right)$, comprising of the 
{\it direct-box, crossed-box} and {\it seagull} amplitudes, contribute to the fractional radiative 
correction, $\delta^{(1)}_{\gamma\gamma}\sim {\mathcal O}\left(\alpha{\cal Q}/M\right)$. The TPE box 
amplitudes are  IR-divergent and their exact analytical evaluation involves a intricate system of 
scalar and tensor {\it three- and four-point integral} functions and their derivatives. In contrast to
the relativistic treatment of the proton propagator within the TPE loops in $d$-dimension, the 
integrals involving the non-relativistic ``heavy nucleon" propagator is a challenge in $d$-dimension. 
To the best of our knowledge an exact analytical evaluation of such ``heavy baryon TPE loop" functions 
in order to isolate the IR divergences has not been pursued in the literature. However, we remark that
efforts are currently underway~\cite{Poonam2021} to analytically isolate the IR singularities from the
box integrals in the context of a cut-off regularization scheme. A direct numerical evaluation of the 
TPE loops without approximation {\it per se} may not be feasible unless the IR-divergent parts are 
first analytically isolated. We therefore rely on an approximate analytical method to evaluate the TPE
box amplitudes in order to project out the IR-singular parts, as outlined in details in the work of 
Ref.~\cite{Talukdar:2019dko}. Notably, the IR-finite TPE seagull amplitude can be straightforwardly 
evaluated analytically without any approximations.  

%\vspace{0.05cm}	

Following the seminal review of Mo and Tsai~\cite{Mo:1968cg}, and as later advocated in the work of 
Koshchii and Afanasev~\cite{Koshchii2017}, Ref.~\cite{Talukdar:2019dko} evaluated the TPE box diagrams 
invoking SPA. A ``less drastic" variant of this approximation was advocated in the work of Maximon and 
Tjon~\cite{Maximon:2000hm}. The use of SPA has the advantage that the seemingly intractable four-point 
functions get reduced to scalar three-point integrals which can be readily evaluated in analytical form. 
The disadvantage of this methodology is that, while the vital IR-divergent parts are evaluated correctly, 
the numerically small finite parts are estimated only partially up to term that preclude the TPE 
kinematical region of simultaneous propagation of two hard photons (see, e.g., Ref.~\cite{Tomalak:2014dja}). 
As demonstrated in Ref.~\cite{Talukdar:2019dko}, using SPA the TPE box amplitudes get factorized into the 
LO Born amplitude $\mathcal{M}^{(0)}_\gamma$. Our NLO$_\alpha$ TPE contribution is given by the sum of the
factorizable IR-singular TPE box amplitudes and the non-factorizable IR-finite ``residual part" of the TPE
seagull amplitude (see Ref.~\cite{Talukdar:2019dko} for details):
\begin{widetext}
\begin{eqnarray}
\mathcal{M}^{lp(1)}_{\gamma\gamma;\, {\rm TPE}}
&=&\,\frac{e^2 Q^2}{16\pi^2 M E}\mathcal{M}^{(0)}_\gamma
\left\{\left[\frac{1}{|\epsilon_{\rm IR}|}+\gamma_E-\ln\left(\frac{4\pi\mu^2}{-Q^2}\right)\right]
\left[\frac{1}{\beta}\ln\sqrt{\frac{1+\beta}{1-\beta}}
+\frac{\eta}{\beta^\prime}\ln\sqrt{\frac{1+\beta^\prime}{1-\beta^\prime}}\,\,\right] \right.
\nonumber\\
&&\left. -\,\frac{1}{\beta}\left[\frac{\pi^2}{2}-\ln^2\sqrt{\frac{1+\beta}{1-\beta}}
-\text{Sp}\left(\frac{2\beta}{1+\beta}\right)\right]
-\frac{\eta}{\beta^\prime}\left[\frac{\pi^2}{2}-\ln^2\sqrt{\frac{1+\beta^\prime}{1-\beta^\prime}} 
-\text{Sp}\left(\frac{2\beta^\prime}{1+\beta^\prime}\right)\right]\,\right\}_{\rm box}  
\nonumber\\
&& +\,\frac{e^4}{16\pi^2 m^2_l M}\Bigg\{{\mathcal N}_1 {\mathcal I}_1 
- {\mathcal N}_2 \left({\mathcal I}_2 + \frac{Q^2}{m^2_l}{\mathcal I}_3\right) 
- {\mathcal N}_3\left({\mathcal I}_6 - \frac{Q^2}{m^2_l}{\mathcal I}_5\right)  
- {\mathcal N}_4\frac{Q^2}{m^2_l}{\mathcal I}_4\Bigg\}_{\rm seagull}\,,
\label{Mlp1_tpe}
\end{eqnarray}
\end{widetext}
where the TPE seagull amplitude is expressed in terms of the non-factorizable amplitudes, 
${\mathcal N_i}\not\propto {\mathcal M}^{(0)}_{\gamma}$ ($i=1,...,4$), namely, 
\begin{eqnarray}
{\mathcal N}_1 &=& \left[\bar u(p^\prime)\gamma^\mu(m_l+\slashed{p})\gamma_\mu u(p)\right]\,
\left[\chi^\dagger(p_p^\prime)\,\chi(p_p)\right]\,,
\nonumber\\
\nonumber\\
{\mathcal N}_2 &=& \left[\bar u(p^\prime)\gamma^\mu(\slashed{p}-\slashed{Q})\gamma_\mu u(p)\right]\,
\left[\chi^\dagger(p_p^\prime)\,\chi(p_p)\right]\,,
\nonumber\\
\nonumber\\
{\mathcal N}_3 &=& \left[\bar u(p^\prime)\gamma^\mu \slashed{Q} \gamma_\mu u(p)\right]\,
\left[\chi^\dagger(p_p^\prime)\,\chi(p_p)\right]\,,
\nonumber\\
\nonumber\\
{\mathcal N}_4 &=& \left[\bar u(p^\prime)\gamma^\mu(2\slashed{Q}-\slashed{p})\gamma_\mu u(p)\right]\,
\left[\chi^\dagger(p_p^\prime)\,\chi(p_p)\right].\,\,\quad
\nonumber\\
\end{eqnarray}
For brevity, the analytic expressions for the $Q^2$ dependent integrals ${\mathcal I}_{\,i=1,...,4}$ are 
given in Ref.~\cite{Talukdar:2019dko}.  

\vspace{0.1cm}

%%%%%%%%%%%%%%%%%%%%%%%%%%%%%%%%%%%%%%%%%%%%%%%%%%%%%%%%%%%%%%
\noindent{\it 4. Complete one-loop Virtual Contribution\,}
%%%%%%%%%%%%%%%%%%%%%%%%%%%%%%%%%%%%%%%%%%%%%%%%%%%%%%%%%%%%%%
Using the result for the TPE amplitude, we find that the total one-loop NLO$_\alpha$ radiative amplitude is 
\begin{eqnarray}
\mathcal{M}_{\gamma\gamma}^{(1)}=\mathcal{M}^{l(1)}_{\gamma\gamma}
+\mathcal{M}^{lp(1)}_{\gamma\gamma;\, {\rm TPE}} \,,
\end{eqnarray}
where $\mathcal{M}^{l(1)}_{\gamma\gamma}$ is determined from the lepton-photon VC and the VP contributions 
at NLO$_\alpha$,  Eqs.~\eqref{Ml1gamgam_1} and \eqref{Ml1gamgam_2}. This yields the {\it lab.}-frame 
one-loop radiative correction to the LO Born differential cross section:
\begin{eqnarray}
\Delta\left[\frac{{\rm d}\sigma^{\rm (NLO)}_{el}(Q^2)}{{\rm d}\Omega^\prime_l}\right]_{\gamma\gamma}\!\!\!
=\,\,\left[\frac{{\rm d}\sigma_{el}(Q^2)}{{\rm d}
\Omega^\prime_l}\right]_{0}\delta^{(1)}_{\gamma\gamma}(Q^2)\,,\quad\,
\nonumber\\
\end{eqnarray}
where the fractional contribution $\delta^{(1)}_{\gamma\gamma}$ including the kinematically suppressed 
${\mathcal O}(\alpha{\cal Q}^2/M^2)$ terms (contributing to the theoretical error) reads
%%%%%%%%%%%%%%%%%%%%%%%%%%%%%%%%%%%%%%%%%%%%%%%%%%%%
\begin{widetext}
\begin{eqnarray}
\delta^{(1)}_{\gamma\gamma}(Q^2) &=& \frac{\sum\limits_{spins}\left[\left|{\mathcal M}^{(0)}_{\gamma\gamma}
+{\mathcal M}^{(1)}_{\gamma\gamma}\right|^2 -|{\mathcal M}^{(1)}_\gamma|^2-2{\mathcal R}\text{e}
\left(\mathcal{M}_\gamma^{(0)\dagger}{\mathcal M}^{(1)}_\gamma
+\mathcal{M}_\gamma^{(0)\dagger}{\mathcal M}^{(0)}_{\gamma\gamma}\right)\right]}{\sum\limits_{spins}
|\mathcal{M}^{(0)}_\gamma|^2}+1
\nonumber\\
&=&\text{\bf IR}^{(1)}_{\gamma\gamma}(Q^2) +\overline{\delta}^{\rm (box)}_{\gamma\gamma}(Q^2)
+\delta^{\rm (seagull)}_{\gamma\gamma}(Q^2) + {\mathscr R}_Q\,\bar{\delta}^{(0)}_{\gamma\gamma}(Q^2) 
\left\{1+\frac{1}{2}(1+\kappa_p)^2\left(\frac{Q^2+4(m^2_l-E^2)}{Q^2+4E^2}\right)\right\} 
\nonumber\\
&& +\, (1+\kappa_p)^2{\mathscr R}_Q\,{\delta}^{(0)}_{\gamma\gamma;2}(Q^2) 
\left\{1-\frac{1}{2}\left(\frac{Q^2+4(m^2_l-E^2)}{Q^2+4E^2}\right)\right\}
+o\left(\alpha\frac{{\cal Q}^2}{M^2}\right)\,.
\label{eq:IR+finite_NLO}
\end{eqnarray}  
Here again, the symbol $o(\alpha{\cal Q}^2/M^2)$ denotes other possible virtual radiative corrections (from 
additional LECs and pion loops) of $1/M^2$ order that are not explicitly accounted for in our analysis. The
insofar obtained IR divergences stemming from the NLO$_\alpha$ one-loop lepton VC and TPE box diagrams are 
contained in 
\begin{eqnarray}
\text{\bf IR}^{(1)}_{\gamma\gamma}(Q^2)&=&\text{\bf IR}^{lp(1)}_{\gamma\gamma;\,{\rm TPE}}(Q^2)
+\text{\bf IR}^{(0)}_{\gamma\gamma}(Q^2)\,{\mathscr R}_Q\,\left\{1+\frac{1}{2}(1+\kappa_p)^2
\left(\frac{Q^2+4(m^2_l-E^2)}{Q^2+4E^2}\right)\right\}\,;\quad\,
\nonumber\\
\nonumber\\
\text{\bf IR}^{lp(1)}_{\gamma\gamma;\,{\rm TPE}} (Q^2)&=&\frac{\alpha Q^2}{2\pi M E\beta}
\left[\frac{1}{|\epsilon_{\rm IR}|}+\gamma_E-\ln\left(\frac{4\pi\mu^2}{-Q^2}\right)\right]
\left[\ln\sqrt{\frac{1+\beta}{1-\beta}}+\frac{\eta\beta}{\beta^\prime}
\ln\sqrt{\frac{1+\beta^\prime}{1-\beta^\prime}}\,\right]\,,
\label{IR1_virt}
\end{eqnarray}
where the term $\text{\bf IR}^{(0)}_{\gamma\gamma}$ is displayed in Eq.~\eqref{IR0_vert}, and the expression 
for $\text{\bf IR}^{lp(1)}_{\gamma\gamma;\,{\rm TPE}}$ is extracted from Ref.~\cite{Talukdar:2019dko}. The 
remaining IR-finite contributions $\overline{\delta}^{\rm (box)}_{\gamma\gamma}$ and 
$\delta^{\rm (seagull)}_{\gamma\gamma}$ originating from the TPE diagrams are respectively given 
as~\cite{Talukdar:2019dko} 
\begin{equation}
\overline{\delta}^{\rm (box)}_{\gamma\gamma}(Q^2)=
-\frac{\alpha Q^2}{2\pi ME\beta}\left\{\left[\frac{\pi^2}{2}-\ln^2\sqrt{\frac{1+\beta}{1-\beta}}
-\text{Sp}\left(\frac{2\beta}{1+\beta}\right)\right]
+\frac{\eta \beta}{\beta^\prime}\left[\frac{\pi^2}{2}-\ln^2\sqrt{\frac{1+\beta^\prime}{1-\beta^\prime}} 
-\text{Sp}\left(\frac{2\beta^\prime}{1+\beta^\prime}\right)\right]\,\right\}\,, 
\label{eq:finite_delta_tpe_box}
\end{equation}
and
\begin{equation}
\delta^{\rm (seagull)}_{\gamma\gamma}(Q^2)= 
-\frac{2\alpha Q^2}{\pi M E}\left[\frac{E^2(1+\eta)}{\eta Q^2+4E^2}\right]
\left\{{\mathcal I}_1(Q^2)+{\mathcal I}_2(Q^2)
+\frac{Q^2}{m^2_l}\left[{\mathcal I}_3(Q^2)-{\mathcal I}_4(Q^2)\right]\right\}\,.
\label{eq:finite_delta_tpe_sea} 
\end{equation}
\end{widetext}
%%%%%%%%%%%%%%%%%%%%%%%%%%%%%%%%%%%%%%%%%%%%%%%%%%%%
As already pointed out, all non-vanishing ${\mathcal O}\left(e^2\alpha{\cal Q}/M\right)$ one-loop 
diagrams, displayed in Figs.~\ref{leptonic:NLO} and \ref{proton_vertex:NLO}, though formally expected
to contribute to the NLO$_\alpha$ virtual corrections, 
$\overline{\delta}^{(1)}_{\gamma\gamma}\sim {\mathcal O}\left(\alpha{\cal Q}/{M}\right)$, are in effect 
kinematically suppressed, contributing at NNLO$_\alpha$. Therefore, such contributions shall be dropped 
in presenting our central results. However, here we prefer to retain the full structure of the 
IR-divergent terms, i.e., including also the $1/M^2$ suppressed contributions proportional to 
$\mathscr{R}_Q$, Eq.~\eqref{IR1_virt}, in order to demonstrate their complete cancellations after 
including the corresponding NLO$_\alpha$ bremsstrahlung counterpart. Thus, at 
${\mathcal O}\left(\alpha{\cal Q}/{M}\right)$, the IR-finite part of the one-loop NLO$_\alpha$ radiative 
corrections, that arises solely from the NLO$_\alpha$ TPE contributions,   %and 
reads 
%%%%%%%%%%%%%%%%%%%%%%%%%%%%%%%%%%%%%%%%%FIGURE%%%%%%%%%%%%%%%%%%%%%%%%%%%%%%%%%%%%%%%%%%%%%	
\begin{figure*}[tbp]
 \centering
        \includegraphics[scale=0.6]{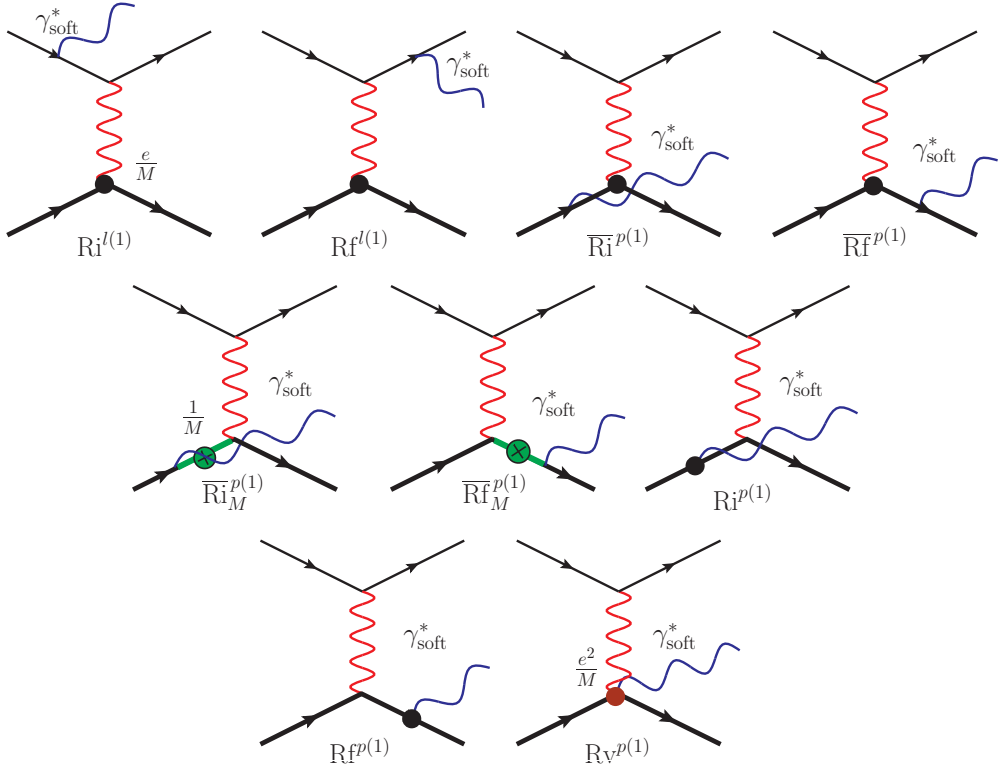}
	    \caption{Soft bremsstrahlung diagrams at NLO$_\alpha$ in HB$\chi$PT [i.e., 
	             ${\mathcal O}\left(e^3{\cal Q}/{M}\right)$], contributing to the radiative 
	             corrections to the elastic $\ell$-p scattering amplitude [see 
	             Eq.~\eqref{M0_gamma}]. The filled blobs represent $1/M$ order proton-photon 
	             vertices. In particular, the two-photon proton vertex (diagram Rv$^{\,p(1)}$) 
	             is of order $e^2/M$. The proton propagators with the crossed blobs ``$\otimes$" 
	             represent $1/M$ order propagator insertions. The amplitudes (diagrams 
	             $\overline{\rm Ri,f}^{\,p(1)}$ and $\overline{\rm Ri,f}^{\,p(1)}_M$) with the 
	             leading order proton-photon vertices trivially vanish.}
        \label{nlobrem}
\end{figure*}
%%%%%%%%%%%%%%%%%%%%%%%%%%%%%%%%%%%%%%%%%%%%%%%%%%%%%%%%%%%%%%%%%%%%%%%%%%%%%%%%%%%%%%%%%%%%
%%%%%%%%%%%%%%%%%%%%%%%%%%%%%%%%%%%%%%%%%%%%%%%%%%%%
\begin{widetext}
\begin{eqnarray}
\overline{\delta}^{(1)}_{\gamma\gamma}(Q^2)
&=&\overline{\delta}^{\rm (box)}_{\gamma\gamma}(Q^2)+\delta^{\rm (seagull)}_{\gamma\gamma}(Q^2)
=-\frac{\alpha Q^2}{\pi ME\beta}\Bigg[\frac{\pi^2}{2}+\ln\left(\frac{-Q^2}{m^2_l}\right)
\ln\sqrt{\frac{1+\beta}{1-\beta}}-\ln^2\sqrt{\frac{1+\beta}{1-\beta}}
\nonumber\\
&&-\,\text{Sp}\left(\frac{2\beta}{1+\beta}\right)
+\frac{4E^2\beta}{Q^2+4E^2}\left\{{\mathcal I}_1(Q^2)+{\mathcal I}_2(Q^2)
+\frac{Q^2}{m^2_l}\Big[{\mathcal I}_3(Q^2)-{\mathcal I}_4(Q^2)\Big]\right\}\Bigg]
+{\mathcal O}\left(\alpha\frac{{\cal Q}^2}{M^2}\right).\,\quad
\label{delta:NLO_tpe}
\end{eqnarray}
\end{widetext}
%%%%%%%%%%%%%%%%%%%%%%%%%%%%%%%%%%%%%%%%%%%%%%%%%%%%
Note that in order to arrive at the above NLO$_\alpha$ expression we removed terms of ${\cal O}(1/M^2)$ by 
replacing $E^\prime\to E$ (i.e, $\eta=1$) and $\beta^\prime\to \beta$. 

%\vspace{0.1cm}

%%%%%%%%%%%%%%%%%%%%%%%%%%%%%%%%%%%%%%%%%%%%%%%%%%%%%%
\subsection{Soft Bremsstrahlung Corrections at NLO$_\alpha$} 
%%%%%%%%%%%%%%%%%%%%%%%%%%%%%%%%%%%%%%%%%%%%%%%%%%%%%%
In HB$\chi$PT, all bremsstrahlung diagrams with leading order photon emission proton vertex, e.g., 
diagrams Ri,f${}^{p(0)}$ in Fig.~\ref{lobrem} and diagrams $\overline{\rm Ri,f}^{\,p(1)}$ and 
$\overline{\rm Ri,f}^{\,p(1)}_M$ in Fig.~\ref{nlobrem} respectively, proportional to $v\cdot\varepsilon^*$ 
vanish at the amplitude level. The NLO$_\alpha$ bremsstrahlung diagrams with a $1/M$ order photon emission 
vertex yield the first non-vanishing contributions. The soft bremsstrahlung corrections at NLO$_\alpha$ 
originate from diagrams with either the lepton or proton emitting a single undetectable soft photon 
($\gamma^*_{\rm soft}$), as illustrated in Fig.~\ref{nlobrem}.  We use the soft photon limit $k\to 0$ to 
evaluate the IR-divergent contributions to the cross section at NLO$_\alpha$. These contributions arise 
from the diagrams, namely, Ri${}^{l(1)}$, Rf${}^{l(1)}$, Ri${}^{p(1)}$ and Rf${}^{p(1)}$, in 
Fig.~\ref{nlobrem} which get factorized into the LO$_\chi$ and NLO$_\chi$ Born amplitudes 
$\mathcal{M}^{(0,1)}_\gamma$. In the $k\to 0$ limit, the amplitudes are respectively given as: 
\begin{eqnarray}
{\mathcal{M}}^{l(1);\,{\rm i}}_{\gamma\gamma^*} \,\,&\stackrel{\gamma^*_{\rm soft}}{\leadsto}&\,\, 
\widetilde{\mathcal{M}^{l(1);\,{\rm i}}_{\gamma\gamma^*}}
=e\mathcal{M}^{(1)}_\gamma\left(\frac{p\cdot\varepsilon^*} {p\cdot k}\right)\,,
\nonumber\\
\nonumber\\
{\mathcal{M}}^{l(1);\,{\rm f}}_{\gamma\gamma^*} \,\,&\stackrel{\gamma^*_{\rm soft}}{\leadsto}&\,\, 
\widetilde{\mathcal{M}^{l(1);\,{\rm f}}_{\gamma\gamma^*}}
=-e\mathcal{M}^{(1)}_\gamma\left(\frac{p^\prime\cdot\varepsilon^*} {p^\prime\cdot k}\right)\,.
\nonumber\\
\nonumber\\
{\mathcal{M}}^{p(1);\,{\rm i}}_{\gamma\gamma^*} \,\,&\stackrel{\gamma^*_{\rm soft}}{\leadsto}&\,\, 
\widetilde{\mathcal{M}^{p(1);\,{\rm i}}_{\gamma\gamma^*}}
=\frac{e}{M}\mathcal{M}^{(0)}_\gamma\left(\frac{p_p\cdot\varepsilon^*} {v\cdot k}\right)=0\,,
\nonumber\\
\nonumber\\
{\mathcal{M}}^{p(1);\,{\rm f}}_{\gamma\gamma^*} \,\,&\stackrel{\gamma^*_{\rm soft}}{\leadsto}&\,\, 
\widetilde{\mathcal{M}^{p(1);\,{\rm f}}_{\gamma\gamma^*}}
=-\frac{e}{M}\mathcal{M}^{(0)}_\gamma\left(\frac{p_p^\prime\cdot\varepsilon^*} {v\cdot k}\right)\,,
\nonumber\\
\end{eqnarray}
where the otherwise non-zero amplitude ${\mathcal{M}}^{p(1);\,{\rm i}}_{\gamma\gamma^*}$ trivially vanishes 
with the initial state proton's residual four-momentum vector as $p_p=0$ in the {\it lab.}-frame. Moreover, 
the proton's spin-dependent terms proportional to the commutator $[S_\mu, S\cdot k]$ also vanishes on the 
soft photon limit. Evidently, restricting to ${\mathcal O}\left(\alpha^3{\cal Q}/{M}\right)$ the bremsstrahlung 
cross section gets contributions only from the interference between the LO$_\alpha$ and NLO$_\alpha$ 
bremsstrahlung amplitudes. Nevertheless, in order to keep track of the systematic uncertainties, we evaluate 
all possible ${\mathcal O}\left(\alpha^3{\cal Q}^2/{M}^2\right)$ terms that may arise from the above amplitudes. 
This requires us to evaluate the squared modulus of the full bremsstrahlung amplitude up-to-and-including 
NLO$_\alpha$, namely, $|\mathcal{M}^{(0)}_{\gamma\gamma^*}+\mathcal{M}^{(1)}_{\gamma\gamma^*}|^2$, and retain 
${\mathcal O}\left(\alpha^3{\cal Q}^2/{M}^2\right)$ terms. Notably, the complete 
${\mathcal O}\left(\alpha^3{\cal Q}^2/{M}^2\right)$ expression for the cross section will need additional 
$1/M^2$ or NNLO$_\alpha$ bremsstrahlung diagrams which we currently exclude commensurate with the NLO$_\alpha$ 
virtual corrections. In contrast to the aforementioned factorizable amplitudes (diagrams Ri,f$^{l(0,1)}$ and 
Ri,f$^{p(1)}$) which potentially give rise to IR divergences, the diagram Rv$^{p(1)}$ with a proton-two-photon 
NLO$_\alpha$ vertex and a non-factorizable amplitude, namely,
\begin{eqnarray}
{\mathcal{M}}^{p(1);\,{\rm v}}_{\gamma\gamma^*}=-\frac{e^3}{M Q^2}\left[{\bar u}_l(p^\prime) 
\slashed{\varepsilon} u_l(p)\right]\left[\chi^\dagger(p_p^\prime) \chi(p_p)\right]\,,\qquad\,
\end{eqnarray} 
yields IR-finite contributions only. The corresponding cross section terms are evaluated without explicitly 
taking the $k\to 0$ limit, as reminiscent of the IR-finite one-loop virtual counterpart arising from the TPE 
seagull diagram having the same NLO$_\alpha$ vertex. To this end, the total bremsstrahlung amplitude using 
the appropriate soft photon limits is given by the sum
\begin{eqnarray}
\widetilde{\mathcal{M}^{(0)}_{\gamma\gamma^*}}+\widetilde{\mathcal{M}^{(1)}_{\gamma\gamma^*}}
&=& \widetilde{\mathcal{M}^{l(0);\,{\rm i}}_{\gamma\gamma^*}} 
+ \widetilde{\mathcal{M}^{l(0);\,{\rm f}}_{\gamma\gamma^*}} 
+\widetilde{\mathcal{M}^{l(1);\,{\rm i}}_{\gamma\gamma^*}}
\nonumber\\
&& +\, \widetilde{\mathcal{M}^{l(1);\,{\rm f}}_{\gamma\gamma^*}} 
+ \widetilde{\mathcal{M}^{p(1);\,{\rm f}}_{\gamma\gamma^*}}
+{\mathcal{M}}^{p(1);\,{\rm v}}_{\gamma\gamma^*}\,.
\nonumber\\
\end{eqnarray}

Next, the relevant contributions to the squared modulus of the full NLO$_\alpha$ amplitude read
%%%%%%%%%%%%%%%%%%%%%%%%%%%%%%%%%%%%%%%%%%%%%%%%%%%%
\begin{widetext}
\begin{eqnarray}
&&\sum_{spins}\Bigg[|\widetilde{\mathcal{M}^{(0)}_{\gamma\gamma^*}}
+\widetilde{\mathcal{M}^{(1)}_{\gamma\gamma^*}}|^2 
- |\widetilde{\mathcal{M}^{(0)}_{\gamma\gamma^*}}|^2\Bigg]\equiv \sum_{spins}
\Bigg[|\widetilde{\mathcal{M}^{(1)}_{\gamma\gamma^*}}|^2 + 2\,{\mathcal R}\text{e}\!
\left(\widetilde{\mathcal{M}^{(0)}_{\gamma\gamma^*}}^{\dagger}\,
\widetilde{\mathcal{M}^{(1)}_{\gamma\gamma^*}}\right)\Bigg]
\nonumber\\
&=& -\,e^2\sum_{spins}|\mathcal{M}_\gamma^{(0)}|^2\Bigg[\left({\mathscr R}_Q
+\frac{\sum\limits_{spins}|{\mathcal M}^{(1)}_\gamma|^2}{\sum\limits_{spins}
|\mathcal{M}^{(0)}_\gamma|^2}\right)\left(\frac{m^2_l}{(p\cdot k)^2}+\frac{m^2_l}{(p^\prime\cdot k)^2}
-\frac{2 p^\prime\cdot p}{(p\cdot k)(p^\prime\cdot k)}\right)
\nonumber\\
&& \hspace{2.8cm}+\,\frac{{Q}^2}{M}\left(\frac{1}{(v\cdot k)(p\cdot k)}
+\frac{1}{(v\cdot k)(p^\prime\cdot k)}\right) 
+ {\mathscr R}_Q\,\frac{2}{(v\cdot k)^2}+{\mathcal O}\left(\frac{{\cal Q}^3}{M^3}\right) \Bigg]
\nonumber\\
&& +\,\sum_{spin} \left[2{\mathcal R}\text{e}\,\left(\mathcal{M}^{(0)}_{\gamma\gamma^*}
+\mathcal{M}^{(1)}_{\gamma\gamma^*}\right)^\dagger\mathcal{M}^{p(1);\,{\rm v}}_{\gamma\gamma^*}
-|\mathcal{M}^{p(1);\,{\rm v}}_{\gamma\gamma^*}|^2\right]\,.
\label{M1gamgam*2}
\end{eqnarray}
\end{widetext}
%%%%%%%%%%%%%%%%%%%%%%%%%%%%%%%%%%%%%%%%%%%%%%%%%%%% 
The first set of terms, namely, those due to the lepton-lepton and lepton-proton bremsstrahlung are 
factorizable being proportional to the squared modulus of the LO Born amplitude $\mathcal{M}_\gamma^{(0)}$. 
The first of these terms are analogous to the LO$_\alpha$ bremsstrahlung contribution, Eq.~\eqref{M0gamgam*2}, 
apart from the extra $1/M^2$ order pre-factors. All such $1/M^2$ order IR-finite terms may be dropped from 
our analytical expressions, as their primary purpose in this work is to contribute to the theoretical error. 
Nevertheless, we prefer to retain all IR-singular contributions up to $1/M^2$ order to demonstrate their 
order by order cancellations with the corresponding IR-divergent one-loop counterparts. The latter set of 
interference terms with the proton-two-photon vertex diagram  lead to IR-finite non-factorizable contributions 
and may be readily evaluated, namely,
%%%%%%%%%%%%%%%%%%%%%%%%%%%%%%%%%%%%%%%%%%%%%%%%%%%%
\begin{widetext}
\begin{eqnarray}
\sum_{spin} && |\mathcal{M}^{p(1);\,{\rm v}}_{\gamma\gamma^*}|^2=
-\frac{32 e^6}{Q^6} M\left\{\frac{\Delta^{(1/M)}_{p(1)\rm v - p(1)\rm v}}{M}\right\}
+ {\mathcal O}\left(\alpha^3\frac{\cal Q}{M}\right)\,,
\nonumber\\
\nonumber\\
\sum_{spin} && 2{\mathcal R}\text{e}\left({\mathcal{M}^{l(0);\,{\rm i}}_{\gamma\gamma^*}}^\dagger
\mathcal{M}^{p(1);\,{\rm v}}_{\gamma\gamma^*}\right)= -\frac{64 e^6}{Q^2q^2 (p\cdot k)} M 
\Bigg\{m^2_l(E+E^\prime) - E (p^\prime \cdot k)+ E^\prime (p\cdot k)+(v\cdot k) \left(\frac{1}{2}Q^2-m^2_l\right)
\nonumber\\
&& \hspace{6.7cm} 
+\, \frac{\Delta^{(1/M)}_{l(0){\rm i}-p(1){\rm v}}}{M} \Bigg\} 
+ {\mathcal O}\left(\!\alpha^3\frac{\cal Q}{M}\!\right),
\nonumber\\
\nonumber\\
\sum_{spin} && 2{\mathcal R}\text{e}\left({\mathcal{M}^{l(0);\,{\rm f}}_{\gamma\gamma^*}}^\dagger
\mathcal{M}^{p(1);\,{\rm v}}_{\gamma\gamma^*}\right)= \frac{64 e^6}{Q^2q^2 (p^\prime \cdot k)} M 
\Bigg\{m^2_l(E+E^\prime) - E (p^\prime \cdot k)+ E^\prime (p\cdot k)-(v\cdot k) \left(\frac{1}{2}Q^2-m^2_l\right)
\nonumber\\
&& \hspace{6.7cm} + \frac{\Delta^{(1/M)}_{l(0){\rm f}-p(1){\rm v}}}{M} \Bigg\} 
+ {\mathcal O}\left(\alpha^3\frac{\cal Q}{M}\right),
\nonumber\\
\nonumber\\
\sum_{spin} && 2{\mathcal R}\text{e}\left({\mathcal{M}^{l(1);\,{\rm i}}_{\gamma\gamma^*}}^\dagger
\mathcal{M}^{p(1);\,{\rm v}}_{\gamma\gamma^*}\right)= 
-\frac{32 e^6}{Q^2 q^2 (p\cdot k)}M \Bigg\{\frac{\Delta^{(1/M)}_{l(1){\rm i}-p(1){\rm v}}}{M}\Bigg\} 
+ {\mathcal O}\left(\alpha^3\frac{\cal Q}{M}\right)\,,
\nonumber\\
\nonumber\\
\sum_{spin} && 2{\mathcal R}\text{e}\left({\mathcal{M}^{l(1);\,{\rm f}}_{\gamma\gamma^*}}^\dagger
\mathcal{M}^{p(1);\,{\rm v}}_{\gamma\gamma^*}\right)= 
-\frac{32 e^6}{Q^2 q^2 (p^\prime\cdot k)}M \Bigg\{\frac{\Delta^{(1/M)}_{l(1){\rm f}-p(1){\rm v}}}{M}\Bigg\} 
+ {\mathcal O}\left(\alpha^3\frac{\cal Q}{M}\right)\,,
\nonumber\\
\nonumber\\
\sum_{spin} && 2{\mathcal R}\text{e}\left({\mathcal{M}^{p(1);\,{\rm i}}_{\gamma\gamma^*}}^\dagger
\mathcal{M}^{p(1);\,{\rm v}}_{\gamma\gamma^*}\right)=
\sum_{spin} 2{\mathcal R}\text{e}\left({\mathcal{M}^{p(1);\,{\rm f}}_{\gamma\gamma^*}}^\dagger
\mathcal{M}^{p(1);\,{\rm v}}_{\gamma\gamma^*}\right)=0\, .
\label{nlobrem_pvamps}
\end{eqnarray}
The last two terms exactly vanish, and the $\Delta^{(1/M)}$s are finite terms contributing to the cross 
section at NNLO$_\alpha$. Hence, in Eq.~\eqref{nlobrem_pvamps}, apart from the interference terms with 
the LO$_\alpha$ lepton bremsstrahlung amplitude, all the remaining terms are needed only to estimate the 
theoretical error incurred in our NLO$_\alpha$ results. We may explicitly spell-out these terms (replacing
$E^\prime\to E$ at this order), namely,
\begin{eqnarray}
\Delta^{(1/M)}_{p(1){\rm v}-p(1){\rm v}}&=& 2Q^2\left(\frac{1}{2}Q^2 + m^2_l\right)\,,
\nonumber\\
\nonumber\\
\Delta^{(1/M)}_{l(0){\rm i}-p(1){\rm v}}&=& -\frac{1}{2}(v\cdot k) E\left[2m^2_l + Q\cdot k\right]
-\frac{1}{2}(v\cdot k)^2 \left(\frac{1}{2}Q^2-m^2_l\right) \,,
\nonumber\\
\nonumber\\
\Delta^{(1/M)}_{l(0){\rm f}-p(1){\rm v}}&=& -\frac{1}{2}(v\cdot k)E\left[2m^2_l + Q\cdot k  \right]
+\frac{1}{2}(v\cdot k)^2 \left(\frac{1}{2}Q^2-m^2_l\right) \,,
\nonumber\\
\nonumber\\
\Delta^{(1/M)}_{l(1){\rm i}-p(1){\rm v}}&=& 
-2(p^\prime\cdot k) \left(\frac{1}{2} Q^2 - m_l^2\right) - 2m_l^2 (Q\cdot k) \,,
\nonumber\\
\nonumber\\
\Delta^{(1/M)}_{l(1){\rm f}-p(1){\rm v}}&=& 
2(p^\prime\cdot k) \left(\frac{1}{2} Q^2 + m_l^2\right) +\frac{1}{2}m_l^2 Q^2 \,.
\label{eq:Delta_M}
\end{eqnarray}
Notably, with $|\mathcal{M}^{(0)}_\gamma|^2\propto M^2$ contributing to the LO Born cross section, for the 
error estimate it is sufficient to retain at the most $M^0$ order terms as displayed above. Next, the 
{\it lab.}-frame soft bremsstrahlung correction to the elastic LO differential cross section at NLO$_\alpha$ 
accuracy in HB$\chi$PT is expressed as
\begin{eqnarray}
\Delta\left[\frac{{\rm d}\sigma^{(\text{NLO})}_{br}(Q^2)}{{\rm d}
\Omega^\prime_l}\right]_{\gamma\gamma^*} &\stackrel{\gamma^*_{\rm soft}}{\leadsto}& 
\frac{\alpha}{2\pi^2}\left[\frac{{\rm d}\sigma_{el}(Q^2)}{{\rm d}\Omega^\prime_l}\right]_0
\nonumber\\
&& \hspace{0.4cm} \times\,\Bigg[\left(-L_{\rm ii} - L_{\rm ff} + L_{\rm if}\right)\mathscr{R}_Q\,
\left\{1+\frac{1}{2}(1+\kappa_p)^2\left(\frac{ Q^2+4(m^2_l-E^2)}{Q^2+4E^2}\right)\right\} 
\nonumber\\
&& \hspace{0.9cm}-\,\frac{Q^2}{M}(L_{\rm i}+L_{\rm f})\Bigg] 
+ \left[\frac{{\rm d}\sigma_{br}(Q^2)}{{\rm d}
\Omega^\prime_l}\right]^{\,lp(1);\,{\rm v}}_{\gamma\gamma^*}\,,\quad
\label{NLO:brem_cross}
\end{eqnarray}
\end{widetext}
%%%%%%%%%%%%%%%%%%%%%%%%%%%%%%%%%%%%%%%%%%%%%%%%%%%%
The IR-divergent integrals, $L_{\rm ii},\,L_{\rm ff}$ and $L_{\rm if}$, are identical to the ones we obtained 
in our LO$_\alpha$ bremsstrahlung evaluations. The two new integrals, $L_{\rm i}$ and $L_{\rm f}$, appearing 
at this order stems from the factorizable lepton-proton interference contribution. All these integrals are 
IR-divergent and conveniently evaluated using DR, as detailed in Appendix B. In particular, to evaluate 
Eq.~\eqref{NLO:brem_cross} at NLO$_\alpha$ accuracy, the exact expression of these integrals displayed in the 
appendix could be approximated with $E^\prime\to E,\, \beta^\prime \to \beta$, leading to 
$L_{\rm ii}=L_{\rm ff}$ and $L_{\rm i}=L_{\rm f}$. The last non-factorizable contribution, namely, 
$\displaystyle\left[\frac{{\rm d}\sigma_{br}}{{\rm d}
\Omega^\prime_l}\right]^{\,lp(1);\,{\rm v}}_{\gamma\gamma^*}$ \!\!\!\! is IR-finite and readily evaluated 
using Eq.~\eqref{nlobrem_pvamps} [see Eq.~\eqref{x_sect:dia_v} in Appendix B]. The resulting expression in the 
{\it lab.}-frame yields the  correction due to soft photon emission with energy less than $\Delta_{\gamma^*}$, 
namely,
\begin{eqnarray}
\Delta\left[\frac{{\rm d}\sigma^{(\text{NLO})}_{br}(Q^2)}{{\rm d}\Omega^\prime_l}
\right]^{(E_{\gamma^*} < \Delta_{\gamma^*})}
\hspace{-0.3cm}=\left[\frac{{\rm d}\sigma_{el}(Q^2)}{{\rm d}\Omega^\prime_l}\right]_0 \!
\delta^{(1)}_{\gamma\gamma^*}(Q^2)\,,
\nonumber\\
\end{eqnarray}
where the fractional NLO$_\alpha$ bremsstrahlung contribution is given as
\begin{eqnarray}
\delta^{(1)}_{\gamma\gamma^*}(Q^2)=\text{\bf IR}^{(1)}_{\gamma\gamma^*}(Q^2)
+\overline{\delta}^{(1)}_{\gamma\gamma^*}(Q^2)\,.
\label{x_sec:brem_NLO}
\end{eqnarray}
Using $\widetilde{L}_{\rm ff}=\widetilde{L}_{\rm ii}$ and $\widetilde{L}_{\rm f}=\widetilde{L}_{\rm i}$ 
at NLO$_\alpha$ accuracy, the finite part of the contribution is expressed as 
%%%%%%%%%%%%%%%%%%%%%%%%%%%%%%%%%%%%%%%%%%%%%%%%%%%%
\begin{widetext}
\begin{eqnarray}
\overline{\delta}^{(1)}_{\gamma\gamma^*}(Q^2)=
\frac{\alpha}{\pi}\Bigg[(-2\widetilde{L}_{\rm ii}+\widetilde{L}_{\rm if}){\mathscr R}_Q\!
\left\{\!1+\frac{1}{2}(1+\kappa_p)^2\left(\frac{Q^2+4(m^2_l-E^2)}{Q^2+4E^2}\right)\!\right\} 
\!-\!\frac{2Q^2}{M}\widetilde{L}_{\rm i}\Bigg] 
\!+\!\delta^{\,lp(1);\,{\rm v}}_{\gamma\gamma^*}(Q^2)\,\,, 
\nonumber
\end{eqnarray}
with
\begin{eqnarray}
\delta^{\,lp(1);\,{\rm v}}_{\gamma\gamma^*}(Q^2)
=\left[\frac{{\rm d}\sigma_{br}(Q^2)}{{\rm d}\Omega^\prime_l}\right]^{\,lp(1);\,{\rm v}}_{\gamma\gamma^*}\!\!\!\!
\Bigg/\left[\frac{{\rm d}\sigma_{el}(Q^2)}{{\rm d}\Omega^\prime_l}\right]_0,
\label{eq:delta_brems} 
\end{eqnarray}
whose explicit expression due to the various interference terms involving the diagram Rv$^{p(1)}$ is worked 
out in Appendix B. The term
\begin{eqnarray}
\text{\bf IR}^{(1)}_{\gamma\gamma^*}(Q^2)&=& 
\text{\bf IR}^{\,lp(1)}_{\gamma\gamma^*}(Q^2)+\text{\bf IR}^{(0)}_{\gamma\gamma^*}(Q^2)\mathscr{R}_Q
\left\{1+\frac{1}{2}(1+\kappa_p)^2\left(\frac{Q^2+4(m^2_l-E^2)}{Q^2+4E^2}\right)\right\} 
\nonumber\\
\nonumber\\
&=&-\,\text{\bf IR}^{\,lp(1)}_{\gamma\gamma;\,{\rm TPE}}(Q^2)-\text{\bf IR}^{(0)}_{\gamma\gamma}(Q^2)\mathscr{R}_Q
\left\{1+\frac{1}{2}(1+\kappa_p)^2\left(\frac{Q^2+4(m^2_l-E^2)}{Q^2+4E^2}\right)\right\}\, 
\end{eqnarray} 
\end{widetext}
%%%%%%%%%%%%%%%%%%%%%%%%%%%%%%%%%%%%%%%%%%%%%%%%%%%%
collects the IR divergences formally arising from the soft bremsstrahlung contributions of the LO$_\alpha$ 
and NLO$_\alpha$ diagrams with $\text{\bf IR}^{(0)}_{\gamma\gamma^*}=-\text{\bf IR}^{(0)}_{\gamma\gamma}$ 
[cf. Eqs.~\eqref{IR0_vert} and \eqref{IR0_brem}], and 
$\text{\bf IR}^{\,lp(1)}_{\gamma\gamma^*}=-\text{\bf IR}^{\,lp(1)}_{\gamma\gamma;\,{\rm TPE}}$, [cf. 
Eq.\eqref{IR1_virt}]. Thus, as anticipated, $\text{\bf IR}^{(1)}_{\gamma\gamma^*}$ is exactly equal in 
magnitude but opposite in sign to $\text{\bf IR}^{(1)}_{\gamma\gamma}$ [cf. Eq.~\eqref{IR1_virt}]. 
Consequently, the sum of the real and virtual radiative corrections at NLO$_\alpha$, namely, 
\begin{eqnarray}
\delta_{2\gamma}^{(1)}(Q^2)&=&\delta^{(1)}_{\gamma\gamma}(Q^2)+\delta^{(1)}_{\gamma\gamma^*}(Q^2)
\nonumber\\
&\equiv&\overline{\delta}^{(1)}_{\gamma\gamma}(Q^2)+\overline{\delta}^{(1)}_{\gamma\gamma^*}(Q^2)\,
\end{eqnarray}
is free of IR divergences, where $\overline{\delta}^{(1)}_{\gamma\gamma}$ represents the finite part of the 
NLO$_\alpha$ TPE contributions [cf. Eq.~\eqref{delta:NLO_tpe}]. Furthermore, the NNLO$_\alpha$ error terms 
that we partially considered is also shown to be IR-free. This, of course, does not preclude the presence of 
further IR-singularities which may arise from various %real and virtual 
NNLO$_\alpha$ contributions not 
included  in this analysis.  

\vspace{-0.05cm}

Having established the complete cancellation of the IR-singularities among the NLO$_\alpha$ virtual and real 
(soft) photon emission diagrams, we explicitly drop all terms of $1/M^{2}$ order, i.e., terms beyond our 
intended order of accuracy. Such excluded terms also include ``implicit" $1/M^2$ order terms proportional 
to $(E-E^\prime)/M$ and $(\beta-\beta^\prime)/M$, that is justified following the replacements, 
$E^\prime\to E$, $\eta\to 1$ and $\beta^\prime\to \beta$ in all the NLO$_\alpha$ expression. This yields the 
finite NLO$_\alpha$ bremsstrahlung contribution, which modifies the total fractional elastic contribution 
and is given by
%%%%%%%%%%%%%%%%%%%%%%%%%%%%%%%%%%%%%%%%%%%%%%%%%%%%
\begin{widetext}
\begin{eqnarray}
\overline{\delta}^{(1)}_{\gamma\gamma^*}(Q^2) 
&=&-\,\frac{\alpha Q^2}{\pi M E \beta }
\Bigg[\ln\left(\frac{4\Delta^{2}_{\gamma^*}}{-Q^2}\right)\ln\sqrt{\frac{1+\beta}{1-\beta}}
+\frac{1}{2}\text{Sp}\left(\frac{2\beta}{\beta+1}\right)
-\frac{1}{2}\text{Sp}\left(\frac{2\beta}{\beta-1}\right)
\nonumber\\
&& \hspace{1.6cm}+\,\frac{32 \beta E^2 \Delta^2_{\gamma^*}}{Q^2(Q^2+4 E^2)}\,\Bigg] 
+ {\mathcal O}\left(\alpha\frac{{\mathcal Q}^2}{M^2}\right)\,.
\label{delta:NLO_brem}
\end{eqnarray}
\end{widetext}
%%%%%%%%%%%%%%%%%%%%%%%%%%%%%%%%%%%%%%%%%%%%%%%%%%%%

%%%%%%%%%%%%%%%%%%%%%%%%%%%%%%%%%%%%%%%%%%%%%%%%%%%%%% 
\subsection{Total Radiative Corrections at NLO$_\alpha$} 
%%%%%%%%%%%%%%%%%%%%%%%%%%%%%%%%%%%%%%%%%%%%%%%%%%%%%%
By using our analytically derived NLO$_\alpha$ expressions for the IR-finite virtual (i.e., TPE) and real 
contributions [cf. Eqs.~\eqref{delta:NLO_tpe} and \eqref{delta:NLO_brem}] , we obtain the total fractional 
radiative corrections to the elastic differential cross section accurate up to order $1/M$, which reads  
\begin{widetext}
\begin{eqnarray}
\delta_{2\gamma}^{(1)}(Q^2)
&=&-\,\frac{\alpha Q^2}{\pi M E \beta }\left[\ln\left(\frac{4\Delta^{2}_{\gamma^*}}{-Q^2}\right)
\ln\sqrt{\frac{1+\beta}{1-\beta}}-\frac{1}{2}\text{Sp}\left(\frac{2\beta}{\beta+1}\right)
-\frac{1}{2}\text{Sp}\left(\frac{2\beta}{\beta-1}\right)\right.
\nonumber\\
&&\hspace{1.6cm} \left.+\,\frac{32 \beta E^2 \Delta^2_{\gamma^*}}{Q^2(Q^2+4 E^2)}+\ln\left(\frac{-Q^2}{m^2_l}\right)
\ln\sqrt{\frac{1+\beta}{1-\beta}}-\ln^2\sqrt{\frac{1+\beta}{1-\beta}}+\frac{\pi^2}{2}\right.
\nonumber\\
&&\hspace{1.6cm} \left.+\,\frac{4E^2\beta}{Q^2+4E^2}\left\{{\mathcal I}_1(Q^2)+{\mathcal I}_2(Q^2)
+\frac{Q^2}{m^2_l}\Big({\mathcal I}_3(Q^2)-{\mathcal I}_4(Q^2)\Big)\right\}\right]
+{\mathcal O}\left(\alpha\frac{{\mathcal Q}^2}{M^2}\right)\,,
\label{delta12gamma}
\end{eqnarray}
\end{widetext}
where ${\mathcal I}_{i=1,...,4}$ are finite integrals given in Ref.~\cite{Talukdar:2019dko}.  

%%%%%%%%%%%%%%%%%%%%%%%%%%%%%%%%%%%%%%%FIGURE%%%%%%%%%%%%%%%%%%%%%%%%%%%%%%%%%%%%%%%%%%%%%%%    
\begin{figure*}[tbp]
 \centering
    \includegraphics[scale=0.53]{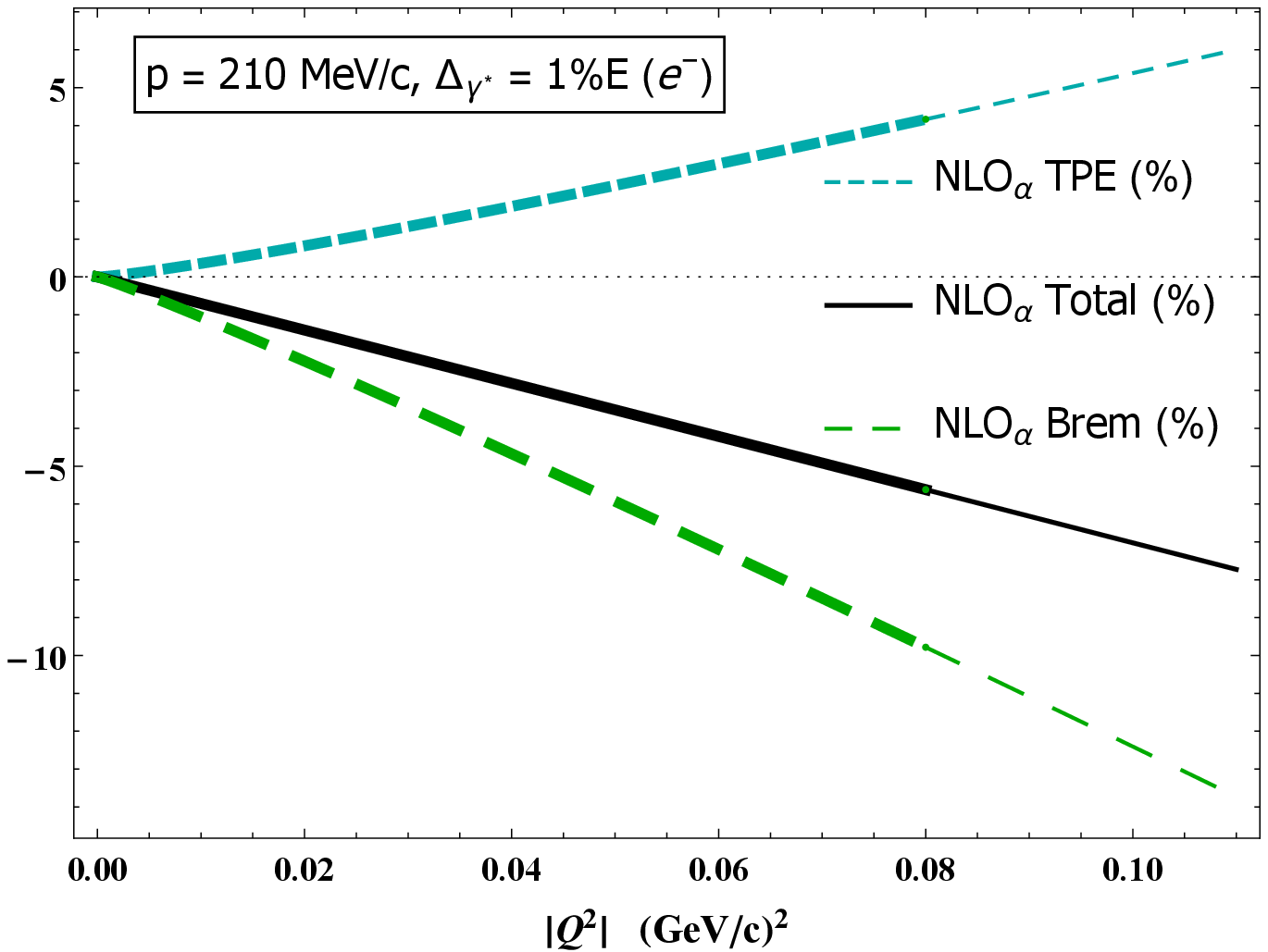} \qquad 
    \includegraphics[scale=0.53]{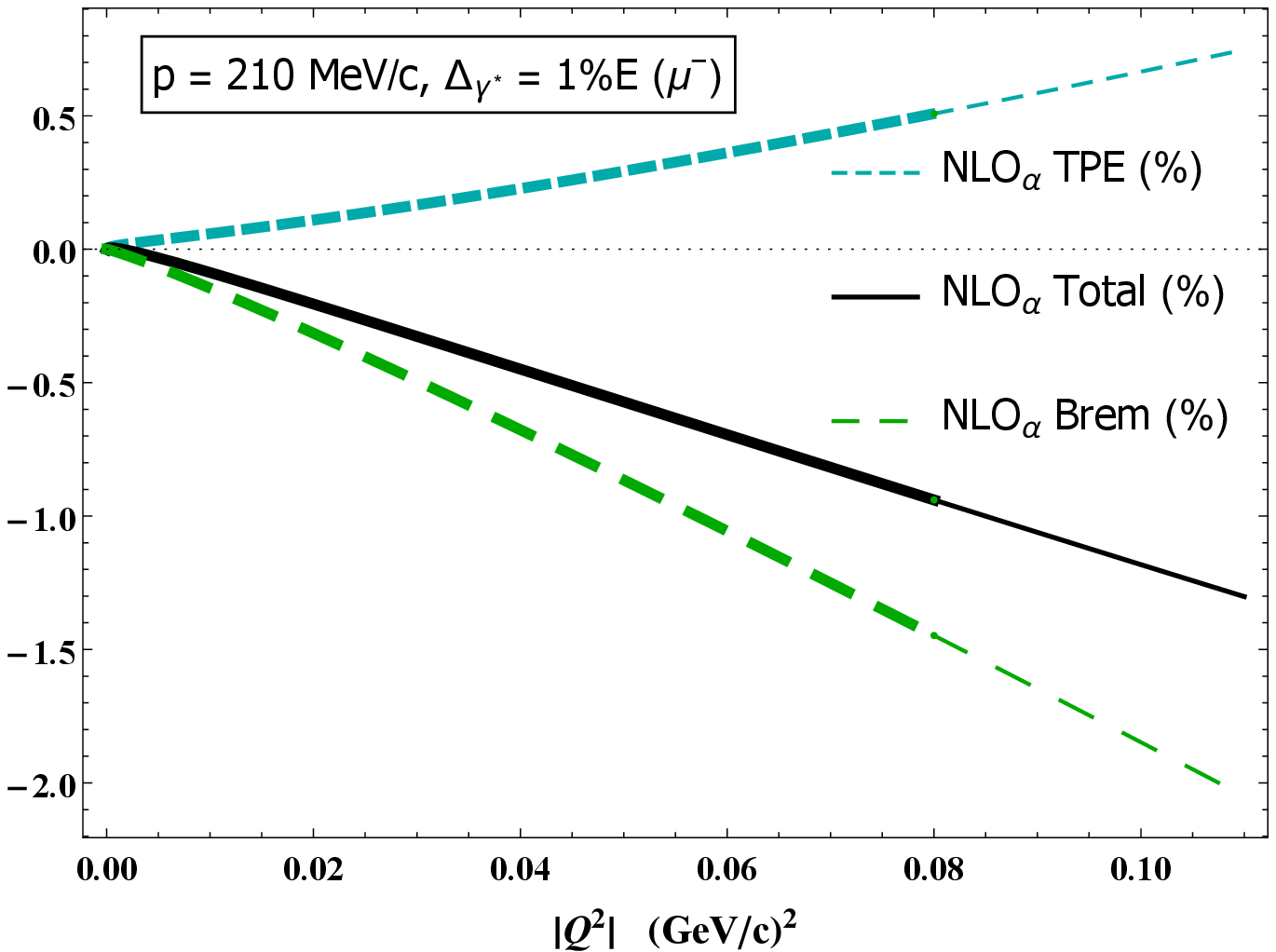}
    
    \vspace{1cm}
    
    \includegraphics[scale=0.53]{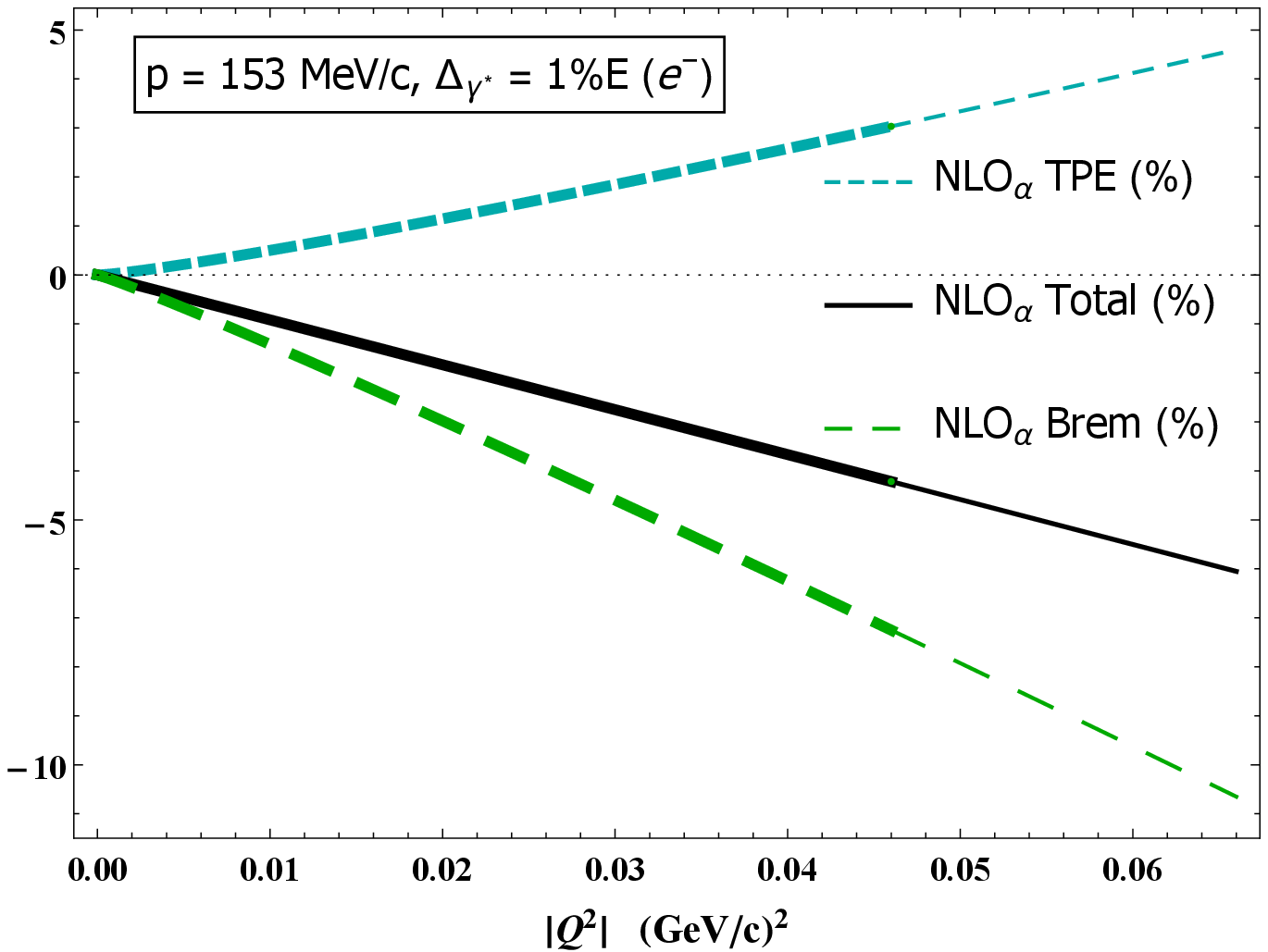} \qquad 
    \includegraphics[scale=0.53]{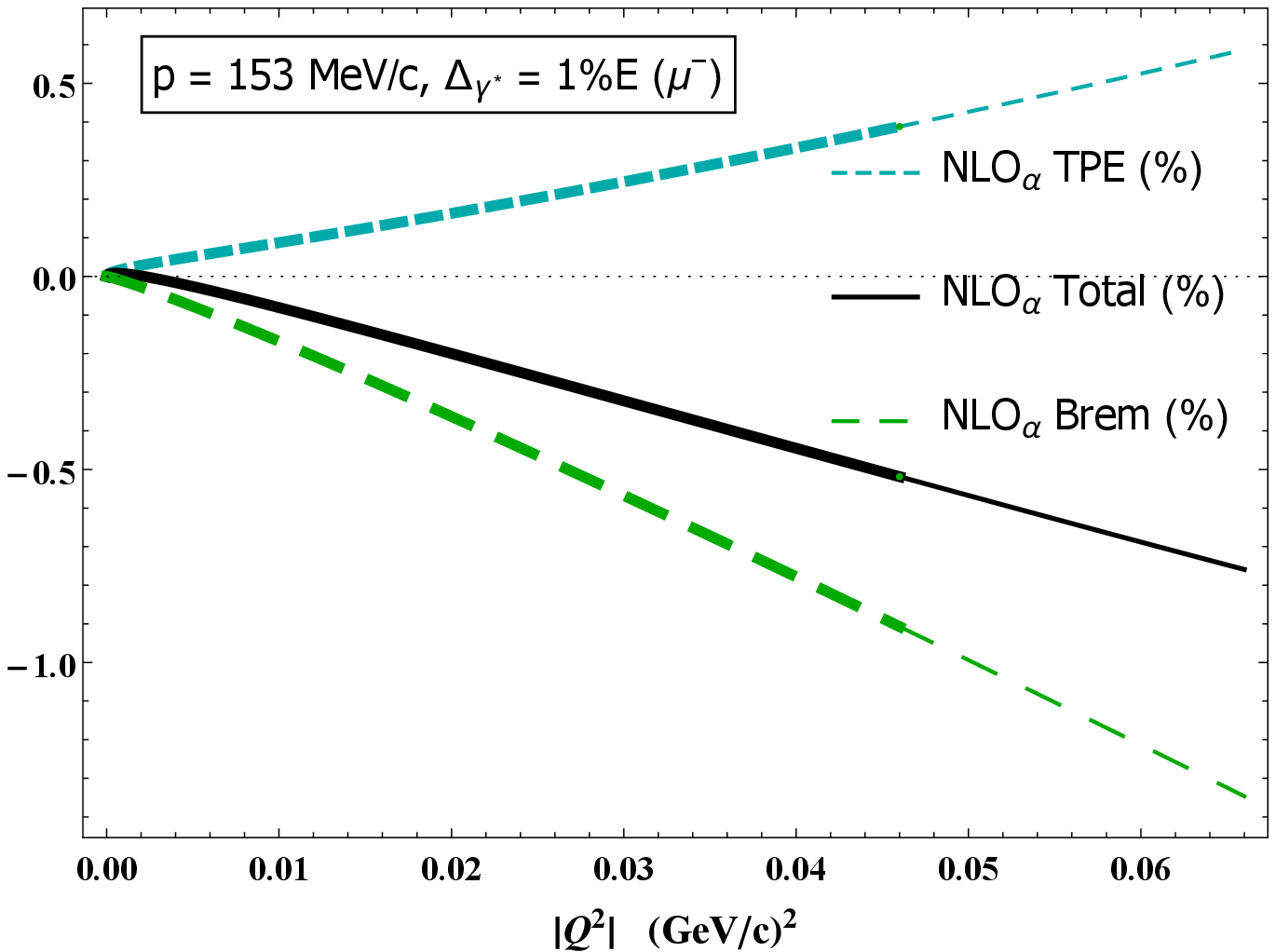}
    
    \vspace{1cm}
    
    \includegraphics[scale=0.53]{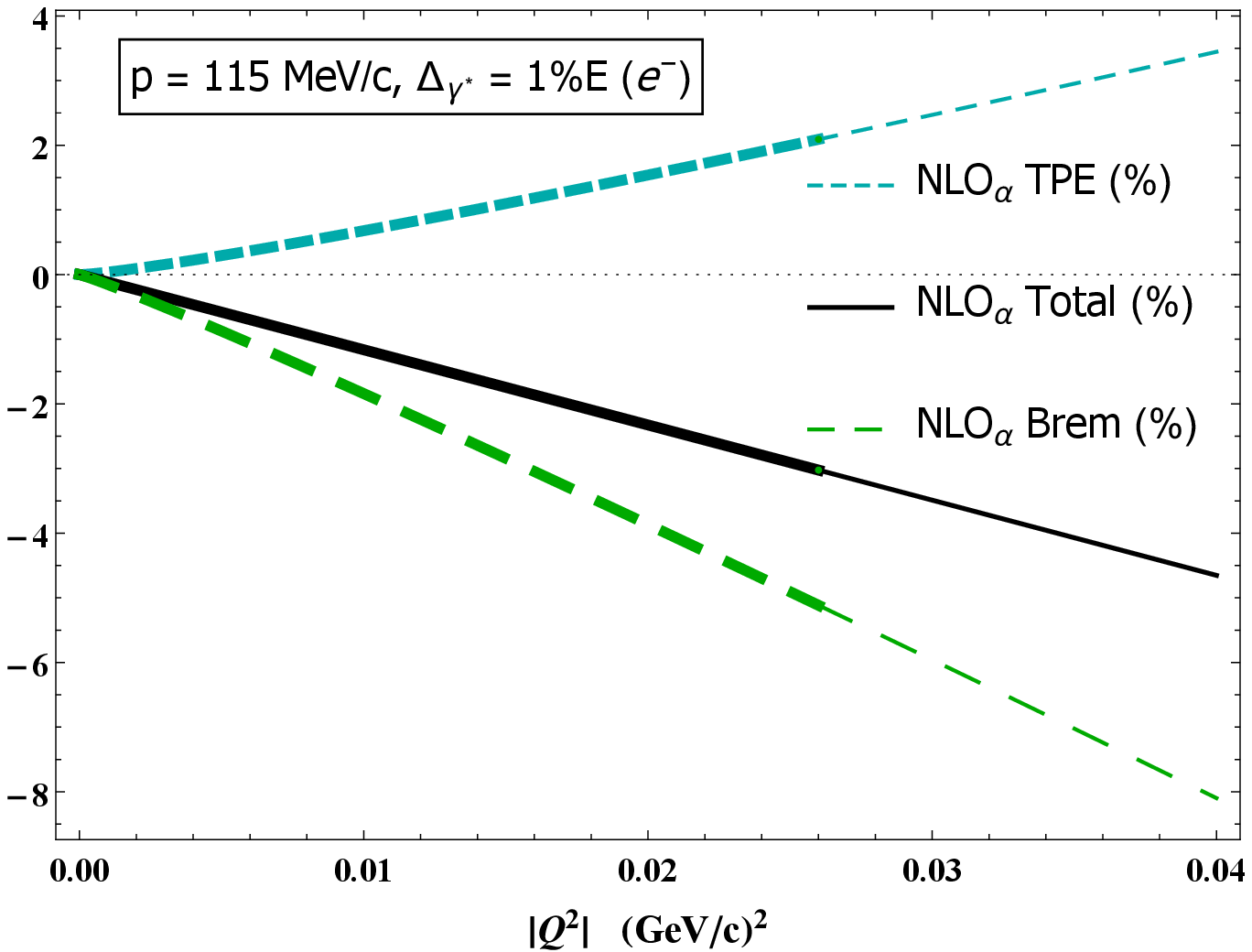} \qquad 
    \includegraphics[scale=0.53]{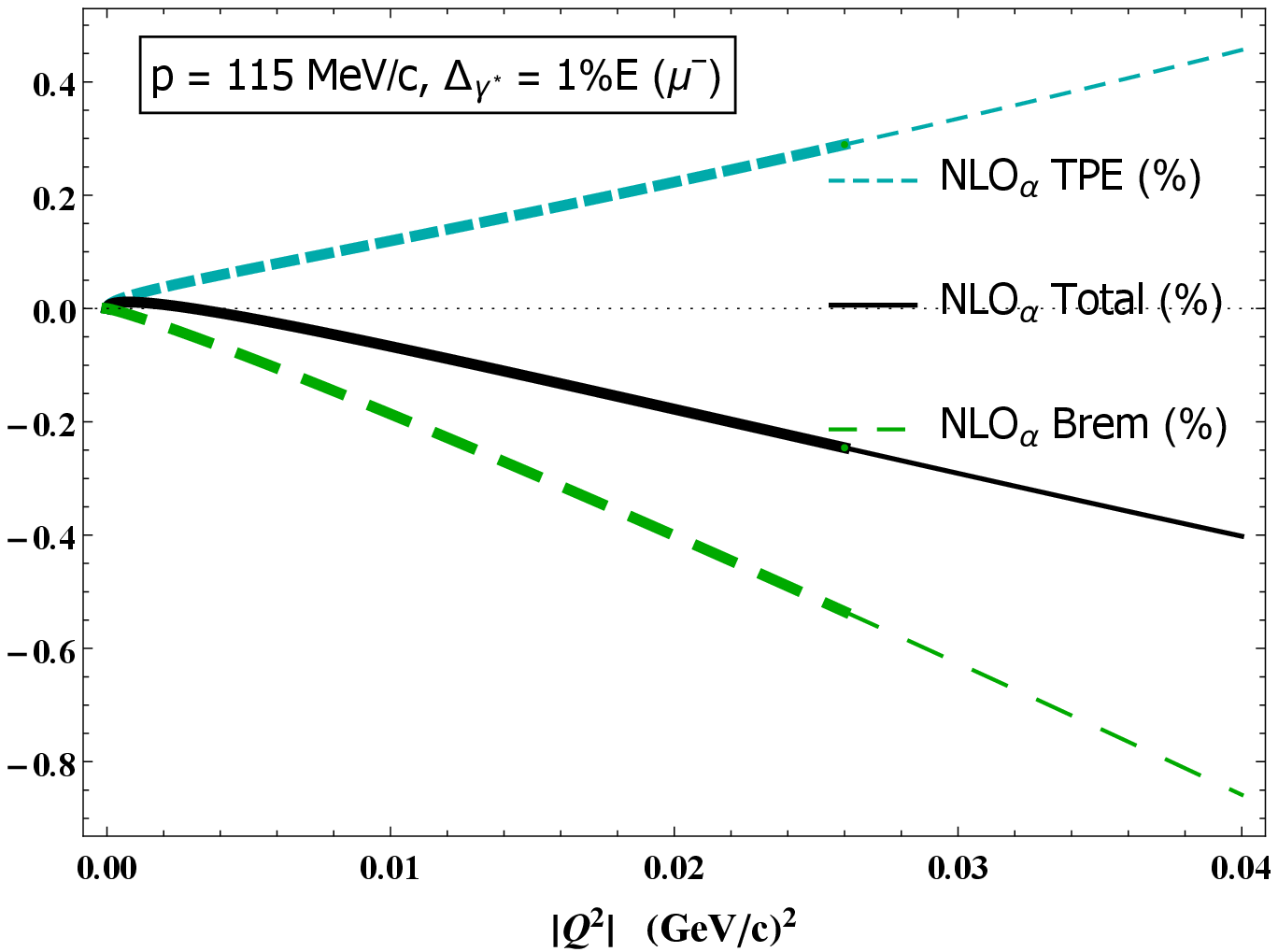}
    \caption{The fractional NLO$_\alpha$  virtual corrections (in percentage) due to the two-photon 
             contributions $\overline{\delta}^{(1)}_{\gamma\gamma}$ (short-dashed curves), the soft 
             photon bremsstrahlung corrections $\overline{\delta}^{(1)}_{\gamma\gamma^*}$ (dashed 
             curves), and their sum $\delta^{(1)}_{2\gamma}$ (solid curves). The left (right) panel 
             displays the results for $e$-p ($\mu$-p) elastic cross section versus the squared 
             four-momentum transfer $|Q^2|$, for the MUSE beam momenta of 
             $|\vec p\,|=p=115,\, 153,\, 210$~MeV/c. Each plot covers the full kinematical scattering 
             range, $0<|Q^2|<|Q^2_{\rm max}|$, when $\theta\in[0,\pi]$. The thickened portion of each 
             curve corresponds to the MUSE kinematic cut, where $\theta\in[20^\circ,100^\circ]$. The 
             {\it lab.}-frame detector acceptance $\Delta_{\gamma^*}$ is taken as $1\%$ of the 
             incident lepton energy $E$.}
    \label{plot:delta_tpe_brem}
\end{figure*}
%%%%%%%%%%%%%%%%%%%%%%%%%%%%%%%%%%%%%%%%%%%%%%%%%%%%%%%%%%%%%%%%%%%%%%%%%%%%%%%%%%%%%%%%%%%%    

%\vspace{0.1cm}

Our calculated radiative corrections depend on the value of the detector resolution parameter 
$\Delta_{\gamma^*}$. Theoretically, $\Delta_{\gamma^*}$ determines the maximal energy for the soft 
bremsstrahlung photon in the {\it lab.}-frame. Anticipating the typical accuracy levels of present day 
experiments, in this work we have chosen a reasonable value, $\Delta_{\gamma^*} = 1\%$ of the incoming 
lepton energy. Figure~\ref{plot:delta_tpe_brem} shows the IR-finite one-loop virtual NLO$_\alpha$ 
radiative (i.e., TPE) corrections $\overline{\delta}^{(1)}_{\gamma\gamma}$ and the soft photon 
bremsstrahlung NLO$_\alpha$ radiative corrections $\overline{\delta}^{(1)}_{\gamma\gamma^*}$, 
Eqs.~\eqref{delta:NLO_tpe} and \eqref{delta:NLO_brem} respectively. The following observations describe 
the features of the NLO$_\alpha$ radiative corrections:
\begin{itemize}
\item The TPE corrections as expected vanish as $Q^2\rightarrow 0$, while the soft bremsstrahlung 
corrections become negative, albeit infinitesimally small, at $Q^2=0$ due to non-vanishing of the 
finite non-factorizable term $\delta^{\,lp(1);\,{\rm v}}_{\gamma\gamma^*}$. 
\item  Both the real and virtual NLO$_\alpha$ corrections display a roughly linear rise with 
increasing momentum transfer $|Q^2|$. 
\item  Like the LO$_\alpha$ chiral order results, the NLO$_\alpha$ corrections do not change 
rapidly with the increasing lepton beam momenta.  
\item Both the NLO$_\alpha$ contributions are comparable in magnitudes but of opposite signs. At 
the {\it largest} MUSE lepton beam momenta we observe about $5\%$ and $1\%$ total radiative 
corrections at NLO$_\alpha$ for electron and muon scatterings respectively. 
\end{itemize}   

%%%%%%%%%%%%%%%%%%%%%%%%%%%%%%%%%%%%%%%%%%%%%%%%%%%%%%%%%%%%%%%%%%%%%%%%%%%%%%%%%%%%%%%%%%%%%%
\section{Numerical Results and Discussion}\label{sec:results}  
%%%%%%%%%%%%%%%%%%%%%%%%%%%%%%%%%%%%%%%%%%%%%%%%%%%%%%%%%%%%%%%%%%%%%%%%%%%%%%%%%%%%%%%%%%%%%% 
In order to determine the total $\ell$-p elastic differential cross section we sum up all the leading 
(LO$_\alpha$) and next-to-leading (NLO$_\alpha$) chiral order radiative corrections, in addition to 
the chirally expanded elastic Born terms (i.e., up to NNLO$_\chi$), as given in 
Eq.~\eqref{eq:delta_chi1}, to yield  
%%%%%%%%%%%%%%%%%%%%%%%%%%%%%%%%%%%%%%%%%%%%%%%%%%%%
\begin{widetext}
\begin{equation}
\frac{{\rm d}\sigma_{el}(Q^2)}{{\rm d} \Omega^\prime_l}\Bigg|_{\rm lab}
=\left[\frac{{\rm d}\sigma_{el}(Q^2)}{{\rm d} \Omega^\prime_l}\right]_0 
\left[1 + \delta_\chi^{(rms)}(Q^2) + \delta_\chi^{(1/M^2)}(Q^2) 
+ \delta^{(0)}_{2\gamma}(Q^2) + \delta^{(1)}_{2\gamma}(Q^2) 
+ \delta^{(2)}_{2\gamma}(Q^2)\right] 
+ {\mathcal O}\left(\frac{Q^3}{M^3}\right) \,, 
\label{eq:full}
\end{equation} 
\end{widetext}
%%%%%%%%%%%%%%%%%%%%%%%%%%%%%%%%%%%%%%%%%%%%%%%%%%%%
where  $\delta^{(0)}_{2\gamma}$ and $\delta^{(1)}_{2\gamma}$ are given in Eq.~\eqref{delta02gamma} and 
Eq.~\eqref{delta12gamma},  respectively. The last term $\delta^{(2)}_{2\gamma}$ includes the interference 
terms between the NNLO$_\chi$ and the LO$_\alpha$ corrections as well as non-factorizable 
NNLO$_\alpha$ terms, e.g., those in Eqs.~\eqref{eq:IR+finite_NLO} and \eqref{eq:delta_brems}, which are 
beyond the intended accuracy of this work. These higher order terms, which constitute the 
$\mathcal O(\alpha{\mathcal Q}^2/M^2)$ fractional chiral-radiative corrections to the LO elastic Born 
cross section, are only partially included in this work and are just used to estimate the systematic error
of our methodology. 

%\vspace{0.1cm} 

By comparing the different contributions in Eq.~(\ref{eq:full}) we can obtain a reasonable estimate of the 
relative magnitudes of the different radiative corrections. To that end, we first discuss the total 
fractional radiative correction up-to-and-including NLO$_\alpha$ accuracy, i.e., 
$\delta_{2\gamma}=\delta^{(0)}_{2\gamma}+\delta^{(1)}_{2\gamma}$ (i.e., excluding NNLO$_\alpha$). 
Table~\ref{tab1} displays the LO$_\alpha$ and NLO$_\alpha$ corrections as well as their sum $\delta_{2\gamma}$, 
for both lepton and anti-lepton scatterings off the proton (i.e., $e^\pm$p and $\mu^\pm$p). As observed in 
the table, we make a comparison of the relative magnitudes of the two chiral order corrections, {\it viz.} 
NLO$_\alpha$ : LO$_\alpha$ (i.e., $\delta^{(1)}_{2\gamma} : \delta^{(0)}_{2\gamma}$). The results for muon 
scattering indicate ratios which change with momentum transfer $|Q^2|$ and incident lepton beam momentum 
$p=|\vec p\,|$, from about $2:3$ for the largest $|Q^2|$ and $p$ values to about $1:110$ for the smallest 
$|Q^2|$ and $p$ values. In comparison, for electron scattering the comparable ratios are about $1:5$ and $1:25$
for the largest and smallest $|Q^2|,\,p$ values respectively. Evidently, there are  drastic changes in the 
NLO$_\alpha$ : LO$_\alpha$ ratio in going from the lowest to the largest possible kinematical limits in case 
of the muon scattering at MUSE. Furthermore, we note that the TPE and the soft bremsstrahlung corrections at 
NLO$_\alpha$ depend on the {\it lepton} ($\ell^\pm$) {\it charge}. Hence, not only do the NLO$_\alpha$ 
corrections $\delta^{(1)}_{2\gamma}$ corresponding to the $\ell^\pm$p scattering processes change sign, the 
total corrections, $\delta_{2\gamma}=\delta^{(0)}_{2\gamma}+\delta^{(1)}_{2\gamma}$, are somewhat smaller 
(larger) for $\mu^-$p ($e^-$p) scattering than the $\mu^+$p ($e^+$p) scattering.
%%%%%%%%%%%%%%%%%%%%%%%%%%%%%%%%%%%%%%%%%%TABLE%%%%%%%%%%%%%%%%%%%%%%%%%%%%%%%%%%%%%%%%%%%HIT
\begin{table*}[tbp]{\footnotesize 
\renewcommand{\arraystretch}{2.0}
\tabcolsep 0.2cm \vspace{0.5cm}
\begin{center}
\begin{tabular}{|c|c||c|c||c|c||c|c||c|c|}
\hline\hline
$p=|\vec{p}\,|$  & $|Q^2|$ & \multicolumn{2}{c||}{LO: $\delta_{2\gamma}^{(0)}$ }&\multicolumn{2}{c||}{NLO: $\delta_{2\gamma}^{(1)}$}&\multicolumn{2}{c||}{LO+NLO: $\delta_{2\gamma}$}&\multicolumn{2}{c|}{LO+NLO: $\delta_{2\gamma}$}\\
\cline{3-10}
GeV/c  & (GeV/c)$^2$ & $e^{\pm}p$ & $\mu^{\pm}p$ & $e^{\pm}p$    & $\mu^{\pm}p$  & $e^{-}p$ & $\mu^{-}p$ & $e^{+}p$  & $\mu^{+}p$ \\ 
\hline\hline
       & 0.005 & -0.1485   & 0.0222      & $\pm 0.0058$ & $\pm 0.0002$ & -0.1543  & 0.0220    & -0.1427  & 0.0224 \\
\cline{2-10}
 0.115   & 0.015 & -0.1671   & 0.0197      & $\pm 0.0174$ & $\pm 0.0012$ & -0.1846  & 0.0185    & -0.1497  & 0.0209 \\
\cline{2-10}
       & 0.025 & -0.1760   & 0.0171      & $\pm 0.0291$ & $\pm 0.0023$ & -0.2050  & 0.0147    & -0.1469  & 0.0194 \\     
\hline
       & 0.01 & -0.1598    & 0.0211       & $\pm 0.0092$ & $\pm 0.0008$ & -0.1689  & 0.0203    & -0.1506  & 0.0219 \\ 
\cline{2-10}
 0.153 & 0.025 & -0.1752   & 0.0171      & $\pm 0.0230$ & $\pm 0.0026$ & -0.1981  & 0.0145    & -0.1523  & 0.0197 \\
\cline{2-10}
       & 0.04 & -0.1831    & 0.0137       & $\pm 0.0367$ & $\pm 0.0045$ & -0.2199  & 0.0092    & -0.1465  & 0.0181 \\     
\hline
       & 0.02 & -0.1709    & 0.0185       & $\pm 0.0140$ & $\pm 0.0020$ & -0.1850  & 0.0165    & -0.1569  & 0.0206 \\
\cline{2-10}
 0.210 & 0.04 & -0.1829    & 0.0137       & $\pm 0.0281$ & $\pm 0.0045$ & -0.2105  & 0.0092    & -0.1543  & 0.0182 \\
\cline{2-10}
      & 0.06 & -0.1892    & 0.0099       & $\pm 0.0421$ & $\pm 0.0069$ & -0.2313  & 0.0030    & -0.1470  & 0.0169 \\     
\hline\hline
\end{tabular}
         \caption{The fractional radiative corrections with respect to the LO elastic Born 
                  cross section, $\delta^{(0)}_{2\gamma}$ at LO$_\alpha$, $\delta^{(1)}_{2\gamma}$ 
                  at NLO$_\alpha$ and their sum 
                  $\delta_{2\gamma}=\delta^{(0)}_{2\gamma}+\delta^{(1)}_{2\gamma}$, evaluated in 
                  HB$\chi$PT for $\ell^\pm$p elastic scattering. The incident lepton beam momenta, 
                  $|\vec p\,|=p= 0.115\,, 0.153\,, 0.210$~GeV/c, at some specific $|Q^2|$ values 
                  within the allowed MUSE kinematic range are used. The above numerical figures 
                  correspond to the {\it lab.}-frame detector acceptance $\Delta_{\gamma^*}=1\%$ 
                  of the incident lepton energy $E$. } 
          \label{tab1}
\end{center}}
\end{table*}
%%%%%%%%%%%%%%%%%%%%%%%%%%%%%%%%%%%%%%%%%%%%%%%%%%%%%%%%%%%%%%%%%%%%%%%%%%%%%%%%%%%%%%%%%%

%\vspace{0.1cm}

In our treatment of the radiative corrections presented hitherto, we only considered the dominant 
${\mathcal O}(\alpha)$ or leading QED contributions. The higher order QED corrections (of negative sign) 
proportional to $\alpha^n$ where $n=2,3,\cdots$ will tend to cause deviations from the leading QED 
predictions, especially at momentum transfers much larger than typical kinematic scale, e.g., in our case
of lepton scattering such enhancements are typically expected for $|Q^2|\gg m^2_l$. In practice, to 
evaluate higher order QED effects involves a very intricate task of calculating multiple photon loop 
diagrams that is evidently beyond the scope of this work. It is, however, well-known that the 
double-logarithmic Sudakov terms are responsible for the largest enhancements, especially in the soft and 
collinear kinematical regions for near-massless particles (such as the electron) in the soft photon limit. 
Schwinger~\cite{Schwinger:1949ra}, based on work by Bloch and Nordsieck~\cite{Bloch:1937pw} (see also 
Refs.~\cite{Yennie:1961ad,Jauch54,Lomon56,Lomon59,Perrin65,Eriksson61,Okubo60,Caianiello:1960pi}) showed 
that one could by and large compensate for such large enhanced negative contributions to {\it all orders} in
$\alpha$ by the exponentiation of their contribution to the elastic cross section. Symbolically, this means 
that if $\delta_{n\gamma}(Q^2)<0$ denotes the $n^{\rm th}$ order photon-loop and soft photon bremsstrahlung 
corrections (not including VP contributions), then the replacement,
$$1+\delta_{n\gamma}(Q^2)\to\exp\left\{\delta_{n\gamma}(Q^2)\right\},$$ leads to an essential suppression of
such ``artificial" enhancements resulting from the truncation of perturbative expansions. Theoretically, a 
tacit assumption in this regard is that the emission and re-absorption of an infinite number of {\it soft} 
photons are statistical independent and that these do not alter the elastic kinematics. Furthermore, as seen 
in our LO$_\alpha$ results, there is significant contribution from the VP corrections which are comparable 
in magnitude but opposite in sign to the photon-loop (vertex) contributions. Consequently, following 
Ref.~\cite{Vanderhaeghen:2000ws}, commensurate with the exponentiation of the radiative corrections arising 
from the photon-loop terms, we find it consistent to include the re-summation of the  one-particle 
irreducible VP diagrams to all orders. This is useful to preserve essential cancellations that can manifest 
itself among the higher order radiative corrections. To this end, the $\ell$-p elastic differential cross 
section reads
\begin{eqnarray}
\frac{{\rm d}\sigma_{el}(Q^2)}{{\rm d} \Omega^\prime_l}\Bigg|_{\rm lab}
&\approx& \left[\frac{{\rm d}\sigma_{el}(Q^2)}{{\rm d} \Omega^\prime_l}\right]_0 
\\
&&\times\,\Big\{1+\delta^{(2)}_\chi (Q^2)+\delta^{el}_{\rm resum}(Q^2)\Big\}\,, 
\nonumber
\label{eq:final_result}
\end{eqnarray}
where $\delta^{(2)}_\chi$ includes the NNLO$_\chi$ chiral corrections (cf. Sec.~\ref{sec:formalism}), while the
{\it modified} fractional QED corrections, taking into account the partial re-summation of all the potentially 
large double logarithm terms, is given by 
\begin{eqnarray*}
\delta^{el}_{\rm resum}(Q^2)=\frac{\exp{\left\{\delta_{2\gamma}(Q^2)-
\delta_{\rm vac}(Q^2)\right\}}}{\left[1-\delta_{\rm vac}(Q^2)/2\right]^2}-1\,;
\end{eqnarray*}
\begin{eqnarray}
\delta_{2\gamma}(Q^2)=\delta^{(0)}_{2\gamma}(Q^2)+\delta^{(1)}_{2\gamma}(Q^2)
+\delta^{(2)}_{2\gamma}(Q^2)\,,
\nonumber\\
\label{eq:delta_resum}
\end{eqnarray} 
with $\delta^{(2)}_{2\gamma}$ (explicit expression not displayed) representing the additional NNLO$_\alpha$ 
error terms. The VP contribution is assigned as $\delta_{\rm vac}\to \delta^{(0)}_{\rm vac}$, the LO$_\alpha$ 
VP correction [cf. Eq.~\eqref{delta0_vac}], in the absence of NLO$_\alpha$ [i.e., 
$\mathcal O(\alpha {\cal Q}/M)$] VP term.\footnote{It is worth noting that the Schwinger's method of 
exponentiating radiative corrections is strictly applicable only for the ``IR-enhanced" double-log terms, 
e.g., ones proportional to $\ln \left(m^2_e/|Q^2|\right)$ or $\ln \left(\Delta^2_{\gamma^*}/|Q^2|\right)$, 
where the so-called {\it Sudakov regions} are clearly defined with the only relevant scale as $|Q^2|\to \infty$. 
However, it is not immediately apparent how to generalize such ultra-relativistic results to low-energies, 
especially with other relevant scales, such as $m^2_\mu, M^2\gtrsim Q^2$, etc., and constitutes a topic certainly 
beyond the scope of the present discussion. Consequently, in a simplified approach, we naively approximate the 
large double-log re-summation by exponentiating our NLO$_\alpha$ result (also including the NNLO$_\alpha$ 
errors) $\delta_{2\gamma}$, save the VP contributions $\delta_{\rm vac}$, which do not contain IR-enhanced 
terms.} While estimating the theoretical error due to NNLO$_\alpha$ corrections, the VP contributions must be 
modified as
\begin{eqnarray}
\delta_{\rm vac} (Q^2)&=&\delta^{(0)}_{\rm vac}(Q^2)\Bigg[1+\mathscr{R}_Q
\left\{1+\frac{1}{2}(1+\kappa_p)^2\right.
\nonumber\\
&&\quad\left.\times\left(\frac{Q^2+4(m^2_l-E^2)}{Q^2+4E^2}\right)\right\}\Bigg]\,.\quad\,
\end{eqnarray}

%\vspace{0.1cm}

\begin{figure*}[tbp]
 \centering
         \includegraphics[scale=0.53]{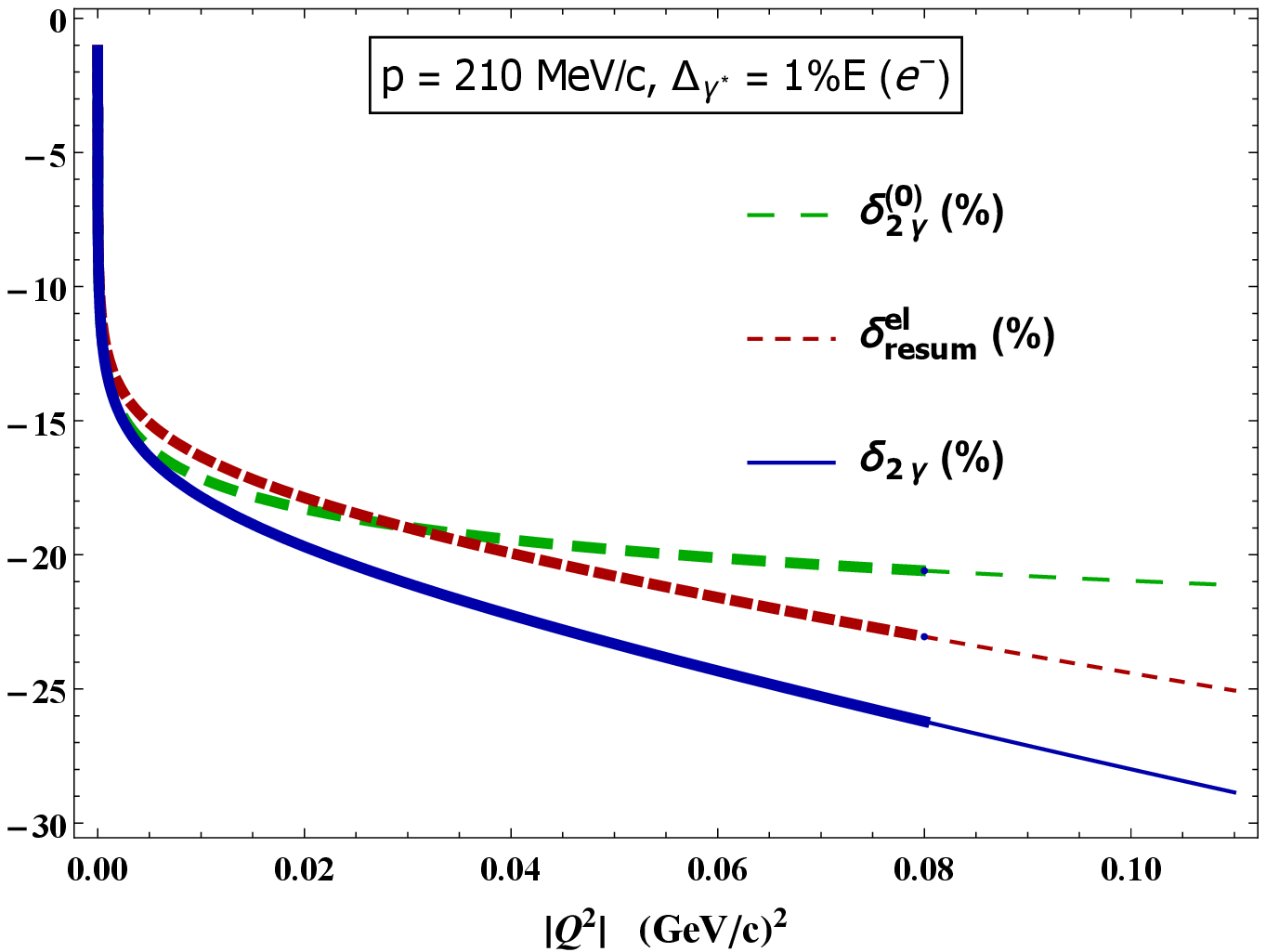} \qquad 
         \includegraphics[scale=0.53]{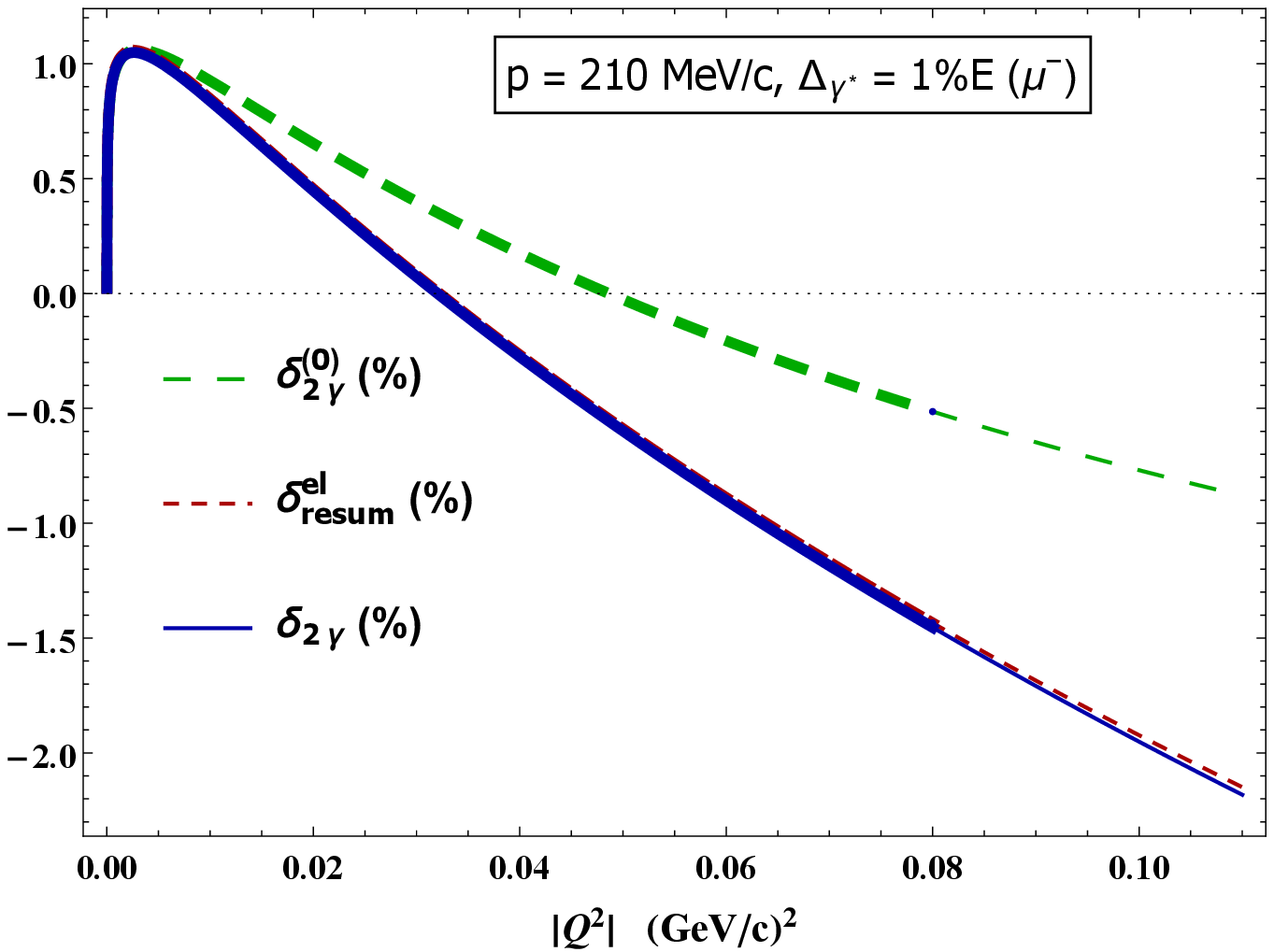}
         
         \vspace{1cm}
         
         \includegraphics[scale=0.53]{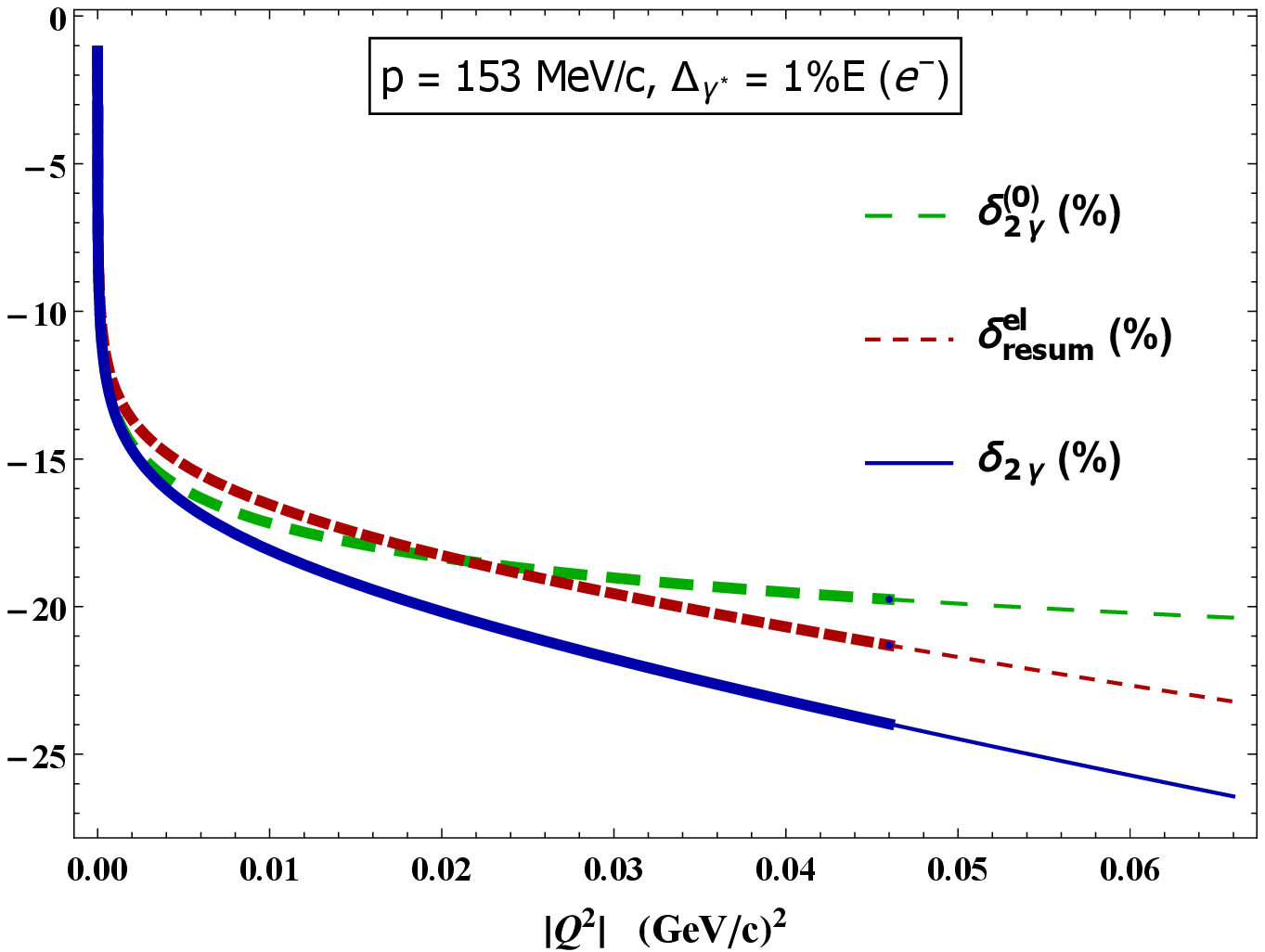} \qquad 
         \includegraphics[scale=0.53]{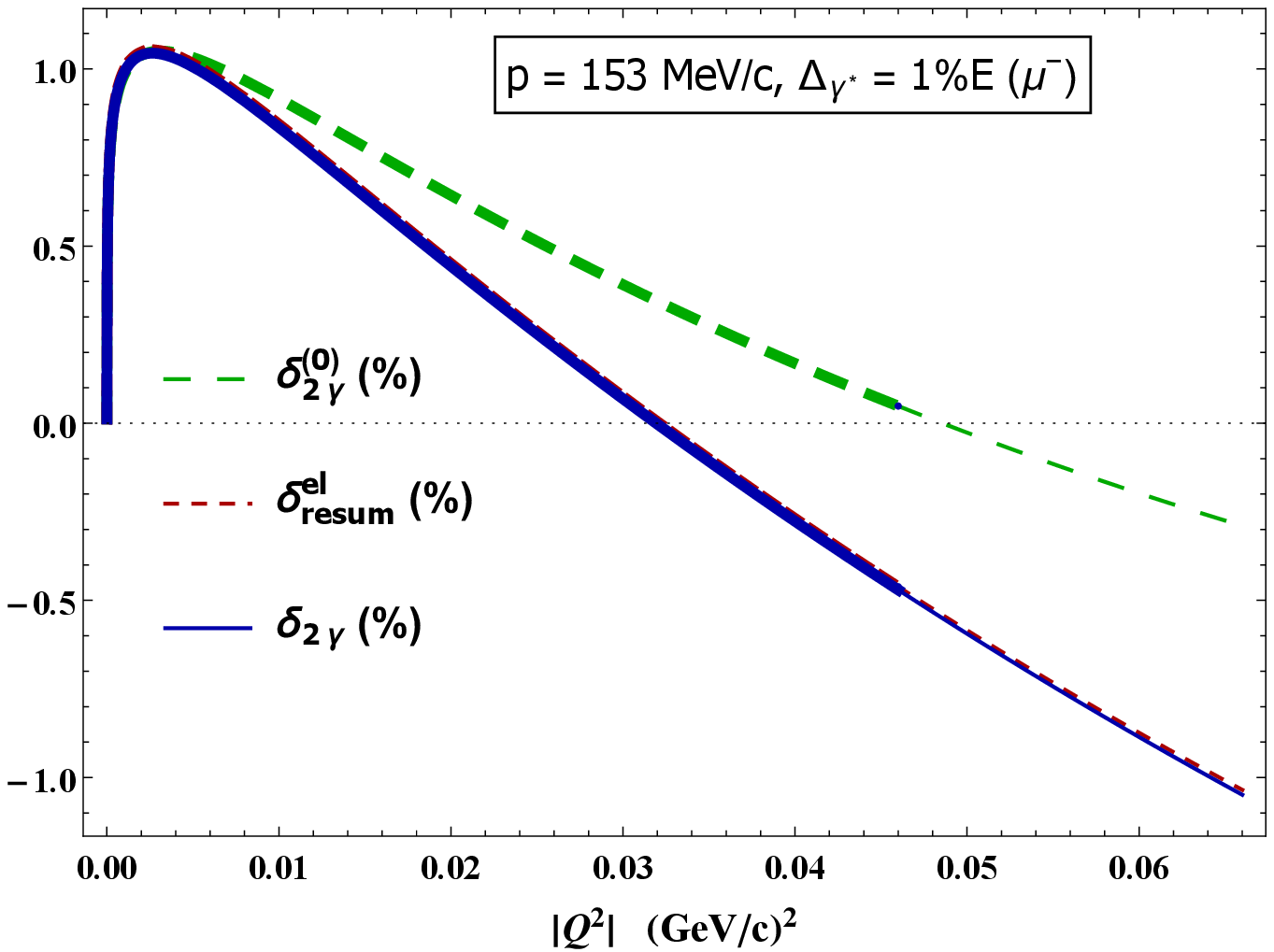}
         
         \vspace{1cm}
         
         \includegraphics[scale=0.53]{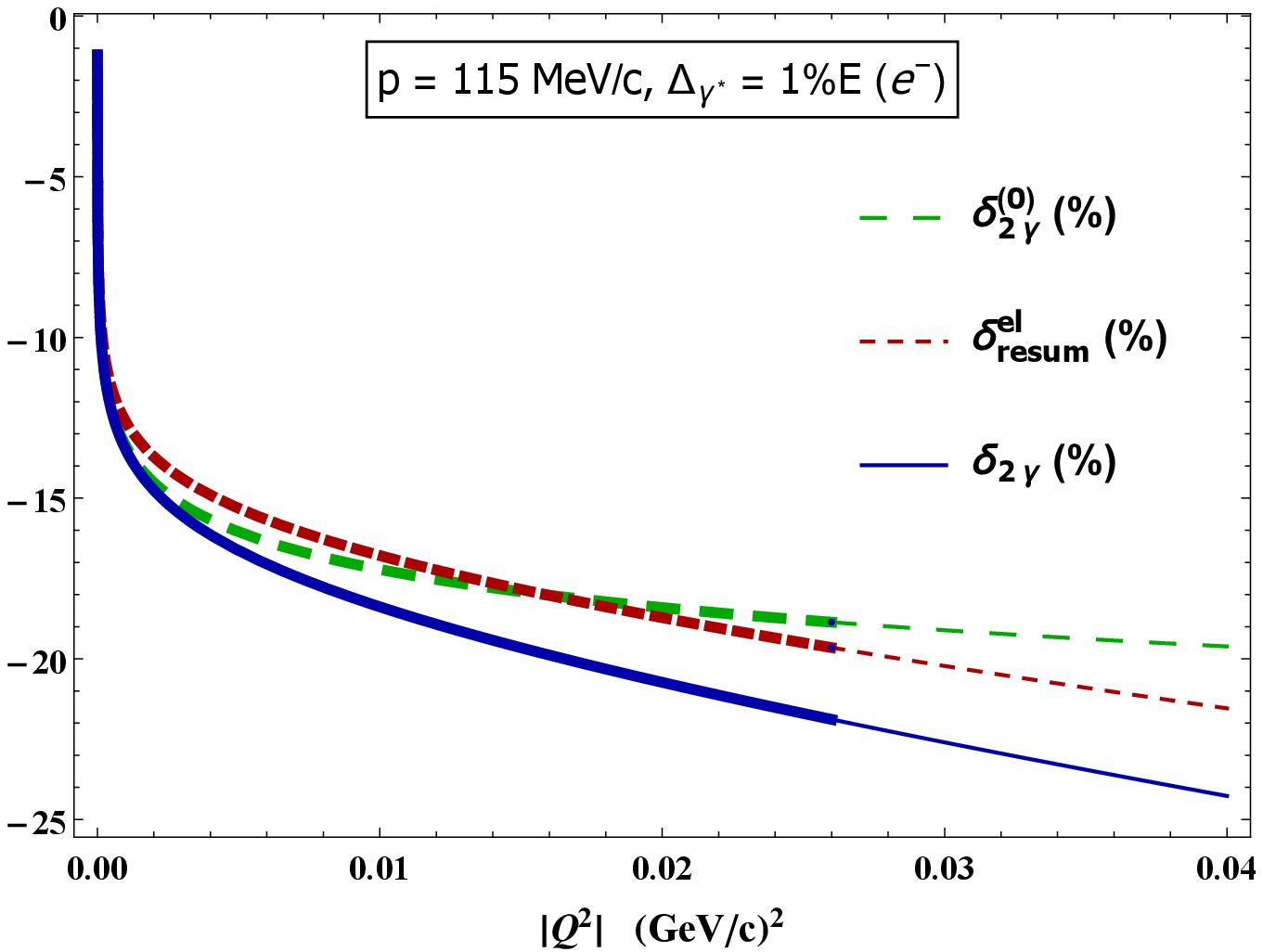} \qquad 
         \includegraphics[scale=0.53]{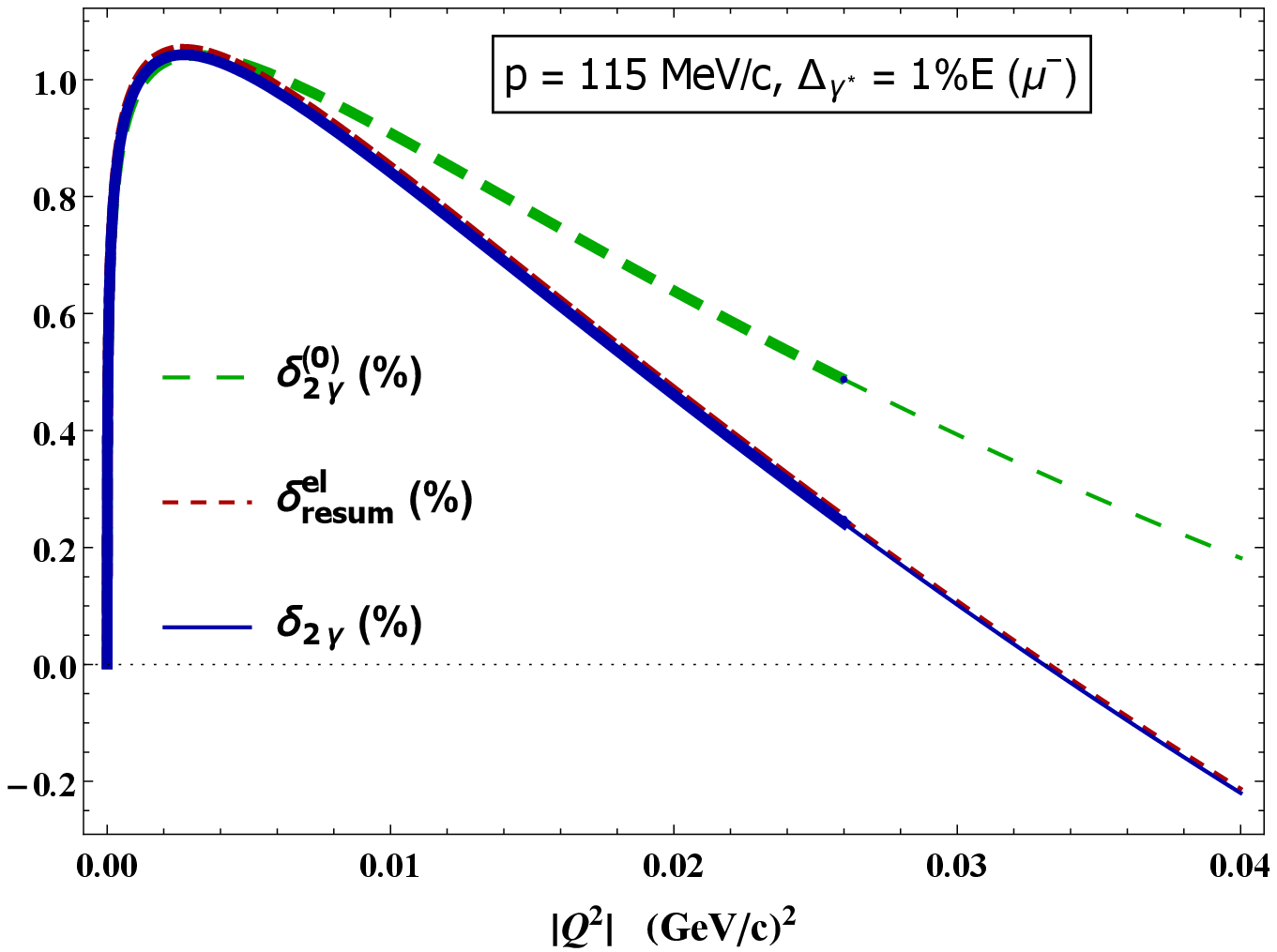}
         \caption{The total fractional radiative corrections (in percentage) at LO$_\alpha$, 
                  $\delta^{(0)}_{2\gamma}$, and up-to-and-including  NLO$_\alpha$, 
                  $\delta_{2\gamma}=\delta^{(0)}_{2\gamma}+\delta^{(1)}_{2\gamma}$, in 
                  HB$\chi$PT for $e$-p (left panel) and $\mu$-p (right panel) elastic cross 
                  sections as a function of $|Q^2|$ for the MUSE beam momenta, 
                  $|\vec p\,|=p=115,\, 153,\, 210$~MeV/c. Each plot covers the full 
                  kinematically allowed scattering range $0<|Q^2|<|Q^2_{\rm max}|$ when 
                  $\theta\in[0,\pi]$. The thickened portion of each curve corresponds to the 
                  MUSE kinematic cut, where $\theta\in[20^\circ,100^\circ]$. The 
                  {\it lab.}-frame detector acceptance $\Delta_{\gamma^*}$ is taken to be 
                  $1\%$ of the incident lepton energy $E$. The corresponding total fractional 
                  re-summed result, $\delta^{el}_{resum}$, Eq.~\eqref{eq:delta_resum}, are 
                  also displayed for comparison.}
         \label{fig:deltaRC_resum}
\end{figure*}

In Fig.~\ref{fig:deltaRC_resum} we plot the total fractional radiative corrections $\delta_{2\gamma}$, 
up-to-and-including NLO$_\alpha$ in HB$\chi$PT, and compare with the LO$_\alpha$ corrections 
$\delta^{(0)}_{2\gamma}$, Eq.~\eqref{delta02gamma}, and the corresponding partially re-summed QED 
results $\delta^{el}_{\rm resum}$, Eq.~\eqref{eq:delta_resum}, for low-energy $\ell$-p scatterings. In
the case of electron scattering with large negative bremsstrahlung contributions, the total radiative 
corrections $\delta_{2\gamma}$ stay negative and the magnitude of $\delta_{2\gamma}$ monotonically 
increases with increasing squared four-momentum transfer $|Q^2|$. The total corrections vary in the 
range of $(22-27)\%$ in the MUSE kinematic range. On the other hand, for muon scattering the total 
radiative corrections reach no larger than $1.5\%$ in same region. However, as distinct from the 
electron scattering case, the $\delta^{(0)}_{2\gamma}$ corrections for muon undergo a sign change at 
these low energies. As seen in the figure, $\delta^{(0)}_{2\gamma}$ is  positive for very small 
momentum transfers, say, $|Q^2|\lesssim 0.04$ (GeV/c)$^2$, due to the dominance of the VP contributions
in that region. However, for larger $|Q^2|$ values, $\delta^{(0)}_{2\gamma}$ turns negative as the VP 
corrections are eventually superseded by the dominant soft bremsstrahlung contributions. Nevertheless, 
for the  lowest MUSE muon beam momentum, $p=|{\vec p}\,|\gtrsim 115$ MeV/c, even the total correction 
$\delta_{2\gamma}$ remains positive due to VP dominance. It is also quite evident that there is no 
significant lepton beam momentum dependence on the individual LO$_\alpha$ and NLO$_\alpha$ components 
of the radiative corrections in the MUSE kinematical range, $115<p<210$~MeV/c. 

%\vspace{0.1cm}

Figure~\ref{fig:deltaRC_resum} also displays the $\delta^{el}_{resum}$ results where the potentially 
large double-logarithms have been effectively iterated to all orders in $\alpha$. The exponentiated 
radiative effects leads to the well-known {\it Sudakov suppression}, as is clearly evident in the 
electron scattering results. This $Q^2$ suppression effect is almost numerically comparable to the 
${\mathcal O}(\alpha)$ NLO$_\alpha$ corrections. In contrast, for muon scattering there is no 
discernible NLO$_\alpha$ suppression. Such contrasting results can be easily anticipated in regard to 
the MUSE kinematics, since the same $|Q^2|$ range that may be identified with typical low-momentum 
transfer dynamics for $\mu$-p scattering, becomes a region of high-momentum transfer in relativistic 
$e$-p scattering. For example, in Ref.~\cite{Talukdar:2018hia} the same reason was attributed to the 
validity of the high-energy {\it peaking approximation} for electron scattering at MUSE, when it fails
for muon scattering. We therefore anticipate such characteristic suppression to manifest itself in the 
radiatively corrected future MUSE data for electron-proton scattering.     

%\vspace{0.1cm}

The theoretical uncertainties involved in our calculations are categorized as the pure chiral hadronic 
corrections and the radiative corrections. The following sources of uncertainties are  identified in 
our treatment of the effective Born cross section. First, the proton's rms charge radius $r_p$ [cf. 
Eq.~\eqref{eq:delta_chi2}] is an essential input to our chirally corrected result at $\nu=2$ order. An 
uncertainty due to the numerical differences in the different proton's rms radius measurements is 
required. These differences in the extracted charge radius from high-precision electronic and muonic 
measurements result in an appreciable error in the value of the chiral correction 
$\delta^{(rms)}_\chi$, namely, $\Delta_{rms} \sim 6.4$\% (i.e., with respect to our central result) 
[cf. Fig.~\ref{plot:deltachi}]. However, given the ongoing contentious radius puzzle
scenario~\cite{Pohl:2010zza,Pohl:2013,Antognini:1900ns,Bernauer:2014,Carlson:2015} it seems not too 
unreasonable to estimate an error of such magnitude.\footnote{For instance, using the measured rms 
radius from the recent PRad Collaboration~\cite{Xiong_2019}, 
$r^{(ep)}_p=0.831 \pm 0.007{\rm (stat)}\pm 0.012 {\rm (syst)}$~fm, and from the erstwhile CREMA 
Collaboration~\cite{Antognini:1900ns}, $r^{(\mu H)}_p=0.84087(39)$~fm, we could obtain an effective 
error estimate due to the input rms radius contributions in the chiral chiral corrections 
$\delta^{(rms)}_\chi$ as
\begin{equation*}
\Delta_{rms}=\sqrt{\Delta^2_{\rm PRad}+\Delta^2_{\rm CREMA}+\Delta^2_{\rm diff}
+ \Delta^2_{\rm NNLO}}\sim 6.4\% \,,
\end{equation*}
where, $\Delta_{\rm NNLO}\sim 1$\%, and 
\begin{eqnarray*}
\Delta_{\rm PRad}&=& \frac{2r^{(ep)}_p\left(\delta r^{(ep)}_p\right)_{\rm exp}}{\left(r^{(ep)}_p\right)^2
-\frac{3\kappa_p}{2M^2}}\sim 5.5\% \,,
\\
\Delta_{\rm CREMA}&=& \frac{2r^{(\mu H)}_p\left(\delta r^{(\mu H)}_p\right)_{\rm exp}}{\left(r^{(\mu H)}_p\right)^2-\frac{3\kappa_p}{2M^2}}\sim 0.1\% \,,
\\
\Delta_{\rm diff} &=&  
\frac{2r^{(\mu H)}_p\left(r^{(ep)}_p-r^{(\mu H)}_p\right) }{\left(r^{(\mu H)}_p\right)^2
-\frac{3\kappa_p}{2M^2}}\sim 3.1\% \,,
\end{eqnarray*}
where $\left(\delta r^{(\mu H)}_p\right)_{\rm exp}$ stands for respective experimental uncertainties. Note 
that the above quoted percentage errors are not relative to the LO Born contributions, but are with respect 
to the central values of $\delta^{(rms)}_{\chi}$.} Second, the hadronic corrections beyond NNLO$_\chi$ 
constitute an important corrections to the LO Born cross section, Eq.~\eqref{dsigmaBornL0}. As mentioned, 
the N$^3$LO$_\chi$ [i.e., ${\mathcal O}({\cal Q}^3/M^3)\sim 0.008$] corrections due to the exclusion of 
$\nu=3$ chiral order hadronic interactions, constitute an uncontrolled error which estimation really lies 
beyond the present scope of this work. Nevertheless, commensurate with our discussions of our EFT power 
counting (cf. Sec.~\ref{intro}), an error of $\Delta_{\rm NNLO}\sim 1$\% on each of the two types of chiral 
corrections, {\it viz.} rms and recoil contributions, [cf. Eq.~\eqref{eq:delta_chi1}] can naively be 
attributed for the low-energy MUSE kinematics [cf. Fig.~\ref{plot:deltachi}]. Third, other sources of hadronic
uncertainties arise from non-perturbative effects due to various resonances and excited nucleon states (see, 
e.g., Refs.~\cite{Bernard:1992qa,Bernard:1995dp,Hemmert:1996xg,Hemmert:1997ye}. Furthermore, non-perturbative 
techniques, such as dispersively improved $\chi$PT~\cite{Lorentz_2015,Lee:2015jqa,Alarcon:2018zbz,Horbatsch:2016ilr} 
are needed for rigorous estimation of the inherent systematics which are not captured in the perturbative 
framework such as our's. However, such contributions also constitute uncontrolled uncertainties that we can 
not readily assess. 

%\vspace{0.1cm}
%%%%%%%%%%%%%%%%%%%%%%%%%%%%%%%%%%%%%%%FIGURE%%%%%%%%%%%%%%%%%%%%%%%%%%%%%%%%%%%%%%%%%%%%%    
%%%%%%%%%%%%%%%%%%%%%%%%%%%%%%%%%%%%%%%%%%%%%%%%%%%%%%%%%%%%%%%%%%%%%%%%%%%%%%%%%%%%%%%%%% 
%%%%%%%%%%%%%%%%%%%%%%%%%%%%%%%%%%%%%%%%FIGURE%%%%%%%%%%%%%%%%%%%%%%%%%%%%%%%%%%%%%%%%%%%%
\begin{figure*}[tbp]
 \centering
           \includegraphics[scale=0.53]{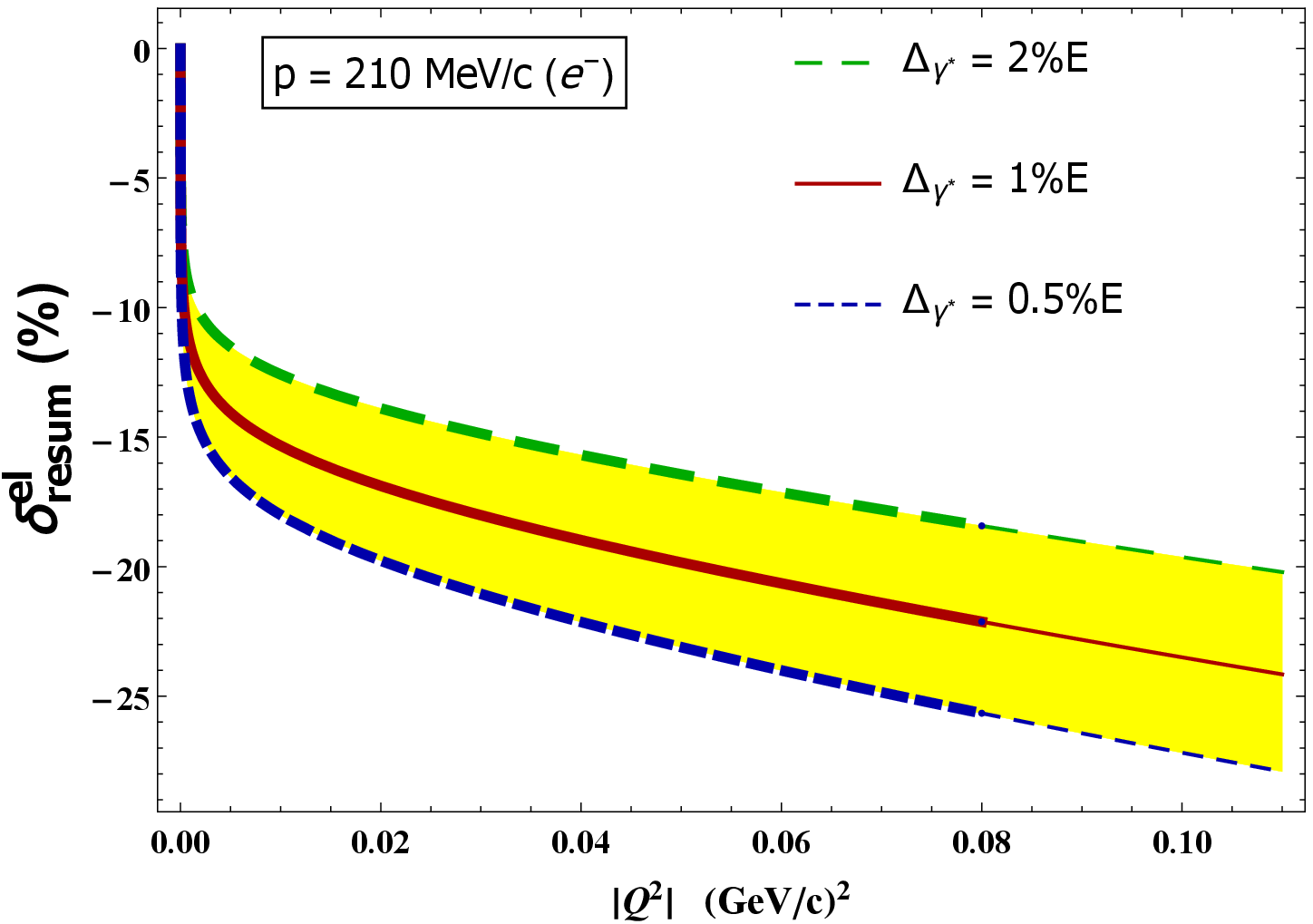} \qquad 
           \includegraphics[scale=0.53]{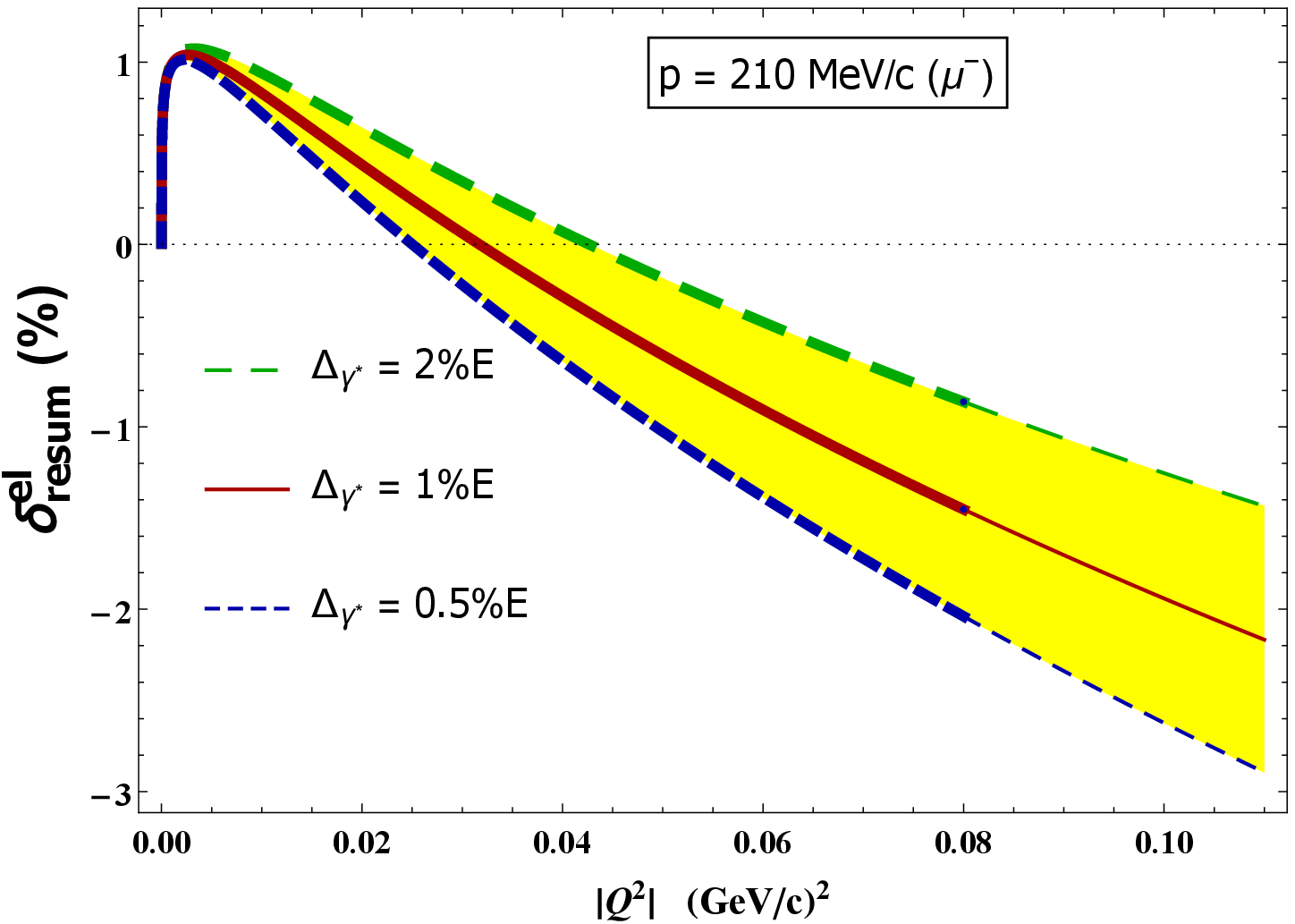}
           
           \vspace{1cm}
           
           \includegraphics[scale=0.53]{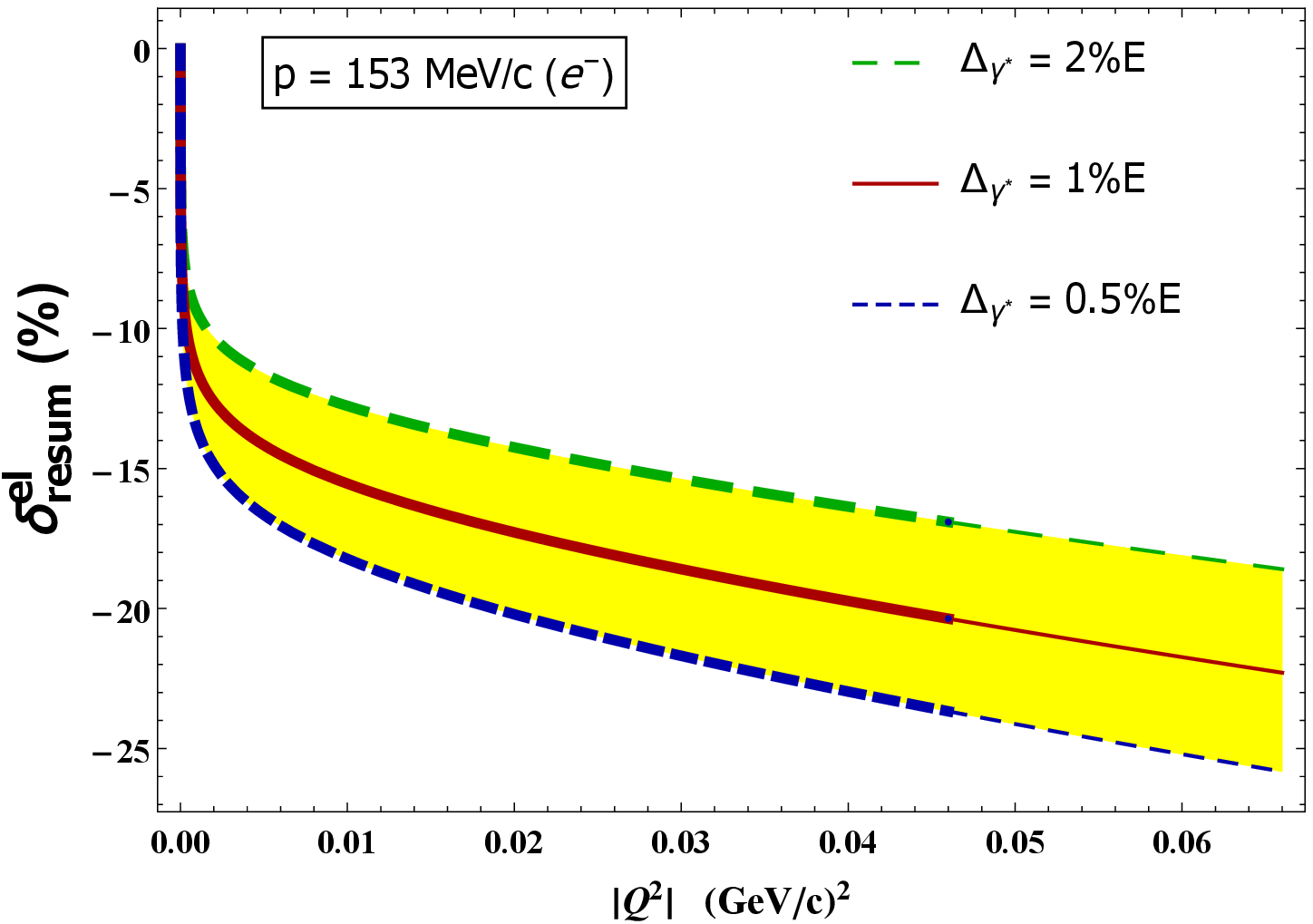} \qquad 
           \includegraphics[scale=0.53]{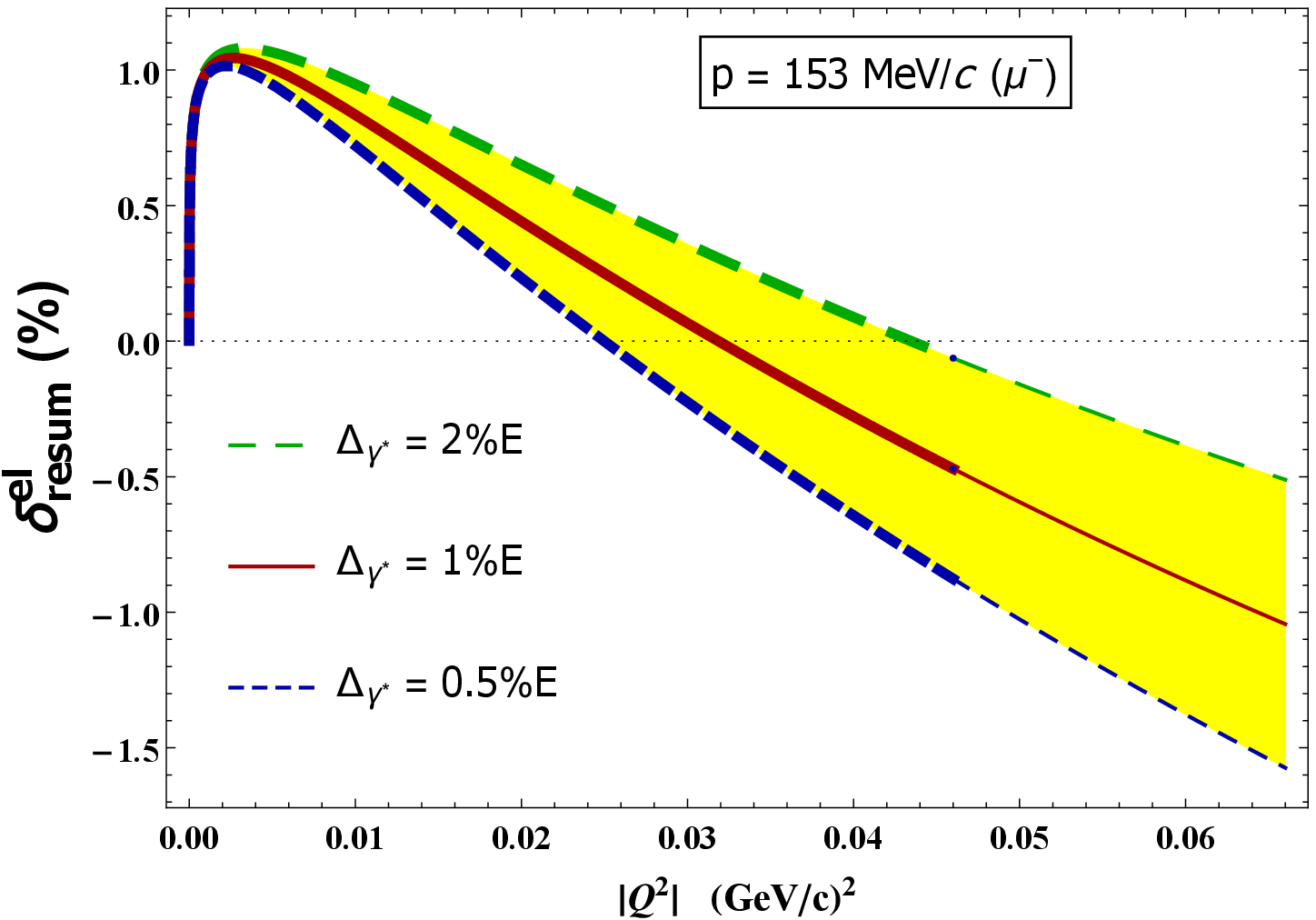}
           
           \vspace{1cm}
           
           \includegraphics[scale=0.53]{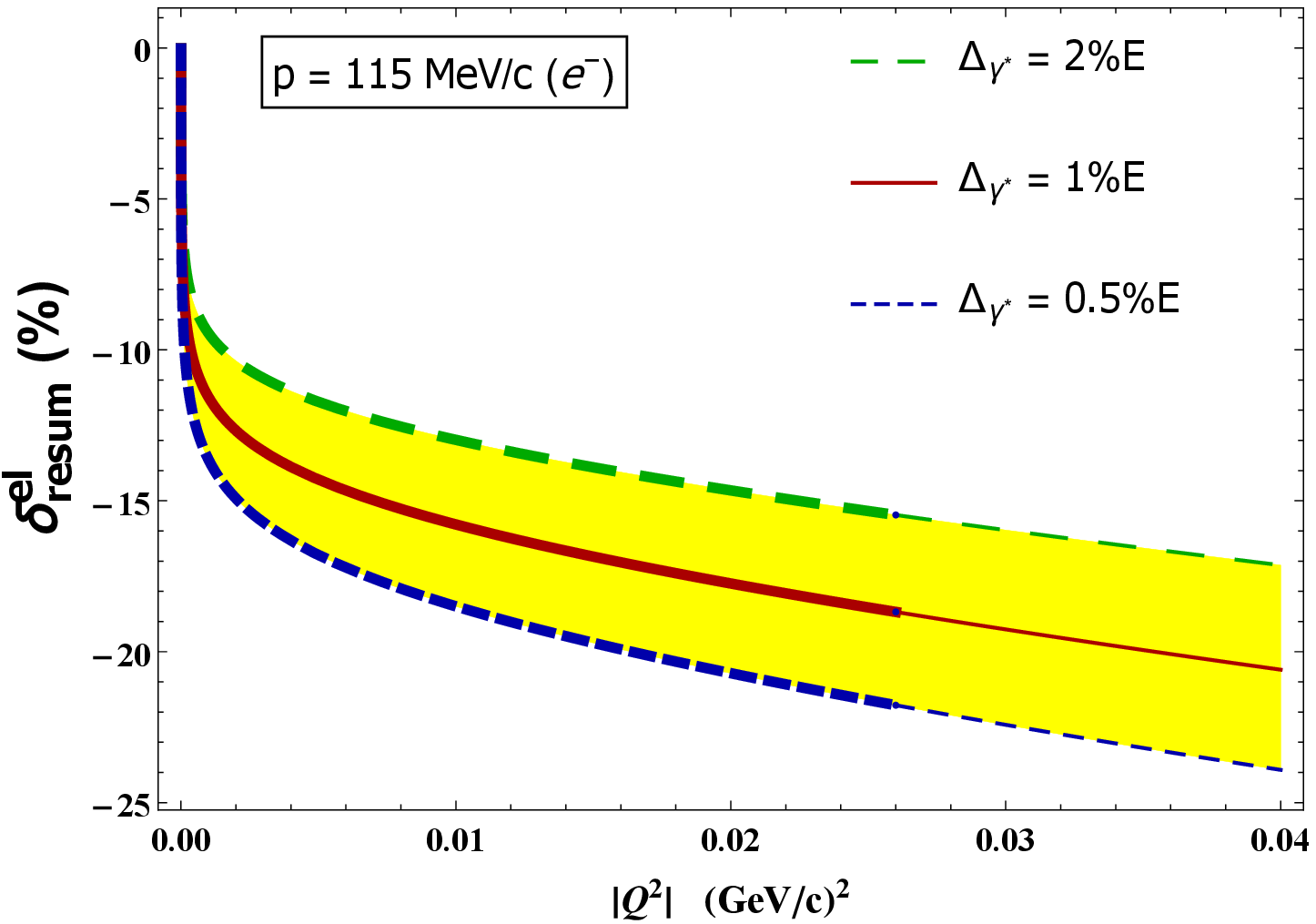} \qquad 
           \includegraphics[scale=0.53]{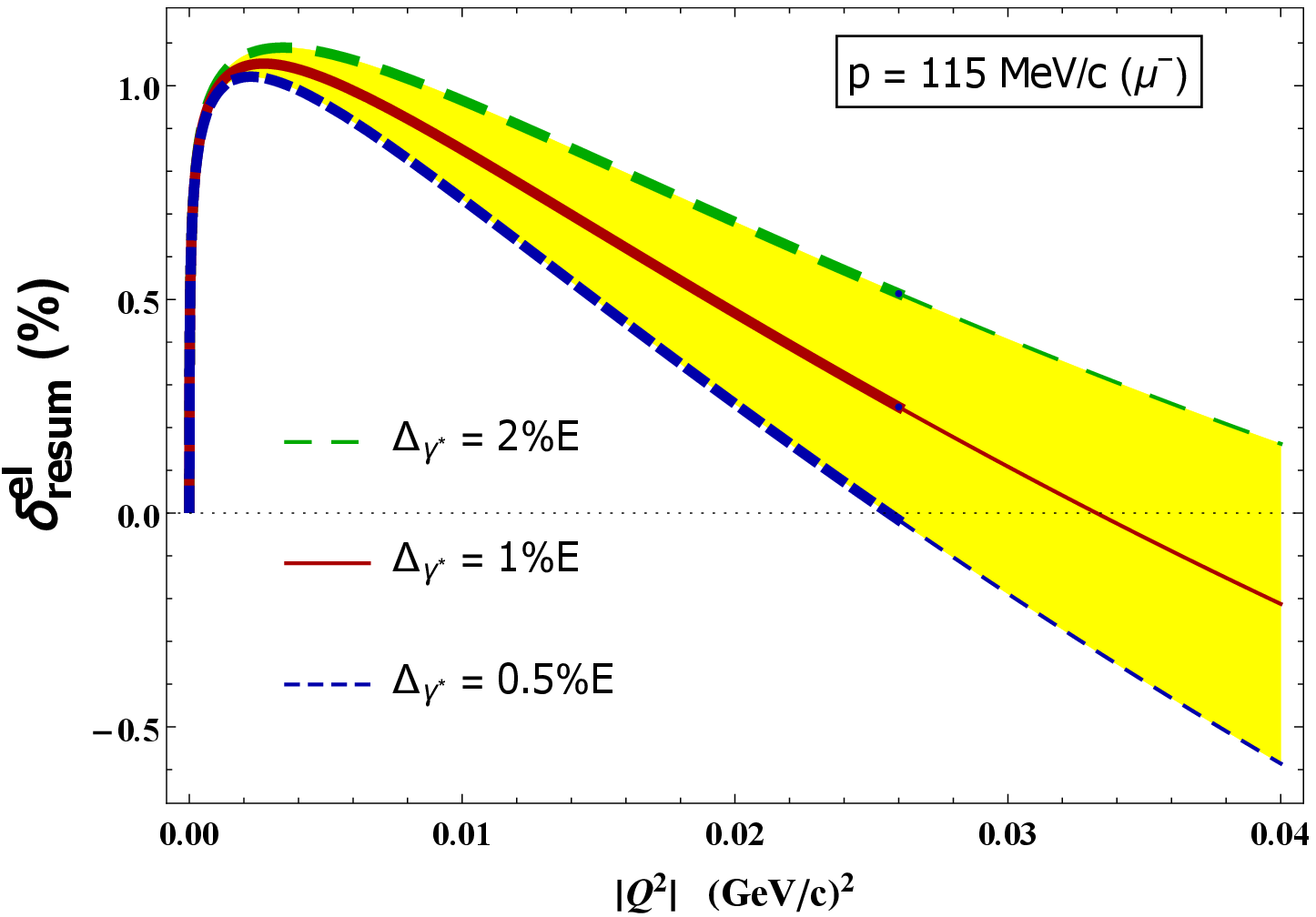}
           \caption{Variation of the partially re-summed fractional radiative correction 
                    up-to-and-including NLO$_\alpha$, $\delta^{el}_{\rm resum}$, in HB$\chi$PT to $e$-p (left 
                    panel) and $\mu$-p (right panel) elastic cross sections as a function of $|Q^2|$ for the 
                    MUSE beam momenta, $|\vec p\,|=p=115,\, 153,\, 210$~MeV/c. Each plot covers the full 
                    kinematic range $0<|Q^2|<|Q^2_{\rm max}|$ when $\theta\in[0,\pi]$. The thickened portion 
                    of each curve corresponds to the MUSE kinematic cut, where $\theta\in[20^\circ,100^\circ]$. 
                    The (yellow) bands correspond to the variation of the results with the {\it lab.}-frame 
                    detector acceptance in the range, $0.5\%<\Delta_{\gamma^*}<2\%$, of the incident lepton 
                    energy $E$.}
            \label{fig:delta_error} 
\end{figure*}
%%%%%%%%%%%%%%%%%%%%%%%%%%%%%%%%%%%%%%%%%%%%%%%%%%%%%%%%%%%%%%%%%%%%%%%%%%%%%%%%%%%%%%%%%% 

\begin{figure*}[tbp]
 \centering
         \includegraphics[scale=0.53]{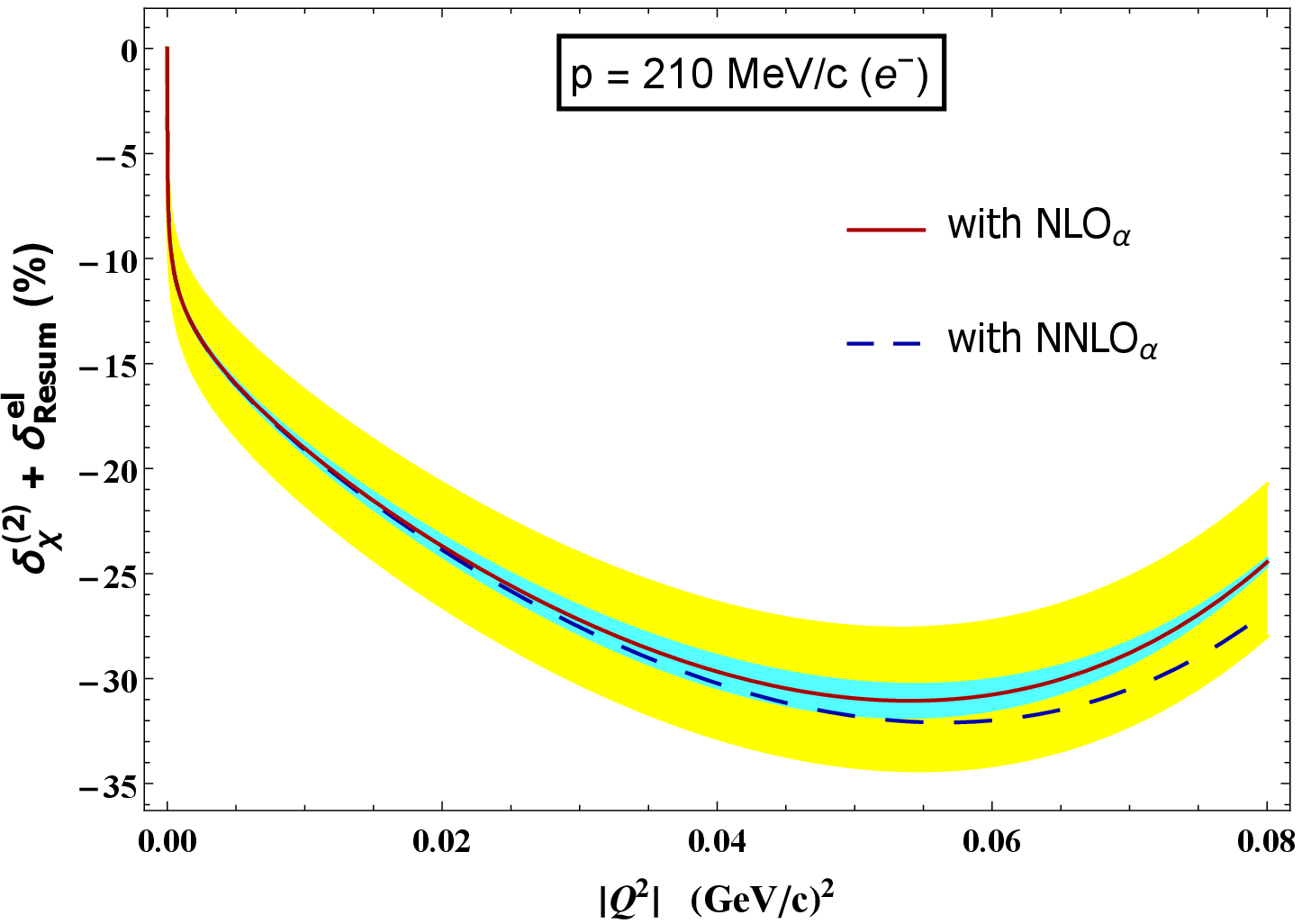} \qquad 
         \includegraphics[scale=0.53]{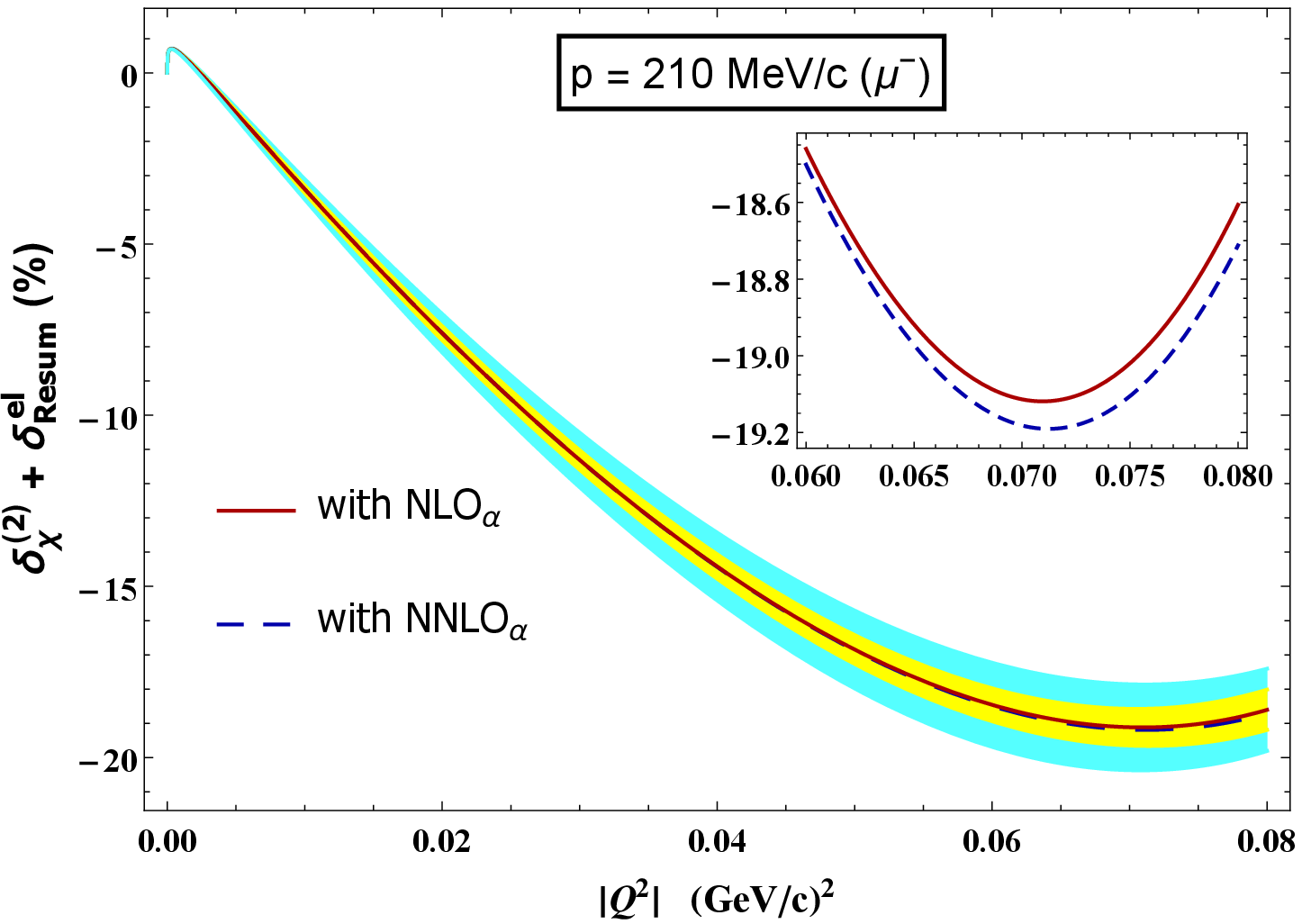}
         
         \vspace{1cm}
         
         \includegraphics[scale=0.53]{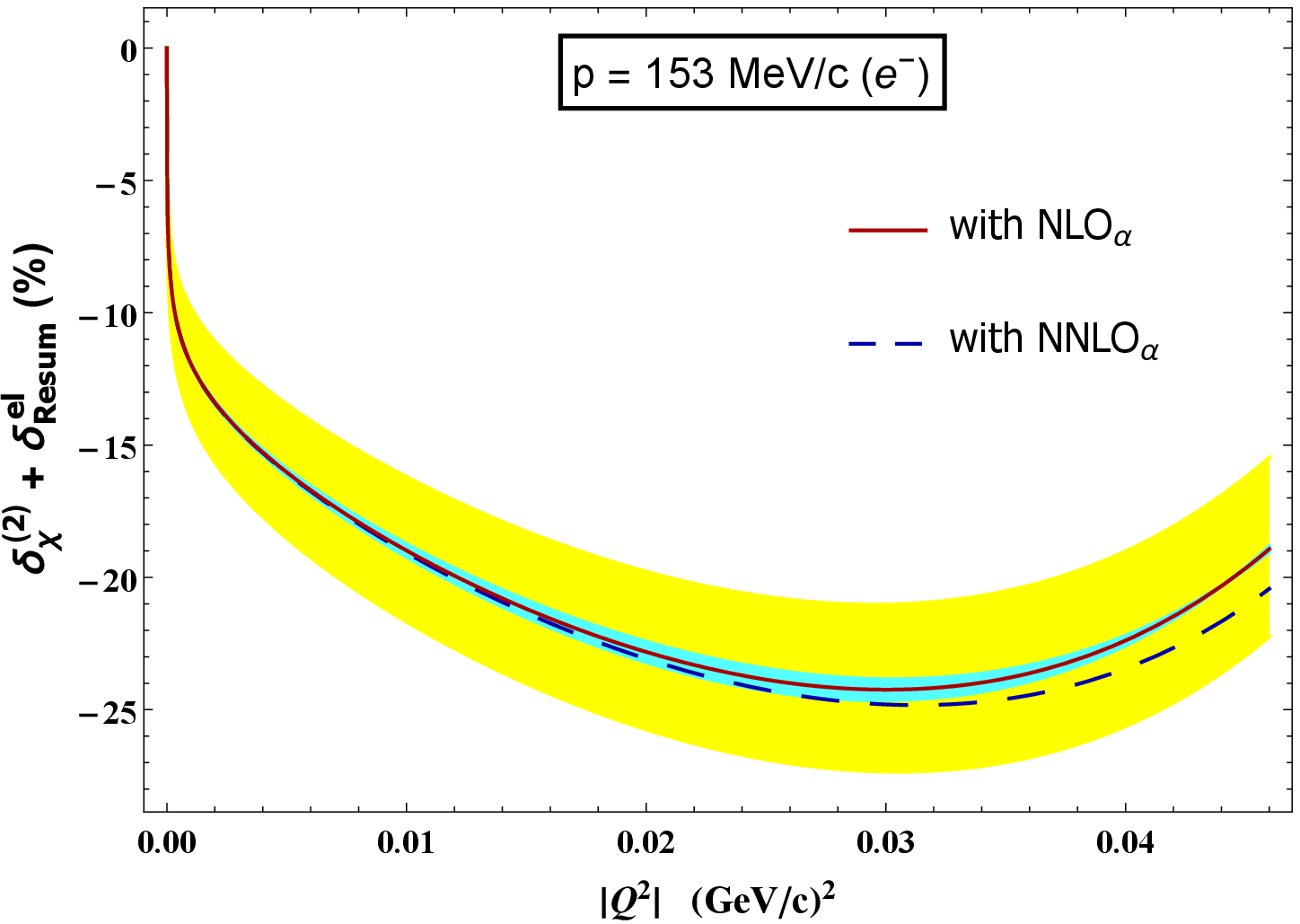} \qquad 
         \includegraphics[scale=0.53]{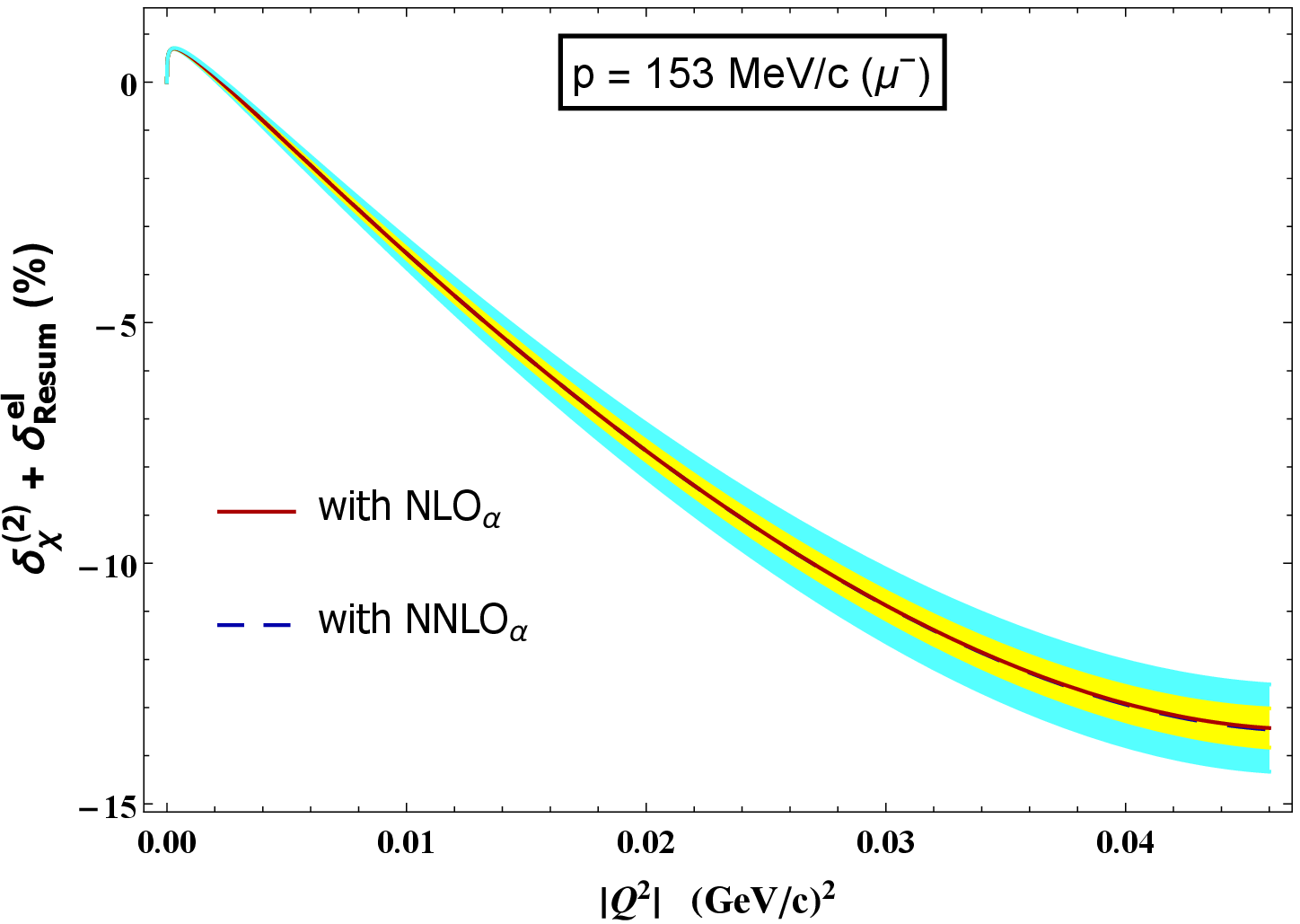}
         
         \vspace{1cm}
         
         \includegraphics[scale=0.53]{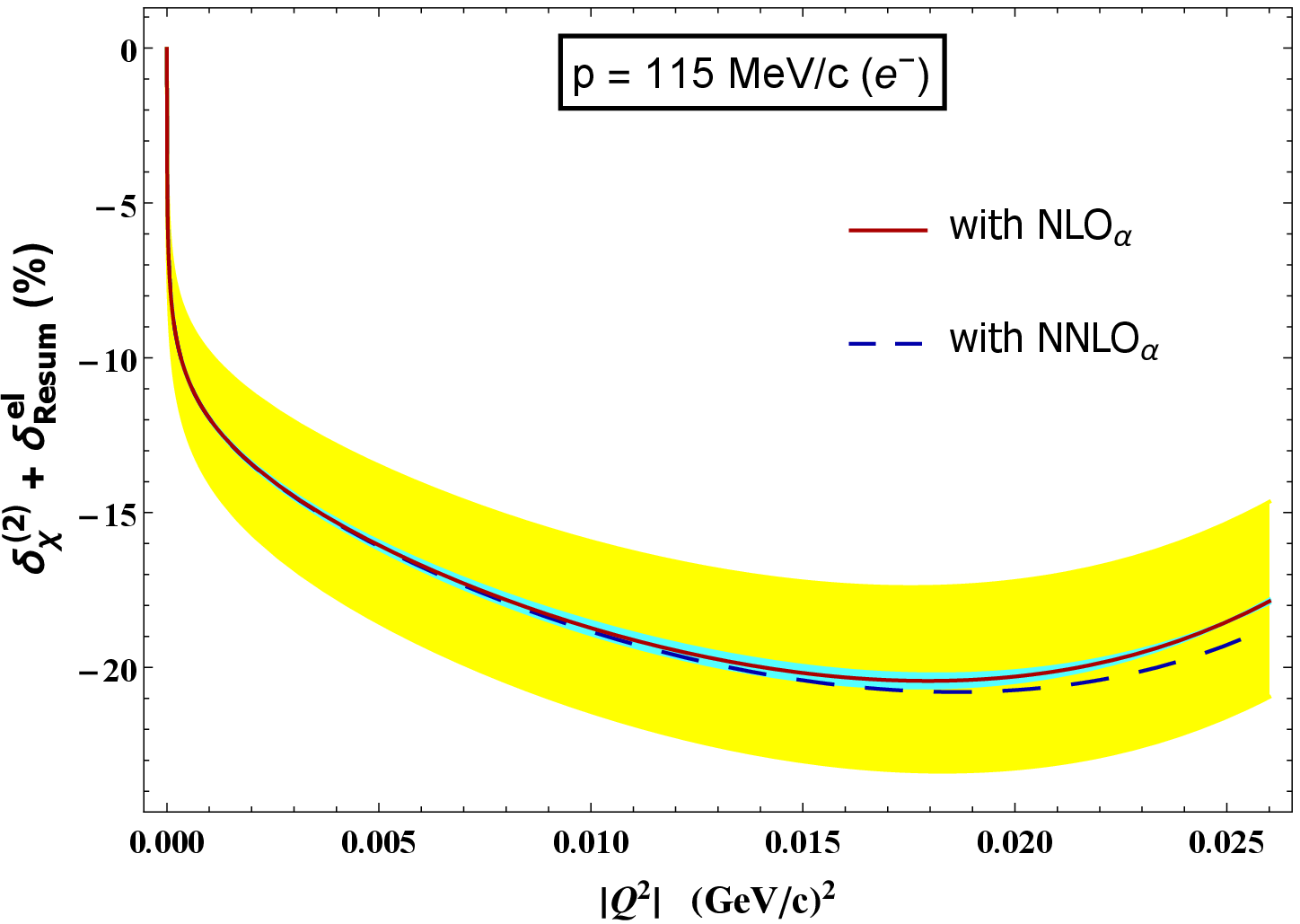} \qquad 
         \includegraphics[scale=0.53]{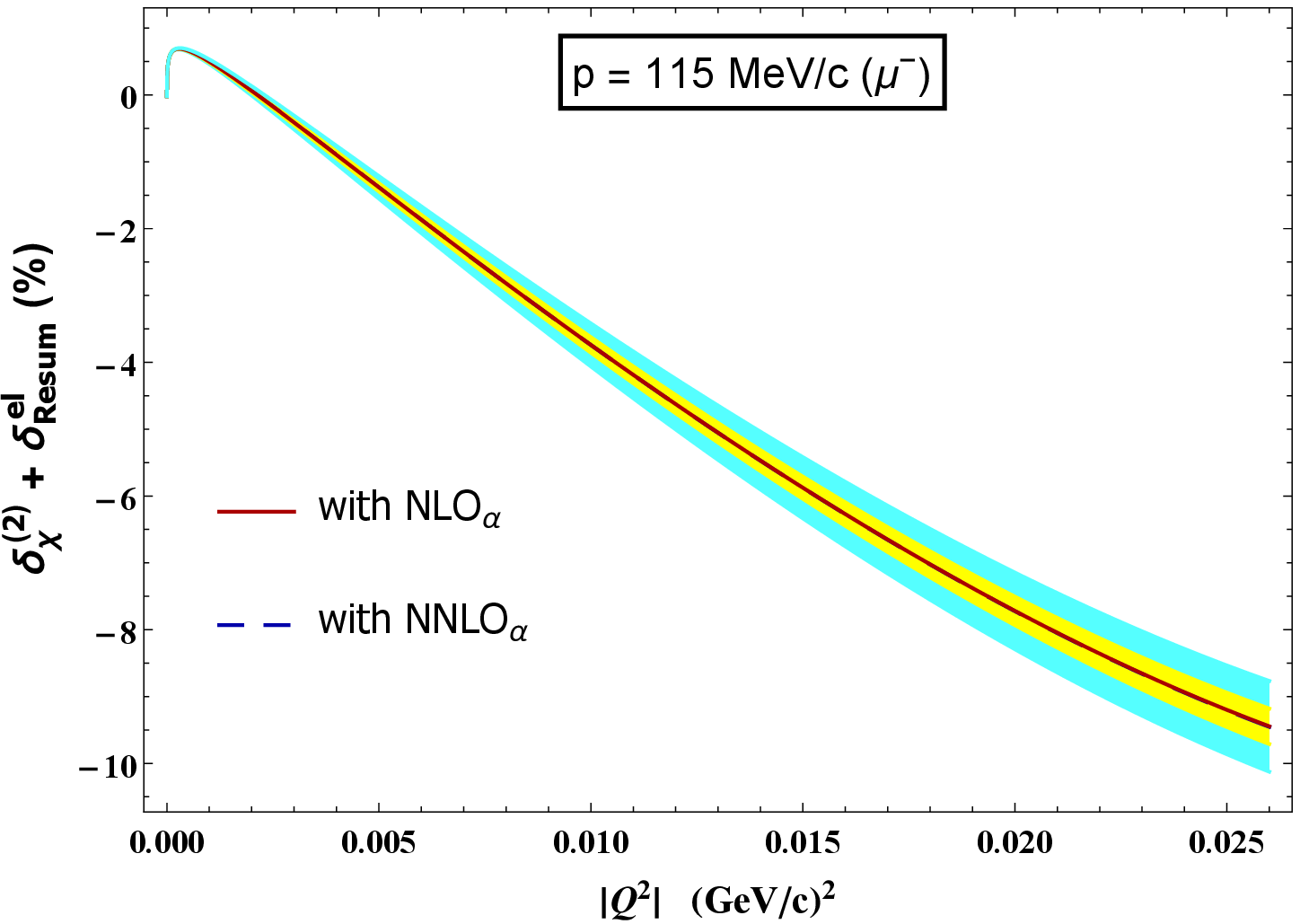}
         \caption{The total fractional corrections considered in this work, i.e. 
                  $\delta^{(2)}_\chi+\delta^{el}_{resum}$, (in percentage) for $e$-p (left panel) 
                  and $\mu$-p (right panel) elastic scattering cross sections as a function of 
                  $|Q^2|$. Each plot covers only the MUSE kinematic range of $|Q^2|$ values where  
                  the scattering angle lies within the range, $\theta\in[20^\circ,100^\circ]$, 
                  at specific incoming lepton momenta, $|\vec p\,|=p=115,\, 153,\, 210$ MeV/c. The
                  solid (dashed) red (blue) curves correspond to the partially re-summed radiative 
                  corrections up-to-and-including NLO$_\alpha$ (including NNLO$_\alpha$), with the 
                  {\it lab.}-frame detector acceptance $\Delta_{\gamma^*}$ is $1\%$ of the 
                  incident lepton energy $E$. The yellow bands correspond to the error in the 
                  radiative corrections due to the variation, $0.5\%<\Delta_{\gamma^*}<2\%$,
                  while the bands in cyan correspond to the error in the pure hadronic chiral 
                  corrections $\delta^{(2)}_\chi$.}
         \label{fig:total_delta_chrad}
\end{figure*}

The uncertainties in our treatment of the radiative corrections can be summarized as follows:    
\begin{enumerate}
\item 
The only free parameter in this work is the soft-photon detector acceptance factor $\Delta_{\gamma^*}$. 
The sensitivity of our result on the the parametric dependence on this cut-off parameter is illustrated 
in Fig.~\ref{fig:delta_error}. This figure depicts our analytical results corresponding to the partially 
re-summed NLO$_\alpha$ radiative corrections $\delta^{el}_{\rm resum}$, as $\Delta_{\gamma^*}$ is varied
in the reasonable range (0.5 - 2)\% of the incident lepton beam energy $E$, with $\Delta_{\gamma^*}=1\%$ 
being our benchmark value of the acceptance. As expected, for electron scattering the radiative 
corrections decrease in magnitude with larger values of the acceptance. However, for muon scattering the 
behavior is somewhat atypical due the change of sign of $\delta^{(0)}_{2\gamma}$ versus $|Q^2|$.  
\item 
Our HB$\chi$PT calculations indicate much larger than expected NLO$_\alpha$ : LO$_\alpha$ relative 
corrections in case of muon scattering close to the upper limit of the MUSE kinematic range. As mentioned, 
for muon scattering $\delta_{2\gamma}$ goes through a zero at some small $Q^2$ value. Especially, when 
$\delta^{(0)}_{2\gamma} \approx 0$, the NNLO$_\alpha$ contributions are needed for a more robust 
evaluation of radiative corrections. We, however, provide a partial assessment of the NNLO$_\alpha$ effects
that reveals a maximal uncertainty of 3\% and 0.1\% respectively, to effect the LO elastic Born cross 
section for electron and muon scatterings at MUSE energies (cf. Fig.~\ref{fig:total_delta_chrad}).  
\item 
One source of uncontrolled systematics afflicting our evaluations is attributed to the inherent differences
in the TPE evaluations with and without invoking SPA. As already mentioned, an exact evaluation of our 
IR-divergent one-loop diagrams with an insertion of a heavy baryon propagator is rather intricate and has 
not been pursued earlier. We already referred to Ref.~\cite{Tomalak:2014dja} where an attempt was made in 
order to include the hard two-photon effects in the TPE contributions. We are, however, unable to currently 
assess the uncertainty due to the missing hard-photon contribution based solely on this work. We simply 
refer to an ongoing effort~\cite{Poonam2021} to analytically evaluate a family of such TPE direct and 
crossed box diagrams at NLO$_\alpha$, wherein the large cancellations amongst them in the $k\to 0$ limit, as
noted in Ref.~\cite{Talukdar:2019dko}, is not explicitly manifest without SPA. A detailed investigation of 
the TPE diagrams without SPA shall be presented in a future publication. 
\end{enumerate}

%%%%%%%%%%%%%%%%%%%%%%%%%%%%%%%%%%%%%%%FIGURE%%%%%%%%%%%%%%%%%%%%%%%%%%%%%%%%%%%%%%%%%%%%%    
%%%%%%%%%%%%%%%%%%%%%%%%%%%%%%%%%%%%%%%%%%%%%%%%%%%%%%%%%%%%%%%%%%%%%%%%%%%%%%%%%%%%%%%%%% 

%\vspace{0.1cm}

Figure~\ref{fig:total_delta_chrad}  displays our HB$\chi$PT results for the total corrections (chiral plus 
radiative) to the $\ell$-p elastic differential cross section, Eq.~\eqref{eq:final_result}, where all 
tractable sources of systematic uncertainties are consolidated. Our central results for the total factional 
corrections, $\delta^{(2)}_\chi+\delta^{el}_{resum}$, as denoted by the solid red (color online) curves, 
correspond to the radiative corrections up-to-and-including NLO$_\alpha$ partially re-summed to all order 
in QED. Likewise, our partially re-summed results including the NNLO$_\alpha$ terms, that contribute to the
theoretical uncertainty, are denoted by the dashed blue (color online) curves. For electron scattering, the
largest conceivable source of theoretical uncertainty evidently stems from the parametric dependence on the 
detector acceptance $\Delta_{\gamma^*}$ (yellow bands) which overwhelms the proton's $\delta^{(2)}_\chi$ 
uncertainty (cyan bands). In contrast, for muon scattering the uncertainty in both $\delta^{el}_{resum}$ and
$\delta^{(2)}_\chi$ are rather moderate with the latter slightly larger than the former. Only for electron 
scattering do we find a moderate difference ($\sim 3$\% of the LO Born cross section) between our complete 
NLO$_\alpha$ and partially included NNLO$_\alpha$ results.

%%%%%%%%%%%%%%%%%%%%%%%%%%%%%%%%%%%%%%%%%%%%%%%%%%%%%%%%%%%%%%%%%%%%%%%%%%%%%%%%%%%%%%%%%%%%%%   
\section{Summary and Outlook}\label{sec:summary} 
%%%%%%%%%%%%%%%%%%%%%%%%%%%%%%%%%%%%%%%%%%%%%%%%%%%%%%%%%%%%%%%%%%%%%%%%%%%%%%%%%%%%%%%%%%%%%%
Within the framework of HB$\chi$PT  we have evaluated up-to-and-including NNLO$_\chi$ hadronic chiral 
corrections as well as the NLO$_\alpha$ radiative corrections to elastic lepton-proton scattering cross 
section. The hadronic chiral corrections include the pion-loop effects and LECs parametrizing the 
proton structure.  We found that in the MUSE kinematic range the NNLO$_\chi$ fractional corrections with
respect to the leading order Born cross section are about $10\%$ and $20\%$ for electron and muon 
scattering processes respectively. We further estimated that the next higher order hadronic corrections 
(N$^3$LO$_\chi$) do not contribute a larger uncertainty than the current experimental discrepancies 
pertaining to the rms radius. Regarding the radiative corrections we included all possible virtual photon 
loops as well as soft photon bremsstrahlung corrections. We demonstrated that the IR divergences 
systematically cancel at each radiative chiral order, i.e.,  at LO$_\alpha$ and NLO$_\alpha$. 

%\vspace{0.1cm}

Only in our analytical calculations of the one-loop TPE box diagrams, do we invoke SPA following our 
evaluations in the previous work outlined in Ref.~\cite{Talukdar:2019dko}. The technical difficulties 
associated with the exact evaluation of the box diagrams in DR, the SPA methodology offers a standard 
analytical procedure in assessing the contribution from the elastic intermediate proton state, as  
advocated in Refs.~\cite{Maximon:2000hm} and \cite{Koshchii2017}.\footnote{To contrast our 
non-relativistic chiral expansion in inverse powers of the proton mass $M$, the analysis of the TPE 
amplitudes, e.g., in Ref~\cite{Koshchii2017}, were evaluated relativistically, meaning that they 
preserve terms to all orders in $1/M$.} Unfortunately, the contribution of the kinematically hard 
two-photon-loops integration regions is left out in the process. It is, however, a well accepted fact 
that the contribution from the hard part of the TPE loops becomes significant only at large-$|Q^2|$ 
values and lepton beam energies as well as in the proximal region of back-scattering. On the other hand,
given the very low-energy dynamics of our calculations, it is conceivable that the hard part (those 
essentially stemming from the inelastic dynamics of partons) would not effect the TPE contributions in 
any significant way (see e.g., Ref.~\cite{Bernauer:2021vbn} for      %rather 
recent discussions). Nevertheless,
the issue of including the two-hard-photon-exchange should be considered in a future investigation in 
order to reduce the systematic errors in the TPE evaluation. In order to minimize possible 
model-dependence, our approach utilizes the existing analytically derived $\chi$PT form
factors~\cite{Bernard:1992qa,Bernard:1995dp,Fearing1998}, consistent with our power counting scheme. By 
directly relying on the input proton's rms radius, we tacitly bypassed the introduction of model form 
factors, $F^p_{1,2}$ at the photon-proton vertices, unlike in the work of, e.g., 
Ref.~\cite{Tomalak:2014dja} where the finite part of the TPE amplitudes were evaluated numerically in a
relativistic framework. What is still not clear is how large are the two-hard-photon contributions 
relative to the SPA they employed in order to isolate the IR-singularity following 
Ref.~\cite{Maximon:2000hm}. For this reason, and since we can not assess the influence the form factors 
on their results, we are at this point unable to estimate the possible systematic uncertainties incurred 
due of the use of SPA in our evaluation.  

%\vspace{0.1cm}

In our estimate of the bremsstrahlung contributions due to the undetected soft photons with energies below
the detector threshold $\Delta_{\gamma^*}$, we use the soft photon momentum limit $k\to 0$ in our 
calculations, a methodology widely used to extract the IR divergences as introduced in 
Ref.~\cite{Yennie:1961ad}.  However, the introduction of the artificial dependence on the free parameter 
$\Delta_{\gamma^*}$ is certainly a demerit of the current methodology that needs to be improved in future, 
e.g., by the inclusion of the hard or detectable part of the elastic radiative tail for a realistic 
estimation of the radiative corrections. All our analytical results presented in this paper depend on 
$\Delta_{\gamma^*}\sim 1\%$, which theoretically complies with the expected lowest bremsstrahlung photon 
energy detectable in present day experiments. We simply note here that in the case of the MUSE set-up with 
a single arm beam-line arrangement only high-energy photons at forward angles can be detected. Thus, from 
a more practical view-point, a modification of our current analysis incorporating the anticipated MUSE 
features must be employed for a more pertinent future data analysis.    

%\vspace{0.1cm}

Notably, in all our evaluations in this work we have explicitly included the masses of the leptons, and also
the often neglected Pauli form factor contributions to the lepton-photon vertex corrections. Our calculations 
revealed that both the TPE and the proton bremsstrahlung process start to contribute to the radiative 
corrections only at NLO$_\alpha$ in HB$\chi$PT. Our work suggests that in the MUSE kinematic range the total 
radiative corrections up-to-and-including NLO$_\alpha$ for electron scattering can be as large as $25\%$, while 
for muon scattering they are no more than $2\%$ (cf. Fig.~\ref{fig:delta_error}). Furthermore, we observe that 
for the muon scattering the LO$_\alpha$ radiative correction $\delta^{(0)}_{2\gamma}$ goes through zero in the
MUSE kinematic range, i.e., in this energy range the NLO$_\alpha$ contributions dominate. This naturally 
indicates the importance of the NNLO$_\alpha$ corrections in order to correctly assess the insofar neglected 
proton's structure dependent chiral-radiative effects. Although such NNLO$_\alpha$ corrections have been 
partially included in this work for the sake of estimating the theoretical error, a complete NNLO$_\alpha$ 
evaluation is relegated to a possible future project. We finally remark that in the work of 
Ref.~\cite{Talukdar:2018hia}, a HB$\chi$PT estimation of the elastic radiative tail distribution was 
considered at NLO$_\alpha$ accuracy, where IR divergences were not explicitly addressed. Thus, a renewed 
low-energy HB$\chi$PT approach is needed in order to explore the prospects of a systematic inclusion of the 
radiative tail effects in a complete radiative ``unfolding" analysis in close analogy to the erstwhile work of 
Ref.~\cite{Vanderhaeghen:2000ws}.

\section{Acknowledgements}
We are thankful to Dipankar Chakrabarti, Poonam Choudhary, Maciej Rybczy\'nski and Steffen Strauch, for 
useful discussions at various stages of this work. PT acknowledges the kind hospitality of the Department 
of Physics, IIT Kanpur, where part of the work was done, and VS acknowledges the financial support through
the Polish National Science Centre (NCN) {\it via} the OPUS project 2019/33/B/ST2/00613.

%%%%%%%%%%%%%%%%%%%%%%%%%%%%%%%%%%%%%%%%%%%%%%%%%%%%%%%%%%%%%%%%%%%%%%%%%%%%%%%%%%%%%%%%%%%%%%  
\section{Appendix}
%%%%%%%%%%%%%%%%%%%%%%%%%%%%%%%%%%%%%%%%%%%%%%%%%%%%%%%%%%%%%%%%%%%%%%%%%%%%%%%%%%%%%%%%%%%%%%
%%%%%%%%%%%%%%%%%%%%%%%%%%%%%%%%%%%%%%%%%%%%%%%%%%%%
\subsection{$S$-frame}
%%%%%%%%%%%%%%%%%%%%%%%%%%%%%%%%%%%%%%%%%%%%%%%%%%%% 
In this appendix, we discuss the transformation relations among kinematical quantities between the laboratory 
frame and the boosted $S$-frame~\cite{Tsai:1961zz,Mo:1968cg,Maximon:2000hm,Vanderhaeghen:2000ws,Bucoveanu:2018soy}. 
The S-frame is defined as the center-of-mass system of the recoil proton and the soft bremsstrahlung photon, 
such that 
\begin{equation}
 {\vec p}^{\,\prime}_p +{\vec k} = {\vec Q}\, \stackrel{S{\rm -frame}}{\longrightarrow} \, {\vec Q}_S = {\vec 0}.  
\end{equation}
Here ${\vec p}^{\,\prime}_p$ and ${\vec k}$ are the respective laboratory frame three-momenta of the recoil proton 
and the emitted soft photon, and ${\vec Q}={\vec p}-{\vec p}^{\,\prime}={\vec p}^{\,\prime}_p-{\vec p}_p$ is the 
three-momentum transferred in the $\ell$-p elastic scattering process in the laboratory frame, i.e., the target 
proton three-momentum ${\vec p}_p=\vec{0}$. The maximum energy of the soft (undetected) photon fixes the upper 
limit of the bremsstrahlung energy integration and is conventionally taken as detector acceptance 
$\Delta_{\gamma^*}$ in the laboratory frame. This in essence corresponds to the maximal deviation of the outgoing 
lepton energy $E^{\prime el}$ from its theoretical elastic limit $E^\prime$ while practically preserving elastic 
conditions, i.e., $E^\prime - E^{\prime\, el}\leq \Delta_{\gamma^*}$. In the ensuing treatment using soft photon 
limit, namely, $k=(E_{\gamma^*},{\vec k})\to 0$, we shall use, 
$E^{\prime el} \approx E^{\prime}$.\footnote{In general, with real photon emissions, $E^{\prime el} \leq E^\prime$. 
The equality only holds for the ``strictly elastic" (non-radiative) kinematics which is evidently unrealistic in a 
given laboratory experiment. In this work, since we are concerned with the ``physical" elastic process that is 
naturally accompanied by soft photon bremsstrahlung, $E^{\prime el} \approx E^{\prime}$ is implicitly understood.} 
In a boosted frame the maximum photon energy limit becomes a frame dependent quantity, which we denote as 
$\Delta_S \neq \Delta_{\gamma^*}$ in the $S$-frame. The phase-space integration for the laboratory frame 
differential cross section for the soft bremsstrahlung process, $\ell{\rm p}\to \ell{\rm p}\gamma^*_{\rm soft}$, 
namely,
\begin{eqnarray}
\left[\widetilde{{\rm d}\sigma}^{\rm (LO_\alpha,\,NLO_\alpha)}_{br}\right]_{\gamma\gamma^*} 
\!\!\!\!\!&=& \frac{(2\pi)\delta^{lab}_k}{8ME^\prime_p E}\,
\frac{{\rm d}^3\vec{p}^{\,\prime}}{(2\pi)^3 2E^\prime}
\frac{{\rm d}^3\vec{k}}{(2\pi)^3 2E_{\gamma^*}}\,
\nonumber\\
&&\!\!\times\,\frac{1}{4}\sum_{spins}
\left|\widetilde{\mathcal{M}}^{\rm (LO_\alpha,\,NLO_\alpha)}_{\gamma\gamma^*}\right|^2\,,\,\,\qquad\,
\label{dsigma_brem_tilde}
\end{eqnarray}
with the LO$_\alpha$ and NLO$_\alpha$ squared bremsstrahlung amplitudes 
$\widetilde{\mathcal{M}}^{\rm (LO_\alpha,\,NLO_\alpha)}_{\gamma\gamma^*}$ in the respective soft photon limits 
[cf. Eqs.~\eqref{M0gamgam*2} and \eqref{M1gamgam*2} in the main text], are complicated by the dependence on the 
photon emission angles present in the energy conserving $\delta$-function, namely, 
$$\delta^{lab}_k\equiv \delta\left(E+M-E^\prime-\sqrt{({\vec Q}-{\vec k})^2+M^2}-E_{\gamma^*}\right),$$ 
appearing in the above expression. Consequently, the emitted photon radiation spectrum in the laboratory frame
becomes anisotropic, being defined over a ellipsoidal integration volume which is difficult to evaluate 
analytically. However, by boosting to the $S$-frame the integration simplifies into a standard spherical one 
(see, e.g., Ref.~\cite{Tsai:1961zz}), with the above $\delta$-function becoming free of the photon angles in the 
soft photon limit. This effectively transforms the $S$-frame kinematics into one akin to a ``reversed" elastic 
scenario in the soft photon limit, denoted by the constraint, 
$$\delta^{S}\equiv\delta\Big(E^S+E^S_p-E^{\prime S}-E^{\prime S}_p\Big),$$ where all $S$-frame quantities are 
denoted by the superscript/subscript ``$S$". In terms of the laboratory frame quantities, the following 
relationships can then be justified:   
\begin{eqnarray}
&\text{(i)}&\,\, E^{\prime S}_p \approx M\,,\hspace{1.06cm} \text{(ii)}\,\, E^S \approx E^{\prime}=\frac{E}{\eta}\,, 
\nonumber\\
&\text{(iii)}&\,\, E^{\prime S} \approx E\,,\hspace{1.1cm} \text{(iv)}\,\, E^S_p \approx E_p^\prime\,, 
\nonumber\\
&\text{(v)}&\,\, \cos\theta_S\approx \cos\theta\,, \hspace{0.35cm} \text{(vi)}\,\, \Delta_S \approx \eta \Delta_{\gamma^*},
\end{eqnarray}
where $\eta=1+2E\sin^2(\theta/2)/M$ is the laboratory frame proton recoil factor. In other words, the energy 
transformations between the two frames are easily effected by simply interchanging the energies between the 
initial and final states of the elastic process. 

%\vspace{0.2cm}

In view of pedagogical interests, we derive these relations between the two frames using the limit of soft photons. 
We make use of the  four-momentum conservation relation for the bremsstrahlung process, namely, 
$p+P-p^\prime=P^\prime+k$. 
\begin{itemize}
\item First, we consider the invariant $(P^\prime + k)^2$ in the $S$-frame:  
\begin{eqnarray*}
\qquad (P^{\prime S} + k^S)^2 &=& M^2 + 2E^{\prime S}_p E^S_{\gamma^*} + 2 (E^S_{\gamma^*})^2 
\\
&\stackrel{\gamma_{\rm soft}}{\leadsto}&  M^2\,.
\end{eqnarray*}
Since ${\vec p}^{\,\prime S}_p+{\vec k}^S=0$, we must have
\begin{eqnarray*}
\qquad (P^{\prime S} + k^S)^2
=(E^{\prime S}_p+E^S_{\gamma^*})^2\stackrel{\gamma_{\rm soft}}{\leadsto} (E^{\prime S}_p)^2\,, 
\end{eqnarray*}
which implies, $\boxed{E^{\prime S}_p\approx M}$\, .
\item Second, we consider the invariant $p\cdot(P^\prime + k)$. In the $S$-frame we have  
\begin{eqnarray*}
\qquad  p^S\!\cdot\!(P^{\prime S} + k^S) \!
=\! E^S (E^{\prime S}_p+E^S_{\gamma^*})\!\stackrel{\gamma_{\rm soft}}{\leadsto}\!  M E^S \,, 
\end{eqnarray*}
while in the laboratory frame we have   
\begin{eqnarray*}
p\cdot(P^\prime + k) &=& p\cdot(P + Q) 
\\
&=& ME +\frac{Q^2}{2}= M E^{\prime}\,. 
\end{eqnarray*}
This implies, $\boxed{E^S\approx E^{\prime}}$.
\item Third, we consider the invariant $p^\prime\cdot(P^\prime + k)$.  
\begin{eqnarray*}
p^\prime\cdot(P^\prime + k) &=& p^{\prime S}\cdot(P^{\prime S} + k^S) 
\\
\hookrightarrow \quad p^\prime\cdot(P + Q) &=& E^{\prime S} (E^{\prime S}_p+E^S_{\gamma^*})
\\
\stackrel{\gamma_{\rm soft}}{\leadsto}\, ME^{\prime}-\frac{Q^2}{2} &\approx &  M E^{\prime S} \,.
\end{eqnarray*}
Since, $E=E^{\prime}-Q^2/(2M)$, the above relation implies, $\boxed{E^{\prime S}\approx E}$. 
\item  Fourth, we consider the energy conservation in the $S$-frame:
\begin{eqnarray*}
\qquad E^S+E_p^S-E^{\prime S} &=& E_p^{\prime S} + E^S_{\gamma^*}\,,
\\
E^S_p &=& E_p^{\prime S} + E^S_{\gamma^*} + E^{\prime S} - E^S
\\
 & \stackrel{\gamma_{\rm soft}}{\leadsto} &  M + E - E^{\prime}\approx E_p^\prime\,,
\end{eqnarray*}
where we have used the relations derived for $E^S,\,E^{\prime S}$ and $E_p^{\prime S}$. Thus, 
$\boxed{E^S_p\approx E_p^\prime}$.
\item Fifth, we use the invariant expression for the squared four-momentum transfer $Q^2=Q^2_S$ in each frame:
\begin{eqnarray*}
\qquad  Q^2 &=&  2m^2_l-2EE^{\prime}(1-\beta\beta^{\prime}\cos\theta) \,,
\\
Q^2_S &=&  2m^2_l-2E^SE^{\prime S}(1-\beta_S\beta^\prime_S\cos\theta_S)
\\
&\stackrel{\gamma_{\rm soft}}{\leadsto}& 2m^2_l-2E^{\prime}E(1-\beta^{\prime}\beta\cos\theta_S)\,,
\end{eqnarray*}
where the incoming and outgoing lepton velocities in the {\it lab.}-frame and $S$-frame are 
($\beta=|{\vec p}\,|/E,\, \beta^\prime= |{\vec p}^{\,\prime}|/E^\prime$) and 
($\beta_S=|{\vec p}^{\,S}|/E^S,\, \beta^\prime= |{\vec p}^{\,\prime S}|/E^{\prime S}$) respectively, and 
$\theta,\,\theta_S$ are the corresponding scattering angles. Using the relations derived for $E^S,\,E^{\prime S}$,
it follows that $\boxed{\cos\theta_S \approx \cos\theta}$. 
\item Finally, squaring the aforementioned four-momentum conservation relation, and then expressing the left 
and right hand sides in terms of the laboratory frame and $S$-frame quantities respectively, yields
\begin{eqnarray*}
\qquad 2m_l^2&-&2p\cdot p^\prime+2M(E-E^{\prime el})=2P^{\prime S}\cdot k^S
\\
&=& 2M\Delta_S\sqrt{1+\left(\frac{\Delta_S}{M}\right)^2}+2\Delta^2_S\,,
\end{eqnarray*}
where $E^\prime -E^{\prime el}\leq \Delta_S$, with $\Delta_S$ being the maximal limit of the emitted soft photon 
energy in the $S$-frame, i.e., $E^S_{\gamma^*}=|{\vec k}^S|\lesssim \Delta_S\ll M$. Next to obtain an estimate for
$\Delta_S$ in the soft photon limit, we further neglect the lepton mass, $m_l \ll M$, such that the above 
relation becomes
\begin{eqnarray*}
M(E-E^{\prime el})&-&E E^{\prime el} (1-\cos\theta)
\\
&=&M\Delta_S\left[1+{\mathcal O}\left(\frac{\Delta_S}{M}\right)\right]\,.
\end{eqnarray*}
Furthermore, in the elastic limit, i.e, with $\Delta_S\to 0$ and $E^{\prime el}\to E^{\prime}$, the above equation 
reduces to 
\begin{eqnarray*}
 M(E-E^{\prime})-E E^{\prime} (1-\cos\theta)=0\,.
\end{eqnarray*}
Then, subtracting the latter relation from the former, yields our desired expression for $\Delta_S$:
\begin{eqnarray*}
 \Delta_S&=&(E^{\prime}-E^{\prime el})\left[1+\frac{2E}{M}\sin^2\left(\frac{\theta}{2}\right)\right]
 \\
 &&+\,{\mathcal O}\left(\frac{\Delta_S}{M}\right) \approx \eta \Delta_{\gamma^*}\,.
\end{eqnarray*}
Thus, the upper limit of the soft photon bremsstrahlung integrals (see Appendix B) in the $S$-frame is taken as 
$\boxed{\Delta_S\leq \eta\Delta_{\gamma^*}}$. 
\end{itemize}

%%%%%%%%%%%%%%%%%%%%%%%%%%%%%%%%%%%%%%%%%%%%%%%%%%%%
\subsection{Soft Bremsstrahlung Integrals}
%%%%%%%%%%%%%%%%%%%%%%%%%%%%%%%%%%%%%%%%%%%%%%%%%%%%
As detailed in this paper, the one-loop virtual radiative corrections posses IR divergences at the amplitude 
level. These  integrals are conveniently evaluated by boosting to the $S$-frame following, e.g., 
Refs.~\cite{Tsai:1961zz,Mo:1968cg,Maximon:2000hm,Vanderhaeghen:2000ws,Bucoveanu:2018soy}, and some details of 
these calculations will be outlined in this appendix. For the bremsstrahlung corrections, similar IR 
divergences have to be extracted at the cross section level, which involve integration over the soft photon 
radiative tail below the detector threshold, i.e., $E_{\gamma^*} < \Delta_{\gamma^*}$. The IR divergences so 
extracted in each case were demonstrated to cancel order by order. Below we demonstrate the process of 
extracting the IR divergences using DR from the phase-phase integration of Eq.~\eqref{dsigma_brem_tilde}. 
Notably, the assumption that soft photon emissions does not effectively alter the elastic kinematics, implies
the approximation that the four-momentum transfer for the bremsstrahlung process, $q=(Q-k)\to Q$, the 
four-momentum transfer for the elastic process. This simplification allows the photon phase-space integration 
to be performed analytically in closed form. Using the LO$_\alpha$ and NLO$_\alpha$ squared bremsstrahlung 
amplitudes [cf. Eqs.~\eqref{M0gamgam*2} and \eqref{M1gamgam*2} in the main text] results in the following type
of integrals:
\begin{eqnarray}
\text{@ LO$_\alpha$ \& NLO$_\alpha$ } &:& \!\!\quad \int\frac{{\rm d}^3\vec{k}}{k}\frac{1}{(p\cdot k)^2}\delta^{lab}_k\,,\quad\,
\end{eqnarray}
\begin{eqnarray}
\text{@ LO$_\alpha$ \& NLO$_\alpha$} &:& \!\!\quad \int\frac{{\rm d}^3\vec{k}}{k}\frac{1}{(p^\prime\cdot k)^2}\delta^{lab}_k\,,\quad\,
\\
\text{@ LO$_\alpha$ \& NLO$_\alpha$} &:& \!\!\quad \int\frac{{\rm d}^3\vec{k}}{k}\frac{p^\prime\cdot p}{(p\cdot k)(p^\prime\cdot k)}\delta^{lab}_k\,,\,\,\,\,\quad\,
\\
\text{@ NLO$_\alpha$} &:& \!\!\quad \int\frac{{\rm d}^3\vec{k}}{k}\frac{l}{(v\cdot k)(p\cdot k)}\delta^{lab}_k\,,\,\,\,\,\quad\,
\\
\text{@ NLO$_\alpha$} &:& \!\!\quad \int\frac{{\rm d}^3\vec{k}}{k}\frac{l}{(v\cdot k)(p^\prime\cdot k)}\delta^{lab}_k\,.\,\,\,\,\quad\,
\end{eqnarray} \\
Especially, the first three types of integrals appearing in both the LO$_\alpha$ and NLO$_\alpha$ 
contributions to the bremsstrahlung process were known from earlier works, e.g., in 
Refs.~\cite{Vanderhaeghen:2000ws,deCalan:1990eb}. The presence of the {\it bremsstrahlung} 
$\delta$-function constraint $\delta^{lab}_k$ (see Appendix A) in the above integrals complicates their 
evaluation in the laboratory frame. However, this apparent hurdle is overcome by employing Tsai's 
technique~\cite{Tsai:1961zz} of boosting to the $S$-frame in the soft photon limit. The corresponding 
{\it elastic} $\delta$-function $\delta^S$ (see Appendix A) becomes $|\vec k|=E_{\gamma^*}$ independent
and can therefore be readily taken outside the integrals. The resulting integrals in the $S$-frame are 
evaluated by analytically continuing to $d-1$ spatial dimensions, where $d-1=3-2\epsilon_{\rm IR}$
with $\epsilon_{\rm IR}<0$. Thus, in terms of laboratory frame variables, the following $S$-frame 
integrals are required to be evaluated {\it via} dimensional regularization (DR):
\begin{eqnarray}
L_{\rm ii}^{(S)} &=& \frac{m^2_l}{2}\int\frac{{\rm d}^3\vec{k}^S}{k^S}\frac{1}{(p^S\cdot k^S)^2}
\nonumber\\
\nonumber\\
&\stackrel{\rm DR}{\longmapsto}& \frac{m^2_l}{2}\int\frac{{\rm d}^{d-1}{k^S}}{k^S}\frac{1}{(p^S\cdot k^S)^2}
\nonumber\\
&=&\frac{m^2_l}{2}(2\pi\mu)^{2\epsilon_{\rm IR}}\int_0^{\eta\Delta_{\gamma^*}} (k^S)^{d-3} {\rm d}k^S
\nonumber\\
&&\times\,\oiint {\rm d}^{d-2}\,\Omega^S_k \frac{1}{(p^S\cdot k^S)^2}
\nonumber\\
&=& \pi\left[\frac{1}{|\epsilon_{\rm IR}|}+\gamma_E-\ln\left(\frac{4\pi\mu^2}{-Q^2}\right)\right]
+2\pi\widetilde{L}^{(S)}_{\rm ii}\,,
\nonumber\\
\end{eqnarray}
where $\mu$ is the subtraction scale, and similarly,
%%%%%%%%%%%%%%%%%%%%%%%%%%%%%%%%%%%%%%%%%%%%%%%%%%%%
\begin{widetext}
\begin{eqnarray}
%%%%%%%%%%%%%%%%%%%%%%%%%%%%%%%%%%%%%%%%%%%%%%
L^{(S)}_{\rm ff}&\stackrel{\rm DR}{\longmapsto}&\frac{m^2_l}{2}
\int\frac{{\rm d}^{d-1}{k^S}}{k^S}\frac{1}{(p^{\prime S}\cdot k^S)^2}
= \pi\left[\frac{1}{|\epsilon_{\rm IR}|}+\gamma_E-\ln\left(\frac{4\pi\mu^2}{-Q^2}\right)\right]
+2\pi\widetilde{L}^{(S)}_{\rm ff}\,,
%%%%%%%%%%%%%%%%%%%%%%%%%%%%%%%%%%%%%%%%%%%%%%
\\
\nonumber\\
%%%%%%%%%%%%%%%%%%%%%%%%%%%%%%%%%%%%%%%%%%%%%%
L^{(S)}_{\rm if}&\stackrel{\rm DR}{\longmapsto}&\int\frac{{\rm d}^{d-1}{k^S}}{k^S}
\frac{(p^{\prime S}\cdot p^S)}{(p^S\cdot k^S)(p^{\prime S}\cdot k^S)}
= \pi\left[\frac{1}{|\epsilon_{\rm IR}|}+\gamma_E-\ln\left(\frac{4\pi\mu^2}{-Q^2}\right)\right] \frac{\nu^2+1}{\nu}
\ln\left[\frac{\nu+1}{\nu-1}\right]+2\pi\widetilde{L}^{(S)}_{\rm if}\,,\,\,\quad
%%%%%%%%%%%%%%%%%%%%%%%%%%%%%%%%%%%%%%%%%%%%%%
\\
\nonumber\\
%%%%%%%%%%%%%%%%%%%%%%%%%%%%%%%%%%%%%%%%%%%%%% 
L_{\rm i}^{(S)} &\stackrel{\rm DR}{\longmapsto}&\frac{1}{2}
\int\frac{{\rm d}^{d-1} {k^S}}{k^S}\frac{1}{(v\cdot k^S)(p^S\cdot k^S)} 
=\frac{\pi}{\beta^\prime E^\prime}\left[\frac{1}{|\epsilon_{\rm IR}|}+\gamma_E-\ln\left(\frac{4\pi\mu^2}{-Q^2}\right)\right]
\ln\sqrt{\frac{1+\beta^\prime}{1-\beta^\prime}}+2\pi\widetilde{L}^{(S)}_{\rm i}\,,\,\,\quad
%%%%%%%%%%%%%%%%%%%%%%%%%%%%%%%%%%%%%%%%%%%%%%
\end{eqnarray}
\begin{eqnarray}
%%%%%%%%%%%%%%%%%%%%%%%%%%%%%%%%%%%%%%%%%%%%%%
L^{(S)}_{\rm f}&\stackrel{\rm DR}{\longmapsto}&\frac{1}{2}
\int\frac{{\rm d}^{d-1} {k^S}}{k^S}\frac{1}{(v\cdot k^S)(p^{\prime S}\cdot k^S)}
=\frac{\pi}{\beta E}\left[\frac{1}{|\epsilon_{\rm IR}|}+\gamma_E-\ln\left(\frac{4\pi\mu^2}{-Q^2}\right)\right]
\ln\sqrt{\frac{1+\beta}{1-\beta}}+2\pi\widetilde{L}^{(S)}_{\rm f}\,.\,\,\quad
%%%%%%%%%%%%%%%%%%%%%%%%%%%%%%%%%%%%%%%%%%%%%%
\end{eqnarray}
\end{widetext} 
%%%%%%%%%%%%%%%%%%%%%%%%%%%%%%%%%%%%%%%%%%%%%%%%%%%%
Here, $k^S=|{\vec k}^S|=E^S_{\gamma^*}$ is the soft photon three-momentum or energy, $\beta=p/E$ and 
$\beta^\prime=p^\prime/E^\prime \approx \beta^{\prime el}=p^{\prime el}/E^{\prime el}$ are the incoming and 
elastically scattered outgoing lepton velocities, and as found ubiquitous in the main text,
$\displaystyle{\nu=\sqrt{1-4m_l^2/Q^2}}$ is an invariant kinematical variable associated with the radiative 
corrections to the lepton scattering. Note that we additionally encounter an integral stemming from 
Eq.~\eqref{M1gamgam*2} (main text), formally contributing to the bremsstrahlung cross section at NLO$_\alpha$ but 
kinematically suppressed to NNLO$_\alpha$ being proportional to ${\mathscr R}_Q$, namely, 
$$2{\mathscr R}_Q\int\frac{{\rm d}^3\vec{k}^S}{(k^S)^3}\,\,
\stackrel{\rm DR}{\longmapsto}\,\,2{\mathscr R}_Q\int\frac{{\rm d}^{d-1}{k^S}}{(k^S)^3}\to 0,$$
which is scaleless and vanishes trivially on using DR. Next, after isolating the finite parts, 
$\widetilde{L}^{(S)}_{\rm ii},\, \widetilde{L}^{(S)}_{\rm ff}$ and $\widetilde{L}^{(S)}_{\rm if}$, from their 
respective IR-divergent parts, we revert back to the laboratory frame. In this case, the corresponding laboratory
frame integrals are readily obtained by substituting the energy transformation relations (see Appendix A), namely,
%%%%%%%%%%%%%%%%%%%%%%%%%%%%%%%%%%%%%%%%%%%%%%%%%%%%
\begin{widetext}
\begin{eqnarray}
%%%%%%%%%%%%%%%%%%%%%%%%%%%%%%%%%%%%%%%%%%%%%%
L_{\rm ii}&=& \pi\left[\frac{1}{|\epsilon_{\rm IR}|}+\gamma_E-\ln\left(\frac{4\pi\mu^2}{-Q^2}\right)\right]+2\pi\widetilde{L}_{\rm ii}
\quad;\quad
\widetilde{L}_{\rm ii} = \frac{1}{2}\ln\left(\frac{4\eta^2\Delta^{2}_{\gamma^*}}{-Q^2}\right)
-\frac{1}{4\beta}\ln\sqrt{\frac{1+\beta}{1-\beta}}\,,
%%%%%%%%%%%%%%%%%%%%%%%%%%%%%%%%%%%%%%%%%%%%%%
\\
\nonumber\\
%%%%%%%%%%%%%%%%%%%%%%%%%%%%%%%%%%%%%%%%%%%%%%
L_{\rm ff}&=& \pi\left[\frac{1}{|\epsilon_{\rm IR}|}+\gamma_E-\ln\left(\frac{4\pi\mu^2}{-Q^2}\right)\right]+2\pi\widetilde{L}_{\rm ff}
\quad;\quad
\widetilde{L}_{\rm ff} = \frac{1}{2}\ln\left(\frac{4\eta^2\Delta^{2}_{\gamma^*}}{-Q^2}\right)
-\frac{1}{4\beta^\prime}\ln\sqrt{\frac{1+\beta^\prime}{1-\beta^\prime}}\,,
%%%%%%%%%%%%%%%%%%%%%%%%%%%%%%%%%%%%%%%%%%%%%%
\\
\nonumber\\
%%%%%%%%%%%%%%%%%%%%%%%%%%%%%%%%%%%%%%%%%%%%%%
L_{\rm if}&=& \pi\left[\frac{1}{|\epsilon_{\rm IR}|}+\gamma_E-\ln\left(\frac{4\pi\mu^2}{-Q^2}\right)\right] \frac{\nu^2+1}{\nu}\ln\left[\frac{\nu+1}{\nu-1}\right]+2\pi\widetilde{L}_{\rm if}\,;
%%%%%%%%%%%%%%%%%%%
\nonumber\\
\nonumber\\
%%%%%%%%%%%%%%%%%%%
\widetilde{L}_{\rm if} &=& \frac{\nu^2+1}{2\nu}\Bigg[\ln\left(\frac{4\eta^2\Delta^{2}_{\gamma^*}}{-Q^2}\right)
\ln\left[\frac{\nu+1}{\nu-1}\right]+\ln^2\sqrt{\frac{1+\beta^\prime}{1-\beta^\prime}}-\ln^2\sqrt{\frac{1+\beta}{1-\beta}}
-\text{Sp}\left(1-\frac{\lambda_\nu E^\prime-E}{(1-\beta^\prime)E^\prime\xi_\nu}\right)
\nonumber\\
&&-\,\text{Sp}\left(1-\frac{\lambda_\nu E^\prime-E}{(1+\beta^\prime)E^\prime\xi_\nu}\right)
+\text{Sp}\left(1-\frac{\lambda_\nu E^\prime-E}{(1-\beta)E\lambda_\nu\xi_\nu}\right)
+\text{Sp}\left(1-\frac{\lambda_\nu E^\prime-E}{(1+\beta)E\lambda_\nu\xi_\nu}\right)\Bigg]\,,
%%%%%%%%%%%%%%%%%%%%%%%%%%%%%%%%%%%%%%%%%%%%%%
\\
\nonumber\\
%%%%%%%%%%%%%%%%%%%%%%%%%%%%%%%%%%%%%%%%%%%%%%
L_{\rm i}&=&\frac{\pi}{\beta E}\left[\frac{1}{|\epsilon_{\rm IR}|}+\gamma_E-\ln\left(\frac{4\pi\mu^2}{-Q^2}\right)\right]\ln\sqrt{\frac{1+\beta}{1-\beta}}+2\pi\widetilde{L}_{\rm i}\,;
%%%%%%%%%%%%%%%%%%%
\nonumber\\
%%%%%%%%%%%%%%%%%%%
\widetilde{L}_{\rm i}&=&\frac{1}{2\beta E}\left[\ln\left(\frac{4\eta^2\Delta^{2}_{\gamma^*}}{-Q^2}\right)\ln\sqrt{\frac{1+\beta}{1-\beta}}+\frac{1}{2}\text{Sp}\left(\frac{2\beta}{\beta+1}\right)-\frac{1}{2}\text{Sp}\left(\frac{2\beta}{\beta-1}\right)\right]\,,
%%%%%%%%%%%%%%%%%%%%%%%%%%%%%%%%%%%%%%%%%%%%%%
\\
\nonumber\\
%%%%%%%%%%%%%%%%%%%%%%%%%%%%%%%%%%%%%%%%%%%%%%
L_{\rm f}&=&\frac{\pi}{\beta^\prime E^\prime}\left[\frac{1}{|\epsilon_{\rm IR}|}+\gamma_E-\ln\left(\frac{4\pi\mu^2}{-Q^2}\right)\right]\ln\sqrt{\frac{1+\beta^\prime}{1-\beta^\prime}}+2\pi\widetilde{L}_{\rm f}\,;
%%%%%%%%%%%%%%%%%%%
\nonumber\\
%%%%%%%%%%%%%%%%%%%
\widetilde{L}_{\rm f}&=&\frac{1}{2\beta^\prime E^\prime}\left[\ln\left(\frac{4\eta^2\Delta^{2}_{\gamma^*}}{-Q^2}\right)\ln\sqrt{\frac{1+\beta^\prime}{1-\beta^\prime}}+\frac{1}{2}\text{Sp}\left(\frac{2\beta^\prime}{\beta^\prime+1}\right)-\frac{1}{2}\text{Sp}\left(\frac{2\beta^\prime}{\beta^\prime-1}\right)\right]\,,
%%%%%%%%%%%%%%%%%%%%%%%%%%%%%%%%%%%%%%%%%%%%%%
\end{eqnarray}
\end{widetext} 
%%%%%%%%%%%%%%%%%%%%%%%%%%%%%%%%%%%%%%%%%%%%%%%%%%%%
where $\displaystyle{\xi_\nu =\frac{2\nu}{(\nu+1)(\nu-1)}}$ and 
$\displaystyle{\lambda_\nu =\frac{3\nu-1}{\nu-1}}$. 

%\vspace{0.1cm}

Finally, we present the expression for the non-factorizable finite part [i.e., not proportional to the 
LO Born contribution, Eq.~\eqref{dsigmaBornL0} in the main text] of the bremsstrahlung differential cross 
sections at NLO$_\alpha$ [cf. Eq.~\eqref{NLO:brem_cross} in the main text] involving an exact analytical 
evaluation ({\it viz.} without considering the soft photon limit) by boosting to the $S$-frame:   
%%%%%%%%%%%%%%%%%%%%%%%%%%%%%%%%%%%%%%%%%%%%%%%%%%%%
\begin{widetext}
\begin{eqnarray}
\left[\frac{{\rm d}\sigma_{br}(Q^2)}{{\rm d}\Omega^\prime_l}\right]^{\,lp(1);\,{\rm v}}_{\gamma\gamma^*}\!
\Bigg|_\text{$S$-frame}\!\!\!\!\!\!\!\!\!\!\!\!&=&
-\,\frac{4\alpha^3}{\pi^2 Q^2 M}\left(\frac{\eta\beta}{\beta^\prime} \right)\int_0^{\eta\Delta_{\gamma^*}} \!\!\!\!\!
\frac{k^S\, {\rm d}k^S}{Q^2-2k^S(E^\prime-E)}\Bigg[4\pi (E+E^\prime) +m_l^2(E+E^\prime)(\text{L}^{(S)}_{\rm{iv}}-\text{L}^{(S)}_{\rm{fv}})
\nonumber\\
&&\hspace{4cm}-\,E^\prime \mathbb{L}^{(S)}_{\rm fv}-E\mathbb{L}^{(S)}_{\rm iv}+\,k^S\left(\frac{1}{2}Q^2-m_l^2\right)(\text{L}^{(S)}_{\rm{iv}}
+\text{L}^{(S)}_{\rm{fv}})\Bigg]\,,
\nonumber\\
&=&-\,\frac{8\alpha^3 \eta\Delta_{\gamma^*}}{\pi Q^4 M}\left(\frac{\eta\beta}{\beta^\prime} \right)\Bigg[\eta\Delta_{\gamma^*}(E+E^\prime)- 
\left\{2m_l^2 (E-E^\prime)-\eta\Delta_{\gamma^*}\left(\frac{1}{2}Q^2-m_l^2\right)\right\}
\nonumber\\
&&\times\,\left\{\frac{1}{E\beta}\ln\sqrt{\frac{1+\beta}{1-\beta}}
-\frac{1}{E^\prime\beta^\prime}\ln\sqrt{\frac{1+\beta^\prime}{1-\beta^\prime}}\,\right\}
-\,\eta\Delta_{\gamma^*}\frac{E^2}{E^\prime\beta^\prime}
\left\{\left(1-\frac{\beta}{\beta^\prime}\right)\ln\sqrt{\!\frac{1+\beta^\prime}{1-\beta^\prime}}-\beta\right\}
\nonumber\\
&&-\,\eta\Delta_{\gamma^*}\frac{E^{\prime 2}}{E\beta}
\left\{\left(1-\frac{\beta^\prime}{\beta}\right)\ln\sqrt{\frac{1+\beta}{1-\beta}}-\beta^\prime\right\}
+o\left(\frac{\left\{{\mathcal Q},\eta\Delta_{\gamma^*}\right\}}{M}\right)\Bigg]\,.
\label{x_sect:dia_v}
\end{eqnarray}
\end{widetext} 
%%%%%%%%%%%%%%%%%%%%%%%%%%%%%%%%%%%%%%%%%%%%%%%%%%%%
Here the evaluated integral is expressed in terms of {\it lab.}-frame 
quantities for convenience. Noting that, $E-E^\prime=-Q^2/2M$, the symbol ``o" represents other possible $1/M$ 
order terms which arise due to the soft bremsstarhlung integration over the $\Delta^{(1/M)}$s [cf. 
Eqs.~\eqref{eq:Delta_M}] and contribute to the NNLO$_\alpha$ theoretical error. Since these terms are eventually 
dropped from our central analytical results intended at NLO$_\alpha$ accuracy, for brevity we refrain from 
displaying such lengthy expressions. They are, nonetheless, evaluated numerically while estimating the error. 
Finally, in the above expression, $\text{L}^{(S)}_{\rm iv},\,\mathbb{L}^{(S)}_{\rm iv},\,\text{L}^{(S)}_{\rm fv}$ 
and $\mathbb{L}^{(S)}_{\rm fv}$ are the finite two-dimensional angular integrals: 
%%%%%%%%%%%%%%%%%%%%%%%%%%%%%%%%%%%%%%%%%%%%%%%%%%%%
\begin{widetext}
\begin{eqnarray}
\text{L}^{(S)}_{\rm iv\,}\!&=&\!\oiint \,{\rm d}\Omega^S_k\, \frac{1}{p^S\cdot k^S}
\!=\!\frac{4\pi}{E^\prime\beta^\prime k^S}\ln\sqrt{\frac{1+\beta^\prime}{1-\beta^\prime}}\,,
\\
\mathbb{L}^{(S)}_{\rm iv}\!&=&\!\oiint \,{\rm d}\Omega^S_k\, \frac{p^{\prime S}\cdot k^S}{p^S\cdot k^S}
=\frac{4\pi E}{E^\prime\beta^\prime}\left[\left(1-\frac{\beta}{\beta^\prime}\right)
\ln\sqrt{\frac{1+\beta^\prime}{1-\beta^\prime}}-\beta\right]\,,
\\
\text{L}^{(S)}_{\rm fv}\!&=&\!\oiint \,{\rm d}\Omega^S_k\, \frac{1}{p^{\prime S}\cdot k^S} 
\!=\!\frac{4\pi}{E\beta k^S}\ln\sqrt{\frac{1+\beta}{1-\beta}}\,,
\\
\mathbb{L}^{(S)}_{\rm fv}\!&=&\!\oiint \,{\rm d}\Omega^S_k\, \frac{p^S\cdot k^S}{p^{\prime S}\cdot k^S}
=\frac{4\pi E^\prime}{E\beta}\left[\left(1-\frac{\beta^\prime}{\beta}\right)
\ln\sqrt{\frac{1+\beta}{1-\beta}}-\beta^\prime\right]\,.
\end{eqnarray}
On reverting back to the laboratory frame from the $S$-frame, we again use the same set of kinematical 
transformations (see Appendix A) to obtain the contribution to the fractional NLO$_\alpha$ 
bremsstrahlung corrections $\delta^{(1)}_{\gamma\gamma^*}$ [cf. Eq.~\eqref{eq:delta_brems} in the 
main text] to the LO elastic (Born) differential cross section:   
\begin{eqnarray}
\delta^{\,lp(1);\,{\rm v}}_{\gamma\gamma^*}(Q^2)
&=&\left[\frac{{\rm d}\sigma_{br}(Q^2)}{{\rm d}\Omega^\prime_l}\right]^{\,lp(1);\,{\rm v}}_{\gamma\gamma^*}\!\!\!\!
\Bigg/\left[\frac{{\rm d}\sigma_{el}(Q^2)}{{\rm d}\Omega^\prime_l}\right]_0
\nonumber\\
&=&-\frac{\alpha}{\pi M\eta}\left( \frac{8\Delta_{\gamma^*}}{\eta Q^2+4E^2} \right)\Bigg[\Delta_{\gamma^*}(E+E^\prime)- 
\left\{2\eta m_l^2 (E-E^\prime)+\Delta_{\gamma^*}\left(\frac{1}{2}Q^2-m_l^2\right)\right\}
\nonumber\\
&&\times\,\left\{\frac{1}{E\beta}\ln\sqrt{\frac{1+\beta}{1-\beta}}
-\frac{1}{E^\prime\beta^\prime}\ln\sqrt{\frac{1+\beta^\prime}{1-\beta^\prime}}\,\right\}
-\,\Delta_{\gamma^*}\frac{E^2}{E^\prime\beta^\prime}
\left\{\left(1-\frac{\beta}{\beta^\prime}\right)\ln\sqrt{\frac{1+\beta^\prime}{1-\beta^\prime}}-\beta\right\}
\nonumber\\
&&\hspace{1.9cm}-\,\Delta_{\gamma^*}\frac{E^{\prime 2}}{E\beta}
\left\{\left(1-\frac{\beta^\prime}{\beta}\right)\ln\sqrt{\frac{1+\beta}{1-\beta}}-\beta^\prime\right\}
\Bigg]+o\left(\alpha\frac{{\mathcal Q}^2}{M^2}\right)\,.
\end{eqnarray}
\end{widetext} 
%%%%%%%%%%%%%%%%%%%%%%%%%%%%%%%%%%%%%%%%%%%%%%%%%%%%

%%%%%%%%%%%%%%%%%%%%%%%%%%%%%%%%%%%%%%%%%%%%%%%%%%%%%%

\end{document}